\documentclass[10pt]{article}
\pdfoutput=1

\input{preamb.sty}

\pgfdeclarelayer{back}
\pgfdeclarelayer{front}
\pgfsetlayers{back,main,front}

\tikzset{->-/.style={decoration={
                markings,
                mark=at position #1 with {\arrow{Triangle[width=3.5pt, length=3.5pt]}}},postaction={decorate}}}

\tikzset{-<-/.style={decoration={
                markings,
                mark=at position #1 with {\arrowreversed{Triangle[width=3.5pt, length=3.5pt]}}},postaction={decorate}}}

\tikzset{
    every picture/.append style={
        line join=round,
        line width = 0.7pt,
        line cap=round
    }
}

\tikzset{
    mathpoly/.style={
        baseline=0ex,        
        x=1.3ex, y=1.3ex,      
        line width=0.11ex,      
        line join=round        
    }
}

\tikzset{mpoint/.style={ transform shape, sloped, 
minimum width=20pt, minimum height=4pt, inner sep=0pt, ultra thin, font=\tiny}}

\tikzset{
    clip even odd rule/.code={\pgfseteorule},
    invclip/.style={clip,insert path=[clip even odd rule]{
    [reset cm](-\maxdimen,-\maxdimen)rectangle(\maxdimen,\maxdimen)
    }}}

\tikzstyle{dot}=[dash pattern= on 0.5pt off 2.5pt]
\tikzstyle{mor}=[circle, minimum size=5pt, inner sep=1pt, fill=gray!20]
\tikzstyle{mps0}=[circle, inner sep=2.5pt, fill=gray!20, draw]
\tikzstyle{mps1}=[circle, inner sep=2.5pt, fill=gray!20, draw]
\tikzstyle{mps2}=[circle, inner sep=1.75pt, fill=gray!20, draw]
\tikzstyle{mps3}=[circle, inner sep=2.5pt, fill=gray!20, draw]
\tikzstyle{hole}=[circle, inner sep=4pt, fill=white, draw=white]
\tikzstyle{mat}=[circle, inner sep=2pt, fill=gray!20, draw]
\tikzstyle{triR}=[regular polygon, regular polygon sides=3 ,inner sep=1.8pt,fill=gray!20, draw, rounded corners=2pt, rotate=30, xshift=-1pt,yshift=.5pt]
\tikzstyle{triL}=[regular polygon, regular polygon sides=3 ,inner sep=1.8pt,fill=gray!20, draw, rounded corners=2pt, rotate=-30, xshift=1pt,yshift=.5pt]
\tikzstyle{cg}=[circle, inner sep=4pt, fill=gray!20, draw]
\tikzstyle{mpo}=[rectangle, rounded corners=2pt, inner sep=3.5pt, fill=gray!20, draw]
\tikzstyle{obj1}=[draw=black]
\tikzstyle{obj2}=[draw=LimeGreen]
\tikzstyle{obj3}=[draw=BrickRed]

\newcommand{\mtriangle}{
    \mathord{
        \tikz[mathpoly] 
            \draw[shift={(0, {cos(60)/(1+cos(60))})}] 
                ({-150}:{1/(1+cos(60))}) 
             -- ({-30}:{1/(1+cos(60))}) 
             -- ({90}:{1/(1+cos(60))}) 
             -- cycle;
    }
}

\newcommand{\msquare}{
    \mathord{
        \tikz[mathpoly] 
            \draw[shift={(0, 1/2)}] 
                ({-135}:{1/(2*cos(45))}) 
             -- ({-45}:{1/(2*cos(45))}) 
             -- ({45}:{1/(2*cos(45))}) 
             -- ({135}:{1/(2*cos(45))}) 
             -- cycle;
    }
}

\newcommand{\mpentagon}{
    \mathord{
        \tikz[mathpoly] 
            \draw[scale=1.1,shift={(0, {cos(36)/(1+cos(36))})}] 
                ({-126}:{1/(1+cos(36))}) 
             -- ({-54}:{1/(1+cos(36))}) 
             -- ({18}:{1/(1+cos(36))}) 
             -- ({90}:{1/(1+cos(36))}) 
             -- ({162}:{1/(1+cos(36))}) 
             -- cycle;
    }
}

\newcommand{\mhexagon}{
    \mathord{
        \tikz[mathpoly] 
            \draw[shift={(0, 1/2)}] 
                ({-120}:{1/(2*cos(30))}) 
             -- ({-60}:{1/(2*cos(30))}) 
             -- ({0}:{1/(2*cos(30))}) 
             -- ({60}:{1/(2*cos(30))}) 
             -- ({120}:{1/(2*cos(30))}) 
             -- ({180}:{1/(2*cos(30))}) 
             -- cycle;
    }
}

\newcommand{\mat}[5]{
    \begin{tikzpicture}[scale=1,baseline={([yshift=-.55ex]current bounding box.center)}]
        \def\l{.7};
        \draw[obj3,-<-=.47] (0,0) --node[pos=1.25]{${\sss #5}$} node[pos=.6,above]{${\sss #3}$} (\l,0);
        \draw[obj1,->-=.73] (0,0) --node[pos=1.25]{${\sss #4}$} node[pos=.6,above]{${\sss #2}$} (-\l,0);
        \node[mps1] at (0,0) {};
        \node[yshift=-12pt] at (0,0) {${\sss #1}$};
    \end{tikzpicture}
}

\newcommand{\matDual}[5]{
    \begin{tikzpicture}[scale=1,baseline={([yshift=-.55ex]current bounding box.center)}]
        \def\l{.7};
        \draw[obj1,-<-=.47] (0,0) --node[pos=1.25]{${\sss #5}$} node[pos=.6,above]{${\sss #3}$} (\l,0);
        \draw[obj3,->-=.73] (0,0) --node[pos=1.25]{${\sss #4}$} node[pos=.6,above]{${\sss #2}$} (-\l,0);
        \node[mps1] at (0,0) {};
        \node[yshift=-12pt] at (0,0) {${\sss #1}$};
    \end{tikzpicture}
}

\newcommand{\MPST}[5]{
    \begin{tikzpicture}[scale=1,baseline={([yshift=-2.15ex]current bounding box.center)}]
        \def\l{.7};
        \draw[obj2,-<-=.47] (0,0) --node[pos=1.25]{${\sss #4}$} node[pos=.6,above]{${\sss #2}$} (\l,0);
        \draw[obj2,->-=.73] (0,0) --node[pos=1.25]{${\sss #3}$} node[pos=.6,above]{${\sss #2}$} (-\l,0);
        \draw[obj1,->-=.73] (0,0) --node[pos=1.25]{${\sss #5}$} (0,\l);
        \node[mps1] at (0,0) {};
        \node[yshift=-9pt] at (0,0) {${\sss #1}$};
    \end{tikzpicture}
}

\newcommand{\TMat}[1]{
    \begin{tikzpicture}[scale=1,baseline={([yshift=-.55ex]current bounding box.center)}]
        \def\l{.7};
        \def\sh{1.2};
        \draw[obj2,-<-=.47] (0,0) -- (\l,0);
        \draw[obj2,->-=.73] (0,0) -- (-\l,0);
        \draw[obj2,->-=.73] (0,\sh*\l) -- (\l,\sh*\l);
        \draw[obj2,-<-=.47] (0,\sh*\l) -- (-\l,\sh*\l);
        \draw[obj1,->-=.73] (0,0) -- (0,\l);
        \node[mps1] at (0,0) {};
        \node[mps1] at (0,\sh*\l) {};
        \node[xshift=0pt, yshift=-9pt] at (0,0) {${\sss #1}$};
        \node[xshift=0pt, yshift=9pt] at (0,\sh*\l) {${\sss \bar #1}$};
    \end{tikzpicture}
}

\newcommand{\TMatOp}[3]{
    \begin{tikzpicture}[scale=1,baseline={([yshift=-.55ex]current bounding box.center)}]
        \def\l{.7};
        \def\sh{1.2};
        \draw[obj2,-<-=.47] (0,0) --node[pos=.6,above]{${\sss #1}$} (\l,0);
        \draw[obj2,->-=.73] (0,0) --node[pos=.6,above]{${\sss #1}$} (-\l,0);
        \draw[obj2,->-=.73] (0,2*\sh*\l) --node[pos=.6,below]{${\sss #1}$} (\l,2*\sh*\l);
        \draw[obj2,-<-=.47] (0,2*\sh*\l) --node[pos=.6,below]{${\sss #1}$} (-\l,2*\sh*\l);
        \draw[obj1,->-=.73] (0,0) -- (0,\l);
        \draw[obj1,->-=.73] (0,\sh*\l) -- (0,\sh*\l+\l);

        \ifthenelse{\equal{#3}{}}{
            \draw[obj3,->-=.73] (0,\sh*\l) --node[pos=.6,above]{${\sss #2}$} (-\l,\sh*\l);
            \draw[obj3,-<-=.47] (0,\sh*\l) --node[pos=.6,above]{${\sss #2}$} (\l,\sh*\l);
            \node[mpo] at (0,\sh*\l) {};
        }{
            \draw[obj3,-<-=.47] (0,\sh*\l) --node[pos=.6,above]{${\sss #2}$} (\l,\sh*\l);
            \node[mps1] at (0,\sh*\l) {};
            \node[xshift=-9pt] at (0,\sh*\l) {${\sss #3}$};
        }
                
        \node[mps1] at (0,0) {};
        \node[mps1] at (0,2*\sh*\l) {};
    \end{tikzpicture}
}

\newcommand{\TMatOpFixed}[4]{
    \begin{tikzpicture}[scale=1,baseline={([yshift=-.55ex]current bounding box.center)}]
        \def\l{.7};
        \def\sh{1.2};

        \ifthenelse{#1=1}{
            \draw[obj2,-<-=.12,-<-=.83,rounded corners=2pt] (0,0) --node[pos=.55,above]{${\sss #2}$} (\sh*\l,0) -- (\sh*\l,2*\sh*\l) --node[pos=.45,below]{${\sss #2}$} (0,2*\sh*\l);
            \draw[obj2,->-=.73] (0,0) --node[pos=.6,above]{${\sss #2}$} (-\l,0);
            \draw[obj2,-<-=.47] (0,2*\sh*\l) --node[pos=.6,below]{${\sss #2}$} (-\l,2*\sh*\l);
            \draw[obj1,->-=.73] (0,0) -- (0,\l);
            \draw[obj1,->-=.73] (0,\sh*\l) -- (0,\sh*\l+\l);
            \draw[obj3,->-=.73] (0,\sh*\l) --node[pos=.6,above]{${\sss #3}$} (-\l,\sh*\l);
            \draw[obj3,-<-=.47] (0,\sh*\l) --node[pos=.6,above]{${\sss #3}$} (\l,\sh*\l);
            \node[mpo] at (0,\sh*\l) {};
                    
            \node[mps1] at (0,0) {};
            \node[mps1] at (\sh*\l,\sh*\l) {};
            \node[xshift=11pt] at (\sh*\l,\sh*\l) {${\sss #4}$};
            \node[mps1] at (0,2*\sh*\l) {};
        }{}

        \ifthenelse{#1=2}{
            \draw[obj2,-<-=.08,-<-=.87, rounded corners=2pt] (0,0) --node[pos=.5,above]{${\sss #2}$} (\l,0) -- (\l,2*\sh*\l) --node[pos=.5,below]{${\sss #2}$} (0,2*\sh*\l);  
            \draw[obj3,->-=.73] (\l,\sh*\l) --node[pos=.6,above]{${\sss #3}$} (0,\sh*\l);
                   
            \node[mps1] at (\l,\sh*\l) {};
            \node[xshift=11pt] at (\l,\sh*\l) {${\sss #4}$};
        }{}

        \ifthenelse{#1=3}{
            \draw[obj2,-<-=.1,-<-=.79,rounded corners=2pt] (0,0) --node[pos=.52,above]{${\sss #2}$} (\sh*\l,0) -- (\sh*\l,2*\sh*\l) --node[pos=.5,below]{${\sss #2}$} (-\l,2*\sh*\l);
            \draw[obj2,->-=.73] (0,0) --node[pos=.6,above]{${\sss #2}$} (-\l,0);
            \draw[obj3,->-=.33, rounded corners=2pt] (\sh*\l,\sh*\l) --node[pos=.5,above]{${\sss #3}$} (0,\sh*\l) -- (0,0);
            \node[mps1] at (0,0) {};
            \node[mps0] at (\sh*\l,\sh*\l) {};
            \node[xshift=11pt] at (\sh*\l,\sh*\l) {${\sss #4}$};
            \node[yshift=-9pt] at (0,0) {${\sss \phi^{r_ahr_a^{-1}\!,a}}$};
        }{}

        \ifthenelse{#1=4}{
            \draw[obj2,-<-=.08,-<-=.86,rounded corners=2pt] (0,0) --node[pos=.5,above]{${\sss #2}$} (\l,0) -- (\l,2*\sh*\l) --node[pos=.5,below]{${\sss #2}$} (0,2*\sh*\l);
            \node[mps0] at (\l,\sh*\l) {};
            \node[xshift=11pt] at (\l,\sh*\l) {${\sss #4}$};
        }{}
    \end{tikzpicture}
}

\newcommand{\TMatProj}[0]{
    \begin{tikzpicture}[scale=1,baseline={([yshift=.4ex]current bounding box.center)}]
        \def\l{.7};
        \def\sh{1.2};
        \def\sp{.12};
        \draw[obj2, -<-=.08,-<-=.87, rounded corners=2pt] (-2*\sp-\sh*\l-\l,-\sh*\l) --node[pos=.45,above]{${\sss a}$} (-2*\sp-\sh*\l,-\sh*\l) -- (-2*\sp-\sh*\l,\sh*\l) --node[pos=.55,below]{${\sss a}$} (-2*\sp-\sh*\l-\l,\sh*\l);
        \draw[obj2,->-=.52, rounded corners=2pt] (0,-\sh*\l) -- (-\sh*\l,-\sh*\l) --node[pos=.5,right]{${\sss a}$} (-\sh*\l,\sh*\l) -- (0,\sh*\l);
        \draw[obj1,-<-=.47] (0,0) -- (0,-\l);
        \draw[obj1,->-=.73] (0,0) -- (0,\l);
        \draw[obj3,-<-=.23, rounded corners=2pt] (0,0) --node[pos=.56,above]{${\sss {}^{r_a}h}$} (\sh*\l,0) -- (\sh*\l,-\sh*\l);
        \draw[obj2,-<-=.47] (0,-\sh*\l) --node[pos=.6,above]{${\sss a}$} (\l,-\sh*\l);
        \draw[obj2,->-=.33, rounded corners=2pt] (2*\sh*\l,0) --node[pos=.55,left]{${\sss a}$} (2*\sh*\l,-\sh*\l) -- (\sh*\l,-\sh*\l);
        \draw[obj2,->-=.38, rounded corners=2pt] (0,\sh*\l) --node[pos=.52,below]{${\sss a}$} (2*\sh*\l,\sh*\l) -- (2*\sh*\l,0);
        \draw[obj2, -<-=.15,-<-=.88, rounded corners=2pt] (2*\sh*\l+2*\sp+2*\l,\sh*\l) --node[pos=.45,below]{${\sss a}$} (2*\sh*\l+2*\sp,\sh*\l) -- (2*\sh*\l+2*\sp,-\sh*\l) --node[pos=.5,above]{${\sss a}$} (2*\sh*\l+2*\sp+\l,-\sh*\l);
        \draw[obj3,-<-=.72, rounded corners=2pt] (2*\sh*\l+2*\sp+\l,-\sh*\l) -- (2*\sh*\l+2*\sp+\l,0) --node[pos=.56,above]{${\sss {}^{r_a}h}$} (2*\sh*\l+2*\sp+2*\l,0);
        \draw[obj2,-<-=.47] (2*\sh*\l+2*\sp+\l,-\sh*\l) --node[pos=.6,above]{${\sss a}$} (2*\sh*\l+2*\sp+2*\l,-\sh*\l);
        
        \node[yshift=-9pt] at (\sh*\l,-\sh*\l) {${\sss \bar \phi^{r_ahr_a^{-1}\!,a}}$};
        \node[xshift=-9pt] at (0,0) {${\sss a_1}$};
        \node[mps1] at (0,0) {};
        \node[mps1] at (0,-\sh*\l) {};
        \node[mps1] at (0,\sh*\l) {};
        \node[mps1] at (2*\sh*\l+2*\sp+\l,-\sh*\l) {};
        \node[yshift=-9pt] at (2*\sh*\l+2*\sp+\l,-\sh*\l) {${\sss \phi^{r_ahr_a^{-1}\!,a}}$};
        \node[mps1] at (\sh*\l,-\sh*\l) {};
        \node[mps2] at (-\sh*\l-2*\sp,0) {};
        \node[xshift=-10pt] at (-\sh*\l-2*\sp,0) {${\sss \mathbb I_{U^*}}$};
    \end{tikzpicture}
}

\newcommand{\productTMat}[1]{
    \begin{tikzpicture}[scale=1,baseline={([yshift=-.55ex]current bounding box.center)}]
        \def\l{.7};
        \def\sh{1.2};        
        \foreach \x in {2,3,6}{
            \draw[obj2,->-=.73] (\x*\l*\sh,0) --node[pos=.6,above]{${\sss #1}$} (\x*\l*\sh-\l,0);
        }
        \draw[obj2,-<-=.47] (6*\sh*\l,0) --node[pos=.6,above]{${\sss #1}$} (6*\sh*\l+\l,0);
        \foreach \x in {2,3,6}{
            \draw[obj2,-<-=.47] (\x*\l*\sh,\sh*\l) --node[pos=.6,below]{${\sss #1}$} (\x*\l*\sh-\l,\sh*\l);
        }
        \draw[obj2,->-=.73] (6*\sh*\l,\sh*\l) --node[pos=.6,below]{${\sss #1}$} (6*\sh*\l+\l,\sh*\l);
        \draw[obj2,-<-=.45] (3*\l*\sh,0) -- (3*\l*\sh+2*\sh*\l/3,0);
        \draw[obj2,->-=.73] (5*\l*\sh,0) -- (5*\l*\sh-2*\sh*\l/3,0);
        \draw[obj2,dot, shorten <= 4pt, shorten >= 4pt] (3*\l*\sh+2*\sh*\l/3,0) -- (5*\l*\sh-2*\sh*\l/3,0);
        \draw[obj2,dot, shorten <= 4pt, shorten >= 4pt] (3*\l*\sh+2*\sh*\l/3,\sh*\l) -- (5*\l*\sh-2*\sh*\l/3,\sh*\l);
        \draw[obj2,->-=.73] (3*\l*\sh,\sh*\l) -- (3*\l*\sh+2*\sh*\l/3,\sh*\l);
        \draw[obj2,-<-=.45] (5*\l*\sh,\sh*\l) -- (5*\l*\sh-2*\sh*\l/3,\sh*\l);
        \foreach \x in {2,3,5,6}{
            \draw[obj1,->-=.73] (\x*\sh*\l,0) -- (\x*\sh*\l,\l);
            \node[mps1] at (\x*\sh*\l,0) {};
            \node[mps1] at (\x*\sh*\l,\sh*\l) {};
        }

    \end{tikzpicture}
}

\newcommand{\TMatRG}[3]{
    \begin{tikzpicture}[scale=1,baseline={([yshift=-.55ex]current bounding box.center)}]
        \def\l{.7};
        \def\sh{1.2};
        \def\sp{.12};
        \draw[obj2, -<-=.08,-<-=.87, rounded corners=2pt] (-\sp-\l,-\sh*\l) --node[pos=.45,above]{${\sss #1}$} (-\sp,-\sh*\l) -- (-\sp,\sh*\l) --node[pos=.55,below]{${\sss #1}$} (-\sp-\l,\sh*\l);
        \draw[obj2, -<-=.08,-<-=.87, rounded corners=2pt] (\sp+\l,\sh*\l) --node[pos=.45,below]{${\sss #1}$} (\sp,\sh*\l) -- (\sp,-\sh*\l) --node[pos=.55,above]{${\sss #1}$} (\sp+\l,-\sh*\l);
        \node[mps2] at (-\sp,0) {};
        \node[mps2] at (\sp,0) {};
        \node[xshift=-8pt] at (-\sp,0) {${\sss #2}$};
        \node[xshift=8pt] at (\sp,0) {${\sss #3}$};
    \end{tikzpicture}
}

\newcommand{\smallTMatRG}[3]{
    \begin{tikzpicture}[scale=1,baseline={([yshift=-.55ex]current bounding box.center)}]
        \def\l{.7};
        \def\sh{1.2};
        \def\sp{.12};
        \draw[obj2, -<-=.1,-<-=.82, rounded corners=2pt] (-\sp-\l,0) --node[pos=.45,above]{${\sss #1}$} (-\sp,0) -- (-\sp,\sh*\l) --node[pos=.55,below]{${\sss #1}$} (-\sp-\l,\sh*\l);
        \draw[obj2, -<-=.1,-<-=.82, rounded corners=2pt] (\sp+\l,\sh*\l) --node[pos=.45,below]{${\sss #1}$} (\sp,\sh*\l) -- (\sp,0) --node[pos=.55,above]{${\sss #1}$} (\sp+\l,0);
        \node[mps2] at (-\sp,.5*\sh*\l) {};
        \node[mps2] at (\sp,.5*\sh*\l) {};
        \node[xshift=-8pt] at (-\sp,.5*\sh*\l) {${\sss #2}$};
        \node[xshift=8pt] at (\sp,.5*\sh*\l) {${\sss #3}$};
    \end{tikzpicture}
}

\newcommand{\RightTMatFixed}[3]{
    \begin{tikzpicture}[scale=1,baseline={([yshift=-.55ex]current bounding box.center)}]
        \def\l{.7};
        \def\sh{1.2};
        \ifthenelse{#1=1}{
            \draw[obj2,-<-=.15,-<-=.79,rounded corners=2pt] (0,0) -- (\sh*\l,0) -- (\sh*\l,\sh*\l) -- (0,\sh*\l);
            \draw[obj2,->-=.73] (0,0) -- (-\l,0);
            \draw[obj2,-<-=.47] (0,\sh*\l) -- (-\l,\sh*\l);
            \draw[obj1,->-=.73] (0,0) -- (0,\l);
            \node[mps1] at (0,0) {};
            \node[mps1] at (0,\sh*\l) {};
            \node[mps0] at (\sh*\l,.5*\sh*\l) {};
            \node[xshift=8pt] at (\sh*\l,.5*\sh*\l) {${\sss #3}$};
            \node[xshift=0pt, yshift=-9pt] at (0,0) {${\sss #2}$};
            \node[xshift=0pt, yshift=9pt] at (0,\sh*\l) {${\sss \bar #2}$};
        }{
            \draw[obj2,-<-=.1,-<-=.82,rounded corners=2pt] (0,0) -- (\l,0) -- (\l,\sh*\l) -- (0,\sh*\l);
            \node[mps0] at (\l,.5*\sh*\l) {};
            \node[xshift=9pt] at (\l,.5*\sh*\l) {${\sss #3}$};
        }
    \end{tikzpicture}
}

\newcommand{\actionSymFixedPoints}[6]{
    \begin{tikzpicture}[scale=1,baseline={([yshift=-.55ex]current bounding box.center)}]
        \def\l{.7};
        \def\sh{1.2};
        \ifthenelse{#1=1}{
            \draw[obj2,-<-=.11,-<-=.84,rounded corners=2pt] (0,0) --node[pos=.52,above]{${\sss #5}$} (\sh*\l,0) -- (\sh*\l,2*\sh*\l) --node[pos=.48,below]{${\sss #5}$} (0,2*\sh*\l);
            \draw[obj2,->-=.73] (0,0) --node[pos=.6,above]{${\sss #6}$} (-\l,0);
            \draw[obj2,-<-=.47] (0,2*\sh*\l) --node[pos=.6,below]{${\sss #6}$} (-\l,2*\sh*\l);
            \draw[obj3,-<-=.47] (0,0) --node[pos=.5,right]{${\sss #4}$} (0,\sh*\l+\l);
            \node[mps1] at (0,0) {};
            \node[mps1] at (0,2*\sh*\l) {};
            \node[mps0] at (\sh*\l,\sh*\l) {};
            \node[xshift=8pt] at (\sh*\l,\sh*\l) {${\sss #3}$};
            \node[xshift=0pt, yshift=-9pt] at (0,0) {${\sss #2}$};
            \node[xshift=0pt, yshift=9pt] at (0,2*\sh*\l) {${\sss \bar #2}$};
        }{}
        
        \ifthenelse{#1=2}{
            \draw[obj2,-<-=.11,-<-=.84,rounded corners=2pt] (0,0) --node[pos=.52,above]{${\sss #5}$} (\sh*\l,0) -- (\sh*\l,2*\sh*\l) --node[pos=.48,below]{${\sss #5}$} (0,2*\sh*\l);
            \draw[obj2,->-=.73] (0,0) --node[pos=.6,above]{${\sss #5}$} (-\l,0);
            \draw[obj2,-<-=.47] (0,2*\sh*\l) --node[pos=.6,below]{${\sss #5}$} (-\l,2*\sh*\l);
            \draw[obj1,->-=.6] (0,0) -- (0,\sh*\l+\l);
            \draw[obj2,->-=.73] (-\sh*\l,0) --node[pos=.6,above]{${\sss #6}$} (-\sh*\l-\l,0);
            \draw[obj2,-<-=.47] (-\sh*\l,2*\sh*\l) --node[pos=.6,below]{${\sss #6}$} (-\sh*\l-\l,2*\sh*\l);
            \draw[obj3,-<-=.47] (-\sh*\l,0) --node[pos=.5,right]{${\sss #4}$} (-\sh*\l,\sh*\l+\l);
            \node[mps1] at (0,0) {};
            \node[mps1] at (0,2*\sh*\l) {};
            \node[mps1] at (-\sh*\l,0) {};
            \node[mps1] at (-\sh*\l,2*\sh*\l) {};
            \node[mps0] at (\sh*\l,\sh*\l) {};
            \node[xshift=8pt] at (\sh*\l,\sh*\l) {${\sss #3}$};
            \node[xshift=0pt, yshift=-9pt] at (-\sh*\l,0) {${\sss #2}$};
            \node[xshift=0pt, yshift=9pt] at (-\sh*\l,2*\sh*\l) {${\sss \bar #2}$};
        }{}

        \ifthenelse{#1=3}{
            \draw[obj2,-<-=.11,-<-=.84,rounded corners=2pt] (0,0) --node[pos=.52,above]{${\sss #5}$} (\sh*\l,0) -- (\sh*\l,2*\sh*\l) --node[pos=.48,below]{${\sss #5}$} (0,2*\sh*\l);
            \draw[obj2,->-=.73] (0,0) --node[pos=.6,above]{${\sss #6}$} (-\l,0);
            \draw[obj2,-<-=.47] (0,2*\sh*\l) --node[pos=.6,below]{${\sss #6}$} (-\l,2*\sh*\l);
            \draw[obj1,->-=.6] (-\sh*\l,0) -- (-\sh*\l,\sh*\l+\l);
            \draw[obj2,->-=.73] (-\sh*\l,0) --node[pos=.6,above]{${\sss #6}$} (-\sh*\l-\l,0);
            \draw[obj2,-<-=.47] (-\sh*\l,2*\sh*\l) --node[pos=.6,below]{${\sss #6}$} (-\sh*\l-\l,2*\sh*\l);
            \draw[obj3,-<-=.47] (0,0) --node[pos=.5,right]{${\sss #4}$} (0,\sh*\l+\l);
            \node[mps1] at (0,0) {};
            \node[mps1] at (0,2*\sh*\l) {};
            \node[mps1] at (-\sh*\l,0) {};
            \node[mps1] at (-\sh*\l,2*\sh*\l) {};
            \node[mps0] at (\sh*\l,\sh*\l) {};
            \node[xshift=8pt] at (\sh*\l,\sh*\l) {${\sss #3}$};
            \node[xshift=0pt, yshift=-9pt] at (0,0) {${\sss #2}$};
            \node[xshift=0pt, yshift=9pt] at (0,2*\sh*\l) {${\sss \bar #2}$};
        }{}

        \ifthenelse{#1=4}{
            \draw[obj2,-<-=.08,-<-=.86,rounded corners=2pt] (0,0) --node[pos=.5,above]{${\sss #6}$} (\l,0) -- (\l,2*\sh*\l) --node[pos=.5,below]{${\sss #6}$} (0,2*\sh*\l);
            \node[mps0] at (\l,\sh*\l) {};
            \node[xshift=13pt] at (\l,\sh*\l) {${\sss #3}$};
        }{}
    \end{tikzpicture}
}

\newcommand{\LeftTMatFixed}[3]{
    \begin{tikzpicture}[scale=1,baseline={([yshift=-.55ex]current bounding box.center)}]
        \def\l{.7};
        \def\sh{1.2};
        \ifthenelse{#1=1}{
            \draw[obj2,-<-=.15,-<-=.79,rounded corners=2pt] (0,\sh*\l) -- (-\sh*\l,\sh*\l) -- (-\sh*\l,0) -- (0,0);
            \draw[obj2,-<-=.47] (0,0) -- (\l,0);
            \draw[obj2,->-=.73] (0,\sh*\l) -- (\l,\sh*\l);
            \draw[obj1,->-=.73] (0,0) -- (0,\l);
            \node[mps1] at (0,0) {};
            \node[mps1] at (0,\sh*\l) {};
            \node[mps0] at (-\sh*\l,.5*\sh*\l) {};
            \node[xshift=-9pt] at (-\sh*\l,.5*\sh*\l) {${\sss #3}$};
            \node[xshift=0pt, yshift=-9pt] at (0,0) {${\sss #2}$};
            \node[xshift=0pt, yshift=9pt] at (0,\sh*\l) {${\sss \bar #2}$};
        }{
            \draw[obj2,-<-=.1,-<-=.82,rounded corners=2pt] (0,\sh*\l) -- (-\l,\sh*\l) -- (-\l,0) -- (0,0);
            \node[mps0] at (-\l,.5*\sh*\l) {};
            \node[xshift=-9pt] at (-\l,.5*\sh*\l) {${\sss #3}$};
        }
    \end{tikzpicture}
}

\newcommand{\MPOT}[6]{
    \begin{tikzpicture}[scale=1,baseline={([yshift=-.55ex]current bounding box.center)}]
        \def\l{.7};
        \draw[obj3,-<-=.47] (0,0) --node[pos=1.25]{${\sss #4}$} node[pos=.6,above]{${\sss #2}$} (\l,0);
        \draw[obj3,->-=.73] (0,0) --node[pos=1.25]{${\sss #3}$} node[pos=.6,above]{${\sss #2}$} (-\l,0);
        \draw[obj1,->-=.73] (0,0) --node[pos=1.25]{${\sss #5}$} (0,\l);
        \draw[obj1,-<-=.47] (0,0) --node[pos=1.25]{${\sss #6}$} (0,-\l);
        \node[mpo] at (0,0) {};
        \node[xshift=6pt, yshift=-6pt] at (0,0) {${\sss #1}$};
    \end{tikzpicture}
}

\newcommand{\MPS}[1]{
    \begin{tikzpicture}[scale=1,baseline={([yshift=-2.3ex]current bounding box.center)}]
        \def\l{.7};
        \def\sh{1.2};
        \draw[obj2,->-=.73] (0,0) -- (-\l,0);
        \draw[obj2,-<-=.47] (8*\sh*\l,0) -- (8*\sh*\l+\l,0);
        \draw[obj2,-<-=.38] (2*\l*\sh,0) -- (3*\l*\sh,0);
        \draw[obj2,-<-=.38] (5*\l*\sh,0) -- (6*\l*\sh,0);
        \foreach \x/\y in {0/2,3/5,6/8}{
            \draw[obj2,-<-=.45] (\x*\l*\sh,0) -- (\x*\l*\sh+2*\sh*\l/3,0);
            \draw[obj2,dot, shorten <= 4pt, shorten >= 4pt] (\x*\l*\sh+2*\sh*\l/3,0) -- (\y*\l*\sh-2*\sh*\l/3,0);
            \draw[obj2,->-=.73] (\y*\l*\sh,0) -- (\y*\l*\sh-2*\sh*\l/3,0);
        }
        \foreach \x in {0,2,3,5,6,8}{
            \draw[obj1,->-=.73] (\x*\l*\sh,0) -- (\x*\l*\sh,\l);
            \node[mps1] at (\x*\l*\sh,0) {};
            \node[xshift=0pt, yshift=-9pt] at (\x*\l*\sh,0) {${\sss #1}$};
        }
        \draw[line width=.5pt] (-\l,-\l/2) -- (-1.3*\l,\l/2);
        \draw[line width=.5pt] (-1.1*\l,-\l/2) -- (-1.4*\l,\l/2);
        \draw[line width=.5pt] (8*\sh*\l+\l,\l/2) -- (8*\sh*\l+1.3*\l,-\l/2);
        \draw[line width=.5pt] (8*\sh*\l+1.1*\l,\l/2) -- (8*\sh*\l+1.4*\l,-\l/2);
    \end{tikzpicture}
}

\newcommand{\MPO}[2]{
    \begin{tikzpicture}[scale=1,baseline={([yshift=-.55ex]current bounding box.center)}]
        \def\l{.7};
        \def\sh{1.2};
        \draw[obj3,->-=.73] (0,0) --node[pos=.6,above]{${\sss #2}$} (-\l,0);
        \draw[obj3,-<-=.47] (8*\sh*\l,0) --node[pos=.6,above]{${\sss #2}$} (8*\sh*\l+\l,0);
        \draw[obj3,-<-=.38] (2*\l*\sh,0) --node[pos=.5,above]{${\sss #2}$} (3*\l*\sh,0);
        \draw[obj3,-<-=.38] (5*\l*\sh,0) --node[pos=.5,above]{${\sss #2}$} (6*\l*\sh,0);
        \foreach \x/\y in {0/2,3/5,6/8}{
            \draw[obj3,-<-=.45] (\x*\l*\sh,0) -- (\x*\l*\sh+2*\sh*\l/3,0);
            \draw[obj3,dot, shorten <= 4pt, shorten >= 4pt] (\x*\l*\sh+2*\sh*\l/3,0) -- (\y*\l*\sh-2*\sh*\l/3,0);
            \draw[obj3,->-=.73] (\y*\l*\sh,0) -- (\y*\l*\sh-2*\sh*\l/3,0);
        }
        \foreach \x in {0,2,3,5,6,8}{
            \draw[obj1,->-=.73] (\x*\l*\sh,0) -- (\x*\l*\sh,\l);
            \draw[obj1,-<-=.47] (\x*\l*\sh,0) -- (\x*\l*\sh,-\l);
            \node[mpo] at (\x*\l*\sh,0) {};
            \node[xshift=0pt, yshift=-9pt] at (\x*\l*\sh,0) {${\sss #1}$};
        }
        \draw[line width=.5pt] (-\l,-\l/2) -- (-1.3*\l,\l/2);
        \draw[line width=.5pt] (-1.1*\l,-\l/2) -- (-1.4*\l,\l/2);
        \draw[line width=.5pt] (8*\sh*\l+\l,\l/2) -- (8*\sh*\l+1.3*\l,-\l/2);
        \draw[line width=.5pt] (8*\sh*\l+1.1*\l,\l/2) -- (8*\sh*\l+1.4*\l,-\l/2);
    \end{tikzpicture}
}

\newcommand{\actionMPO}[8]{
    \begin{tikzpicture}[scale=1,baseline={([yshift=-.55ex]current bounding box.center)}]
        \def\l{.7};
        \def\sh{1.2};
        \ifthenelse{#1=1}{
            \draw[obj2,-<-=.47] (0,0) --node[pos=0.6,above]{${\sss #3}$} (\l,0);
            \draw[obj2,->-=.73] (0,0) --node[pos=0.6,above]{${\sss #3}$} (-\l,0);
            \draw[obj3,-<-=.47] (0,\sh*\l) --node[pos=0.6,above]{${\sss #6}$} (\l,\sh*\l);
            \draw[obj3,->-=.73] (0,\sh*\l) --node[pos=0.6,above]{${\sss #6}$} (-\l,\sh*\l);
            \draw[obj1,->-=.73] (0,0) -- (0,\l);
            \draw[obj1,->-=.73] (0,\sh*\l) -- (0,\sh*\l+\l);
            \node[mps1] at (0,0) {};
            \node[mpo] at (0,\sh*\l) {};
            \node[yshift=-9pt] at (0,0) {${\sss #2}$};
        }{}
        \ifthenelse{#1=2}{
            \draw[obj2,-<-=.47] (0,0) --node[pos=0.6,above]{${\sss #4}$} (\l,0);
            \draw[obj2,-<-=.47] (\sh*\l,0) --node[pos=0.6,above]{${\sss #3}$} (\sh*\l+\l,0);
            \draw[obj2,->-=.73] (0,0) --node[pos=0.6,above]{${\sss #4}$} (-\l,0);
            \draw[obj2,->-=.73] (-\sh*\l,0) --node[pos=0.6,above]{${\sss #3}$} (-\sh*\l-\l,0);
            \draw[obj3,-<-=.72, rounded corners=2pt] (\sh*\l,0) -- (\sh*\l,\sh*\l) --node[pos=.53,above]{${\sss #6}$} (\sh*\l+\l,\sh*\l);
            \draw[obj3,->-=.85, rounded corners=2pt] (-\sh*\l,0) -- (-\sh*\l,\sh*\l) --node[pos=.53,above]{${\sss #6}$} (-\sh*\l-\l,\sh*\l);
            \draw[obj1,->-=.73] (0,0) -- (0,\l);
            \node[mps1] at (0,0) {};
            \node[mps1] at (\sh*\l,0) {};
            \node[mps1] at (-\sh*\l,0) {};
            \node[yshift=-9pt] at (0,0) {${\sss #2}$};
            \node[yshift=-9pt] at (\sh*\l,0) {${\sss #7}$};
            \node[yshift=-9pt] at (-\sh*\l,0) {${\sss #8}$};
        }{}
    \end{tikzpicture}
}

\newcommand{\actionMPORepG}[8]{
    \begin{tikzpicture}[scale=1,baseline={([yshift=-.55ex]current bounding box.center)}]
        \def\l{.7};
        \def\sh{1.2};
        \ifthenelse{#1=1}{
            \draw[obj2,-<-=.47] (0,0) --node[pos=0.6,above]{${\sss #3}$} (\l,0);
            \draw[obj2,->-=.73] (0,0) --node[pos=0.6,above]{${\sss #3}$} (-\l,0);
            \draw[obj3,-<-=.47] (0,\sh*\l) --node[pos=0.6,above]{${\sss #6}$} (\l,\sh*\l);
            \draw[obj3,->-=.73] (0,\sh*\l) --node[pos=0.6,above]{${\sss #6}$} (-\l,\sh*\l);
            \draw[obj1,->-=.73] (0,0) -- (0,\l);
            \draw[obj1,->-=.73] (0,\sh*\l) -- (0,\sh*\l+\l);
            \node[mps1] at (0,0) {};
            \node[mpo] at (0,\sh*\l) {};
            \node[yshift=-9pt] at (0,0) {${\sss #2}$};
        }{}
        \ifthenelse{#1=2}{
            \draw[obj2,-<-=.47] (0,0) --node[pos=0.6,above]{${\sss #4}$} (\l,0);
            \draw[obj2,-<-=.47] (\sh*\l,0) --node[pos=0.6,above]{${\sss #3}$} (\sh*\l+\l,0);
            \draw[obj2,->-=.73] (0,0) --node[pos=0.6,above]{${\sss #4}$} (-\l,0);
            \draw[obj2,->-=.73] (-\sh*\l,0) --node[pos=0.6,above]{${\sss #3}$} (-\sh*\l-\l,0);
            \draw[obj3,-<-=.72, rounded corners=2pt] (\sh*\l,0) -- (\sh*\l,\sh*\l) --node[pos=.53,above]{${\sss #6}$} (\sh*\l+\l,\sh*\l);
            \draw[obj3,->-=.85, rounded corners=2pt] (-\sh*\l,0) -- (-\sh*\l,\sh*\l) --node[pos=.53,above]{${\sss #6}$} (-\sh*\l-\l,\sh*\l);
            \draw[obj1,->-=.73] (0,0) -- (0,\l);
            \node[mps1] at (0,0) {};
            \node[mps1] at (\sh*\l,0) {};
            \node[mps1] at (-\sh*\l,0) {};
            \node[yshift=-9pt] at (0,0) {${\sss #2}$};
            \node[yshift=-12pt] at (\sh*\l,0) {${\sss #7}$};
            \node[yshift=-12pt] at (-\sh*\l,0) {${\sss #8}$};
        }{}
    \end{tikzpicture}
}

\newcommand{\fusionT}[8]{
    \begin{tikzpicture}[scale=1,baseline={([yshift=-.55ex]current bounding box.center)}]
        \def\l{.7};
        \def\sh{1.2};
        \ifthenelse{#1=1}{
            \draw[obj3,-<-=.62, rounded corners=2pt] (0,0) -- (0,\sh*\l/2) -- node[pos=1.3]{${\sss #6}$} node[pos=.53,above]{${\sss #3}$} (\l,\sh*\l/2);
            \draw[obj3,-<-=.62, rounded corners=2pt] (0,0) -- (0,-\sh*\l/2) -- node[pos=1.3]{${\sss #7}$} node[pos=.53,below]{${\sss #4}$} (\l,-\sh*\l/2);
            \draw[obj3,->-=.73] (0,0) --node[pos=1.25]{${\sss #8}$} node[pos=.58,above]{${\sss #5}$} (-\l,0);
            \node[mps1] at (0,0) {};
            \node[xshift=16pt] at (0,0) {${\sss #2}$};
        }{}
        \ifthenelse{#1=2}{
            \draw[obj3,->-=.78, rounded corners=2pt] (0,0) -- (0,\sh*\l/2) -- node[pos=1.25]{${\sss #6}$} node[pos=.53,above]{${\sss #3}$} (-\l,\sh*\l/2);
            \draw[obj3,->-=.78, rounded corners=2pt] (0,0) -- (0,-\sh*\l/2) -- node[pos=1.25]{${\sss #7}$} node[pos=.53,below]{${\sss #4}$} (-\l,-\sh*\l/2);
            \draw[obj3,-<-=.47] (0,0) --node[pos=1.25]{${\sss #8}$} node[pos=.58,above]{${\sss #5}$} (\l,0);
            \node[mps1] at (0,0) {};
            \node[xshift=-16pt] at (0,0) {${\sss #2}$};
        }{}
    \end{tikzpicture}
}

\newcommand{\fusionMPO}[6]{
    \begin{tikzpicture}[scale=1,baseline={([yshift=-.55ex]current bounding box.center)}]
        \def\l{.7};
        \def\sh{1.2};
        \ifthenelse{#1=1}{
            \draw[obj3,-<-=.47] (0,0) --node[pos=.6,above]{${\sss #3}$} (\l,0);
            \draw[obj3,->-=.73] (0,0) --node[pos=.6,above]{${\sss #3}$} (-\l,0);
            \draw[obj3,-<-=.47] (0,\sh*\l) --node[pos=.6,above]{${\sss #2}$} (\l,\sh*\l);
            \draw[obj3,->-=.73] (0,\sh*\l) --node[pos=.6,above]{${\sss #2}$} (-\l,\sh*\l);
            \draw[obj1,->-=.73] (0,0) -- (0,\l);
            \draw[obj1,->-=.73] (0,\sh*\l) -- (0,\sh*\l+\l);
            \draw[obj1,-<-=.47] (0,0) -- (0,-\l);
            \node[mpo] at (0,0) {};
            \node[mpo] at (0,\sh*\l) {};
        }{}
        \ifthenelse{#1=2}{
            \draw[obj1,->-=.73] (0,0) -- (0,\l);
            \draw[obj1,-<-=.47] (0,0) -- (0,-\l);
            \draw[obj3,-<-=.47] (0,0) --node[pos=.6,above]{${\sss #4}$} (\l,0);
            \draw[obj3,->-=.73] (0,0) --node[pos=.6,above]{${\sss #4}$}  (-\l,0);
            \draw[obj3,-<-=.62, rounded corners=2pt] (\sh*\l,0) -- (\sh*\l,\sh*\l/2) --node[pos=.53,above]{${\sss #2}$} (\sh*\l+\l,\sh*\l/2);
            \draw[obj3,-<-=.62, rounded corners=2pt] (\sh*\l,0) -- (\sh*\l,-\sh*\l/2) --node[pos=.53,below]{${\sss #3}$} (\sh*\l+\l,-\sh*\l/2);
            \draw[obj3,->-=.78, rounded corners=2pt] (-\sh*\l,0) -- (-\sh*\l,\sh*\l/2) --node[pos=.53,above]{${\sss #2}$} (-\sh*\l-\l,\sh*\l/2);
            \draw[obj3,->-=.78, rounded corners=2pt] (-\sh*\l,0) -- (-\sh*\l,-\sh*\l/2) --node[pos=.53,below]{${\sss #3}$} (-\sh*\l-\l,-\sh*\l/2);
            \node[mpo] at (0,0) {};
            \node[mps1] at (\sh*\l,0) {};
            \node[mps1] at (-\sh*\l,0) {};
            \node[xshift=16pt] at (\sh*\l,0) {${\sss #5}$};
            \node[xshift=-16pt] at (-\sh*\l,0) {${\sss #6}$};
        }{}
    \end{tikzpicture}
}

\newcommand{\fusionAssoc}[1]{
    \begin{tikzpicture}[scale=1,baseline={([yshift=-.55ex]current bounding box.center)}]
        \def\l{.7};
        \def\sh{1.2};
        \ifthenelse{#1=1}{
            \draw[obj3,->-=.73] (0,.75*\sh*\l) --node[above,pos=.8]{${\sss g_1g_2g_3}$} (-\l,.75*\sh*\l);
            \draw[obj3,->-=.4, rounded corners=2pt] (\sh*\l+\l,0) --node[pos=.5,below]{${\sss g_3}$} (0,0) -- (0,.75*\sh*\l);
            \draw[obj3,-<-=.62, rounded corners=2pt] (\sh*\l,1.5*\sh*\l) -- (\sh*\l,2*\sh*\l) --node[pos=.53,above]{${\sss g_1}$} (\sh*\l+\l,2*\sh*\l);
            \draw[obj3,-<-=.62, rounded corners=2pt] (\sh*\l,1.5*\sh*\l) -- (\sh*\l,\sh*\l) --node[pos=.53,below]{${\sss g_2}$} (\sh*\l+\l,\sh*\l);
            \draw[obj3,-<-=.62, rounded corners=2pt] (0,.75*\sh*\l) -- (0,1.5*\sh*\l) --node[pos=.47,above]{${\sss g_1g_2}$} (\sh*\l,1.5*\sh*\l);
            \node[mps1] at (\sh*\l,1.5*\sh*\l) {};
            \node[mps1] at (0,.75*\sh*\l) {};
        }{}
        \ifthenelse{#1=2}{
            \draw[obj3,->-=.73] (0,1.25*\sh*\l) --node[above,pos=.8]{${\sss g_1g_2g_3}$} (-\l,1.25*\sh*\l);
            \draw[obj3,->-=.4, rounded corners=2pt] (\sh*\l+\l,0) --node[pos=.5,below]{${\sss g_3}$} (\sh*\l,0) -- (\sh*\l,.5*\sh*\l);
            \draw[obj3,->-=.4, rounded corners=2pt] (\sh*\l+\l,2*\sh*\l) --node[pos=.5,above]{${\sss g_1}$} (0,2*\sh*\l) -- (0,1.25*\sh*\l);
            \draw[obj3,-<-=.62, rounded corners=2pt] (\sh*\l,.5*\sh*\l) -- (\sh*\l,\sh*\l) --node[pos=.53,above]{${\sss g_2}$} (\sh*\l+\l,\sh*\l);
            \draw[obj3,-<-=.62, rounded corners=2pt] (0,1.25*\sh*\l) -- (0,.5*\sh*\l) --node[pos=.47,below]{${\sss g_1g_2}$} (\sh*\l,.5*\sh*\l);
            \node[mps1] at (\sh*\l,.5*\sh*\l) {};
            \node[mps1] at (0,1.25*\sh*\l) {};
        }{}
    \end{tikzpicture}
}

\newcommand{\fusionAssocRepG}[1]{
    \begin{tikzpicture}[scale=1,baseline={([yshift=-.55ex]current bounding box.center)}]
        \def\l{.7};
        \def\sh{1.2};
        \ifthenelse{#1=1}{
            \draw[obj3,->-=.73] (0,.75*\sh*\l) --node[above,pos=.8]{${\sss V_4}$} (-\l,.75*\sh*\l);
            \draw[obj3,->-=.4, rounded corners=2pt] (\sh*\l+\l,0) --node[pos=.5,below]{${\sss V_3}$} (0,0) -- (0,.75*\sh*\l);
            \draw[obj3,-<-=.62, rounded corners=2pt] (\sh*\l,1.5*\sh*\l) -- (\sh*\l,2*\sh*\l) --node[pos=.53,above]{${\sss V_1}$} (\sh*\l+\l,2*\sh*\l);
            \draw[obj3,-<-=.62, rounded corners=2pt] (\sh*\l,1.5*\sh*\l) -- (\sh*\l,\sh*\l) --node[pos=.53,below]{${\sss V_2}$} (\sh*\l+\l,\sh*\l);
            \draw[obj3,-<-=.62, rounded corners=2pt] (0,.75*\sh*\l) -- (0,1.5*\sh*\l) --node[pos=.47,above]{${\sss V_5}$} (\sh*\l,1.5*\sh*\l);
            \node[xshift=17pt] at (\sh*\l,1.5*\sh*\l) {${\sss \varphi^{V_1\!V_2}_{V_5,i_1}}$};
            \node[xshift=17pt, yshift=-4pt] at (0,.75*\sh*\l) {${\sss \varphi^{V_5\!V_3}_{V_4,i_2}}$}; 
            \node[mps1] at (\sh*\l,1.5*\sh*\l) {};
            \node[mps1] at (0,.75*\sh*\l) {};
        }{}
        \ifthenelse{#1=2}{
            \draw[obj3,->-=.73] (0,1.25*\sh*\l) --node[above,pos=.8]{${\sss V_4}$} (-\l,1.25*\sh*\l);
            \draw[obj3,->-=.4, rounded corners=2pt] (\sh*\l+\l,0) --node[pos=.5,below]{${\sss V_3}$} (\sh*\l,0) -- (\sh*\l,.5*\sh*\l);
            \draw[obj3,->-=.4, rounded corners=2pt] (\sh*\l+\l,2*\sh*\l) --node[pos=.5,above]{${\sss V_1}$} (0,2*\sh*\l) -- (0,1.25*\sh*\l);
            \draw[obj3,-<-=.62, rounded corners=2pt] (\sh*\l,.5*\sh*\l) -- (\sh*\l,\sh*\l) --node[pos=.53,above]{${\sss V_2}$} (\sh*\l+\l,\sh*\l);
            \draw[obj3,-<-=.62, rounded corners=2pt] (0,1.25*\sh*\l) -- (0,.5*\sh*\l) --node[pos=.47,below]{${\sss V_6}$} (\sh*\l,.5*\sh*\l);
            \node[xshift=17pt] at (\sh*\l,.5*\sh*\l) {${\sss \varphi^{V_2\!V_3}_{V_5,i_3}}$};
            \node[xshift=17pt, yshift=4pt] at (0,1.25*\sh*\l) {${\sss \varphi^{V_1\!V_6}_{V_4,i_4}}$};   
            \node[mps1] at (\sh*\l,.5*\sh*\l) {};
            \node[mps1] at (0,1.25*\sh*\l) {};
        }{}
    \end{tikzpicture}
}

\newcommand{\orthoFusion}[0]{
    \begin{tikzpicture}[scale=1,baseline={([yshift=-.55ex]current bounding box.center)}]
        \def\l{.7};
        \def\sh{1.2};
        \def\r{2};
        \draw[obj3,-<-=.46, rounded corners=2pt] (0,0) -- (0,\sh*\l/2) --node[pos=.5,above]{${\sss g_1}$} (\r*\sh*\l,\sh*\l/2) -- (\r*\sh*\l,0);
        \draw[obj3,-<-=.46, rounded corners=2pt] (0,0) -- (0,-\sh*\l/2) --node[pos=.5,below]{${\sss g_2}$} (\r*\sh*\l,-\sh*\l/2) -- (\r*\sh*\l,0);
        \draw[obj3,->-=.73] (0,0) --node[pos=.6,above]{${\sss g_1g_2}$} (-\l,0);
        \draw[obj3,-<-=.47] (\r*\l+.5*\l,0) --node[pos=.58,above]{${\sss g_1g_2}$} (\r*\sh*\l+\l,0);
        \node[mps1] at (0,0) {};
        \node[mps1] at (\r*\sh*\l,0) {};
    \end{tikzpicture}
}

\newcommand{\orthoAction}[6]{
    \begin{tikzpicture}[scale=1,baseline={([yshift=-2.87ex]current bounding box.center)}]
        \def\l{.7};
        \def\sh{1.2};
        \def\r{2};
        \draw[obj3,-<-=.46, rounded corners=2pt] (0,0) -- (0,\sh*\l) --node[pos=.5,above]{${\sss #2}$} (\r*\sh*\l,\sh*\l) -- (\r*\sh*\l,0);
        \draw[obj2,-<-=.46] (0,0) --node[pos=0.5,above]{${\sss #1}$} (\r*\sh*\l,0);
        \draw[obj2,->-=.73] (0,0) --node[pos=0.6,above]{${\sss #3}$} (-\l,0);
        \draw[obj2,-<-=.47] (\r*\sh*\l,0) --node[pos=0.6,above]{${\sss #3}$} (\r*\sh*\l+\l,0);
        \node[mps1] at (0,0) {};
        \node[mps1] at (\r*\sh*\l,0) {};
        \node[yshift=-9pt] at (0,0) {${\sss #5}$};
        \node[yshift=-9pt] at (\r*\sh*\l,0) {${\sss #6}$};
    \end{tikzpicture}
}

\newcommand{\orthoActionRepG}[6]{
    \begin{tikzpicture}[scale=1,baseline={([yshift=-2.5ex]current bounding box.center)}]
        \def\l{.7};
        \def\sh{1.2};
        \def\r{2};
        \draw[obj3,-<-=.46, rounded corners=2pt] (0,0) -- (0,\sh*\l) --node[pos=.5,above]{${\sss #2}$} (\r*\sh*\l,\sh*\l) -- (\r*\sh*\l,0);
        \draw[obj2,-<-=.46] (0,0) --node[pos=0.5,above]{${\sss #1}$} (\r*\sh*\l,0);
        \draw[obj2,->-=.73] (0,0) --node[pos=0.6,above]{${\sss #3}$} (-\l,0);
        \draw[obj2,-<-=.47] (\r*\sh*\l,0) --node[pos=0.6,above]{${\sss #4}$} (\r*\sh*\l+\l,0);
        \node[mps1] at (0,0) {};
        \node[mps1] at (\r*\sh*\l,0) {};
        \node[yshift=-12pt] at (0,0) {${\sss #5}$};
        \node[yshift=-12pt] at (\r*\sh*\l,0) {${\sss #6}$};
    \end{tikzpicture}
}

\newcommand{\assocAction}[1]{
    \begin{tikzpicture}[scale=1,baseline={([yshift=-5.5ex]current bounding box.center)}]
        \def\l{.7};
        \def\sh{1.2};
        \ifthenelse{#1=1}{
            \draw[obj2,-<-=.47] (0,0) -- (\l,0);
            \draw[obj2,->-=.73] (0,0) -- (-\l,0);
            \draw[obj2,-<-=.47] (\sh*\l,0) --node[pos=.52,above]{${\sss a}$} (2*\sh*\l+\l,0);
            \draw[obj2,->-=.57] (-\sh*\l,0) --node[pos=.52,above]{${\sss a}$}  (-2*\sh*\l-\l,0);
            \draw[obj3,-<-=.62, rounded corners=2pt] (2*\sh*\l,1.5*\sh*\l) -- (2*\sh*\l,2*\sh*\l) --node[pos=.53,above]{${\sss g_1}$} (2*\sh*\l+\l,2*\sh*\l);
            \draw[obj3,-<-=.62, rounded corners=2pt] (2*\sh*\l,1.5*\sh*\l) -- (2*\sh*\l,\sh*\l) --node[pos=.53,below]{${\sss g_2}$} (2*\sh*\l+\l,\sh*\l);
            \draw[obj3,->-=.78, rounded corners=2pt] (-2*\sh*\l,1.5*\sh*\l) -- (-2*\sh*\l,\sh*\l) --node[pos=.53,below]{${\sss g_2}$} (-2*\sh*\l-\l,\sh*\l);
            \draw[obj3,->-=.78, rounded corners=2pt] (-2*\sh*\l,1.5*\sh*\l) -- (-2*\sh*\l,2*\sh*\l) --node[pos=.53,above]{${\sss g_1}$} (-2*\sh*\l-\l,2*\sh*\l);
            \draw[obj3,-<-=.73, rounded corners=2pt] (\sh*\l,0) -- (\sh*\l,1.5*\sh*\l) --node[pos=.47,above]{${\sss g_1g_2}$} (2*\sh*\l,1.5*\sh*\l);
            \draw[obj3,->-=.82, rounded corners=2pt] (-\sh*\l,0) -- (-\sh*\l,1.5*\sh*\l) --node[pos=.45,above]{${\sss g_1g_2}$} (-2*\sh*\l,1.5*\sh*\l);
            \draw[obj1,->-=.73] (0,0) -- (0,\l);
            \node[mps1] at (0,0) {};
            \node[mps1] at (2*\sh*\l,1.5*\sh*\l) {};
            \node[mps1] at (-2*\sh*\l,1.5*\sh*\l) {};
            \node[mps1] at (\sh*\l,0) {};
            \node[mps1] at (-\sh*\l,0) {};
            \node[yshift=-9pt] at (\sh*\l,0) {${\sss \phi^{g_1 \! g_2,a}}$};
            \node[yshift=-9pt] at (-\sh*\l,0) {${\sss \bar \phi^{g_1 \! g_2,a}}$};
        }{}
        \ifthenelse{#1=2}{
            \draw[obj2,-<-=.47] (0,0) -- (\l,0);
            \draw[obj2,-<-=.47] (2*\sh*\l,0) --node[pos=.6,above]{${\sss a}$} (2*\sh*\l+\l,0);
            \draw[obj2,->-=.73] (-2*\sh*\l,0) --node[pos=.6,above]{${\sss a}$} (-2*\sh*\l-\l,0);
            \draw[obj2,-<-=.47] (\sh*\l,0) --node[pos=.6,above]{${\sss g_2(a)}$} (\sh*\l+\l,0);
            \draw[obj2,->-=.73] (0,0) --(-\l,0);
            \draw[obj2,->-=.73] (-\sh*\l,0) --node[pos=.6,above]{${\sss g_2(a)}$} (-\sh*\l-\l,0);
            \draw[obj3,-<-=.72, rounded corners=2pt] (2*\sh*\l,0) -- (2*\sh*\l,\sh*\l) --node[pos=.53,above]{${\sss g_2}$} (2*\sh*\l+\l,\sh*\l);
            \draw[obj3,->-=.85, rounded corners=2pt] (-2*\sh*\l,0) -- (-2*\sh*\l,\sh*\l) --node[pos=.53,above]{${\sss g_2}$} (-2*\sh*\l-\l,\sh*\l);
            \draw[obj3,-<-=.2, rounded corners=2pt] (-2*\sh*\l-\l,2*\sh*\l) --node[pos=.49,above]{${\sss g_1}$} (-\sh*\l,2*\sh*\l) -- (-\sh*\l,0);
            \draw[obj3,->-=.26, rounded corners=2pt] (2*\sh*\l+\l,2*\sh*\l) --node[pos=.49,above]{${\sss g_1}$} (\sh*\l,2*\sh*\l) -- (\sh*\l,0);
            \draw[obj1,->-=.73] (0,0) -- (0,\l);
            \node[mps1] at (0,0) {};
            \node[mps1] at (\sh*\l,0) {};
            \node[mps1] at (-\sh*\l,0) {};
            \node[mps1] at (2*\sh*\l,0) {};
            \node[mps1] at (-2*\sh*\l,0) {};
            \node[yshift=-9pt, xshift=-2pt] at (\sh*\l,0) {${\sss \phi^{g_1 \! ,g_2(a)}}$};
            \node[yshift=-9pt, xshift=2pt] at (-\sh*\l,0) {${\sss \bar \phi^{g_1\! ,g_2(a)}}$};
            \node[yshift=-9pt, xshift=2pt] at (2*\sh*\l,0) {${\sss \phi^{g_2,a}}$};
            \node[yshift=-9pt, xshift=-2pt] at (-2*\sh*\l,0) {${\sss \bar \phi^{g_2,a}}$};
        }{}
    \end{tikzpicture}
}

\newcommand{\assocActionRepG}[1]{
    \begin{tikzpicture}[scale=1,baseline={([yshift=-5.2ex]current bounding box.center)}]
        \def\l{.7};
        \def\sh{1.2};
        \ifthenelse{#1=1}{
            \draw[obj2,-<-=.47] (0,0) --node[pos=.6,above]{${\sss U_2}$} (\l,0);
            \draw[obj2,->-=.73] (0,0) --node[pos=.6,above]{${\sss U_2}$} (-\l,0);
            \draw[obj2,-<-=.47] (\sh*\l,0) --node[pos=.52,above]{${\sss U_1}$} (2*\sh*\l+\l,0);
            \draw[obj2,->-=.57] (-\sh*\l,0) --node[pos=.52,above]{${\sss U_1}$}  (-2*\sh*\l-\l,0);
            \draw[obj3,-<-=.62, rounded corners=2pt] (2*\sh*\l,1.5*\sh*\l) -- (2*\sh*\l,2*\sh*\l) --node[pos=.53,above]{${\sss V_1}$} (2*\sh*\l+\l,2*\sh*\l);
            \draw[obj3,-<-=.62, rounded corners=2pt] (2*\sh*\l,1.5*\sh*\l) -- (2*\sh*\l,\sh*\l) --node[pos=.53,below]{${\sss V_2}$} (2*\sh*\l+\l,\sh*\l);
            \draw[obj3,->-=.78, rounded corners=2pt] (-2*\sh*\l,1.5*\sh*\l) -- (-2*\sh*\l,\sh*\l) --node[pos=.53,below]{${\sss V_2}$} (-2*\sh*\l-\l,\sh*\l);
            \draw[obj3,->-=.78, rounded corners=2pt] (-2*\sh*\l,1.5*\sh*\l) -- (-2*\sh*\l,2*\sh*\l) --node[pos=.53,above]{${\sss V_1}$} (-2*\sh*\l-\l,2*\sh*\l);
            \draw[obj3,-<-=.73, rounded corners=2pt] (\sh*\l,0) -- (\sh*\l,1.5*\sh*\l) --node[pos=.47,above]{${\sss V_3}$} (2*\sh*\l,1.5*\sh*\l);
            \draw[obj3,->-=.82, rounded corners=2pt] (-\sh*\l,0) -- (-\sh*\l,1.5*\sh*\l) --node[pos=.45,above]{${\sss V_3}$} (-2*\sh*\l,1.5*\sh*\l);
            \draw[obj1,->-=.73] (0,0) -- (0,\l);
            \node[mps1] at (0,0) {};
            \node[mps1] at (2*\sh*\l,1.5*\sh*\l) {};
            \node[mps1] at (-2*\sh*\l,1.5*\sh*\l) {};
            \node[mps1] at (\sh*\l,0) {};
            \node[mps1] at (-\sh*\l,0) {};
            \node[xshift=17pt] at (2*\sh*\l,1.5*\sh*\l) {${\sss \varphi^{V_1\!V_2}_{V_3,i_1}}$};
            \node[xshift=-16pt] at (-2*\sh*\l,1.5*\sh*\l) {${\sss \bar \varphi^{V_1\!V_2}_{V_3,i_1}}$};
            \node[yshift=-12pt] at (\sh*\l,0) {${\sss \phi^{V_3 U_1}_{U_2,i_2}}$};
            \node[yshift=-12pt] at (-\sh*\l,0) {${\sss \bar \phi^{V_3  U_1}_{U_2,i_2}}$};
        }{}
        \ifthenelse{#1=2}{
            \draw[obj2,-<-=.47] (0,0) --node[pos=.6,above]{${\sss U_2}$} (\l,0);
            \draw[obj2,-<-=.47] (2*\sh*\l,0) --node[pos=.6,above]{${\sss U_1}$} (2*\sh*\l+\l,0);
            \draw[obj2,->-=.73] (-2*\sh*\l,0) --node[pos=.6,above]{${\sss U_1}$} (-2*\sh*\l-\l,0);
            \draw[obj2,-<-=.47] (\sh*\l,0) --node[pos=.6,above]{${\sss U_3}$} (\sh*\l+\l,0);
            \draw[obj2,->-=.73] (0,0) --node[pos=.6,above]{${\sss U_2}$} (-\l,0);
            \draw[obj2,->-=.73] (-\sh*\l,0) --node[pos=.6,above]{${\sss U_3}$} (-\sh*\l-\l,0);
            \draw[obj3,-<-=.72, rounded corners=2pt] (2*\sh*\l,0) -- (2*\sh*\l,\sh*\l) --node[pos=.53,above]{${\sss V_2}$} (2*\sh*\l+\l,\sh*\l);
            \draw[obj3,->-=.85, rounded corners=2pt] (-2*\sh*\l,0) -- (-2*\sh*\l,\sh*\l) --node[pos=.53,above]{${\sss V_2}$} (-2*\sh*\l-\l,\sh*\l);
            \draw[obj3,-<-=.2, rounded corners=2pt] (-2*\sh*\l-\l,2*\sh*\l) --node[pos=.49,above]{${\sss V_1}$} (-\sh*\l,2*\sh*\l) -- (-\sh*\l,0);
            \draw[obj3,->-=.26, rounded corners=2pt] (2*\sh*\l+\l,2*\sh*\l) --node[pos=.49,above]{${\sss V_1}$} (\sh*\l,2*\sh*\l) -- (\sh*\l,0);
            \draw[obj1,->-=.73] (0,0) -- (0,\l);
            \node[mps1] at (0,0) {};
            \node[mps1] at (\sh*\l,0) {};
            \node[mps1] at (-\sh*\l,0) {};
            \node[mps1] at (2*\sh*\l,0) {};
            \node[mps1] at (-2*\sh*\l,0) {};
            \node[yshift=-12pt, xshift=-2pt] at (\sh*\l,0) {${\sss \phi^{V_1 U_3}_{U_2,i_4}}$};
            \node[yshift=-12pt, xshift=2pt] at (-\sh*\l,0) {${\sss \bar \phi^{V_1 U_3}_{U_2,i_4}}$};
            \node[yshift=-12pt, xshift=2pt] at (2*\sh*\l,0) {${\sss \phi^{V_2 U_1}_{U_3,i_3}}$};
            \node[yshift=-12pt, xshift=-2pt] at (-2*\sh*\l,0) {${\sss \bar \phi^{V_2 U_1}_{U_3,i_3}}$};
        }{}
    \end{tikzpicture}
}

\newcommand{\moduleAssoc}[1]{
    \begin{tikzpicture}[scale=1,baseline={([yshift=-5.6ex]current bounding box.center)}]
        \def\l{.7};
        \def\sh{1.2};
        \ifthenelse{#1=1}{
            \draw[obj2,->-=.73] (0,0) --node[above,pos=.9]{${\sss (g_1g_2)(a)}$} (-\l,0);
            \draw[obj2,-<-=.46] (0,0) --node[pos=.52,above]{${\sss a}$}  (\sh*\l+\l,0);
            \draw[obj3,-<-=.62, rounded corners=2pt] (\sh*\l,1.5*\sh*\l) -- (\sh*\l,2*\sh*\l) --node[pos=.53,above]{${\sss g_1}$} (\sh*\l+\l,2*\sh*\l);
            \draw[obj3,-<-=.62, rounded corners=2pt] (\sh*\l,1.5*\sh*\l) -- (\sh*\l,\sh*\l) --node[pos=.53,below]{${\sss g_2}$} (\sh*\l+\l,\sh*\l);
            \draw[obj3,-<-=.73, rounded corners=2pt] (0,0) -- (0,1.5*\sh*\l) --node[pos=.47,above]{${\sss g_1g_2}$} (\sh*\l,1.5*\sh*\l);
            \node[mps1] at (\sh*\l,1.5*\sh*\l) {};
            \node[mps1] at (0,0) {};
            \node[yshift=-9pt] at (0,0) {${\sss \phi^{g_1 \! g_2,a}}$};
        }{}
        \ifthenelse{#1=2}{
            \draw[obj2,->-=.73] (0,0) --node[above,pos=.9]{${\sss g_1(g_2(a))}$} (-\l,0);
            \draw[obj2,-<-=.47] (0,0) --node[above,pos=.6]{${\sss g_2(a)}$} (\l,0);
            \draw[obj2,-<-=.47] (\sh*\l,0) --node[pos=.6,above]{${\sss a}$} (\sh*\l+\l,0);
            \draw[obj3,-<-=.72, rounded corners=2pt] (\sh*\l,0) -- (\sh*\l,\sh*\l) --node[pos=.53,above]{${\sss g_2}$} (\sh*\l+\l,\sh*\l);
            \draw[obj3,->-=.26, rounded corners=2pt] (\sh*\l+\l,2*\sh*\l) --node[pos=.49,above]{${\sss g_1}$} (0,2*\sh*\l) -- (0,0);
            \node[mps1] at (0,0) {};
            \node[mps1] at (\sh*\l,0) {};
            \node[yshift=-9pt, xshift=-2pt] at (0,0) {${\sss \phi^{g_1 \! ,g_2(a)}}$};
            \node[yshift=-9pt, xshift=2pt] at (\sh*\l,0) {${\sss \phi^{g_2,a}}$};
        }{}
    \end{tikzpicture}
}

\newcommand{\moduleAssocRepG}[1]{
    \begin{tikzpicture}[scale=1,baseline={([yshift=-5.2ex]current bounding box.center)}]
        \def\l{.7};
        \def\sh{1.2};
        \ifthenelse{#1=1}{
            \draw[obj2,->-=.73] (0,0) --node[above,pos=.6]{${\sss U_2}$} (-\l,0);
            \draw[obj2,-<-=.46] (0,0) --node[pos=.52,above]{${\sss U_1}$}  (\sh*\l+\l,0);
            \draw[obj3,-<-=.62, rounded corners=2pt] (\sh*\l,1.5*\sh*\l) -- (\sh*\l,2*\sh*\l) --node[pos=.53,above]{${\sss V_1}$} (\sh*\l+\l,2*\sh*\l);
            \draw[obj3,-<-=.62, rounded corners=2pt] (\sh*\l,1.5*\sh*\l) -- (\sh*\l,\sh*\l) --node[pos=.53,below]{${\sss V_2}$} (\sh*\l+\l,\sh*\l);
            \draw[obj3,-<-=.73, rounded corners=2pt] (0,0) -- (0,1.5*\sh*\l) --node[pos=.47,above]{${\sss V_3}$} (\sh*\l,1.5*\sh*\l);
            \node[mps1] at (\sh*\l,1.5*\sh*\l) {};
            \node[mps1] at (0,0) {};
            \node[yshift=-12pt] at (0,0) {${\sss \phi^{V_3 U_1}_{U_2,i_2}}$};
            \node[xshift=17pt] at (\sh*\l,1.5*\sh*\l) {${\sss \varphi^{V_1 \! V_2}_{V_3,i_1}}$};
        }{}
        \ifthenelse{#1=2}{
            \draw[obj2,->-=.73] (0,0) --node[above,pos=.6]{${\sss U_2}$} (-\l,0);
            \draw[obj2,-<-=.47] (0,0) --node[above,pos=.6]{${\sss U_3}$} (\l,0);
            \draw[obj2,-<-=.47] (\sh*\l,0) --node[pos=.6,above]{${\sss U_1}$} (\sh*\l+\l,0);
            \draw[obj3,-<-=.72, rounded corners=2pt] (\sh*\l,0) -- (\sh*\l,\sh*\l) --node[pos=.53,above]{${\sss V_2}$} (\sh*\l+\l,\sh*\l);
            \draw[obj3,->-=.26, rounded corners=2pt] (\sh*\l+\l,2*\sh*\l) --node[pos=.49,above]{${\sss V_1}$} (0,2*\sh*\l) -- (0,0);
            \node[mps1] at (0,0) {};
            \node[mps1] at (\sh*\l,0) {};
            \node[yshift=-12pt, xshift=-2pt] at (0,0) {${\sss \phi^{V_1 U_3}_{U_2,i_4}}$};
            \node[yshift=-12pt, xshift=2pt] at (\sh*\l,0) {${\sss \phi^{V_2 U_1}_{U_3,i_3}}$};
        }{}
    \end{tikzpicture}
}

\newcommand{\Fsym}[1]{
    \begin{tikzpicture}[scale=1,baseline={([yshift=-5.4
    ex]current bounding box.center)}]
        \def\l{.7};
        \def\sh{1.2};
        \ifthenelse{#1=1}{
            \draw[obj2,->-=.73] (0,0) --node[above,pos=.9]{${\sss g_1(g_2(a))}$} (-\l,0);
            \draw[obj2,-<-=.46] (0,0) --node[above,pos=.5]{${\sss a}$} (\sh*\l+2*\l,0);
            \draw[obj2,-<-=.47] (\sh*\l+2*\l,0) --node[above,pos=.6]{${\sss g_2(a)}$} (\sh*\l+3*\l,0);
            \draw[obj2,-<-=.47] (2*\sh*\l+2*\l,0) --node[above,pos=.95]{${\sss g_1(g_2(a))}$} (2*\sh*\l+3*\l,0);
            \draw[obj3,-<-=.33, rounded corners=2pt] (\sh*\l,1.5*\sh*\l) -- (\sh*\l,2*\sh*\l) --node[pos=.5,above]{${\sss g_1}$} (2*\sh*\l+2*\l,2*\sh*\l) -- (2*\sh*\l+2*\l,0);
            \draw[obj3,-<-=.4, rounded corners=2pt] (\sh*\l,1.5*\sh*\l) -- (\sh*\l,\sh*\l) --node[pos=.53,above]{${\sss g_2}$} (\sh*\l+2*\l,\sh*\l) -- (\sh*\l+2*\l,0);
            \draw[obj3,-<-=.73, rounded corners=2pt] (0,0) -- (0,1.5*\sh*\l) --node[pos=.47,above]{${\sss g_1g_2}$} (\sh*\l,1.5*\sh*\l);
            \node[mps1] at (2*\sh*\l+2*\l,0) {};
            \node[mps1] at (\sh*\l+2*\l,0) {};
            \node[mps1] at (\sh*\l,1.5*\sh*\l) {};
            \node[mps1] at (0,0) {};
            \node[yshift=-9pt] at (0,0) {${\sss \phi^{g_1 \! g_2,a}}$};
            \node[yshift=-9pt, xshift=-2pt] at (\sh*\l+2*\l,0) {${\sss \bar \phi^{g_2,a}}$};
            \node[yshift=-9pt, xshift=2pt] at (2*\sh*\l+2*\l,0) {${\sss \bar \phi^{g_1 \! ,g_2(a)}}$};
        }{
            \draw[obj2,-<-=.47] (0,0) --node[above,pos=.9]{${\sss g_1(g_2(a))}$} (\l,0);
            \draw[obj2,->-=.54] (0,0) --node[above,pos=.5]{${\sss a}$} (-\sh*\l-2*\l,0);
            \draw[obj2,->-=.73] (-\sh*\l-2*\l,0) --node[above,pos=.6]{${\sss g_2(a)}$} (-\sh*\l-3*\l,0);
            \draw[obj2,->-=.73] (-2*\sh*\l-2*\l,0) --node[above,pos=.95]{${\sss g_1(g_2(a))}$} (-2*\sh*\l-3*\l,0);
            \draw[obj3,->-=.38, rounded corners=2pt] (-\sh*\l,1.5*\sh*\l) -- (-\sh*\l,2*\sh*\l) --node[pos=.5,above]{${\sss g_1}$} (-2*\sh*\l-2*\l,2*\sh*\l) -- (-2*\sh*\l-2*\l,0);
            \draw[obj3,->-=.47, rounded corners=2pt] (-\sh*\l,1.5*\sh*\l) -- (-\sh*\l,\sh*\l) --node[pos=.53,above]{${\sss g_2}$} (-\sh*\l-2*\l,\sh*\l) -- (-\sh*\l-2*\l,0);
            \draw[obj3,->-=.83, rounded corners=2pt] (0,0) -- (0,1.5*\sh*\l) --node[pos=.47,above]{${\sss g_1g_2}$} (-\sh*\l,1.5*\sh*\l);
            \node[mps1] at (-2*\sh*\l-2*\l,0) {};
            \node[mps1] at (-\sh*\l-2*\l,0) {};
            \node[mps1] at (-\sh*\l,1.5*\sh*\l) {};
            \node[mps1] at (0,0) {};
            \node[yshift=-9pt] at (0,0) {${\sss \bar \phi^{g_1 \! g_2,a}}$};
            \node[yshift=-9pt, xshift=2pt] at (-\sh*\l-2*\l,0) {${\sss  \phi^{g_2,a}}$};
            \node[yshift=-9pt, xshift=-2pt] at (-2*\sh*\l-2*\l,0) {${\sss \phi^{g_1 \! ,g_2(a)}}$};
        }
    \end{tikzpicture}
}

\newcommand{\FsymRepG}[0]{
    \begin{tikzpicture}[scale=1,baseline={([yshift=-5.2
    ex]current bounding box.center)}]
        \def\l{.7};
        \def\sh{1.2};

        \draw[obj2,-<-=.47] (0,0) --node[above,pos=.6]{${\sss U_4}$} (\l,0);
        \draw[obj2,->-=.54] (0,0) --node[above,pos=.5]{${\sss U_1}$} (-\sh*\l-2*\l,0);
        \draw[obj2,->-=.73] (-\sh*\l-2*\l,0) --node[above,pos=.6]{${\sss U_3}$} (-\sh*\l-3*\l,0);
        \draw[obj2,->-=.73] (-2*\sh*\l-2*\l,0) --node[above,pos=.6]{${\sss U_2}$} (-2*\sh*\l-3*\l,0);
        \draw[obj3,->-=.38, rounded corners=2pt] (-\sh*\l,1.5*\sh*\l) -- (-\sh*\l,2*\sh*\l) --node[pos=.5,above]{${\sss V_1}$} (-2*\sh*\l-2*\l,2*\sh*\l) -- (-2*\sh*\l-2*\l,0);
        \draw[obj3,->-=.47, rounded corners=2pt] (-\sh*\l,1.5*\sh*\l) -- (-\sh*\l,\sh*\l) --node[pos=.53,below]{${\sss V_2}$} (-\sh*\l-2*\l,\sh*\l) -- (-\sh*\l-2*\l,0);
        \draw[obj3,->-=.83, rounded corners=2pt] (0,0) -- (0,1.5*\sh*\l) --node[pos=.47,above]{${\sss V_3}$} (-\sh*\l,1.5*\sh*\l);
        \node[mps1] at (-2*\sh*\l-2*\l,0) {};
        \node[mps1] at (-\sh*\l-2*\l,0) {};
        \node[mps1] at (-\sh*\l,1.5*\sh*\l) {};
        \node[mps1] at (0,0) {};
        \node[yshift=-12pt] at (0,0) {${\sss \bar \phi^{V_3 U_1}_{U_2,i_2}}$};
        \node[yshift=-12pt, xshift=2pt] at (-\sh*\l-2*\l,0) {${\sss  \phi^{V_2 U_1}_{U_3,i_3}}$};
        \node[yshift=-12pt, xshift=-2pt] at (-2*\sh*\l-2*\l,0) {${\sss \phi^{V_1 U_3}_{U_2,i_4}}$};
        \node[xshift=-16pt] at (-\sh*\l,1.5*\sh*\l) {${\sss \bar \varphi^{V_1\!V_2}_{V_3,i_1}}$};
    \end{tikzpicture}
}

\newcommand{\localOp}[1]{
    \begin{tikzpicture}[scale=1,baseline={([yshift=-.55ex]current bounding box.center)}]
        \def\l{.7};
        \draw[obj1,-<-=.47] (0,0) -- (0,-\l);
        \draw[obj1,->-=.73] (0,0) -- (0,\l);
        \node[mps1] at (0,0) {};
        \node[xshift=-9pt] at (0,0) {${\sss #1}$};
    \end{tikzpicture}
}

\newcommand{\localOpExt}[3]{
    \begin{tikzpicture}[scale=1,baseline={([yshift=-.55ex]current bounding box.center)}]
        \def\l{.7};
        \draw[obj1,-<-=.47] (0,0) -- (0,-\l);
        \draw[obj1,->-=.73] (0,0) -- (0,\l);
        \draw[obj1,-<-=.47] (0,0) --node[pos=.6,above]{${\sss #2}$} node[pos=1.25]{${\sss #3}$} (\l,0);

        \node[mps1] at (0,0) {};
        \node[xshift=-13pt] at (0,0) {${\sss #1}$};
    \end{tikzpicture}
}

\newcommand{\localOpTube}[5]{
    \begin{tikzpicture}[scale=1,baseline={([yshift=-.55ex]current bounding box.center)}]
        \def\l{.7};
        \def\sh{1.2};
        \ifthenelse{#1=1}{
            \draw[obj1,-<-=.47] (0,0) -- (0,-\l);
            \draw[obj1,-<-=.47] (0,-\sh*\l) -- (0,-\sh*\l-\l);
            \draw[obj1,->-=.73] (0,0) -- (0,\l);
            \draw[obj1,->-=.73] (0,\sh*\l) -- (0,\sh*\l+\l);
            \draw[obj3,-<-=.48, rounded corners=2pt] (0,-\sh*\l) -- (-\sh*\l,-\sh*\l) --node[pos=.5,left]{${\sss #3}$} (-\sh*\l,\sh*\l) -- (0,\sh*\l);
            \draw[obj3,->-=.52, rounded corners=2pt] (0,-\sh*\l) -- (\sh*\l,-\sh*\l) --node[pos=.5,right]{${\sss #3}$} (\sh*\l,\sh*\l) -- (0,\sh*\l);
            \node[mps1] at (0,0) {};
            \node[mpo] at (0,-\sh*\l) {};
            \node[mpo] at (0,\sh*\l) {};
            \node[xshift=-9pt] at (0,0) {${\sss #2}$};
        }{
            \draw[obj1,-<-=.47] (0,0) -- (0,-\l);
            \draw[obj1,->-=.73] (0,0) -- (0,\l);
            \draw[obj3,->-=.51, rounded corners=2pt] (\sh*\l,0) -- (\sh*\l,-\sh*\l) -- (3*\sh*\l,-\sh*\l) --node[pos=.5,right]{${\sss #3}$} (3*\sh*\l,\sh*\l) -- (\sh*\l,\sh*\l) -- (\sh*\l,0);
            \draw[obj1,-<-=.47] (0,0) --node[pos=.6,above]{${\sss #4}$} (\l,0);
            \draw[obj1,-<-=.47] (\sh*\l,0) --node[pos=.6,above]{${\sss #4}$} node[pos=1.25]{${\sss #5}$} (\sh*\l+\l,0);
            \node[mps1] at (0,0) {};
            \node[mpo] at (\sh*\l,0) {};
            \node[xshift=-13pt] at (0,0) {${\sss #2}$};
        }
    \end{tikzpicture}
}

\newcommand{\twistedLocalOp}[3]{
    \begin{tikzpicture}[scale=1,baseline={([yshift=-.55ex]current bounding box.center)}]
        \def\l{.7};
        \draw[obj1,-<-=.47] (0,0) -- (0,-\l);
        \draw[obj1,->-=.73] (0,0) -- (0,\l);
        \draw[obj3,-<-=.47] (0,0) --node[pos=.6,above]{${\sss #2}$} node[pos=1.25]{${\sss #3}$} (\l,0);
        \node[mps1] at (0,0) {};
        \node[xshift=-9pt] at (0,0) {${\sss #1}$};
    \end{tikzpicture}
}

\newcommand{\twistedLocalOpDual}[2]{
    \begin{tikzpicture}[scale=1,baseline={([yshift=-.55ex]current bounding box.center)}]
        \def\l{.7};
        \draw[obj1,-<-=.47] (0,0) -- (0,-\l);
        \draw[obj1,->-=.73] (0,0) -- (0,\l);
        \draw[obj3,->-=.73] (0,0) --node[pos=.6,above]{${\sss #2}$} (-\l,0);
        \node[mps1] at (0,0) {};
        \node[xshift=8pt] at (0,0) {${\sss #1}$};
    \end{tikzpicture}
}

\newcommand{\twistedLocalOpExt}[4]{
    \begin{tikzpicture}[scale=1,baseline={([yshift=-.55ex]current bounding box.center)}]
        \def\l{.7};
        \def\sh{1.2};
        \draw[obj1,-<-=.47] (0,0) -- (0,-\l);
        \draw[obj1,->-=.73] (0,0) -- (0,\l);
        \draw[obj1,-<-=.47] (0,0) --node[pos=.6,above]{${\sss #2}$} (\l,0);
        \draw[obj3,-<-=.47] (\sh*\l,0) --node[pos=.6,above]{${\sss #4}$} (\sh*\l+\l,0);
        \node[mps1] at (0,0) {};
        \node[mps1] at (\sh*\l,0) {};
        \node[xshift=-15pt] at (0,0) {${\sss #1}$};
        \node[yshift=-12pt] at (\sh*\l,0) {${\sss #3}$};
    \end{tikzpicture}
}

\newcommand{\twistedLocalOpDualExt}[4]{
    \begin{tikzpicture}[scale=1,baseline={([yshift=-.55ex]current bounding box.center)}]
        \def\l{.7};
        \def\sh{1.2};
        \draw[obj1,-<-=.47] (0,0) -- (0,-\l);
        \draw[obj1,->-=.73] (0,0) -- (0,\l);
        \draw[obj1,->-=.73] (0,0) --node[pos=.6,above]{${\sss #2}$} (-\l,0);
        \draw[obj3,->-=.73] (-\sh*\l,0) --node[pos=.6,above]{${\sss #4}$} (-\sh*\l-\l,0);
        \node[mps1] at (0,0) {};
        \node[mps1] at (-\sh*\l,0) {};
        \node[xshift=15pt] at (0,0) {${\sss #1}$};
        \node[yshift=-12pt] at (-\sh*\l,0) {${\sss #3}$};
    \end{tikzpicture}
}

\newcommand{\fusionTwistedLocalOp}[5]{
    \begin{tikzpicture}[scale=1,baseline={([yshift=-.55ex]current bounding box.center)}]
        \def\l{.7};
        \def\sh{1.2};
        \draw[obj1,-<-=.47] (0,0) -- (0,-\l);
        \draw[obj1,->-=.73] (0,0) -- (0,\l);
        \draw[obj1,->-=.73] (0,\sh*\l) -- (0,\sh*\l+\l);
        \draw[obj3,-<-=.47] (\sh*\l,.5*\sh*\l) --node[pos=.6,above]{${\sss #3}$} (\sh*\l+\l,.5*\sh*\l);
        \draw[obj3,->-=.78, rounded corners=2pt] (\sh*\l,.5*\sh*\l) -- (\sh*\l,0) -- node[pos=.53,below]{${\sss #2}$} (\sh*\l-\l,0);
        \draw[obj3,->-=.78, rounded corners=2pt] (\sh*\l,.5*\sh*\l) -- (\sh*\l,\sh*\l) -- node[pos=.53,above]{${\sss #1}$} (\sh*\l-\l,\sh*\l);
        \node[mps1] at (0,0) {};
        \node[mps1] at (\sh*\l,.5*\sh*\l) {};
        \node[mps1] at (0,\sh*\l) {};
        \node[xshift=-9pt] at (0,0) {${\sss #4}$};
        \node[xshift=-9pt] at (0,\sh*\l) {${\sss #5}$};
    \end{tikzpicture}
}

\newcommand{\Tube}[6]{
    \begin{tikzpicture}[scale=1,baseline={([yshift=-.55ex]current bounding box.center)}]
        \def\l{.7};
        \def\sh{1.2};
        \draw[obj1,-<-=.45] (0,0) -- (0,-\sh*\l);
        \draw[obj1,->-=.65] (0,0) -- (0,\sh*\l);
        \draw[obj1,->-=.73] (0,\sh*\l) -- (0,\sh*\l+\l);
        \draw[obj1,-<-=.47] (0,-\sh*\l) -- (0,-\sh*\l-\l);
        \draw[obj3,-<-=.47] (0,0) --node[pos=.56,above]{${\sss #4}$} (\sh*\l,0);
        \draw[obj3,-<-=.23, rounded corners=2pt] (\sh*\l,0) --node[pos=.56,above]{${\sss #3}$} (2*\sh*\l,0) -- (2*\sh*\l,-\sh*\l);
        \draw[obj3,-<-=.47] (2*\sh*\l,-\sh*\l) --node[pos=.6, right]{${\sss #2}$} (2*\sh*\l,-\sh*\l-\l);
        \draw[obj3,-<-=.43] (2*\sh*\l,-\sh*\l) -- node[pos=.49,below]{${\sss #1}$} (0,-\sh*\l);
        \draw[obj3,-<-=.48, rounded corners=2pt] (0,-\sh*\l) -- (-\sh*\l,-\sh*\l) --node[pos=.5,left]{${\sss #1}$} (-\sh*\l,\sh*\l) -- (0,\sh*\l);
        \draw[obj3,-<-=.23, rounded corners=2pt] (0,\sh*\l) --node[pos=.56,above]{${\sss #1}$} (\sh*\l,\sh*\l) -- (\sh*\l,0);
        \node[yshift=-12pt] at (\sh*\l,0) {${\sss #5}$};
        \node[xshift=17pt] at (2*\sh*\l,-\sh*\l) {${\sss #6}$};

        \node[hole] at (0,0) {};
        \node[mpo] at (0,-\sh*\l) {};
        \node[mpo] at (0,\sh*\l) {};
        \node[mps1] at (\sh*\l,0) {};
        \node[mps1] at (2*\sh*\l,-\sh*\l) {};
    \end{tikzpicture}
}

\newcommand{\TubeDag}[6]{
    \begin{tikzpicture}[scale=1,baseline={([yshift=-.55ex]current bounding box.center)}]
        \def\l{.7};
        \def\sh{1.2};
        \draw[obj1,-<-=.45] (0,0) -- (0,-\sh*\l);
        \draw[obj1,->-=.65] (0,0) -- (0,\sh*\l);
        \draw[obj1,->-=.73] (0,\sh*\l) -- (0,\sh*\l+\l);
        \draw[obj1,-<-=.47] (0,-\sh*\l) -- (0,-\sh*\l-\l);
        \draw[obj3,-<-=.47] (0,0) --node[pos=.56,below]{${\sss #2}$} (\sh*\l,0);
        \draw[obj3,-<-=.47] (\sh*\l,0) --node[pos=.6,below]{${\sss #3}$} (\sh*\l+\l,0);
        \draw[obj3,-<-=.5] (2*\sh*\l,0) --node[pos=.55, right]{${\sss #4}$} (2*\sh*\l,-\sh*\l-\l);
        \draw[obj3,-<-=.24, rounded corners=2pt] (0,-\sh*\l) -- node[pos=.55,below]{${\sss #1}$} (\sh*\l,-\sh*\l) -- (\sh*\l,0);
        \draw[obj3,->-=.52, rounded corners=2pt] (0,-\sh*\l) -- (-\sh*\l,-\sh*\l) --node[pos=.5,left]{${\sss #1}$} (-\sh*\l,\sh*\l) -- (0,\sh*\l);
        \draw[obj3,->-=.4, rounded corners=2pt] (0,\sh*\l) --node[pos=.56,above]{${\sss #1}$} (2*\sh*\l,\sh*\l) -- (2*\sh*\l,0);
        
        \node[yshift=12pt] at (\sh*\l,0) {${\sss #5}$};
        \node[xshift=17pt] at (2*\sh*\l,0) {${\sss #6}$};

        \node[hole] at (0,0) {};
        \node[mpo] at (0,-\sh*\l) {};
        \node[mpo] at (0,\sh*\l) {};
        \node[mps1] at (\sh*\l,0) {};
        \node[mps1] at (2*\sh*\l,0) {};
    \end{tikzpicture}
}

\newcommand{\TubeMult}[7]{
    \begin{tikzpicture}[scale=1,baseline={([yshift=-.55ex]current bounding box.center)}]
        \def\l{.7};
        \def\sh{1.2};
        \draw[obj1,-<-=.45] (0,0) -- (0,-\sh*\l);
        \draw[obj1,->-=.65] (0,0) -- (0,\sh*\l);
        \draw[obj1,->-=.73] (0,\sh*\l) -- (0,\sh*\l+\l);
        \draw[obj1,-<-=.47] (0,-\sh*\l) -- (0,-\sh*\l-\l);
        \draw[obj1,->-=.73] (0,2*\sh*\l) -- (0,2*\sh*\l+\l);
        \draw[obj1,-<-=.47] (0,-2*\sh*\l) -- (0,-2*\sh*\l-\l);
        
        \draw[obj3,-<-=.47] (0,0) --node[pos=.56,above]{${\sss #7}$} (\sh*\l,0);
        \draw[obj3,-<-=.23, rounded corners=2pt] (\sh*\l,0) --node[pos=.56,above]{${\sss #6}$} (2*\sh*\l,0) -- (2*\sh*\l,-\sh*\l);
        \draw[obj3,-<-=.47] (4*\sh*\l,-2*\sh*\l) --node[pos=.6, right]{${\sss #3}$} (4*\sh*\l,-2*\sh*\l-\l);
        \draw[obj3,-<-=.43] (2*\sh*\l,-\sh*\l) -- node[pos=.49,below]{${\sss #1}$} (0,-\sh*\l);
        \draw[obj3,-<-=.48, rounded corners=2pt] (0,-\sh*\l) -- (-\sh*\l,-\sh*\l) --node[pos=.5,left]{${\sss #1}$} (-\sh*\l,\sh*\l) -- (0,\sh*\l);
        \draw[obj3,-<-=.48, rounded corners=2pt] (0,-2*\sh*\l) -- (-2*\sh*\l,-2*\sh*\l) --node[pos=.5,left]{${\sss #2}$} (-2*\sh*\l,2*\sh*\l) -- (0,2*\sh*\l);
        \draw[obj3,-<-=.23, rounded corners=2pt] (0,\sh*\l) --node[pos=.56,above]{${\sss #1}$} (\sh*\l,\sh*\l) -- (\sh*\l,0);
        \draw[obj3,-<-=.47] (2*\sh*\l,-\sh*\l) --node[pos=.6,above]{${\sss #5}$} (2*\sh*\l+\l,-\sh*\l);
        \draw[obj3,-<-=.23, rounded corners=2pt] (3*\sh*\l,-\sh*\l) --node[pos=.56,above]{${\sss #4}$} (4*\sh*\l,-\sh*\l) -- (4*\sh*\l,-2*\sh*\l);
        \draw[obj3,-<-=.23, rounded corners=2pt] (0,2*\sh*\l) --node[pos=.56,above]{${\sss #2}$} (3*\sh*\l,2*\sh*\l) -- (3*\sh*\l,-\sh*\l);
        \draw[obj3,->-=.5] (0,-2*\sh*\l) --node[pos=.48,below]{${\sss #2}$} (4*\sh*\l,-2*\sh*\l);
        
        \node[hole] at (0,0) {};
        \node[mpo] at (0,-\sh*\l) {};
        \node[mpo] at (0,\sh*\l) {};
        \node[mpo] at (0,-2*\sh*\l) {};
        \node[mpo] at (0,2*\sh*\l) {};
        \node[mps1] at (\sh*\l,0) {};
        \node[mps1] at (2*\sh*\l,-\sh*\l) {};
        \node[mps1] at (3*\sh*\l,-\sh*\l) {};
        \node[mps1] at (4*\sh*\l,-2*\sh*\l) {};
    \end{tikzpicture}
}

\newcommand{\TubeMultRepG}[0]{
    \begin{tikzpicture}[scale=1,baseline={([yshift=-.55ex]current bounding box.center)}]
        \def\l{.7};
        \def\sh{1.2};
        \draw[obj1,-<-=.45] (0,0) -- (0,-\sh*\l);
        \draw[obj1,->-=.65] (0,0) -- (0,\sh*\l);
        \draw[obj1,->-=.73] (0,\sh*\l) -- (0,\sh*\l+\l);
        \draw[obj1,-<-=.47] (0,-\sh*\l) -- (0,-\sh*\l-\l);
        \draw[obj1,->-=.73] (0,2*\sh*\l) -- (0,2*\sh*\l+\l);
        \draw[obj1,-<-=.47] (0,-2*\sh*\l) -- (0,-2*\sh*\l-\l);
        
        \draw[obj3,-<-=.47] (0,0) --node[pos=.56,above]{${\sss S_1}$} (\sh*\l,0);
        \draw[obj3,-<-=.23, rounded corners=2pt] (\sh*\l,0) --node[pos=.56,above]{${\sss S_2}$} (2*\sh*\l,0) -- (2*\sh*\l,-\sh*\l);
        \draw[obj3,-<-=.47] (4*\sh*\l,-2*\sh*\l) --node[pos=.6, right]{${\sss S_6}$} (4*\sh*\l,-2*\sh*\l-\l);
        \draw[obj3,-<-=.43] (2*\sh*\l,-\sh*\l) -- node[pos=.49,below]{${\sss V_2}$} (0,-\sh*\l);
        \draw[obj3,-<-=.48, rounded corners=2pt] (0,-\sh*\l) -- (-\sh*\l,-\sh*\l) --node[pos=.5,left]{${\sss V_2}$} (-\sh*\l,\sh*\l) -- (0,\sh*\l);
        \draw[obj3,-<-=.48, rounded corners=2pt] (0,-2*\sh*\l) -- (-2*\sh*\l,-2*\sh*\l) --node[pos=.5,left]{${\sss V_1}$} (-2*\sh*\l,2*\sh*\l) -- (0,2*\sh*\l);
        \draw[obj3,-<-=.23, rounded corners=2pt] (0,\sh*\l) --node[pos=.56,above]{${\sss V_2}$} (\sh*\l,\sh*\l) -- (\sh*\l,0);
        \draw[obj3,-<-=.47] (2*\sh*\l,-\sh*\l) --node[pos=.6,above]{${\sss S_3}$} (2*\sh*\l+\l,-\sh*\l);
        \draw[obj3,-<-=.23, rounded corners=2pt] (3*\sh*\l,-\sh*\l) --node[pos=.56,above]{${\sss S_5}$} (4*\sh*\l,-\sh*\l) -- (4*\sh*\l,-2*\sh*\l);
        \draw[obj3,-<-=.23, rounded corners=2pt] (0,2*\sh*\l) --node[pos=.56,above]{${\sss V_1}$} (3*\sh*\l,2*\sh*\l) -- (3*\sh*\l,-\sh*\l);
        \draw[obj3,->-=.5] (0,-2*\sh*\l) --node[pos=.48,below]{${\sss V_1}$} (4*\sh*\l,-2*\sh*\l);

        \node[yshift=-12pt] at (\sh*\l,0) {${\sss \bar \varphi^{V_2 S_1}_{S_2,i_1} }$};
        \node[yshift=-12pt,xshift=-1pt] at (2*\sh*\l,-\sh*\l) {${\sss \varphi^{S_3 V_2}_{S_2,i_2} }$};
        \node[yshift=-12pt, xshift=1pt] at (3*\sh*\l,-\sh*\l) {${\sss \bar \varphi^{V_1 S_3}_{S_5,i_3} }$};
        \node[xshift=17pt] at (4*\sh*\l,-2*\sh*\l) {${\sss \varphi^{S_6 V_1}_{S_5,i_4} }$};
        
        \node[hole] at (0,0) {};
        \node[mpo] at (0,-\sh*\l) {};
        \node[mpo] at (0,\sh*\l) {};
        \node[mpo] at (0,-2*\sh*\l) {};
        \node[mpo] at (0,2*\sh*\l) {};
        \node[mps1] at (\sh*\l,0) {};
        \node[mps1] at (2*\sh*\l,-\sh*\l) {};
        \node[mps1] at (3*\sh*\l,-\sh*\l) {};
        \node[mps1] at (4*\sh*\l,-2*\sh*\l) {};
    \end{tikzpicture}
}

\newcommand{\twistedLocalOpTube}[8]{
    \begin{tikzpicture}[scale=1,baseline={([yshift=-.55ex]current bounding box.center)}]
        \def\l{.7};
        \def\sh{1.2};
        \ifthenelse{#1=1}{
            \draw[obj1,-<-=.47] (0,0) -- (0,-\l);
            \draw[obj1,->-=.73] (0,0) -- (0,\l);
            \draw[obj1,->-=.73] (0,\sh*\l) -- (0,\sh*\l+\l);
            \draw[obj1,-<-=.47] (0,-\sh*\l) -- (0,-\sh*\l-\l);
            \draw[obj3,-<-=.47] (0,0) --node[pos=.56,above]{${\sss #6}$} (\l,0);
            \draw[obj3,-<-=.23, rounded corners=2pt] (\sh*\l,0) --node[pos=.56,above]{${\sss #5}$} (2*\sh*\l,0) -- (2*\sh*\l,-\sh*\l);
            \draw[obj3,-<-=.47] (2*\sh*\l,-\sh*\l) --node[pos=.6, right]{${\sss #4}$} (2*\sh*\l,-\sh*\l-\l);
            \draw[obj3,-<-=.43] (2*\sh*\l,-\sh*\l) -- node[pos=.49,below]{${\sss #3}$} (0,-\sh*\l);
            \draw[obj3,-<-=.48, rounded corners=2pt] (0,-\sh*\l) -- (-\sh*\l,-\sh*\l) --node[pos=.5,left]{${\sss #3}$} (-\sh*\l,\sh*\l) -- (0,\sh*\l);
            \draw[obj3,-<-=.23, rounded corners=2pt] (0,\sh*\l) --node[pos=.56,above]{${\sss #3}$} (\sh*\l,\sh*\l) -- (\sh*\l,0);
            \node[xshift=-9pt] at (0,0) {${\sss #2}$};
            \node[yshift=-12pt] at (\sh*\l,0) {${\sss #7}$};
            \node[xshift=17pt] at (2*\sh*\l,-\sh*\l) {${\sss #8}$};
    
            \node[mps1] at (0,0) {};
            \node[mpo] at (0,-\sh*\l) {};
            \node[mpo] at (0,\sh*\l) {};
            \node[mps1] at (\sh*\l,0) {};
            \node[mps1] at (2*\sh*\l,-\sh*\l) {};
        }{}
        \ifthenelse{#1=2}{
            \path (0,0) -- (0,\sh*\l+\l);
            \draw[obj1,-<-=.47] (0,0) -- (0,-\l);
            \draw[obj1,->-=.73] (0,0) -- (0,\l);
            \draw[obj1,-<-=.47] (0,0) --node[pos=.6,above]{${\sss #7}$} (\l,0);
            \draw[obj3,-<-=.47] (2*\sh*\l,0) --node[pos=.6, above]{${\sss #6}$} (2*\sh*\l+\l,0);
            \draw[obj1,-<-=.47] (\sh*\l,0) --node[pos=.56,above]{${\sss #7}$} (\sh*\l+\l,0);
            \draw[obj3,-<-=.23, rounded corners=2pt] (3*\sh*\l,0) --node[pos=.56,above]{${\sss #5}$} (4*\sh*\l,0) -- (4*\sh*\l,-\sh*\l);
            \draw[obj3,-<-=.47] (4*\sh*\l,-\sh*\l) --node[pos=.6, right]{${\sss #4}$} (4*\sh*\l,-\sh*\l-\l);
            \draw[obj3,-<-=.34, rounded corners=2pt] (4*\sh*\l,-\sh*\l) -- node[pos=.49,below]{${\sss #3}$} (\sh*\l,-\sh*\l) -- (\sh*\l,0);
            \draw[obj3,-<-=.48, rounded corners=2pt] (\sh*\l,0) -- (\sh*\l,\sh*\l) --node[pos=.51,above]{${\sss #3}$} (3*\sh*\l,\sh*\l) -- (3*\sh*\l,0);
            \node[xshift=-13pt] at (0,0) {${\sss #8}$};
            \node[yshift=-11pt] at (2*\sh*\l,0) {${\sss #2}$};
    
            \node[mps1] at (0,0) {};
            \node[mpo] at (\sh*\l,0) {};
            \node[mps1] at (2*\sh*\l,0) {};
            \node[mps1] at (3*\sh*\l,0) {};
            \node[mps1] at (4*\sh*\l,-\sh*\l) {};
        }{}
        \ifthenelse{#1=3}{
            \path (0,0) -- (0,\sh*\l+\l);
            \draw[obj1,-<-=.47] (0,0) -- (0,-\l);
            \draw[obj1,->-=.73] (0,0) -- (0,\l);
            \draw[obj1,-<-=.47] (0,0) --node[pos=.6,above]{${\sss #7}$} (\l,0);
            \draw[obj3,-<-=.47] (2*\sh*\l,0) --node[pos=.6, above]{${\sss #6}$} (2*\sh*\l+\l,0);
            \draw[obj1,-<-=.47] (\sh*\l,0) --node[pos=.56,above]{${\sss #7}$} (\sh*\l+\l,0);
            \draw[obj3,-<-=.23, rounded corners=2pt] (3*\sh*\l,0) --node[pos=.56,above]{${\sss #5}$} (4*\sh*\l,0) -- (4*\sh*\l,-\sh*\l);
            \draw[obj3,-<-=.47] (4*\sh*\l,-\sh*\l) --node[pos=.6, right]{${\sss #4}$} (4*\sh*\l,-\sh*\l-\l);
            \draw[obj3,-<-=.34, rounded corners=2pt] (4*\sh*\l,-\sh*\l) -- node[pos=.49,below]{${\sss #3}$} (\sh*\l,-\sh*\l) -- (\sh*\l,0);
            \draw[obj3,-<-=.48, rounded corners=2pt] (\sh*\l,0) -- (\sh*\l,\sh*\l) --node[pos=.51,above]{${\sss #3}$} (3*\sh*\l,\sh*\l) -- (3*\sh*\l,0);
            \node[xshift=17pt] at (4*\sh*\l,-\sh*\l) {${\sss \varphi^{S_3 V}_{S_2,i_2}}$};
            \node[yshift=-12pt, xshift=-1pt] at (2*\sh*\l,0) {${\sss \bar \varphi^{W}_{S_1,j_2}}$};
            \node[yshift=-12pt, xshift=1pt] at (3*\sh*\l,0) {${\sss \bar \varphi^{V S_1}_{S_2,i_1}}$};
    
            \node[mps1] at (0,0) {};
            \node[mpo] at (\sh*\l,0) {};
            \node[mps1] at (2*\sh*\l,0) {};
            \node[mps1] at (3*\sh*\l,0) {};
            \node[mps1] at (4*\sh*\l,-\sh*\l) {};
        }{}
    \end{tikzpicture}
}

\newcommand{\ModuleCsts}[0]{
    \begin{tikzpicture}[scale=1,baseline={([yshift=-.55ex]current bounding box.center)}]
        \def\l{.7};
        \def\sh{1.2};
        \draw[obj1,-<-=.47] (5*\sh*\l,0) --node[pos=.6,above]{${\sss W}$} (5*\sh*\l+\l,0);
        \draw[obj1,->-=.73] (\sh*\l,0) --node[pos=.6,above]{${\sss W}$} (\sh*\l-\l,0);
        \draw[obj3,-<-=.47] (2*\sh*\l,0) --node[pos=.6, above]{${\sss S_1}$} (2*\sh*\l+\l,0);
        \draw[obj1,-<-=.47] (\sh*\l,0) --node[pos=.56,above]{${\sss W}$} (\sh*\l+\l,0);
        \draw[obj3,-<-=.47] (3*\sh*\l,0) --node[pos=.56,above]{${\sss S_2}$} (3*\sh*\l+\l,0);
        \draw[obj3,-<-=.47] (4*\sh*\l,0) --node[pos=.6, below]{${\sss S_3}$} (4*\sh*\l+\l,0);
        \draw[obj3,-<-=.48, rounded corners=2pt] (4*\sh*\l,0) -- (4*\sh*\l,-\sh*\l) --node[pos=.5,below]{${\sss V}$} (\sh*\l,-\sh*\l) -- (\sh*\l,0);
        \draw[obj3,-<-=.48, rounded corners=2pt] (\sh*\l,0) -- (\sh*\l,\sh*\l) --node[pos=.51,above]{${\sss V}$} (3*\sh*\l,\sh*\l) -- (3*\sh*\l,0);
        
        \node[yshift=12pt,xshift=5pt] at (4*\sh*\l,0) {${\sss \varphi^{S_3 V}_{S_2,i_2}}$};
        \node[yshift=-12pt, xshift=-1pt] at (2*\sh*\l,0) {${\sss \bar \varphi^{W}_{S_1,j_2}}$};
        \node[yshift=-12pt, xshift=1pt] at (3*\sh*\l,0) {${\sss \bar \varphi^{V S_1}_{S_2,i_1}}$};
        \node[yshift=-12pt, xshift=6pt] at (5*\sh*\l,0) {${\sss \varphi^{W}_{S_34,j_1}}$};

        \node[mpo] at (\sh*\l,0) {};
        \node[mps1] at (5*\sh*\l,0) {};
        \node[mps1] at (2*\sh*\l,0) {};
        \node[mps1] at (3*\sh*\l,0) {};
        \node[mps1] at (4*\sh*\l,0) {};
    \end{tikzpicture}
}

\newcommand{\RepGTubeModStr}[0]{
    \begin{tikzpicture}[scale=1,baseline={([yshift=-.55ex]current bounding box.center)}]
        \def\l{.7};
        \def\sh{1.2};
        \draw[obj1,-<-=.47] (7*\sh*\l,0) --node[pos=.6,above]{${\sss W}$} (7*\sh*\l+\l,0);
        \draw[obj1,->-=.73] (0,0) --node[pos=.6,above]{${\sss W}$} (-\l,0);
        \draw[obj1,-<-=.47] (0,0) --node[pos=.6,above]{${\sss W}$} (\l,0);
        \draw[obj3,-<-=.47] (2*\sh*\l,0) --node[pos=.6, above]{${\sss S_1}$} (2*\sh*\l+\l,0);
        \draw[obj1,-<-=.47] (\sh*\l,0) --node[pos=.56,above]{${\sss W}$} (\sh*\l+\l,0);
        \draw[obj3,-<-=.47] (3*\sh*\l,0) --node[pos=.56,above]{${\sss S_2}$} (3*\sh*\l+\l,0);
        \draw[obj3,-<-=.47] (4*\sh*\l,0) --node[pos=.6, below]{${\sss S_3}$} (4*\sh*\l+\l,0);
        \draw[obj3,-<-=.47] (5*\sh*\l,0) --node[pos=.6, above]{${\sss S_5}$} (5*\sh*\l+\l,0);
        \draw[obj3,-<-=.47] (6*\sh*\l,0) --node[pos=.6, below]{${\sss S_6}$} (6*\sh*\l+\l,0);
        
        \draw[obj3,-<-=.48, rounded corners=2pt] (4*\sh*\l,0) -- (4*\sh*\l,-\sh*\l) --node[pos=.5,below]{${\sss V_2}$} (\sh*\l,-\sh*\l) -- (\sh*\l,0);
        \draw[obj3,-<-=.48, rounded corners=2pt] (\sh*\l,0) -- (\sh*\l,\sh*\l) --node[pos=.51,above]{${\sss V_2}$} (3*\sh*\l,\sh*\l) -- (3*\sh*\l,0);
        \draw[obj3,-<-=.48, rounded corners=2pt] (6*\sh*\l,0) -- (6*\sh*\l,-2*\sh*\l) --node[pos=.5,below]{${\sss V_1}$} (0,-2*\sh*\l) -- (0,0);
        \draw[obj3,-<-=.48, rounded corners=2pt] (0,0) -- (0,2*\sh*\l) --node[pos=.51,above]{${\sss V_1}$} (5*\sh*\l,2*\sh*\l) -- (5*\sh*\l,0);
        
        \node[yshift=12pt,xshift=5pt] at (4*\sh*\l,0) {${\sss \varphi^{S_3 V_2}_{S_2,i_2}}$};
        \node[yshift=-12pt, xshift=-1pt] at (2*\sh*\l,0) {${\sss \bar \varphi^{W}_{S_1,j_3}}$};
        \node[yshift=-12pt, xshift=1pt] at (3*\sh*\l,0) {${\sss \bar \varphi^{V_2 S_1}_{S_2,i_1}}$};
        \node[yshift=-12pt, xshift=5pt] at (5*\sh*\l,0) {${\sss \bar \varphi^{V_1,S_3}_{S_5,i_3}}$};
        \node[yshift=12pt, xshift=5pt] at (6*\sh*\l,0) {${\sss  \varphi^{S_6 V_1}_{S_5,i_4}}$};
        \node[yshift=-12pt, xshift=6pt] at (7*\sh*\l,0) {${\sss \varphi^{W}_{S_6,j_1}}$};

        \node[mpo] at (\sh*\l,0) {};
        \node[mpo] at (0,0) {};
        \node[mps1] at (7*\sh*\l,0) {};
        \node[mps1] at (2*\sh*\l,0) {};
        \node[mps1] at (3*\sh*\l,0) {};
        \node[mps1] at (4*\sh*\l,0) {};
        \node[mps1] at (5*\sh*\l,0) {};
        \node[mps1] at (6*\sh*\l,0) {};
    \end{tikzpicture}
}

\newcommand{\StringOp}[3]{
    \begin{tikzpicture}[scale=1,baseline={([yshift=-.55ex]current bounding box.center)}]
        \def\l{.7};
        \def\sh{1.2};
        \draw[obj3,-<-=.38] (2*\l*\sh,0) --node[pos=.5,above]{${\sss #1}$} (3*\l*\sh,0);
        \draw[obj3,-<-=.38] (5*\l*\sh,0) --node[pos=.5,above]{${\sss #1}$} (6*\l*\sh,0);
        \foreach \x/\y in {3/5}{
            \draw[obj3,-<-=.45] (\x*\l*\sh,0) -- (\x*\l*\sh+2*\sh*\l/3,0);
            \draw[obj3,dot, shorten <= 4pt, shorten >= 4pt] (\x*\l*\sh+2*\sh*\l/3,0) -- (\y*\l*\sh-2*\sh*\l/3,0);
            \draw[obj3,->-=.73] (\y*\l*\sh,0) -- (\y*\l*\sh-2*\sh*\l/3,0);
        }
        \foreach \x in {2,3,5,6}{
            \draw[obj1,->-=.73] (\x*\l*\sh,0) -- (\x*\l*\sh,\l);
            \draw[obj1,-<-=.47] (\x*\l*\sh,0) -- (\x*\l*\sh,-\l);
            \ifthenelse{\x = 3 \OR \x = 5}{
                \node[mpo] at (\x*\l*\sh,0) {};
            }{
                \node[mps1] at (\x*\l*\sh,0) {};
            }
        }
        \node[xshift=-9pt] at (2*\sh*\l,0) {${\sss #2}$};
        \node[xshift=9pt] at (6*\sh*\l,0) {${\sss #3}$};
    \end{tikzpicture}
}

\newcommand{\StringOpExpVal}[4]{
    \begin{tikzpicture}[scale=1,baseline={([yshift=-.55ex]current bounding box.center)}]
        \def\l{.7};
        \def\sh{1.2};
        \draw[obj3,-<-=.38] (2*\l*\sh,0) --node[pos=.5,above]{${\sss #2}$} (3*\l*\sh,0);
        \draw[obj3,-<-=.38] (5*\l*\sh,0) --node[pos=.5,above]{${\sss #2}$} (6*\l*\sh,0);
        \foreach \x/\y in {3/5}{
            \draw[obj3,-<-=.45] (\x*\l*\sh,0) -- (\x*\l*\sh+2*\sh*\l/3,0);
            \draw[obj3,dot, shorten <= 4pt, shorten >= 4pt] (\x*\l*\sh+2*\sh*\l/3,0) -- (\y*\l*\sh-2*\sh*\l/3,0);
            \draw[obj3,->-=.73] (\y*\l*\sh,0) -- (\y*\l*\sh-2*\sh*\l/3,0);
        }
        \foreach \x in {2,3,5,6}{
            \draw[obj1,->-=.73] (\x*\l*\sh,0) -- (\x*\l*\sh,\l);
            \draw[obj1,-<-=.47] (\x*\l*\sh,0) -- (\x*\l*\sh,-\l);
            \ifthenelse{\x = 3 \OR \x = 5}{
                \node[mpo] at (\x*\l*\sh,0) {};
            }{
                \node[mps1] at (\x*\l*\sh,0) {};
            }
        }
        \node[xshift=-9pt] at (2*\sh*\l,0) {${\sss #3}$};
        \node[xshift=9pt] at (6*\sh*\l,0) {${\sss #4}$};

        \draw[obj1,->-=.55] (0,-\sh*\l) -- (0,\sh*\l);
        \draw[obj1,->-=.55] (8*\sh*\l,-\sh*\l) -- (8*\sh*\l,\sh*\l);
        
        \draw[obj2,->-=.73] (0,-\sh*\l) --node[pos=.6,above]{${\sss #1}$} (-\l,-\sh*\l);
        \draw[obj2,-<-=.47] (8*\sh*\l,-\sh*\l) --node[pos=.6,above]{${\sss #1}$} (8*\sh*\l+\l,-\sh*\l);
        \draw[obj2,-<-=.47] (2*\l*\sh,-\sh*\l) --node[pos=.6,above]{${\sss #1}$} (2*\l*\sh+\l,-\sh*\l);
        \draw[obj2,-<-=.47] (5*\l*\sh,-\sh*\l) --node[pos=.6,above]{${\sss #1}$} (5*\l*\sh+\l,-\sh*\l);
        \draw[obj2,-<-=.47] (0,\sh*\l) --node[pos=.6,below]{${\sss #1}$} (-\l,\sh*\l);
        \draw[obj2,->-=.73] (8*\sh*\l,\sh*\l) --node[pos=.6,below]{${\sss #1}$} (8*\sh*\l+\l,\sh*\l);
        \draw[obj2,->-=.73] (2*\l*\sh,\sh*\l) --node[pos=.6,below]{${\sss #1}$} (2*\l*\sh+\l,\sh*\l);
        \draw[obj2,->-=.73] (5*\l*\sh,\sh*\l) --node[pos=.6,below]{${\sss #1}$} (5*\l*\sh+\l,\sh*\l);
        \foreach \x/\y in {0/2,3/5,6/8}{
            \draw[obj2,-<-=.45] (\x*\l*\sh,-\sh*\l) -- (\x*\l*\sh+2*\sh*\l/3,-\sh*\l);
            \draw[obj2,->-=.73] (\y*\l*\sh,-\sh*\l) -- (\y*\l*\sh-2*\sh*\l/3,-\sh*\l);
        }
        \foreach \x/\y in {0/2,3/5,6/8}{
            \draw[obj2,->-=.73] (\x*\l*\sh,\sh*\l) -- (\x*\l*\sh+2*\sh*\l/3,\sh*\l);
            \draw[obj2,-<-=.45] (\y*\l*\sh,\sh*\l) -- (\y*\l*\sh-2*\sh*\l/3,\sh*\l);
        }
        \foreach \s in {+1,-1}{
            \foreach \x/\y in {0/2,3/5,6/8}{
                \draw[obj2,dot, shorten <= 4pt, shorten >= 4pt] (\x*\l*\sh+2*\sh*\l/3,-\s*\sh*\l) -- (\y*\l*\sh-2*\sh*\l/3,-\s*\sh*\l);
            }
            \foreach \x in {0,2,3,5,6,8}{
                \node[mps1] at (\x*\l*\sh,-\s*\sh*\l) {};
            }
            \draw[line width=.5pt] (-\l,-\s*\sh*\l-\l/2) -- (-1.3*\l,-\s*\sh*\l+\l/2);
            \draw[line width=.5pt] (-1.1*\l,-\s*\sh*\l-\l/2) -- (-1.4*\l,-\s*\sh*\l+\l/2);
            \draw[line width=.5pt] (8*\sh*\l+\l,-\s*\sh*\l+\l/2) -- (8*\sh*\l+1.3*\l,-\s*\sh*\l-\l/2);
            \draw[line width=.5pt] (8*\sh*\l+1.1*\l,-\s*\sh*\l+\l/2) -- (8*\sh*\l+1.4*\l,-\s*\sh*\l-\l/2);
        }
    \end{tikzpicture}
}

\newcommand{\StringOpExpValLimL}[5]{
    \begin{tikzpicture}[scale=1,baseline={([yshift=.4ex]current bounding box.center)}]
        \def\l{.7};
        \def\sh{1.2};
        \draw[obj2,->-=.52, rounded corners=2pt] (0,-\sh*\l) -- (-\sh*\l,-\sh*\l) --node[pos=.5,left]{${\sss #1}$} (-\sh*\l,\sh*\l) -- (0,\sh*\l);
        \draw[obj1,-<-=.47] (0,0) -- (0,-\l);
        \draw[obj1,->-=.73] (0,0) -- (0,\l);
        \draw[obj3,-<-=.23, rounded corners=2pt] (0,0) --node[pos=.56,above]{${\sss #2}$} (\sh*\l,0) -- (\sh*\l,-\sh*\l);
        \draw[obj2,-<-=.47] (0,-\sh*\l) --node[pos=.6,above]{${\sss #1}$} (\l,-\sh*\l);
        \draw[obj2,->-=.33, rounded corners=2pt] (2*\sh*\l,0) --node[pos=.55,right]{${\sss #1}$} (2*\sh*\l,-\sh*\l) -- (\sh*\l,-\sh*\l);
        \draw[obj2,->-=.38, rounded corners=2pt] (0,\sh*\l) --node[pos=.52,above]{${\sss #1}$} (2*\sh*\l,\sh*\l) -- (2*\sh*\l,0);
        \node[yshift=-9pt] at (\sh*\l,-\sh*\l) {${\sss #3}$};
        \node[xshift=-9pt] at (0,0) {${\sss #4}$};
        \node[mps0] at (2*\sh*\l,0) {};
        \node[xshift=12pt] at (2*\sh*\l,0) {${\sss #5}$};
        \node[mps1] at (0,0) {};
        \node[mps1] at (0,-\sh*\l) {};
        \node[mps1] at (0,\sh*\l) {};
        \node[mps1] at (\sh*\l,-\sh*\l) {};
    \end{tikzpicture}
}

\newcommand{\StringOpExpValLimR}[5]{
    \begin{tikzpicture}[scale=1,baseline={([yshift=.4ex]current bounding box.center)}]
        \def\l{.7};
        \def\sh{1.2};
        \draw[obj2,->-=.42, rounded corners=2pt] (-\sh*\l,-\sh*\l) -- (-2*\sh*\l,-\sh*\l) --node[pos=.5,left]{${\sss #1}$} (-2*\sh*\l,\sh*\l) -- (0,\sh*\l);
        \draw[obj1,-<-=.47] (0,0) -- (0,-\l);
        \draw[obj1,->-=.73] (0,0) -- (0,\l);
        \draw[obj3,->-=.33, rounded corners=2pt] (0,0) --node[pos=.56,above]{${\sss #2}$} (-\sh*\l,0) -- (-\sh*\l,-\sh*\l);
        \draw[obj2,->-=.73] (0,-\sh*\l) --node[pos=.6,above]{${\sss #1}$} (-\l,-\sh*\l);
        \draw[obj2,->-=.33, rounded corners=2pt] (\sh*\l,0) --node[pos=.55,right]{${\sss #1}$} (\sh*\l,-\sh*\l) -- (0,-\sh*\l);
        \draw[obj2,->-=.33, rounded corners=2pt] (0,\sh*\l) --node[pos=.55,above]{${\sss #1}$} (\sh*\l,\sh*\l) -- (\sh*\l,0);
        \node[yshift=-9pt] at (-\sh*\l,-\sh*\l) {${\sss #3}$};
        \node[xshift=9pt] at (0,0) {${\sss #4}$};
        \node[mps0] at (\sh*\l,0) {};
        \node[xshift=12pt] at (\sh*\l,0) {${\sss #5}$};
        \node[mps1] at (0,0) {};
        \node[mps1] at (0,-\sh*\l) {};
        \node[mps1] at (0,\sh*\l) {};
        \node[mps1] at (-\sh*\l,-\sh*\l) {};
    \end{tikzpicture}
}

\newcommand{\TwisedLocalOpEnv}[4]{
    \begin{tikzpicture}[scale=1,baseline={([yshift=.4ex]current bounding box.center)}]
        \def\l{.7};
        \def\sh{1.2};
        \draw[obj2,->-=.42, rounded corners=2pt] (-\sh*\l,-\sh*\l) -- (-2*\sh*\l,-\sh*\l) --node[pos=.5,left]{${\sss #1}$} (-2*\sh*\l,\sh*\l) -- (0,\sh*\l);
        \draw[obj1,-<-=.45] (0,0) -- (0,-\sh*\l);
        \draw[obj1,->-=.65] (0,0) -- (0,\sh*\l);
        \draw[obj3,->-=.36, rounded corners=2pt] (0,0) --node[pos=.6,above]{${\sss #2}$} (-\sh*\l,0) -- (-\sh*\l,-\sh*\l);
        \draw[obj2,->-=.73] (0,-\sh*\l) --node[pos=.6,above]{${\sss #1}$} (-\l,-\sh*\l);
        \draw[obj2,->-=.33, rounded corners=2pt] (\sh*\l,0) --node[pos=.55,right]{${\sss #1}$} (\sh*\l,-\sh*\l) -- (0,-\sh*\l);
        \draw[obj2,->-=.33, rounded corners=2pt] (0,\sh*\l) --node[pos=.55,above]{${\sss #1}$} (\sh*\l,\sh*\l) -- (\sh*\l,0);
        \node[yshift=-9pt] at (-\sh*\l,-\sh*\l) {${\sss #3}$};
        \node[xshift=12pt] at (\sh*\l,0) {${\sss #4}$};
        \node[mps0] at (\sh*\l,0) {};
        \node[hole] at (0,0) {};
        \node[mps1] at (0,-\sh*\l) {};
        \node[mps1] at (0,\sh*\l) {};
        \node[mps1] at (-\sh*\l,-\sh*\l) {};
    \end{tikzpicture}
}

\newcommand{\TubeModEnvA}[9]{
    \begin{tikzpicture}[scale=1,baseline={([yshift=.9ex]current bounding box.center)}]
        \def\l{.7};
        \def\sh{1.2};
        \draw[obj1,-<-=.45] (0,0) -- (0,-\sh*\l);
        \draw[obj1,->-=.65] (0,0) -- (0,\sh*\l);
        \draw[obj1,->-=.73] (0,\sh*\l) -- (0,\sh*\l+\l);
        \draw[obj1,-<-=.47] (0,-\sh*\l) -- (0,-\sh*\l-\l);
        \draw[obj3,->-=.65] (0,0) --node[pos=.56,below]{${\sss #2}$} (-\sh*\l,0);
        \draw[obj3,->-=.73, rounded corners=2pt] (-\sh*\l,0) --node[pos=.6,below]{${\sss #3}$} (-\sh*\l-\l,0);
        \draw[obj3,-<-=.43] (-2*\sh*\l,-2*\sh*\l) --node[pos=.5, left]{${\sss #4}$} (-2*\sh*\l,0);
        \draw[obj3,-<-=.47, rounded corners=2pt] (0,-\sh*\l) -- (\sh*\l,-\sh*\l) --node[pos=.5,right]{${\sss #1}$} (\sh*\l,\sh*\l) -- (0,\sh*\l);
        \draw[obj3,->-=.33, rounded corners=2pt] (0,-\sh*\l) --node[pos=.56,below]{${\sss #1}$} (-\sh*\l,-\sh*\l) -- (-\sh*\l,0);
        \draw[obj3,-<-=.31, rounded corners=2pt] (0,\sh*\l) --node[pos=.52,above]{${\sss #1}$} (-2*\sh*\l,\sh*\l) -- (-2*\sh*\l,0);
        \draw[obj2,-<-=.44] (-2*\sh*\l,-2*\sh*\l) --node[pos=.49,above]{${\sss #5}$} (0,-2*\sh*\l);
        \draw[obj2,-<-=.23, rounded corners=2pt] (0,-2*\sh*\l) --node[pos=.52,above]{${\sss #5}$} (2*\sh*\l,-2*\sh*\l) -- (2*\sh*\l,0);
        \draw[obj2,->-=.29, rounded corners=2pt] (0,2*\sh*\l) --node[pos=.52,below]{${\sss #5}$} (2*\sh*\l,2*\sh*\l) -- (2*\sh*\l,0);
        \draw[obj2,->-=.385,rounded corners=2pt] (-2*\sh*\l,-2*\sh*\l) -- (-3*\sh*\l,-2*\sh*\l) --node[pos=.5,left]{${\sss #5}$} (-3*\sh*\l,2*\sh*\l) -- (0,2*\sh*\l);

        \node[yshift=-9pt] at (-2*\sh*\l,-2*\sh*\l) {${\sss #6}$};
        \node[xshift=-11pt,yshift=12pt] at (-2*\sh*\l,0) {${\sss #7}$};
        \node[yshift=12pt] at (-\sh*\l,0) {${\sss #8}$};
        \node[xshift=12pt] at (2*\sh*\l,0) {${\sss #9}$};
        \node[hole] at (0,0) {};
        \node[mpo] at (0,-\sh*\l) {};
        \node[mpo] at (0,\sh*\l) {};
        \node[mps1] at (-\sh*\l,0) {};
        \node[mps1] at (-2*\sh*\l,0) {};
        \node[mps1] at (-2*\sh*\l,-2*\sh*\l) {};
        \node[mps1] at (0,-2*\sh*\l) {};
        \node[mps1] at (0,2*\sh*\l) {};
        \node[mps0] at (2*\sh*\l,0) {};
    \end{tikzpicture}
}

\newcommand{\TubeModEnvB}[9]{
    \begin{tikzpicture}[scale=1,baseline={([yshift=.9ex]current bounding box.center)}]
        \def\l{.7};
        \def\sh{1.2};
        \draw[obj1,-<-=.44] (0,0) -- (0,-2*\sh*\l);
        \draw[obj1,->-=.55] (0,0) -- (0,2*\sh*\l);
        \draw[obj3,->-=.65] (0,0) --node[pos=.56,below]{${\sss #2}$} (-\sh*\l,0);
        \draw[obj3,->-=.73, rounded corners=2pt] (-\sh*\l,0) --node[pos=.6,below]{${\sss #3}$} (-\sh*\l-\l,0);
        \draw[obj2,->-=.37, rounded corners=2pt] (\sh*\l,0) --node[pos=.5,right]{${\sss #6}$} (\sh*\l,-2*\sh*\l) -- (0,-2*\sh*\l);
        \draw[obj2,->-=.69, rounded corners=2pt] (0,2*\sh*\l) -- (\sh*\l,2*\sh*\l) --node[pos=.5,right]{${\sss #6}$} (\sh*\l,0);
        \draw[obj2,->-=.73] (0,-2*\sh*\l) --node[above,pos=.6]{${\sss #6}$} (-\l,-2*\sh*\l);
        \draw[obj2,->-=.73] (-\sh*\l,-2*\sh*\l) --node[above,pos=.6]{${\sss #5}$} (-\sh*\l-\l,-2*\sh*\l);
        \draw[obj2,->-=.73] (-2*\sh*\l,-2*\sh*\l) --node[above,pos=.6]{${\sss #5}$} (-2*\sh*\l-\l,-2*\sh*\l);
        \draw[obj2,->-=.345, rounded corners=2pt] (-3*\sh*\l,-2*\sh*\l) -- (-4*\sh*\l,-2*\sh*\l) --node[left,pos=.5]{${\sss #6}$}(-4*\sh*\l,2*\sh*\l) -- (0,2*\sh*\l);
        \draw[obj3,-<-=.33,rounded corners=2pt] (-3*\sh*\l,-2*\sh*\l) --node[pos=.5,left]{${\sss #1}$} (-3*\sh*\l,0) -- (-2*\sh*\l,0);
        \draw[obj3,-<-=.43] (-2*\sh*\l,-2*\sh*\l) --node[pos=.5, left]{${\sss #4}$} (-2*\sh*\l,0);
        \draw[obj3,->-=.52] (-\sh*\l,-2*\sh*\l) --node[pos=.5, left]{${\sss #1}$} (-\sh*\l,0);

        \node[yshift=-9pt] at (-3*\sh*\l,-2*\sh*\l) {${\sss #7}$};
        \node[yshift=-9pt] at (-2*\sh*\l,-2*\sh*\l) {${\sss #8}$};
        \node[yshift=-9pt] at (-\sh*\l,-2*\sh*\l) {${\sss #9}$};
        \node[xshift=16pt] at (\sh*\l,0) {${\sss \Xi_{g(a)}}$};
        \node[hole] at (0,0) {};
        \node[mps0] at (\sh*\l,0) {};
        \node[mps1] at (0,-2*\sh*\l) {};
        \node[mps1] at (0,2*\sh*\l) {};
        \node[mps1] at (-\sh*\l,-2*\sh*\l) {}; 
        \node[mps1] at (-2*\sh*\l,-2*\sh*\l) {};
        \node[mps1] at (-3*\sh*\l,-2*\sh*\l) {};
        \node[mps1] at (-\sh*\l,0) {};
        \node[mps1] at (-2*\sh*\l,0) {};
    \end{tikzpicture}
}

\newcommand{\RepGTubeModEnvB}[0]{
    \begin{tikzpicture}[scale=1,baseline={([yshift=.9ex]current bounding box.center)}]
        \def\l{.7};
        \def\sh{1.2};
        \draw[obj1,-<-=.44] (0,0) -- (0,-2*\sh*\l);
        \draw[obj1,->-=.55] (0,0) -- (0,2*\sh*\l);
        \draw[obj3,->-=.65] (0,0) --node[pos=.56,below]{${\sss S_3}$} (-\sh*\l,0);
        \draw[obj3,->-=.73, rounded corners=2pt] (-\sh*\l,0) --node[pos=.6,below]{${\sss S_2}$} (-\sh*\l-\l,0);
        \draw[obj2,->-=.37, rounded corners=2pt] (\sh*\l,0) --node[pos=.5,right]{${\sss U_2}$} (\sh*\l,-2*\sh*\l) -- (0,-2*\sh*\l);
        \draw[obj2,->-=.69, rounded corners=2pt] (0,2*\sh*\l) -- (\sh*\l,2*\sh*\l) --node[pos=.5,right]{${\sss U_2}$} (\sh*\l,0);
        \draw[obj2,->-=.73] (0,-2*\sh*\l) --node[above,pos=.6]{${\sss U_2}$} (-\l,-2*\sh*\l);
        \draw[obj2,->-=.73] (-\sh*\l,-2*\sh*\l) --node[above,pos=.6]{${\sss U_1}$} (-\sh*\l-\l,-2*\sh*\l);
        \draw[obj2,->-=.73] (-2*\sh*\l,-2*\sh*\l) --node[above,pos=.6]{${\sss U_1}$} (-2*\sh*\l-\l,-2*\sh*\l);
        \draw[obj2,->-=.345, rounded corners=2pt] (-3*\sh*\l,-2*\sh*\l) -- (-4*\sh*\l,-2*\sh*\l) --node[left,pos=.5]{${\sss U_2}$}(-4*\sh*\l,2*\sh*\l) -- (0,2*\sh*\l);
        \draw[obj3,-<-=.33,rounded corners=2pt] (-3*\sh*\l,-2*\sh*\l) --node[pos=.5,left]{${\sss V}$} (-3*\sh*\l,0) -- (-2*\sh*\l,0);
        \draw[obj3,-<-=.43] (-2*\sh*\l,-2*\sh*\l) --node[pos=.5, left]{${\sss S_1}$} (-2*\sh*\l,0);
        \draw[obj3,->-=.52] (-\sh*\l,-2*\sh*\l) --node[pos=.5, left]{${\sss V}$} (-\sh*\l,0);

        \node[yshift=-12pt,xshift=-4pt] at (-3*\sh*\l,-2*\sh*\l) {${\sss \phi^{VU_1}_{U_2,j_4}}$};
        \node[yshift=-12pt] at (-2*\sh*\l,-2*\sh*\l) {${\sss \phi^{S_1U_1}_{U_1,j_2}}$};
        \node[yshift=-12pt,xshift=4pt] at (-\sh*\l,-2*\sh*\l) {${\sss \bar \phi^{VU_1}_{U_2,j_4}}$};
        \node[yshift=12pt,xshift=2pt] at (-\sh*\l,0) {${\sss \varphi^{S_3V}_{S_2,i_2}}$};
        \node[yshift=12pt,xshift=-2pt] at (-2*\sh*\l,0) {${\sss \bar \varphi^{VS_1}_{S_2,i_1}}$};
        \node[xshift=15pt] at (\sh*\l,0) {${\sss \Xi_{U_2}}$};
        \node[hole] at (0,0) {};
        \node[mps0] at (\sh*\l,0) {};
        \node[mps1] at (0,-2*\sh*\l) {};
        \node[mps1] at (0,2*\sh*\l) {};
        \node[mps1] at (-\sh*\l,-2*\sh*\l) {}; 
        \node[mps1] at (-2*\sh*\l,-2*\sh*\l) {};
        \node[mps1] at (-3*\sh*\l,-2*\sh*\l) {};
        \node[mps1] at (-\sh*\l,0) {};
        \node[mps1] at (-2*\sh*\l,0) {};
    \end{tikzpicture}
}

\newcommand{\sesameTrick}[1]{
    \begin{tikzpicture}[scale=1,baseline={([yshift=-.55ex]current bounding box.center)}]
        \def\l{.7};
        \def\sh{1.2};
        
        \draw[obj3,-<-=.47] (2*\sh*\l,0) --node[pos=.6, above]{${\sss S_1}$} (2*\sh*\l+\l,0);
        \draw[obj1,-<-=.47] (\sh*\l,0) --node[pos=.56,above]{${\sss W}$} (\sh*\l+\l,0);
        \draw[obj3,-<-=.47] (3*\sh*\l,0) --node[pos=.56,above]{${\sss S_2}$} (3*\sh*\l+\l,0);
        \draw[obj3,-<-=.35, rounded corners=2pt] (4*\sh*\l,0) -- (4*\sh*\l,-\sh*\l) --node[pos=.3,below]{${\sss V_2}$} (\sh*\l,-\sh*\l) -- (\sh*\l,0);
        \draw[obj3,-<-=.48, rounded corners=2pt] (\sh*\l,0) -- (\sh*\l,\sh*\l) --node[pos=.51,above]{${\sss V_2}$} (3*\sh*\l,\sh*\l) -- (3*\sh*\l,0);

        \ifthenelse{#1=1}{
            \draw[obj3,-<-=.35, rounded corners=2pt] (4*\sh*\l,0) -- (5*\sh*\l,0) --node[pos=.52, left]{${\sss S_3}$} (5*\sh*\l,-2*\sh*\l) -- (2.5*\sh*\l,-2*\sh*\l);
            \draw[obj1,->-=.39, rounded corners=2pt] (\sh*\l,0) -- (0,0) --node[pos=.52,right]{${\sss W}$} (0,-2*\sh*\l) -- (2.5*\sh*\l,-2*\sh*\l);
            \draw[obj1,->-=.27, rounded corners=2pt] (2.5*\sh*\l,-3*\sh*\l) --node[pos=.534,above]{${\sss W}$} (-\sh*\l,-3*\sh*\l) -- (-\sh*\l,0) -- (-\sh*\l-\l,0);
            \draw[obj3,-<-=.24, rounded corners=2pt] (2.5*\sh*\l,-3*\sh*\l) --node[pos=.54,above]{${\sss S_3}$} (6*\sh*\l,-3*\sh*\l) -- (6*\sh*\l,0) -- (6*\sh*\l+\l,0);
            \node[yshift=12pt] at (2.5*\sh*\l,-2*\sh*\l) {${\sss \varphi^{W}_{S_3,j_2}}$};
            \node[yshift=-12pt] at (2.5*\sh*\l,-3*\sh*\l) {${\sss \bar \varphi^{W}_{S_3,j_2}}$};
            \node[mps1] at (2.5*\sh*\l,-2*\sh*\l) {};
            \node[mps1] at (2.5*\sh*\l,-3*\sh*\l) {};
        }{}
        \ifthenelse{#1=2}{
            \draw[obj1,->-=.72] (\sh*\l,0) --node[pos=.6,above]{${\sss W}$} (\sh*\l-\l,0);
            \draw[obj3,-<-=.47] (4*\sh*\l,0) --node[pos=.6,below]{${\sss S_3}$} (4*\sh*\l+\l,0);
        }{}
        
        \node[yshift=12pt,xshift=5pt] at (4*\sh*\l,0) {${\sss \varphi^{S_3 V_2}_{S_2,i_2}}$};
        \node[yshift=-12pt, xshift=-1pt] at (2*\sh*\l,0) {${\sss \bar \varphi^{W}_{S_1,j_3}}$};
        \node[yshift=-12pt, xshift=1pt] at (3*\sh*\l,0) {${\sss \bar \varphi^{V_2 S_1}_{S_2,i_1}}$};

        \node[mpo] at (\sh*\l,0) {};
        \node[mps1] at (2*\sh*\l,0) {};
        \node[mps1] at (3*\sh*\l,0) {};
        \node[mps1] at (4*\sh*\l,0) {};
    \end{tikzpicture}
}

\newcommand{\halfBraidingRepG}[1]{
    \begin{tikzpicture}[scale=1,baseline={([yshift=-.55ex]current bounding box.center)}]
        \def\l{.7};
        \def\sh{1.2};
        \ifthenelse{#1=1}{
            \draw[obj3,-<-=.47] (0,0) --node[pos=.6,right]{${\sss V}$} (0,\l);
            \draw[obj3,->-=.72] (0,0) --node[pos=.6,left]{${\sss V}$} (0,-\l);
            \draw[obj1,->-=.72] (0,0) --node[pos=.6,above]{${\sss W}$} (-\l,0);
            \draw[obj1,-<-=.47] (0,0) --node[pos=.6,below]{${\sss W}$} (\l,0);
            \node[mpo] at (0,0) {};
        }{
            \draw[obj3,-<-=.47] (2*\sh*\l,0) --node[pos=.6,left]{${\sss V}$} (2*\sh*\l,\l);
            \draw[obj3,->-=.72] (\sh*\l,0) --node[pos=.6,right]{${\sss V}$} (\sh*\l,-\l);
            \draw[obj3,-<-=.47] (0,0) --node[pos=.6,below]{${\sss S_3}$} (\l,0);
            \draw[obj3,-<-=.47] (\sh*\l,0) --node[pos=.6,above]{${\sss S_2}$} (\sh*\l+\l,0);
            \draw[obj3,-<-=.47] (2*\sh*\l,0) --node[pos=.6,above]{${\sss S_1}$} (2*\sh*\l+\l,0);
            \draw[obj1,->-=.72] (0,0) --node[pos=.6,above]{${\sss W}$} (-\l,0);
            \draw[obj1,-<-=.47] (3*\sh*\l,0) --node[pos=.6,below]{${\sss W}$} (3*\sh*\l+\l,0);

            \node[yshift=-12pt] at (0,0) {${\sss \bar \varphi^W_{S_3,j_1}}$};
            \node[yshift=12pt,xshift=-5pt] at (\sh*\l,0) {${\sss \bar \varphi^{S_3V}_{S_2,i_2}}$};
            \node[yshift=-12pt] at (2*\sh*\l,0) {${\sss \varphi^{VS_1}_{S_2,i_1}}$};
            \node[yshift=12pt,xshift=5pt] at (3*\sh*\l,0) {${\sss \varphi^W_{S_1,j_2}}$};
            \node[mps1] at (0,0) {};
            \node[mps1] at (\sh*\l,0) {};
            \node[mps1] at (2*\sh*\l,0) {};
            \node[mps1] at (3*\sh*\l,0) {};
        }
    \end{tikzpicture}
}

\newcommand{\halfBraidingCstsRepG}[0]{
    \begin{tikzpicture}[scale=1,baseline={([yshift=-.55ex]current bounding box.center)}]
        \def\l{.7};
        \def\sh{1.2};
        \draw[obj3,-<-=.47, rounded corners=2pt] (0,0) -- (0,\sh*\l) --node[pos=.5,above]{${\sss V}$} (2*\sh*\l,\sh*\l) -- (2*\sh*\l,0);
        \draw[obj3,->-=.52, rounded corners=2pt] (0,0) -- (0,-\sh*\l) --node[pos=.5,below]{${\sss V}$} (-2*\sh*\l,-\sh*\l) -- (-2*\sh*\l,0);
        \draw[obj1,->-=.72] (0,0) --node[pos=.6,above]{${\sss W}$} (-\l,0);
        \draw[obj1,-<-=.47] (0,0) --node[pos=.6,below]{${\sss W}$} (\l,0);
        \draw[obj3,-<-=.47] (\sh*\l,0) --node[pos=.6,above]{${\sss S_1}$} (\sh*\l+\l,0);
        \draw[obj3,->-=.72] (-\sh*\l,0) --node[pos=.6,below]{${\sss S_3}$} (-\sh*\l-\l,0);
        \draw[obj3,-<-=.47] (2*\sh*\l,0) --node[pos=.6,above]{${\sss S_2}$} (2*\sh*\l+\l,0);
        \draw[obj3,->-=.72] (-2*\sh*\l,0) --node[pos=.6,below]{${\sss S_2}$} (-2*\sh*\l-\l,0);

        \node[yshift=12pt,xshift=-5pt] at (\sh*\l,0) {${\sss \bar \varphi^W_{S_1,j_2}}$};
        \node[yshift=-12pt] at (2*\sh*\l,0) {${\sss \bar \varphi^{VS_1}_{S_2,i_1}}$};
        \node[yshift=12pt] at (-2*\sh*\l,0) {${\sss \varphi^{S_3V}_{S_2,i_2}}$};
        \node[yshift=-12pt,xshift=5pt] at (-\sh*\l,0) {${\sss \varphi^W_{S_3,j_1}}$};
        \node[mpo] at (0,0) {};
        \node[mps1] at (-\sh*\l,0) {};
        \node[mps1] at (\sh*\l,0) {};
        \node[mps1] at (2*\sh*\l,0) {};
        \node[mps1] at (-2*\sh*\l,0) {};
    \end{tikzpicture}
}

\newcommand{\halfBraidingVecG}[1]{
    \begin{tikzpicture}[scale=1,baseline={([yshift=-.55ex]current bounding box.center)}]
        \def\l{.7};
        \def\sh{1.2};
        \ifthenelse{#1=1}{
            \draw[obj3,-<-=.47] (0,0) --node[pos=.6,right]{${\sss g}$} (0,\l);
            \draw[obj3,->-=.72] (0,0) --node[pos=.6,left]{${\sss g}$} (0,-\l);
            \draw[obj1,->-=.72] (0,0) --node[pos=.6,above]{${\sss W}$} (-\l,0);
            \draw[obj1,-<-=.47] (0,0) --node[pos=.6,below]{${\sss W}$} (\l,0);
            \node[mpo] at (0,0) {};
        }{
            \draw[obj3,-<-=.47] (2*\sh*\l,0) --node[pos=.6,right]{${\sss g}$} (2*\sh*\l,\l);
            \draw[obj3,->-=.72] (\sh*\l,0) --node[pos=.6,right]{${\sss g}$} (\sh*\l,-\l);
            \draw[obj3,-<-=.47] (0,0) --node[pos=.6,above]{${\sss {}^gx}$} (\l,0);
            \draw[obj3,-<-=.47] (\sh*\l,0) --node[pos=.6,above]{${\sss gx}$} (\sh*\l+\l,0);
            \draw[obj3,-<-=.47] (2*\sh*\l,0) --node[pos=.6,above]{${\sss x}$} (2*\sh*\l+\l,0);
            \draw[obj1,->-=.72] (0,0) --node[pos=.6,above]{${\sss W}$} (-\l,0);
            \draw[obj1,-<-=.47] (3*\sh*\l,0) --node[pos=.6,below]{${\sss W}$} (3*\sh*\l+\l,0);

            \node[yshift=-12pt] at (3*\sh*\l,0) {${\sss \psi^W_{\mathbb C_x,c_2}}$};
            \node[yshift=-12pt] at (0,0) {${\sss \bar \psi^W_{\mathbb C_{{}^g \! x},c_1}}$};

            \node[mps1] at (0,0) {};
            \node[mps1] at (\sh*\l,0) {};
            \node[mps1] at (2*\sh*\l,0) {};
            \node[mps1] at (3*\sh*\l,0) {};
        }
    \end{tikzpicture}
}

\newcommand{\halfBraidingCstsVecG}[0]{
    \begin{tikzpicture}[scale=1,baseline={([yshift=-.55ex]current bounding box.center)}]
        \def\l{.7};
        \def\sh{1.2};
        \draw[obj3,-<-=.47, rounded corners=2pt] (0,0) -- (0,\sh*\l) --node[pos=.5,above]{${\sss g}$} (2*\sh*\l,\sh*\l) -- (2*\sh*\l,0);
        \draw[obj3,->-=.52, rounded corners=2pt] (0,0) -- (0,-\sh*\l) --node[pos=.5,below]{${\sss g}$} (-2*\sh*\l,-\sh*\l) -- (-2*\sh*\l,0);
        \draw[obj1,->-=.72] (0,0) --node[pos=.6,above]{${\sss W}$} (-\l,0);
        \draw[obj1,-<-=.47] (0,0) --node[pos=.6,below]{${\sss W}$} (\l,0);
        \draw[obj3,-<-=.47] (\sh*\l,0) --node[pos=.6,above]{${\sss x}$} (\sh*\l+\l,0);
        \draw[obj3,->-=.72] (-\sh*\l,0) --node[pos=.6,above]{${\sss {}^gx}$} (-\sh*\l-\l,0);
        \draw[obj3,-<-=.47] (2*\sh*\l,0) --node[pos=.6,above]{${\sss gx}$} (2*\sh*\l+\l,0);
        \draw[obj3,->-=.72] (-2*\sh*\l,0) --node[pos=.6,below]{${\sss gx}$} (-2*\sh*\l-\l,0);

        \node[yshift=12pt,xshift=-5pt] at (\sh*\l,0) {${\sss \bar \psi^W_{\mathbb C_x,c_2}}$};
        \node[yshift=-12pt,xshift=5pt] at (-\sh*\l,0) {${\sss \psi^W_{\mathbb C_{{}^g \! x},c_1}}$};
        \node[mpo] at (0,0) {};
        \node[mps1] at (-\sh*\l,0) {};
        \node[mps1] at (\sh*\l,0) {};
        \node[mps1] at (2*\sh*\l,0) {};
        \node[mps1] at (-2*\sh*\l,0) {};
    \end{tikzpicture}
}

\newcommand{\DOT}[5]{
    \begin{tikzpicture}[scale=1,baseline={([yshift=-.55ex]current bounding box.center)}]
        \def\l{.7};
        \draw[obj2,-<-=.47] (0,0) --node[pos=1.25]{${\sss #3}$} node[pos=.6,above]{${\sss #1}$} (\l,0);
        \draw[obj2,->-=.73] (0,0) --node[pos=1.25]{${\sss #2}$} node[pos=.6,above]{${\sss #1}$} (-\l,0);
        \draw[obj1,->-=.73] (0,0) --node[pos=1.25]{${\sss #4}$} (0,\l);
        \draw[obj1,-<-=.47] (0,0) --node[pos=1.25]{${\sss #5}$} (0,-\l);
        \node[mpo] at (0,0) {};
    \end{tikzpicture}
}

\newcommand{\DO}[2]{
    \begin{tikzpicture}[scale=1,baseline={([yshift=-.55ex]current bounding box.center)}]
        \def\l{.7};
        \def\sh{1.2};
        \draw[obj2,->-=.73] (0,0) --node[pos=.6,above]{${\sss #2}$} (-\l,0);
        \draw[obj2,-<-=.47] (8*\sh*\l,0) --node[pos=.6,above]{${\sss #2}$} (8*\sh*\l+\l,0);
        \draw[obj2,-<-=.38] (2*\l*\sh,0) --node[pos=.5,above]{${\sss #2}$} (3*\l*\sh,0);
        \draw[obj2,-<-=.38] (5*\l*\sh,0) --node[pos=.5,above]{${\sss #2}$} (6*\l*\sh,0);
        \foreach \x/\y in {0/2,3/5,6/8}{
            \draw[obj2,-<-=.45] (\x*\l*\sh,0) -- (\x*\l*\sh+2*\sh*\l/3,0);
            \draw[obj2,dot, shorten <= 4pt, shorten >= 4pt] (\x*\l*\sh+2*\sh*\l/3,0) -- (\y*\l*\sh-2*\sh*\l/3,0);
            \draw[obj2,->-=.73] (\y*\l*\sh,0) -- (\y*\l*\sh-2*\sh*\l/3,0);
        }
        \foreach \x in {0,2,3,5,6,8}{
            \draw[obj1,->-=.73] (\x*\l*\sh,0) -- (\x*\l*\sh,\l);
            \draw[obj1,-<-=.47] (\x*\l*\sh,0) -- (\x*\l*\sh,-\l);
            \node[mpo] at (\x*\l*\sh,0) {};
            \node[xshift=0pt, yshift=-9pt] at (\x*\l*\sh,0) {${\sss #1}$};
        }
        \draw[line width=.5pt] (-\l,-\l/2) -- (-1.3*\l,\l/2);
        \draw[line width=.5pt] (-1.1*\l,-\l/2) -- (-1.4*\l,\l/2);
        \draw[line width=.5pt] (8*\sh*\l+\l,\l/2) -- (8*\sh*\l+1.3*\l,-\l/2);
        \draw[line width=.5pt] (8*\sh*\l+1.1*\l,\l/2) -- (8*\sh*\l+1.4*\l,-\l/2);
    \end{tikzpicture}
}

\newcommand{\DOvevO}[2]{
    \begin{tikzpicture}[scale=1,baseline={([yshift=-.55ex]current bounding box.center)}]
        \def\l{.7};
        \def\sh{1.2};
        \draw[obj2,->-=.73] (0,0) --node[pos=.6,above]{${\sss #2}$} (-\l,0);
        \draw[obj2,-<-=.47] (8*\sh*\l,0) --node[pos=.6,above]{${\sss #2}$} (8*\sh*\l+\l,0);
        \draw[obj2,-<-=.38] (2*\l*\sh,0) --node[pos=.5,above]{${\sss #2}$} (3*\l*\sh,0);
        \draw[obj2,-<-=.38] (5*\l*\sh,0) --node[pos=.5,above]{${\sss #2}$} (6*\l*\sh,0);
        \foreach \x/\y in {0/2,3/5,6/8}{
            \draw[obj2,-<-=.45] (\x*\l*\sh,0) -- (\x*\l*\sh+2*\sh*\l/3,0);
            \draw[obj2,dot, shorten <= 4pt, shorten >= 4pt] (\x*\l*\sh+2*\sh*\l/3,0) -- (\y*\l*\sh-2*\sh*\l/3,0);
            \draw[obj2,->-=.73] (\y*\l*\sh,0) -- (\y*\l*\sh-2*\sh*\l/3,0);
        }
        
        \draw[obj1,->-=.73] (3*\l*\sh,0) -- (3*\l*\sh,\l);
        \draw[obj1,->-=.73] (5*\l*\sh,0) -- (5*\l*\sh,\l);
        \draw[obj1,->-=.73] (3*\sh*\l,\sh*\l) -- (3*\sh*\l,\sh*\l+\l);
        \draw[obj1,->-=.73] (5*\sh*\l,\sh*\l) -- (5*\sh*\l,\sh*\l+\l);
        \node[mpo] at (4*\sh*\l,\sh*\l) {\q\q\q\q\;\;\;};
        \node[yshift=-10pt] at (4*\sh*\l,\sh*\l) {${\sss \mc O}$};
        \draw[obj1,->-=.58] (0,0) -- (0,\sh*\l+\l); 
        \draw[obj1,->-=.58] (2*\sh*\l,0) -- (2*\sh*\l,\sh*\l+\l); 
        \draw[obj1,->-=.58] (6*\sh*\l,0) -- (6*\sh*\l,\sh*\l+\l); 
        \draw[obj1,->-=.58] (8*\sh*\l,0) -- (8*\sh*\l,\sh*\l+\l); 
        
        \foreach \x in {0,2,3,5,6,8}{
            \draw[obj1,-<-=.47] (\x*\l*\sh,0) -- (\x*\l*\sh,-\l);
            \node[mpo] at (\x*\l*\sh,0) {};
            \node[xshift=0pt, yshift=-9pt] at (\x*\l*\sh,0) {${\sss #1}$};
            \draw[line width=.5pt] (\x*\sh*\l+\l/2,\sh*\l+\l) -- (\x*\sh*\l-\l/2,\sh*\l+1.3*\l);
            \draw[line width=.5pt] (\x*\sh*\l+\l/2,\sh*\l+1.1*\l) -- (\x*\sh*\l-\l/2,\sh*\l+1.4*\l);
            \draw[line width=.5pt] (\x*\sh*\l-\l/2,-\l) -- (\x*\sh*\l+\l/2,-1.3*\l);
            \draw[line width=.5pt] (\x*\sh*\l-\l/2,-1.1*\l) -- (\x*\sh*\l+\l/2,-1.4*\l);
        }
        
        \draw[line width=.5pt] (-\l,-\l/2) -- (-1.3*\l,\l/2);
        \draw[line width=.5pt] (-1.1*\l,-\l/2) -- (-1.4*\l,\l/2);
        \draw[line width=.5pt] (8*\sh*\l+\l,\l/2) -- (8*\sh*\l+1.3*\l,-\l/2);
        \draw[line width=.5pt] (8*\sh*\l+1.1*\l,\l/2) -- (8*\sh*\l+1.4*\l,-\l/2);
    \end{tikzpicture}
}

\newcommand{\productDOTM}[1]{
    \begin{tikzpicture}[scale=1,baseline={([yshift=-.55ex]current bounding box.center)}]
        \def\l{.7};
        \def\sh{1.2};
        \draw[obj2,->-=.73] (2*\sh*\l,0) -- (2*\sh*\l-\l,0);
        \draw[obj2,->-=.73] (6*\sh*\l+\l,0) -- (6*\sh*\l,0);
        \draw[obj2,-<-=.38] (2*\l*\sh,0) --node[pos=.5,above]{${\sss #1}$} (3*\l*\sh,0);
        \draw[obj2,-<-=.38] (5*\l*\sh,0) --node[pos=.5,above]{${\sss #1}$} (6*\l*\sh,0);
        \draw[obj2,-<-=.45] (3*\l*\sh,0) -- (3*\l*\sh+2*\sh*\l/3,0);
        \draw[obj2,dot, shorten <= 4pt, shorten >= 4pt] (3*\l*\sh+2*\sh*\l/3,0) -- (5*\l*\sh-2*\sh*\l/3,0);
        \draw[obj2,->-=.73] (5*\l*\sh,0) -- (5*\l*\sh-2*\sh*\l/3,0);
        
        \foreach \x in {2,3,5,6}{
            \draw[obj1,->-=.73] (\x*\l*\sh,0) -- (\x*\l*\sh,\l);
            \draw[obj1,-<-=.47] (\x*\l*\sh,0) -- (\x*\l*\sh,-\l);
            \node[mpo] at (\x*\l*\sh,0) {};
            \draw[line width=.5pt] (\x*\sh*\l+\l/2,\l) -- (\x*\sh*\l-\l/2,1.3*\l);
            \draw[line width=.5pt] (\x*\sh*\l+\l/2,1.1*\l) -- (\x*\sh*\l-\l/2,1.4*\l);
            \draw[line width=.5pt] (\x*\sh*\l-\l/2,-\l) -- (\x*\sh*\l+\l/2,-1.3*\l);
            \draw[line width=.5pt] (\x*\sh*\l-\l/2,-1.1*\l) -- (\x*\sh*\l+\l/2,-1.4*\l);
        }

    \end{tikzpicture}
}

\newcommand{\actionDOStrong}[8]{
    \begin{tikzpicture}[scale=1,baseline={([yshift=-.55ex]current bounding box.center)}]
        \def\l{.7};
        \def\sh{1.2};
        \ifthenelse{#1=1}{
            \draw[obj2,-<-=.47] (0,0) --node[pos=0.6,above]{${\sss #3}$} (\l,0);
            \draw[obj2,->-=.73] (0,0) --node[pos=0.6,above]{${\sss #3}$} (-\l,0);
            \draw[obj3,-<-=.47] (0,\sh*\l) --node[pos=0.6,above]{${\sss #6}$} (\l,\sh*\l);
            \draw[obj3,->-=.73] (0,\sh*\l) --node[pos=0.6,above]{${\sss #6}$} (-\l,\sh*\l);
            \draw[obj1,->-=.73] (0,0) -- (0,\l);
            \draw[obj1,-<-=.47] (0,0) -- (0,-\l);
            \draw[obj1,->-=.73] (0,\sh*\l) -- (0,\sh*\l+\l);
            \node[mpo] at (0,0) {};
            \node[mpo] at (0,\sh*\l) {};
            \node[yshift=-9pt] at (0,0) {${\sss #2}$};
        }{}
        \ifthenelse{#1=2}{
            \draw[obj1,-<-=.47] (0,0) -- (0,-\l);
            \draw[obj2,-<-=.47] (0,0) --node[pos=0.6,above]{${\sss #4}$} (\l,0);
            \draw[obj2,-<-=.47] (\sh*\l,0) --node[pos=0.6,above]{${\sss #3}$} (\sh*\l+\l,0);
            \draw[obj2,->-=.73] (0,0) --node[pos=0.6,above]{${\sss #4}$} (-\l,0);
            \draw[obj2,->-=.73] (-\sh*\l,0) --node[pos=0.6,above]{${\sss #3}$} (-\sh*\l-\l,0);
            \draw[obj3,-<-=.72, rounded corners=2pt] (\sh*\l,0) -- (\sh*\l,\sh*\l) --node[pos=.53,above]{${\sss #6}$} (\sh*\l+\l,\sh*\l);
            \draw[obj3,->-=.85, rounded corners=2pt] (-\sh*\l,0) -- (-\sh*\l,\sh*\l) --node[pos=.53,above]{${\sss #6}$} (-\sh*\l-\l,\sh*\l);
            \draw[obj1,->-=.73] (0,0) -- (0,\l);
            \node[mpo] at (0,0) {};
            \node[mps1] at (\sh*\l,0) {};
            \node[mps1] at (-\sh*\l,0) {};
            \node[yshift=-9pt] at (0,0) {${\sss #2}$};
            \node[yshift=-9pt] at (\sh*\l,0) {${\sss #7}$};
            \node[yshift=-9pt] at (-\sh*\l,0) {${\sss #8}$};
        }{}
    \end{tikzpicture}
}

\newcommand{\actionDOWeak}[8]{
    \begin{tikzpicture}[scale=1,baseline={([yshift=-.55ex]current bounding box.center)}]
        \def\l{.7};
        \def\sh{1.2};
        \ifthenelse{#1=1}{
            \draw[obj2,-<-=.47] (0,0) --node[pos=0.6,above]{${\sss #3}$} (\l,0);
            \draw[obj2,->-=.73] (0,0) --node[pos=0.6,above]{${\sss #3}$} (-\l,0);
            \draw[obj3,-<-=.47] (0,\sh*\l) --node[pos=0.6,above]{${\sss #6}$} (\l,\sh*\l);
            \draw[obj3,->-=.73] (0,\sh*\l) --node[pos=0.6,above]{${\sss #6}$} (-\l,\sh*\l);
            \draw[obj3,->-=.73] (0,-\sh*\l) --node[pos=0.6,below]{${\sss #6}$} (\l,-\sh*\l);
            \draw[obj3,-<-=.47] (0,-\sh*\l) --node[pos=0.6,below]{${\sss #6}$} (-\l,-\sh*\l);
            \draw[obj1,->-=.73] (0,0) -- (0,\l);
            \draw[obj1,-<-=.47] (0,0) -- (0,-\l);
            \draw[obj1,-<-=.47] (0,-\sh*\l) -- (0,-\sh*\l-\l);
            \draw[obj1,->-=.73] (0,\sh*\l) -- (0,\sh*\l+\l);
            \node[mpo] at (0,0) {};
            \node[mpo] at (0,\sh*\l) {};
            \node[mpo] at (0,-\sh*\l) {};
            \node[yshift=-9pt] at (0,0) {${\sss #2}$};
        }{}
        \ifthenelse{#1=2}{
            \draw[obj2,-<-=.47] (0,0) --node[pos=0.6,above]{${\sss #4}$} (\l,0);
            \draw[obj2,-<-=.47] (\sh*\l,0) --node[pos=0.6,above]{${\sss #3}$} (\sh*\l+\l,0);
            \draw[obj2,->-=.73] (0,0) --node[pos=0.6,above]{${\sss #4}$} (-\l,0);
            \draw[obj2,->-=.73] (-\sh*\l,0) --node[pos=0.6,above]{${\sss #3}$} (-\sh*\l-\l,0);
            \draw[obj3,-<-=.72, rounded corners=2pt] (\sh*\l,0) -- (\sh*\l,\sh*\l) --node[pos=.53,above]{${\sss #6}$} (\sh*\l+\l,\sh*\l);
            \draw[obj3,->-=.85, rounded corners=2pt] (-\sh*\l,0) -- (-\sh*\l,\sh*\l) --node[pos=.53,above]{${\sss #6}$} (-\sh*\l-\l,\sh*\l);
            \draw[obj3,->-=.85, rounded corners=2pt] (\sh*\l,0) -- (\sh*\l,-\sh*\l) --node[pos=.53,below]{${\sss #6}$} (\sh*\l+\l,-\sh*\l);
            \draw[obj3,-<-=.72, rounded corners=2pt] (-\sh*\l,0) -- (-\sh*\l,-\sh*\l) --node[pos=.53,below]{${\sss #6}$} (-\sh*\l-\l,-\sh*\l);
            \draw[obj1,->-=.73] (0,0) -- (0,\l);
            \draw[obj1,-<-=.47] (0,0) -- (0,-\l);
            \node[mpo] at (0,0) {};
            \node[mps1] at (\sh*\l,0) {};
            \node[mps1] at (-\sh*\l,0) {};
            \node[yshift=-9pt] at (0,0) {${\sss #2}$};
            \node[xshift=10pt,yshift=-9pt] at (\sh*\l,0) {${\sss #7}$};
            \node[xshift=-10pt,yshift=-9pt] at (-\sh*\l,0) {${\sss #8}$};
        }{}
    \end{tikzpicture}
}

\newcommand{\orthoActionWeak}[6]{
    \begin{tikzpicture}[scale=1,baseline={([yshift=-.55ex]current bounding box.center)}]
        \def\l{.7};
        \def\sh{1.2};
        \def\r{2};
        \draw[obj3,-<-=.46, rounded corners=2pt] (0,0) -- (0,\sh*\l) --node[pos=.5,above]{${\sss #2}$} (\r*\sh*\l,\sh*\l) -- (\r*\sh*\l,0);
        \draw[obj3,->-=.52, rounded corners=2pt] (0,0) -- (0,-\sh*\l) --node[pos=.5,below]{${\sss #2}$} (\r*\sh*\l,-\sh*\l) -- (\r*\sh*\l,0);
        \draw[obj2,-<-=.46] (0,0) --node[pos=0.5,above]{${\sss #1}$} (\r*\sh*\l,0);
        \draw[obj2,->-=.73] (0,0) --node[pos=0.6,above]{${\sss #3}$} (-\l,0);
        \draw[obj2,-<-=.47] (\r*\sh*\l,0) --node[pos=0.6,above]{${\sss #3}$} (\r*\sh*\l+\l,0);
        \node[mps1] at (0,0) {};
        \node[mps1] at (\r*\sh*\l,0) {};
        \node[yshift=-9pt,xshift=10pt] at (0,0) {${\sss #5}$};
        \node[yshift=-9pt,xshift=-10pt] at (\r*\sh*\l,0) {${\sss #6}$};
    \end{tikzpicture}
}

\newcommand{\moduleAssocStrong}[1]{
    \begin{tikzpicture}[scale=1,baseline={([yshift=-5.7ex]current bounding box.center)}]
        \def\l{.7};
        \def\sh{1.2};
        \ifthenelse{#1=1}{
            \draw[obj2,->-=.73] (0,0) --node[above,pos=.9]{${\sss a}$} (-\l,0);
            \draw[obj2,-<-=.46] (0,0) --node[pos=.52,above]{${\sss a}$}  (\sh*\l+\l,0);
            \draw[obj3,-<-=.62, rounded corners=2pt] (\sh*\l,1.5*\sh*\l) -- (\sh*\l,2*\sh*\l) --node[pos=.53,above]{${\sss {}^{r_a}n_1}$} (\sh*\l+\l,2*\sh*\l);
            \draw[obj3,-<-=.62, rounded corners=2pt] (\sh*\l,1.5*\sh*\l) -- (\sh*\l,\sh*\l) --node[pos=.53,below]{${\sss {}^{r_a}n_2}$} (\sh*\l+\l,\sh*\l);
            \draw[obj3,-<-=.73, rounded corners=2pt] (0,0) -- (0,1.5*\sh*\l) --node[pos=.3,above]{${\sss {}^{r_a}n_1n_2}$} (\sh*\l,1.5*\sh*\l);
            \node[mps1] at (\sh*\l,1.5*\sh*\l) {};
            \node[mps1] at (0,0) {};
            \node[yshift=-9pt] at (0,0) {${\sss \phi^{r_an_1n_2r_a^{-1}\!,a}}$};
        }{}
        \ifthenelse{#1=2}{
            \draw[obj2,->-=.73] (0,0) --node[above,pos=.9]{${\sss a}$} (-\l,0);
            \draw[obj2,-<-=.47] (0,0) --node[above,pos=.6]{${\sss a}$} (\l,0);
            \draw[obj2,-<-=.47] (\sh*\l,0) --node[pos=.6,above]{${\sss a}$} (\sh*\l+\l,0);
            \draw[obj3,-<-=.72, rounded corners=2pt] (\sh*\l,0) -- (\sh*\l,\sh*\l) --node[pos=.53,above]{${\sss {}^{r_a}n_2}$} (\sh*\l+\l,\sh*\l);
            \draw[obj3,->-=.26, rounded corners=2pt] (\sh*\l+\l,2*\sh*\l) --node[pos=.49,above]{${\sss {}^{r_a}n_1}$} (0,2*\sh*\l) -- (0,0);
            \node[mps1] at (0,0) {};
            \node[mps1] at (\sh*\l,0) {};
            \node[yshift=-10pt, xshift=-10pt] at (0,0) {${\sss \phi^{r_an_1r_a^{-1}\! ,a}}$};
            \node[yshift=-10pt, xshift=10pt] at (\sh*\l,0) {${\sss \phi^{r_an_2r_a^{-1}\!,a}}$};
        }{}
    \end{tikzpicture}
}

\newcommand{\assocDO}[1]{
    \begin{tikzpicture}[scale=1,baseline={([yshift=-3.5ex]current bounding box.center)}]
        \def\l{.7};
        \def\sh{1.2};
        \ifthenelse{#1=1}{
            \draw[obj2,->-=.73] (0,0) --node[above,pos=.6]{${\sss g(a)}$} (-\l,0);
            \draw[obj2,-<-=.47] (0,0) --node[above,pos=.6]{${\sss a}$} (\l,0);
            \draw[obj2,-<-=.47] (\sh*\l,0) --node[pos=.6,above]{${\sss a}$} (\sh*\l+\l,0);
            \draw[obj3,-<-=.72, rounded corners=2pt] (\sh*\l,0) -- (\sh*\l,\sh*\l) --node[pos=.53,above]{${\sss {}^{r_a}n}$} (\sh*\l+\l,\sh*\l);
            \draw[obj3,->-=.26, rounded corners=2pt] (\sh*\l+\l,2*\sh*\l) --node[pos=.49,above]{${\sss g}$} (0,2*\sh*\l) -- (0,0);
            \draw[obj3,->-=.75, rounded corners=2pt] (0,0) -- (0,-\sh*\l) --node[pos=.53,below]{${\sss g}$} (\sh*\l+\l,-\sh*\l);
            
            \node[mps1] at (0,0) {};
            \node[mps1] at (\sh*\l,0) {};
            \node[yshift=-9pt, xshift=-10pt] at (0,0) {${\sss \eta^{g,a}}$};
            \node[yshift=-9pt] at (\sh*\l,0) {${\sss \phi^{r_anr_a^{-1}\!,a}}$};
        }{}
        \ifthenelse{#1=2}{
            \draw[obj2,->-=.73] (0,0) --node[above,pos=.6]{${\sss g(a)}$} (-\l,0);
            \draw[obj2,-<-=.47] (0,0) --node[above,pos=.6]{${\sss g(a)}$} (\l,0);
            \draw[obj2,-<-=.5] (\sh*\l,0) --node[above,pos=.55]{${\sss a}$} (2*\sh*\l+\l,0);
            \draw[obj3,-<-=.47] (2*\sh*\l,\sh*\l) --node[above,pos=.6]{${\sss {}^{r_a}n}$} (2*\sh*\l+\l,\sh*\l);
            \draw[obj3,->-=.75, rounded corners=2pt] (\sh*\l,0) -- (\sh*\l,-\sh*\l) --node[pos=.53,below]{${\sss g}$} (2*\sh*\l+\l,-\sh*\l);
            \draw[obj3,-<-=.72, rounded corners=2pt] (2*\sh*\l,\sh*\l) -- (2*\sh*\l,2*\sh*\l) --node[pos=.53,above]{${\sss g}$} (2*\sh*\l+\l,2*\sh*\l);
            \draw[obj3,-<-=.47] (\sh*\l,\sh*\l) --node[above,pos=.61]{${\sss g{}^{r_a}n}$} (\sh*\l+\l,\sh*\l);
            \draw[obj3,->-=.35, rounded corners=2pt] (\sh*\l,\sh*\l) --node[pos=.55,above]{${\sss {}^{gr_a}n}$} (0,\sh*\l) -- (0,0);
            \draw[obj3,-<-=.47] (\sh*\l,0) --node[right,pos=.6]{${\sss g}$} (\sh*\l,\l);

            \node[mps1] at (0,0) {};
            \node[mps1] at (\sh*\l,0) {};
            \node[mps1] at (\sh*\l,\sh*\l) {};
            \node[mps1] at (2*\sh*\l,\sh*\l) {};
            \node[yshift=-9pt,xshift=-9pt] at (0,0) {${\sss \phi^{gr_an(gr_a)^{-1}\!,g(a)}}$};
        }{}
    \end{tikzpicture}
}

\newcommand{\DOTM}[1]{
    \begin{tikzpicture}[scale=1,baseline={([yshift=-.55ex]current bounding box.center)}]
        \def\l{.7};
        \def\sh{1.2};
        \draw[obj2,-<-=.47] (0,0) -- node[pos=.6,above]{${\sss #1}$} (\l,0);
        \draw[obj2,->-=.73] (0,0) -- node[pos=.6,above]{${\sss #1}$} (-\l,0);
        \draw[obj1,->-=.73] (0,0) -- (0,\l);
        \draw[obj1,-<-=.47] (0,0) -- (0,-\l);
        \node[mpo] at (0,0) {};

        \draw[line width=.5pt] (\l/2,\l) -- (-\l/2,1.3*\l);
        \draw[line width=.5pt] (\l/2,1.1*\l) -- (-\l/2,1.4*\l);
        \draw[line width=.5pt] (-\l/2,-\l) -- (\l/2,-1.3*\l);
        \draw[line width=.5pt] (-\l/2,-1.1*\l) -- (\l/2,-1.4*\l);
    \end{tikzpicture}
}

\newcommand{\DOFptSym}[1]{
    \begin{tikzpicture}[scale=1,baseline={([yshift=-.55ex]current bounding box.center)}]
        \def\l{.7};
        \def\sh{1.2};
        \ifthenelse{#1=1}{
            \draw[obj2,-<-=.47] (0,0) -- node[pos=.6,above]{${\sss a}$} (\l,0);
            \draw[obj2,->-=.73] (0,0) -- node[pos=.6,above]{${\sss g(a)}$} (-\l,0);
            \draw[obj3,-<-=.47] (0,0) --node[pos=.6,right]{${\sss g}$} (0,\l);
            \draw[obj3,->-=.73] (0,0) --node[pos=.6,right]{${\sss g}$} (0,-\l);
            \node[mps1] at (0,0) {};
            \draw[obj2,-<-=.47] (\sh*\l,0) -- node[pos=.6,above]{${\sss a}$} (\sh*\l+\l,0);
            \draw[obj1,->-=.73] (\sh*\l,0) -- (\sh*\l,\l);
            \draw[obj1,-<-=.47] (\sh*\l,0) -- (\sh*\l,-\l);
            \node[mps1] at (\sh*\l,0) {};
            \node[yshift=-9pt, xshift=-10pt] at (0,0) {${\sss \eta^{g,a}}$};

            \draw[line width=.5pt] (\sh*\l+\l/2,\l) -- (\sh*\l-\l/2,1.3*\l);
            \draw[line width=.5pt] (\sh*\l+\l/2,1.1*\l) -- (\sh*\l-\l/2,1.4*\l);
            \draw[line width=.5pt] (\sh*\l-\l/2,-\l) -- (\sh*\l+\l/2,-1.3*\l);
            \draw[line width=.5pt] (\sh*\l-\l/2,-1.1*\l) -- (\sh*\l+\l/2,-1.4*\l);
            \draw[line width=.5pt] (\l/2,\l) -- (-\l/2,1.3*\l);
            \draw[line width=.5pt] (\l/2,1.1*\l) -- (-\l/2,1.4*\l);
            \draw[line width=.5pt] (-\l/2,-\l) -- (\l/2,-1.3*\l);
            \draw[line width=.5pt] (-\l/2,-1.1*\l) -- (\l/2,-1.4*\l);
            \node[triR] at (2*\sh*\l,0) {};
            \node[xshift=13pt] at (2*\sh*\l,0) {${\sss \Xi_{{\rm r},a}}$};
        }{}

        \ifthenelse{#1=2}{
            \draw[obj2,-<-=.47] (0,0) -- node[pos=.6,above]{${\sss a}$} (\l,0);
            \draw[obj2,->-=.73] (0,0) -- node[pos=.6,above]{${\sss g(a)}$} (-\l,0);
            \draw[obj3,-<-=.67, rounded corners=2pt] (0,0) -- (0,\sh*\l) --node[pos=.45,above]{${\sss g}$} (\sh*\l,\sh*\l);
            \draw[obj3,->-=.75, rounded corners=2pt] (2*\sh*\l,\sh*\l+\l) -- (2*\sh*\l,\sh*\l) --node[pos=.45,above]{${\sss g}$} (\sh*\l,\sh*\l);
            \draw[obj3,->-=.75, rounded corners=2pt] (0,0) -- (0,-\sh*\l) --node[pos=.45,above]{${\sss g}$} (\sh*\l,-\sh*\l);
            \draw[obj3,-<-=.64, rounded corners=2pt] (2*\sh*\l,-\sh*\l-\l) -- (2*\sh*\l,-\sh*\l) --node[pos=.45,above]{${\sss g}$} (\sh*\l,-\sh*\l);
            \node[mps1] at (0,0) {};
            \draw[obj2,-<-=.5] (\sh*\l,0) -- node[pos=.55,above]{${\sss a}$} (2*\sh*\l+\l,0);
            \draw[obj1,->-=.73] (\sh*\l,0) -- (\sh*\l,\l);
            \draw[obj1,-<-=.47] (\sh*\l,0) -- (\sh*\l,-\l);
            \draw[obj1,->-=.73] (\sh*\l,\sh*\l) -- (\sh*\l,\sh*\l+\l);
            \draw[obj1,-<-=.47] (\sh*\l,-\sh*\l) -- (\sh*\l,-\sh*\l-\l);
            \node[mps1] at (\sh*\l,0) {};
            \node[yshift=-9pt, xshift=-10pt] at (0,0) {${\sss \eta^{g,a}}$};

            \node[mpo] at (\sh*\l,\sh*\l) {};
            \node[mpo] at (\sh*\l,-\sh*\l) {};

            \draw[line width=.5pt] (\sh*\l+\l/2,\sh*\l+\l) -- (\sh*\l-\l/2,\sh*\l+1.3*\l);
            \draw[line width=.5pt] (\sh*\l+\l/2,\sh*\l+1.1*\l) -- (\sh*\l-\l/2,\sh*\l+1.4*\l);
            \draw[line width=.5pt] (\sh*\l-\l/2,-\sh*\l-\l) -- (\sh*\l+\l/2,-\sh*\l-1.3*\l);
            \draw[line width=.5pt] (\sh*\l-\l/2,-\sh*\l-1.1*\l) -- (\sh*\l+\l/2,-\sh*\l-1.4*\l);
            \draw[line width=.5pt] (2*\sh*\l+\l/2,\sh*\l+\l) -- (2*\sh*\l-\l/2,\sh*\l+1.3*\l);
            \draw[line width=.5pt] (2*\sh*\l+\l/2,\sh*\l+1.1*\l) -- (2*\sh*\l-\l/2,\sh*\l+1.4*\l);
            \draw[line width=.5pt] (2*\sh*\l-\l/2,-\sh*\l-\l) -- (2*\sh*\l+\l/2,-\sh*\l-1.3*\l);
            \draw[line width=.5pt] (2*\sh*\l-\l/2,-\sh*\l-1.1*\l) -- (2*\sh*\l+\l/2,-\sh*\l-1.4*\l);
            \node[triR] at (3*\sh*\l,0) {};
            \node[xshift=13pt] at (3*\sh*\l,0) {${\sss \Xi_{{\rm r},a}}$};
        }{}

        \ifthenelse{#1=3}{
            \draw[obj2,-<-=.47] (0,0) -- node[pos=.6,above]{${\sss g(a)}$} (\l,0);
            \draw[obj2,->-=.73] (0,0) -- node[pos=.6,above]{${\sss g(a)}$} (-\l,0);
            \draw[obj3,-<-=.47] (\sh*\l,0) --node[pos=.6,right]{${\sss g}$} (\sh*\l,\l);
            \draw[obj3,->-=.73] (\sh*\l,0) --node[pos=.6,right]{${\sss g}$} (\sh*\l,-\l);
            \draw[obj2,-<-=.47] (\sh*\l,0) -- node[pos=.6,above]{${\sss a}$} (\sh*\l+\l,0);
            \draw[obj1,->-=.73] (0,0) -- (0,\l);
            \draw[obj1,-<-=.47] (0,0) -- (0,-\l);
            \node[mps1] at (\sh*\l,0) {};
            \node[mps1] at (0,0) {};
            \node[yshift=-9pt, xshift=-10pt] at (\sh*\l,0) {${\sss \eta^{g,a}}$};

            \draw[line width=.5pt] (\sh*\l+\l/2,\l) -- (\sh*\l-\l/2,1.3*\l);
            \draw[line width=.5pt] (\sh*\l+\l/2,1.1*\l) -- (\sh*\l-\l/2,1.4*\l);
            \draw[line width=.5pt] (\sh*\l-\l/2,-\l) -- (\sh*\l+\l/2,-1.3*\l);
            \draw[line width=.5pt] (\sh*\l-\l/2,-1.1*\l) -- (\sh*\l+\l/2,-1.4*\l);
            \draw[line width=.5pt] (\l/2,\l) -- (-\l/2,1.3*\l);
            \draw[line width=.5pt] (\l/2,1.1*\l) -- (-\l/2,1.4*\l);
            \draw[line width=.5pt] (-\l/2,-\l) -- (\l/2,-1.3*\l);
            \draw[line width=.5pt] (-\l/2,-1.1*\l) -- (\l/2,-1.4*\l);
            \node[triR] at (2*\sh*\l,0) {};
            \node[xshift=13pt] at (2*\sh*\l,0) {${\sss \Xi_{{\rm r},a}}$};
        }{}
    \end{tikzpicture}
}

\newcommand{\DOFptR}[6]{
    \begin{tikzpicture}[scale=1,baseline={([yshift=-.55ex]current bounding box.center)}]
        \def\l{.7};
        \def\sh{1.2};
        \draw[obj2,-<-=.47] (0,0) -- node[pos=.6,above]{${\sss #2}$} (\l,0);
        \draw[obj2,->-=.73] (0,0) -- node[pos=.6,above]{${\sss #3}$} (-\l,0);
        \ifthenelse{#1=1}{
            \draw[obj1,->-=.73] (0,0) -- (0,\l);
            \draw[obj1,-<-=.47] (0,0) -- (0,-\l);
            \node[mpo] at (0,0) {};
        }{
            \draw[obj3,-<-=.47] (0,0) --node[pos=.6,right]{${\sss #4}$} (0,\l);
            \draw[obj3,->-=.73] (0,0) --node[pos=.6,right]{${\sss #4}$} (0,-\l);
            \node[mps1] at (0,0) {};
            \node[yshift=-9pt, xshift=-10pt] at (0,0) {${\sss #6}$};
        }
        
        \draw[line width=.5pt] (\l/2,\l) -- (-\l/2,1.3*\l);
        \draw[line width=.5pt] (\l/2,1.1*\l) -- (-\l/2,1.4*\l);
        \draw[line width=.5pt] (-\l/2,-\l) -- (\l/2,-1.3*\l);
        \draw[line width=.5pt] (-\l/2,-1.1*\l) -- (\l/2,-1.4*\l);
        \node[triR] at (\sh*\l,0) {};
        \node[xshift=11pt] at (\sh*\l,0) {${\sss #5}$};
    \end{tikzpicture}
}

\newcommand{\DOFptRT}[2]{
    \begin{tikzpicture}[scale=1,baseline={([yshift=-.6ex]0,0)}]
        \def\l{.7};
        \def\sh{1.2};
        \draw[obj2,-<-=.39] (0,0) -- node[pos=.5,above]{${\sss #1}$} (\l,0);
        \node[triR] at (\sh*\l,0) {};
        \node[xshift=11pt] at (\sh*\l,0) {${\sss #2}$};
    \end{tikzpicture}
}

\newcommand{\DOFptLR}[3]{
    \begin{tikzpicture}[scale=1,baseline={([yshift=-.6ex]0,0)}]
        \def\l{.7};
        \def\sh{1.2};
        \draw[obj2,-<-=.39] (0,0) -- node[pos=.5,above]{${\sss #1}$} (\l,0);
        \draw[obj2,-<-=.39] (\sh*\l+.6*\l,0) -- node[pos=.5,above]{${\sss #1}$} (\sh*\l+.6*\l+\l,0);
        \node[triR] at (\sh*\l,0) {};
        \node[triL] at (\sh*\l+.6*\l,0) {};
        \node[yshift=-11pt] at (\sh*\l,0) {${\sss #2}$};
        \node[yshift=-11pt] at (\sh*\l+.6*\l,0) {${\sss #3}$};
    \end{tikzpicture}
}

\newcommand{\DOFptLT}[2]{
    \begin{tikzpicture}[scale=1,baseline={([yshift=-.6ex]0,0)}]
        \def\l{.7};
        \def\sh{1.2};
        \draw[obj2,-<-=.39] (-\l,0) -- node[pos=.5,above]{${\sss #1}$} (0,0);
        \node[triL] at (-\sh*\l,0) {};
        \node[xshift=-11pt] at (-\sh*\l,0) {${\sss #2}$};
    \end{tikzpicture}
}

\newcommand{\DOFptL}[5]{
    \begin{tikzpicture}[scale=1,baseline={([yshift=-.55ex]current bounding box.center)}]
        \def\l{.7};
        \def\sh{1.2};
        \draw[obj2,-<-=.47] (0,0) -- node[pos=.6,above]{${\sss #2}$} (\l,0);
        \draw[obj2,->-=.73] (0,0) -- node[pos=.6,above]{${\sss #3}$} (-\l,0);
        \ifthenelse{#1=1}{
            \draw[obj1,->-=.73] (0,0) -- (0,\l);
            \draw[obj1,-<-=.47] (0,0) -- (0,-\l);
        }{
            \draw[obj3,-<-=.47] (0,0) --node[pos=.6,right]{${\sss #4}$} (0,\l);
            \draw[obj3,->-=.73] (0,0) --node[pos=.6,right]{${\sss #4}$} (0,-\l);
        }
        
        \node[mpo] at (0,0) {};
        \draw[line width=.5pt] (\l/2,\l) -- (-\l/2,1.3*\l);
        \draw[line width=.5pt] (\l/2,1.1*\l) -- (-\l/2,1.4*\l);
        \draw[line width=.5pt] (-\l/2,-\l) -- (\l/2,-1.3*\l);
        \draw[line width=.5pt] (-\l/2,-1.1*\l) -- (\l/2,-1.4*\l);
        \node[triL] at (-\sh*\l,0) {};
        \node[xshift=-11pt] at (-\sh*\l,0) {${\sss #5}$};
    \end{tikzpicture}
}

\newcommand{\MixedStringOpExpValLimL}[6]{
    \begin{tikzpicture}[scale=1,baseline={([yshift=-.6ex]0,0)}]
        \def\l{.7};
        \def\sh{1.2};
        \draw[obj1,-<-=.47] (0,\sh*\l) -- (0,\sh*\l-\l);
        \draw[obj1,->-=.73] (0,\sh*\l) -- (0,\sh*\l+\l);
        \draw[obj1,-<-=.47] (0,0) -- (0,-\l);
        \draw[obj2,-<-=.47] (0,0) --node[pos=.6,above]{${\sss #1}$} (\l,0);
        \draw[obj2,-<-=.47] (\sh*\l,0) --node[pos=.6,above]{${\sss #1}$} (\sh*\l+\l,0);
        \draw[obj2,->-=.73] (0,0) --node[pos=.6,above]{${\sss #1}$} (-\l,0);
        \draw[obj3,-<-=.23, rounded corners=2pt] (0,\sh*\l) --node[pos=.56,above]{${\sss #2}$} (\sh*\l,\sh*\l) -- (\sh*\l,0);
        \node[yshift=-9pt] at (\sh*\l,0) {${\sss #3}$};
        \node[xshift=-9pt] at (0,\sh*\l) {${\sss #4}$};
        \node[triR] at (2*\sh*\l,0) {};
        \node[xshift=11pt] at (2*\sh*\l,0) {${\sss #6}$};
        \node[triL] at (-\sh*\l,0) {};
        \node[xshift=-11pt] at (-\sh*\l,0) {${\sss #5}$};
        \node[mpo] at (0,0) {};
        \node[mps1] at (0,\sh*\l) {};
        \node[mps1] at (\sh*\l,0) {};
        \draw[line width=.5pt] (\l/2,\sh*\l+\l) -- (-\l/2,\sh*\l+1.3*\l);
        \draw[line width=.5pt] (\l/2,\sh*\l+1.1*\l) -- (-\l/2,\sh*\l+1.4*\l);
        \draw[line width=.5pt] (-\l/2,-\l) -- (\l/2,-1.3*\l);
        \draw[line width=.5pt] (-\l/2,-1.1*\l) -- (\l/2,-1.4*\l);
    \end{tikzpicture}
}

\newcommand{\MixedStringOpExpValLimR}[6]{
    \begin{tikzpicture}[scale=1,baseline={([yshift=-.6ex]0,0)}]
        \def\l{.7};
        \def\sh{1.2};
        \draw[obj1,-<-=.47] (0,\sh*\l) -- (0,\sh*\l-\l);
        \draw[obj1,->-=.73] (0,\sh*\l) -- (0,\sh*\l+\l);
        \draw[obj1,-<-=.47] (0,0) -- (0,-\l);
        \draw[obj2,-<-=.47] (0,0) --node[pos=.6,above]{${\sss #1}$} (\l,0);
        \draw[obj2,->-=.73] (-\sh*\l,0) --node[pos=.6,above]{${\sss #1}$} (-\sh*\l-\l,0);
        \draw[obj2,->-=.73] (0,0) --node[pos=.6,above]{${\sss #1}$} (-\l,0);
        \draw[obj3,->-=.3, rounded corners=2pt] (0,\sh*\l) --node[pos=.56,above]{${\sss #2}$} (-\sh*\l,\sh*\l) -- (-\sh*\l,0);
        \node[yshift=-9pt] at (-\sh*\l,0) {${\sss #3}$};
        \node[xshift=9pt] at (0,\sh*\l) {${\sss #4}$};
        \node[triR] at (\sh*\l,0) {};
        \node[xshift=11pt] at (\sh*\l,0) {${\sss #6}$};
        \node[triL] at (-2*\sh*\l,0) {};
        \node[xshift=-11pt] at (-2*\sh*\l,0) {${\sss #5}$};
        \node[mpo] at (0,0) {};
        \node[mps1] at (0,\sh*\l) {};
        \node[mps1] at (-\sh*\l,0) {};
        \draw[line width=.5pt] (\l/2,\sh*\l+\l) -- (-\l/2,\sh*\l+1.3*\l);
        \draw[line width=.5pt] (\l/2,\sh*\l+1.1*\l) -- (-\l/2,\sh*\l+1.4*\l);
        \draw[line width=.5pt] (-\l/2,-\l) -- (\l/2,-1.3*\l);
        \draw[line width=.5pt] (-\l/2,-1.1*\l) -- (\l/2,-1.4*\l);
    \end{tikzpicture}
}

\newcommand{\MixedTwistedLocalOpEnv}[6]{
    \begin{tikzpicture}[scale=1,baseline={([yshift=-.6ex]0,0)}]
        \def\l{.7};
        \def\sh{1.2};
        \draw[obj1,-<-=.55] (0,\sh*\l) -- (0,\sh*\l-\l);
        \draw[obj1,->-=.78] (0,\sh*\l) -- (0,\sh*\l+\l);
        \draw[obj1,-<-=.47] (0,0) -- (0,-\l);
        \draw[obj2,-<-=.47] (0,0) --node[pos=.6,above]{${\sss #1}$} (\l,0);
        \draw[obj2,->-=.73] (-\sh*\l,0) --node[pos=.6,above]{${\sss #1}$} (-\sh*\l-\l,0);
        \draw[obj2,->-=.73] (0,0) --node[pos=.6,above]{${\sss #1}$} (-\l,0);
        \draw[obj3,->-=.3, rounded corners=2pt] (0,\sh*\l) --node[pos=.56,above]{${\sss #2}$} (-\sh*\l,\sh*\l) -- (-\sh*\l,0);
        \node[yshift=-10pt] at (-\sh*\l,0) {${\sss #3}$};
        \node[xshift=9pt] at (0,\sh*\l) {${\sss #4}$};
        \node[triR] at (\sh*\l,0) {};
        \node[xshift=11pt] at (\sh*\l,0) {${\sss #6}$};
        \node[triL] at (-2*\sh*\l,0) {};
        \node[xshift=-11pt] at (-2*\sh*\l,0) {${\sss #5}$};
        \node[mpo] at (0,0) {};
        \node[hole] at (0,\sh*\l) {};
        \node[mps1] at (-\sh*\l,0) {};
        \draw[line width=.5pt] (\l/2,\sh*\l+\l) -- (-\l/2,\sh*\l+1.3*\l);
        \draw[line width=.5pt] (\l/2,\sh*\l+1.1*\l) -- (-\l/2,\sh*\l+1.4*\l);
        \draw[line width=.5pt] (-\l/2,-\l) -- (\l/2,-1.3*\l);
        \draw[line width=.5pt] (-\l/2,-1.1*\l) -- (\l/2,-1.4*\l);
    \end{tikzpicture}
}

\newcommand{\MixedTubeModEnvA}[9]{
    \begin{tikzpicture}[scale=1,baseline={([yshift=-.6ex]0,-1.68)}]
        \def\l{.7};
        \def\sh{1.2};
        \draw[obj1,-<-=.45] (0,0) -- (0,-\sh*\l);
        \draw[obj1,->-=.65] (0,0) -- (0,\sh*\l);
        \draw[obj1,->-=.73] (0,\sh*\l) -- (0,\sh*\l+\l);
        \draw[obj1,-<-=.47] (0,-\sh*\l) -- (0,-\sh*\l-\l);
        \draw[obj1,-<-=.47] (0,-2*\sh*\l) -- (0,-2*\sh*\l-\l);
        \draw[obj3,->-=.65] (0,0) --node[pos=.56,below]{${\sss #2}$} (-\sh*\l,0);
        \draw[obj3,->-=.73, rounded corners=2pt] (-\sh*\l,0) --node[pos=.6,below]{${\sss #3}$} (-\sh*\l-\l,0);
        \draw[obj3,-<-=.43] (-2*\sh*\l,-2*\sh*\l) --node[pos=.5, left]{${\sss #4}$} (-2*\sh*\l,0);
        \draw[obj3,-<-=.47, rounded corners=2pt] (0,-\sh*\l) -- (\sh*\l,-\sh*\l) --node[pos=.5,right]{${\sss #1}$} (\sh*\l,\sh*\l) -- (0,\sh*\l);
        \draw[obj3,->-=.33, rounded corners=2pt] (0,-\sh*\l) --node[pos=.56,below]{${\sss #1}$} (-\sh*\l,-\sh*\l) -- (-\sh*\l,0);
        \draw[obj3,-<-=.31, rounded corners=2pt] (0,\sh*\l) --node[pos=.52,above]{${\sss #1}$} (-2*\sh*\l,\sh*\l) -- (-2*\sh*\l,0);
        \draw[obj2,-<-=.44] (-2*\sh*\l,-2*\sh*\l) --node[pos=.49,above]{${\sss #5}$} (0,-2*\sh*\l);
        \draw[obj2,-<-=.47] (0,-2*\sh*\l) --node[pos=.6,above]{${\sss #5}$} (\l,-2*\sh*\l);
        \draw[obj2,->-=.73] (-2*\sh*\l,-2*\sh*\l) --node[pos=.6,above]{${\sss #5}$} (-2*\sh*\l-\l,-2*\sh*\l);

        \draw[line width=.5pt] (\l/2,\sh*\l+\l) -- (-\l/2,\sh*\l+1.3*\l);
        \draw[line width=.5pt] (\l/2,\sh*\l+1.1*\l) -- (-\l/2,\sh*\l+1.4*\l);
        \draw[line width=.5pt] (-\l/2,-2*\sh*\l-\l) -- (\l/2,-2*\sh*\l-1.3*\l);
        \draw[line width=.5pt] (-\l/2,-2*\sh*\l-1.1*\l) -- (\l/2,-2*\sh*\l-1.4*\l);

        \node[yshift=-9pt] at (-2*\sh*\l,-2*\sh*\l) {${\sss #6}$};
        \node[xshift=-11pt,yshift=12pt] at (-2*\sh*\l,0) {${\sss #7}$};
        \node[xshift=-11pt] at (-3*\sh*\l,-2*\sh*\l) {${\sss #8}$};
        \node[xshift=11pt] at (\sh*\l,-2*\sh*\l) {${\sss #9}$};
        \node[hole] at (0,0) {};
        \node[mpo] at (0,-\sh*\l) {};
        \node[mpo] at (0,\sh*\l) {};
        \node[mps1] at (-\sh*\l,0) {};
        \node[mps1] at (-2*\sh*\l,0) {};
        \node[mps1] at (-2*\sh*\l,-2*\sh*\l) {};
        \node[mpo] at (0,-2*\sh*\l) {};
        \node[triR] at (\sh*\l,-2*\sh*\l) {};
        \node[triL] at (-3*\sh*\l,-2*\sh*\l) {};
    \end{tikzpicture}
}

\newcommand{\MixedTubeModEnvB}[9]{
    \begin{tikzpicture}[scale=1,baseline={([yshift=-.6ex]0,-1.68)}]
        \def\l{.7};
        \def\sh{1.2};
        \draw[obj1,-<-=.45] (0,0) -- (0,-\sh*\l);
        \draw[obj1,->-=.74] (0,0) -- (0,\l);
        \draw[obj1,-<-=.47] (0,-\sh*\l) -- (0,-\sh*\l-\l);
        \draw[obj1,-<-=.47] (0,-2*\sh*\l) -- (0,-2*\sh*\l-\l);
        \draw[obj1,-<-=.47] (0,-3*\sh*\l) -- (0,-3*\sh*\l-\l);
        \draw[obj3,->-=.65] (0,0) --node[pos=.56,below]{${\sss #2}$} (-\sh*\l,0);
        \draw[obj3,->-=.73, rounded corners=2pt] (-\sh*\l,0) --node[pos=.6,below]{${\sss #3}$} (-\sh*\l-\l,0);
        \draw[obj3,-<-=.43] (-2*\sh*\l,-2*\sh*\l) --node[pos=.5, left]{${\sss #4}$} (-2*\sh*\l,0);
        \draw[obj3,-<-=.64, rounded corners=2pt] (0,-\sh*\l) -- (\sh*\l,-\sh*\l) --node[pos=.5,right]{${\sss #1}$} (\sh*\l,\l);
        \draw[obj3,->-=.33, rounded corners=2pt] (0,-\sh*\l) --node[pos=.56,below]{${\sss #1}$} (-\sh*\l,-\sh*\l) -- (-\sh*\l,0);
        \draw[obj3,->-=.73] (-2*\sh*\l,0) --node[pos=.6,left]{${\sss #1}$} (-2*\sh*\l,\l);
        \draw[obj3,->-=.84, rounded corners=2pt] (0,-3*\sh*\l) -- (\sh*\l,-3*\sh*\l) --node[pos=.5,right]{${\sss #1}$} (\sh*\l,-3*\sh*\l-\l);
        \draw[obj3,-<-=.82, rounded corners=2pt] (0,-3*\sh*\l) -- (-2*\sh*\l,-3*\sh*\l) --node[pos=.5,left]{${\sss #1}$} (-2*\sh*\l,-3*\sh*\l-\l);

        \draw[obj2,-<-=.44] (-2*\sh*\l,-2*\sh*\l) --node[pos=.49,above]{${\sss #5}$} (0,-2*\sh*\l);
        \draw[obj2,-<-=.47] (0,-2*\sh*\l) --node[pos=.6,above]{${\sss #5}$} (\l,-2*\sh*\l);
        \draw[obj2,->-=.73] (-2*\sh*\l,-2*\sh*\l) --node[pos=.6,above]{${\sss #5}$} (-2*\sh*\l-\l,-2*\sh*\l);

        \foreach \x in {-2,0,1}{
            \draw[line width=.5pt] (\x*\sh*\l+\l/2,\l) -- (\x*\sh*\l-\l/2,1.3*\l);
            \draw[line width=.5pt] (\x*\sh*\l+\l/2,1.1*\l) -- (\x*\sh*\l-\l/2,1.4*\l);
            \draw[line width=.5pt] (\x*\sh*\l-\l/2,-3*\sh*\l-\l) -- (\x*\sh*\l+\l/2,-3*\sh*\l-1.3*\l);
            \draw[line width=.5pt] (\x*\sh*\l-\l/2,-3*\sh*\l-1.1*\l) -- (\x*\sh*\l+\l/2,-3*\sh*\l-1.4*\l);
        }

        \node[yshift=-9pt] at (-2*\sh*\l,-2*\sh*\l) {${\sss #6}$};
        \node[xshift=-11pt,yshift=12pt] at (-2*\sh*\l,0) {${\sss #7}$};
        \node[xshift=-11pt] at (-3*\sh*\l,-2*\sh*\l) {${\sss #8}$};
        \node[xshift=11pt] at (\sh*\l,-2*\sh*\l) {${\sss #9}$};
        \node[hole] at (0,0) {};
        \node[mpo] at (0,-\sh*\l) {};
        \node[mpo] at (0,-3*\sh*\l) {};
        \node[mps1] at (-\sh*\l,0) {};
        \node[mps1] at (-2*\sh*\l,0) {};
        \node[mps1] at (-2*\sh*\l,-2*\sh*\l) {};
        \node[mpo] at (0,-2*\sh*\l) {};
        \node[triR] at (\sh*\l,-2*\sh*\l) {};
        \node[triL] at (-3*\sh*\l,-2*\sh*\l) {};
    \end{tikzpicture}
}

\newcommand{\MixedTubeModEnvC}[9]{
    \begin{tikzpicture}[scale=1,baseline={([yshift=-.6ex]0,-1.68)}]
        \def\l{.7};
        \def\sh{1.2};
        \draw[obj1,-<-=.49] (0,0) -- (0,-\sh*\l-\l);
        \draw[obj1,->-=.74] (0,0) -- (0,\l);
        \draw[obj1,-<-=.49] (0,-2*\sh*\l) -- (0,-3*\sh*\l-\l);
        \draw[obj3,->-=.65] (0,0) --node[pos=.56,below]{${\sss #2}$} (-\sh*\l,0);
        \draw[obj3,->-=.73, rounded corners=2pt] (-\sh*\l,0) --node[pos=.6,below]{${\sss #3}$} (-\sh*\l-\l,0);
        \draw[obj3,-<-=.43] (-2*\sh*\l,-2*\sh*\l) --node[pos=.5, left]{${\sss #4}$} (-2*\sh*\l,0);
        \draw[obj3,-<-=.45] (-\sh*\l,0) --node[pos=.5,left]{${\sss #1}$} (-\sh*\l,-2*\sh*\l);
        \draw[obj3,->-=.73] (-2*\sh*\l,0) --node[pos=.6,left]{${\sss #1}$} (-2*\sh*\l,\l);
        \draw[obj3,->-=.51, rounded corners=2pt] (-2*\sh*\l,-3*\sh*\l-\l) -- (-2*\sh*\l,-3*\sh*\l) --node[pos=.5,below]{${\sss g}$} (-\sh*\l,-3*\sh*\l) -- (-\sh*\l,-2*\sh*\l);

        \draw[obj2,->-=.73] (0,-2*\sh*\l) --node[pos=.6,above]{${\sss g(a)}$} (-\l,-2*\sh*\l);
        \draw[obj2,->-=.73] (-\sh*\l,-2*\sh*\l) --node[pos=.6,above]{${\sss #5}$} (-\sh*\l-\l,-2*\sh*\l);
        \draw[obj2,-<-=.47] (0,-2*\sh*\l) --node[pos=.6,above]{${\sss g(a)}$} (\l,-2*\sh*\l);
        \draw[obj2,->-=.73] (-2*\sh*\l,-2*\sh*\l) --node[pos=.6,above]{${\sss #5}$} (-2*\sh*\l-\l,-2*\sh*\l);

        \foreach \x in {-2,0}{
            \draw[line width=.5pt] (\x*\sh*\l+\l/2,\l) -- (\x*\sh*\l-\l/2,1.3*\l);
            \draw[line width=.5pt] (\x*\sh*\l+\l/2,1.1*\l) -- (\x*\sh*\l-\l/2,1.4*\l);
            \draw[line width=.5pt] (\x*\sh*\l-\l/2,-3*\sh*\l-\l) -- (\x*\sh*\l+\l/2,-3*\sh*\l-1.3*\l);
            \draw[line width=.5pt] (\x*\sh*\l-\l/2,-3*\sh*\l-1.1*\l) -- (\x*\sh*\l+\l/2,-3*\sh*\l-1.4*\l);
        }

        \node[yshift=-9pt] at (-2*\sh*\l,-2*\sh*\l) {${\sss #6}$};
        \node[xshift=-11pt,yshift=12pt] at (-2*\sh*\l,0) {${\sss #7}$};
        \node[xshift=-11pt] at (-3*\sh*\l,-2*\sh*\l) {${\sss #8}$};
        \node[xshift=11pt] at (\sh*\l,-2*\sh*\l) {${\sss #9}$};
        \node[hole] at (0,0) {};
        \node[mps1] at (-\sh*\l,0) {};
        \node[mps1] at (-\sh*\l,-2*\sh*\l) {};
        \node[mps1] at (-2*\sh*\l,0) {};
        \node[mps1] at (-2*\sh*\l,-2*\sh*\l) {};
        \node[mpo] at (0,-2*\sh*\l) {};
        \node[triR] at (\sh*\l,-2*\sh*\l) {};
        \node[triL] at (-3*\sh*\l,-2*\sh*\l) {};
        \node[yshift=-9pt, xshift=10pt] at (-\sh*\l,-2*\sh*\l) {${\sss \bar \eta^{g,a}}$};
    \end{tikzpicture}
}

\title{\boldmath Algebras of order parameters in one-dimensional \par spin systems}

\author[\mtriangle]{Ameya Chavda,}
\author[\msquare]{Clement Delcamp,}
\author[\mpentagon]{Alex Turzillo,}
\author[\,\mhexagon]{Minyoung You}

\affiliation[\normalfont \mtriangle]{Center for Theoretical Physics, Department of Physics, Columbia University, New York, U.S.A}
\affiliation[\normalfont \msquare]{Laboratoire Alexander Grothendieck, Institut des Hautes \'Etudes Scientifiques \& CNRS, \\ Bures-sur-Yvette, France}
\affiliation[\normalfont \mpentagon]{Department of Applied Mathematics and Theoretical Physics, University of Cambridge, \\ Cambridge, United Kingdom}
\affiliation[\normalfont \mhexagon]{Yukawa Institute for Theoretical Physics, Kyoto University, Kyoto, Japan}

\emailAdd{ameya.chavda@columbia.edu}
\emailAdd{delcamp@ihes.fr}
\emailAdd{amt201@cam.ac.uk}
\emailAdd{miyou849@gmail.com}

\abstract{\\~\\ We study order parameters in one-dimensional quantum lattice models with finite invertible or non-invertible symmetry. We investigate what properties a string operator must satisfy in order to acquire a non-vanishing expectation value in a given gapped phase. We deduce that multiplets of string order parameters organise into a Lagrangian algebra in the Drinfel'd centre of the symmetry category. In particular, we highlight the role of the multiplication rule as governing the fusion of the twisted sector local operators that constitute the string operator in the infrared limit. Our derivations exploit the tensor network approach to the classification of gapped phases and its reformulation in terms of module categories over the symmetry category. Within this framework, a gapped phase is associated with a pattern of spontaneous symmetry breaking wherein a Morita class of algebras of topological lines is preserved in the ground state subspace. The crux of the proof is to show that the expectation value of any string operator explicitly depends on the tube algebra module associated with the Lagrangian algebra, which is realised as the full centre of the corresponding module category. Finally, we demonstrate that these techniques extend to phases of symmetric mixed states. 
}

\begin{document} 
	\vspace*{-2em}
        \maketitle
	\flushbottom
	\newpage
	
\section{Introduction}

In condensed matter theory, arguably the most interesting aspect of the notion of \emph{symmetry} is the absence thereof, or more precisely, the idea that a physical state---typically a ground state or a thermal equilibrium state---possesses less symmetry than the theory describing it. We distinguish two situations:  the system possesses either a unique symmetric ground state that is left invariant by all symmetry operators, or a space of degenerate and orthogonal ground states that are related by the action of these operators. Within the Landau paradigm, distinct gapped phases of matter are classified by such patterns of \emph{spontaneous symmetry breaking}.

Closely related to spontaneous symmetry breaking is the concept of an \emph{order parameter}. Fundamentally, an order parameter is a physical observable used to detect \emph{phase transitions}. It typically consists of a local operator whose vacuum expectation value vanishes in the symmetric phase but acquires a non-zero value in the symmetry-broken one. A stricter definition requires this local operator to have a unique expectation value for each symmetry-broken state. In spite of being a ubiquitous tool across theoretical physics, the contours of this concept continue to evolve. For example, the discovery of \emph{symmetry-protected topological} (SPT) phases in one-dimensional quantum lattice models during the 1980s eventually challenged this paradigm \cite{PhysRevLett.50.1153}. Famously, (non-trivial) SPT phases cannot be detected by local order parameters; rather, they require non-local \emph{string order parameters} \cite{PhysRevB.40.4709,PhysRevLett.100.167202,PhysRevB.86.125441}. The existence of such SPT phases stems from the fact that there may be multiple (inequivalent) ways for a state to be left invariant by all the symmetry operators. A well know manifestation of this phenomenon is the presence of \emph{edge modes} that transform projectively under the symmetry group, leading to a classification scheme based on \emph{group cohomology} \cite{Kennedy:1992ifl,PhysRevB.80.155131,PhysRevB.83.035107,PhysRevB.84.165139}. 

The concept of symmetry itself has also been evolving. Indeed, reformulating symmetries in the language of \emph{topological defects} has led to generalisations of the concept, commonly referred to as \emph{generalised symmetries} \cite{Gaiotto:2014kfa,Freed:2022qnc}. In particular, a (global internal) symmetry may be defined in terms of transformations of the degrees of freedom that cannot be `undone', leading to the concept of a \emph{non-invertible symmetry}. This formalises an idea that has long been anticipated by the study, for instance, of \emph{Verlinde lines} \cite{Verlinde:1988sn, PETKOVA2001157} and related topological defects \cite{Oshikawa:1996dj,Oshikawa:1996ww,Frohlich:2004ef,Frohlich:2006ch} in (1+1)d rational conformal field theories (CFTs). Another common source of examples are ordinary global symmetries with respect to non-abelian groups, which upon \emph{gauging} produce non-invertible symmetries \cite{Tachikawa:2017gyf, Bhardwaj:2017xup, Delcamp:2021szr,Bhardwaj:2022lsg,Bartsch:2022mpm,Bhardwaj:2022yxj}. While ordinary symmetries are captured by group theory, these generalised symmetries necessitate higher mathematical structures. In (1+1)d, assuming finiteness and semi-simplicity, the relevant mathematical framework is provided by \emph{(spherical) fusion category} theory \cite{Bhardwaj:2017xup,Chang:2018iay,Thorngren:2019iar,PhysRevResearch.2.043086,Schafer-Nameki:2023jdn,Shao:2023gho}. Naturally, this new notion of symmetry has necessitated a rethinking of the traditional concepts of gapped phases, spontaneous symmetry breaking, charges, and order parameters, as well as the identification of the relevant mathematical tools. Remarkably, these developments have shed light on the invertible case as well. In that spirit, this manuscript aims to provide a framework that refines our understanding of order parameters for gapped phases with invertible symmetries, while simultaneously tackling the more challenging case of non-invertible symmetries.

\bigskip \noindent
A lot of the intuition surrounding the physics of non-invertible symmetries stems from a picture that has been developed alongside it, namely that a theory with a given finite symmetry can be realised as a boundary theory of a higher-dimensional fully extended \emph{topological quantum field theory} (TQFT) hosting a gauged version of this symmetry. An early incarnation of this mechanism within the context of two-dimensional rational CFTs was explored in ref.~\cite{Fuchs:2002cm, FUCHS2004277, FUCHS2004511, FUCHS2005539}. More recently, this idea has born multiple names, namely ‘strange correlators’ \cite{PhysRevLett.112.247202,PhysRevLett.121.177203}, ‘topological holography’ \cite{KONG201762,PhysRevResearch.2.043086,PhysRevResearch.2.033417,Ji:2021esj,Chatterjee:2022kxb,Huang:2023pyk,Huang:2024ror}, ‘Symmetry Topological Field Theory (SymTFT)’ \cite{Apruzzi:2021nmk,Kaidi:2022cpf,Kaidi:2023maf}, or the ‘sandwich construction’ \cite{Freed:2022qnc}. The upshot is that all properties of a theory pertaining to its symmetry can be understood in terms of this higher-dimensional TQFT. This picture can be made especially explicit on the lattice \cite{Aasen:2016dop,Freed:2018cec,PhysRevLett.121.177203,Aasen:2020jwb,Vanhove:2021nav,Delcamp:2024cfp}. In the context of quantum lattice models, it boils down to the so-called \emph{anyonic chain} construction \cite{Feiguin:2006ydp,PhysRevB.87.235120,Buican:2017rxc,Inamura:2021szw,Lootens:2021tet,Lootens:2022avn}, and higher-dimensional generalisations thereof \cite{Delcamp:2023kew,Inamura:2023qzl,Moradi:2023dan,Choi:2024rjm,Vancraeynest-DeCuiper:2025wkh,Eck:2025ldx}. 

One important insight this picture suggests is that symmetric gapped phases are classified by topological \emph{boundary conditions} of the corresponding SymTFT, yielding a classification of (1+1)d symmetric gapped phases in terms \emph{indecomposable module categories} over the symmetry fusion category \cite{Thorngren:2019iar,Komargodski:2020mxz}. In this formulation, vacua are in one-to-one correspondence with simple objects in this module category. In the invertible case, this readily recovers the ordinary classification of gapped phases of matter.
This picture also sheds light on the notion of (generalised) charge for non-invertible symmetries. In general, a charge with respect to a non-invertible symmetry is a \emph{twisted sector local operator}, i.e. a local operator living at the endpoint of topological symmetry line. Multiplets of twisted sector local operators form modules over the so-called \emph{tube algebra} $\Tu(\mc C)$ of the symmetry (spherical) fusion category $\mc C$---a construction with an illustrious history going back to subfactor theory \cite{ocneanu1994chirality,ocneanu2001operator,Izumi2000,MUGER2003159,Neshveyev_Yamashita_2018}---via a process sometimes referred as the `lasso' action of the symmetry lines \cite{Lin:2022dhv,Bhardwaj:2023wzd,Bartsch:2023wvv}. Crucially, the category $\Mod(\Tu(\mc C))$ of modules over the tube algebra is equivalent to the \emph{Drinfel'd centre} $\mc Z(\mc C)$ of the fusion category $\mc C$ . In general, for every symmetry fusion category, the Drinfel'd centre is a \emph{non-degenerate braided fusion category}, and it encodes the bulk operators of the corresponding SymTFT \cite{Turaev:1992hq,Barrett:1993ab}. This classification of the multiplets of twisted sector local operators coincide with that of the topological sectors of theory, i.e. charge sectors of the various twisted Hilbert spaces. This is a task that can be performed very explicitly on the lattice within the anyonic chain framework \cite{Aasen:2020jwb,Lin:2022dhv,Lootens:2022avn}.

In this language, string operators can be thought as pairs of twisted sector local operators connected by a common topological line. We wish to understand what kind of properties string operators---or rather the corresponding twisted sector local operators---need to satisfy in order to be candidate order parameters for a given gapped symmetric phase. More specifically, we are interested in the space of string operators 
that acquire non-vanishing expectation values in the ground state subspace. 
Crystallising earlier ideas, a general statement has recently emerged in the literature \cite{Moradi:2022lqp,Bhardwaj:2023idu}, stating that order parameters form a specific structure within the Drinfel'd centre $\mc Z(\mc C)$ of the symmetry fusion category $\mc C$ known as a \emph{Lagrangian} algebra \cite{Davydov2013}.

Lagrangian algebras in $\mc Z(\mc C)$ are in one-to-one correspondence with indecomposable module categories over $\mc C$. In the SymTFT picture, they encode which collections of bulk line operators \emph{condense} on the corresponding gapped boundary, as well as they way in which they condense \cite{Bais:2008ni,kongBdries,Kong:2013aya}. Because bulk line operators in the Lagrangian algebra terminate on the gapped boundary, one expects the corresponding string operators to  acquire a non-vanishing expectation value in the ground state subspace of the corresponding gapped phase. 
Conversely, the operators that do not condense---those that are \emph{confined} to the boundary---form a fusion category known as the Morita dual $\mc C^*_\mc M$ of the symmetry fusion category $\mc C$ with respect to the module category $\mc M$. These operators produce the \emph{excitations} on top of the ground states of the gapped phase so that the expectation values of the corresponding string operators necessarily vanish. Notably, most recent studies have focused on the meaning of the object underlying the Lagrangian algebra, the role of the multiplication having been largely ignored. However, it is known that a single object in $\mc Z(\mc C)$ can be equipped with distinct Lagrangian algebra structures. Within our context, this implies the existence of string order parameters that cannot be specified solely by the way they transform \cite{PhysRevB.86.125441,Kobayashi:2025ykb,Kobayashi:2025pxs}. The goal of this manuscript is to provide further evidence in support of this conjecture and highlight the role of the multiplication rule. 

\bigskip \noindent
Our focus is on one-dimensional quantum lattice models with \emph{non-anomalous} symmetries. We demonstrate the following results. Given an arbitrary representative of a symmetric gapped phase, one proves that in order for the expectation value of a generic string operator not to vanish in the ground state subspace, the corresponding twisted sector local operators must transform in specific ways. More precisely, we show that any irreducible multiplet of twisted sector local operators with non-vanishing expectation values must be associated with a simple object in $\mc Z(\mc C)$ that is a direct summand of the Lagrangian algebra associated with the gapped phase. We do so by showing that, up to terms that decay exponentially in the length of the string, the expectation values explicitly involve the module of the tube algebra associated with this Lagrangian algebra. Practically, we exploit the \emph{tensor network} approach to the classification of gapped phases \cite{Perez-Garcia:2006nqo,PhysRevB.84.165139,Garre-Rubio:2022uum}, allowing us to pick an arbitrary gapped phase representative in the form of a \emph{block-injective matrix product state} (MPS), and its reformulation in terms of indecomposable module categories over the symmetry fusion category. 

While these derivations are relatively straightforward in the invertible case, non-invertible symmetries present greater computational challenges. Whenever the symmetry is described by the fusion category $\Rep(G)$ of representations of a finite group $G$, one can conveniently exploit the fact that the Lagrangian algebras are known very explicitly, which allows for instance to decompose them into simple objects \cite{DAVYDOV2017149}. Beyond this specific scenario, explicit expressions are largely unavailable. However, one can rely on Davydov’s \emph{full centre} construction \cite{DAVYDOV2010319,Davydov:2011kb}.  Viewing the Lagrangian algebra as the full centre of the relevant module category makes its \emph{half-braiding natural isomorphism} explicit in terms of the module category associator. Since objects of $\mc Z(\mc C)$ are equivalent to $\Tu(\mc C)$-modules, this immediately yields the $\Tu(\mc C)$-module structure of the corresponding object in the same form that appears in our derivations.

This result raises interesting conceptual points. Even when a non-invertible symmetry is fully spontaneously broken, it  can still admit some non-local string operators whose expectation values in the ground state are non-zero, in sharp contrast with the invertible case. For instance, for a $\Rep(G)$ symmetry, this occurs whenever ground states are associated with higher-dimensional representations. Because these ground states possess varying degrees of entanglement, they are not strictly equivalent \cite{Bhardwaj:2023idu,Lootens:2024gfp,Chung:2025ulc}. By construction, the topological lines entering the definition of such string order parameters are preserved in these ground states. Crucially, the collection of all such lines do not form a fusion subcategory as they are typically not even closed under the monoidal structure of the symmetry fusion category. This highlights an important aspect of spontaneous symmetry breaking, namely that the symmetry that is preserved in a given ground state is captured by an \emph{algebra of lines} in the symmetry category. More precisely, the algebra of lines that is preserved in a given ground state is the \emph{internal hom} of the associated simple object in the corresponding module category. Therefore, the symmetry of the theory and the symmetry of the ground states are not on the same level of mathematical abstraction. While different ground states generally preserve non-isomorphic algebras, these algebras are \emph{Morita equivalent}, which is sufficient to ensure that they are all compatible with the same module category labelling the gapped phase \cite{Ostrik:2001xnt}.

In general, for a given gapped phase representative, many irreducible multiplets of twisted sector local operators associated with a direct summand of the corresponding Lagrangian algebra may be defined. Not every multiplet of string operators will acquire an expectation value in the vacuum.  
In particular, this may happen when several Lagrangian algebra structures can be defined on the same underlying object in the Drinfel'd centre. 
For \emph{renormalisation group fixed points} of the gapped phase, we argue that given string operators that yield non-vanishing expectation values, when projecting them into the \emph{infrared}---an operation that the tensor network formalism allows us to perform rather straightforwardly---we obtain a multiplet of twisted sector local operators that fuse according to the multiplication rule of the Lagrangian algebra. In other words, a (generic) string operator whose projection into the infrared does not reproduce the expected multiplication rule would have a vanishing expectation value. We verify this statement very explicitly in the invertible case. We posit that a more systematic study of this aspect beyond renormalisation group fixed points would require operator algebraic techniques \cite{cmp/1104249404,Nachtergaele1996,Ogata:2020ofz,Rubio:2024tsw,Evans:2025msy}.

\bigskip \noindent
\textbf{Organisation of the manuscript:} In sec.~\ref{sec:VecG_phases}, we review the tensor network description of one-dimensional gapped phases with finite invertible symmetry and the construction of renormalisation group fixed point representatives for these phases. In sec.~\ref{sec:VecG_operators}, we define string operators in terms of twisted-sector local operators and explain how their multiplets transform as modules over the tube algebra, or equivalently as objects of the Drinfel'd centre. We then show that only those irreducible multiplets appearing as direct summands of the associated Lagrangian algebra can acquire non-vanishing expectation values. At fixed points, we further show that the order parameters which survive in the infrared fuse according to the multiplication of this Lagrangian algebra. Explicit examples are included to illustrate how this algebraic structure refines the usual notion of string order. Sec.~\ref{sec:NonInv} extends the construction to non-invertible symmetries and demonstrates the analogous result that string order parameters again furnish direct summands of the Lagrangian algebra associated with the relevant module category. We clarify that the symmetry preserved in a given ground state is encoded into the internal hom of the corresponding simple objects in the module category. In sec.~\ref{sec:mixed}, we quickly show how our methods also apply to phases of symmetric mixed states  for invertible symmetries. Various mathematical definitions and results are collected in appendices.

\vfill \noindent
\textbf{Acknowledgements:} C.~D. is grateful to Lea Bottini, Laurens Lootens and Giovanni Rizi for insightful discussions. The authors would like to thank the Isaac Newton Institute for Mathematical Sciences, Cambridge, for support and hospitality during the programme \emph{Quantum field theory with boundaries, impurities, and defects}, where work on this paper was undertaken. This work was supported by \emph{EPSRC grant EP/Z000580/1}. A.~C. gratefully acknowledges the support and hospitality of the Institut des Hautes Études Scientifiques during the completion of part of this work. A.~C. is supported by the National Science Foundation Graduate Research Fellowship under \emph{Grant No. DGE-2036197}. A.~T. acknowledges funding from the EU’s Horizon 2020 research and innovation programme under the \emph{Marie Sk\l{}odowska-Curie grant agreement No. 101273289} as well as from the \emph{UKRI grant EP/Z003342/1}. 

\newpage
\section{Symmetric gapped phases\label{sec:VecG_phases}}

\emph{Given a finite group $G$, we review the classification of one-dimensional $G$-symmetric gapped phases in terms of block-injective tensor networks, which we reinterpret in terms of indecomposable module categories over the fusion category $\Vect_G$. This formalism is then exploited to construct families of renormalisation group fixed points for each gapped phase. Relevant category theoretic concepts are introduced in app~\ref{app:algebras}.}

\subsection{Microscopic Hilbert space \label{sec:VecG_Hilbert}}

Consider a one-dimensional quantum theory with a finite non-anomalous invertible (0-form) symmetry, i.e., a finite `ordinary' symmetry.\footnote{Exploiting the anyonic chain framework developed in ref.~\cite{Feiguin:2006ydp,PhysRevB.87.235120,Buican:2017rxc,Lootens:2021tet,Lootens:2022avn}, we could easily generalise our results to the case of an anomalous symmetry. However, this would force us to define the microscopic Hilbert space in such a way that making contact with the existing literature on order parameters would not be as enlightening.} Let $G$ be the finite group of symmetry transformations. The corresponding topological lines are encoded into the (pivotal unitary) fusion category $\Vect_G$ of finite-dimensional $G$-graded vector spaces and grading preserving linear maps. Recall that representatives of isomorphism classes of simple objects in $\Vect_G$ are provided by the one-dimensional vector spaces $\mathbb C_g = \mathbb C\{e_g\}$, for every $g \in G$, such that ${(\mathbb C_{g})}_{x} = \delta_{g,x} \, \mathbb C$ and $\Hom_{\Vect_G}(\mathbb C_{g}, \mathbb C_{x}) \cong \delta_{g,x} \, \mathbb C$, for every $g,x \in G$. The tensor product of two simple objects is provided by the multiplication rule of $G$ according to $\mathbb C_{g_1} \otimes \mathbb C_{g_2} \cong \mathbb C_{g_1g_2}$, for every $g_1,g_2 \in G$. 

There are many ways to realise a symmetry $G$ (or rather $\Vect_G$) in one-dimensional quantum lattice models  \cite{Lootens:2021tet,Lootens:2022avn,Jones:2024lws}. Throughout this manuscript, we work within the following setup:
Let $\Rep(G)$ be the category of finite-dimensional representation of $G$. Let $V = \mathbb C\{v_d\}_{d=1,\ldots,\dim_\mathbb C V}$ be an object in $\Rep(G)$ with algebra homomorphism $\rho \colon \mathbb C[G]\to \End_\mathbb C(V)$ between the group algebra $\mathbb C[G] = \mathbb C\{e_g\}_{g \in G}$ and the algebra $\End_\mathbb C(V) = V \otimes V^*$ of endomorphisms of $V$.\footnote{Throughout, all the (linear and projective) representations are assumed to be \emph{unitary}.} We write the basis of the dual vector space as $V^* := \Hom(V,\mathbb C)  = \mathbb C\{v^d\}_{d=1,\ldots,\dim_\mathbb C V}$. Pick a finite subset $\Lambda$ of the lattice $\mathbb Z$. To every `site' $\msf i \in \Lambda$, we assign a copy of the vector space $V$ together with a copy of the algebra ${\rm Mat}_\mathbb C(\dim_\mathbb C V)$ of complex $\dim_\mathbb C V \times \dim_\mathbb C V$ matrices. We identify $\mc A_\Lambda := \bigotimes_{\msf i \in \Lambda} {\rm Mat}_\mathbb C(\dim_\mathbb C V)_{\{\msf i\}}$ with the algebra of bounded operators acting on the microscopic Hilbert space $\mc H_\Lambda := \bigotimes_{\msf i \in \Lambda} V_{\{\msf i\}}$. To every topological line labelled by $\mathbb C_g \in \Vect_G$, we then assign the bounded operator $\bigotimes_{\msf i \in \Lambda} \rho(g)_{\{\msf i\}} = \prod_{\msf i \in \Lambda} \rho(g)_\msf i$, where, as customary, $\rho(g)_\msf i \equiv \rho(e_g)_\msf i$\footnote{Throughout, we freely conflate group homomorphisms and the corresponding linear extensions.} denotes the embedding of $\rho(g) \in {\rm Mat}_\mathbb C(\dim_\mathbb C V)_{\{\msf i\}}$ into $\mc A_\Lambda$ by tensoring with the identity operator. Requiring $V \in \Rep(G)$ to be a \emph{faithful} representation guarantees that the symmetry $\Vect_G$ acts faithfully on $\mc H_\Lambda$. Finally, every $G$-equivariant map $\varphi \colon V^{\otimes N} \to V^{\otimes N}$ gives rise to a symmetric operator $\varphi_{\msf i, \ldots,\msf i+N-1} \in \mc A_{\Lambda}$, where the copies of $V$ are here assigned to sites $\msf i,\ldots, \msf i+N-1 \in \Lambda$. 

Any state in $\mc H_\Lambda$ can be expressed as a Matrix Product State (MPS) (see ref.~\cite{RevModPhys.93.045003} for review of tensor network techniques). In general, one can define a \emph{translation invariant} MPS in $\mc H_{\Lambda}$ via a choice of linear map $\theta : M \to V \otimes M$, where $M = \mathbb C\{m_l\}_{l=1,\ldots,\dim_\mathbb C M}$ is some finite-dimensional vector space. It is customary to refer to $V$ as the `physical' vector space and $M$ as a `virtual' vector space. Writing $M^* =  \Hom(M,\mathbb C) = \mathbb C\{m^l\}_{l=1,\ldots,\dim_\mathbb C M}$, we can express the linear map $\theta$ as the following tensor:
\begin{equation}
    \theta \equiv \sum_{l_1,l_2=1}^{\dim_\mathbb C M} \theta^{l_1}_{l_2} \otimes  m_{l_1} \otimes m^{l_2} \equiv \sum_{l_1,l_2 = 1}^{\dim_\mathbb C M} \sum_{d=1}^{\dim_\mathbb C V} (\theta^{l_1}_{l_2})^d \; v_d \otimes m_{l_1} \otimes m^{l_2}
    \, ,   
\end{equation}
such that $\theta^{l_1}_{l_2} \in V$, for every $l_1,l_2 \in \{1,\ldots,\dim_
\mathbb C M\}$, and $(\theta^{l_1}_{l_1})^d \in \mathbb C$, for every $d \in \{1,\ldots,\dim_\mathbb C V\}$. As such, $\theta$ is typically referred to as the `MPS tensor' in the literature.
As is customary, we shall use graphical notation. We depict the MPS tensor $\theta$ as\footnote{Typically, since one only deals with one MPS tensor at a time, we can omit the corresponding label for visual clarity.}
\begin{equation}
    \theta \equiv \!\!\! \MPST{\theta}{}{}{}{} \!\!\!
    \equiv 
    \sum_{l_1,l_2 = 1}^{\dim_\mathbb C M}
    \sum_{d=1}^{\dim_\mathbb C V} \! \MPST{\theta}{}{l_1}{l_2}{d} v_d \otimes m_{l_1} \otimes m^{l_2} \, .
\end{equation}
Assuming \emph{closed periodic} boundary conditions, the MPS $\Psi(\theta) \in \mc H_\Lambda$ associated with the map $\theta$ reads
\begin{equation}
    \Psi(\theta) := \sum_{l_1,\ldots,l_{|\Lambda|}} \theta^{l_1}_{l_2} \otimes \theta^{l_2}_{l_3} \otimes \cdots \otimes \theta^{l_{|\Lambda|}}_{l_1} \, .
\end{equation}
Alternatively, defining $\theta^d := \sum_{l_1,l_2=1}^{\dim_\mathbb C M} (\theta^{l_1}_{l_2})^d \; m_{l_1} \otimes m^{l_2} \in \End_\mathbb C(M)$, for every $d \in \{1,\ldots,\dim_\mathbb C V\}$, the same MPS can be written as 
\begin{equation}
    \Psi(\theta) = \sum_{d_1,\ldots,d_{|\Lambda|}} {\rm tr}_M \big[\theta^{d_1} \, \theta^{d_2} \cdots \theta^{d_{|\Lambda|}} \big] \; v_{d_1} \otimes v_{d_2} \otimes \cdots \otimes v_{d_{|\Lambda|}} \, . 
\end{equation}
Graphically,
\begin{equation}
    \Psi(\theta) \equiv \raisebox{-5pt}{\MPS{\theta}} \! .
\end{equation}
A special class of MPSs consists of \emph{injective} MPSs. An MPS is said to be injective if the linear map $\End_\mathbb C(M) \to V, m_{l_1} \otimes m^{l_2} \mapsto \theta^{l_1}_{l_2}$ is injective;\footnote{In words, the tensor $\theta$, interpreted as a map from the virtual space to the physical space, is injective.} equivalently, if the set $\{\theta^d\}_{d=1,\ldots,\dim_\mathbb C V}$ spans the matrix algebra $\text{Mat}_\mathbb C(\dim_\mathbb C M)$. An important object is the \emph{transfer matrix} $\mathbb E_\theta$ of the MPS tensor $\theta$, which is defined as
\begin{equation}
\begin{split}
    \mathbb E_\theta := \sum_{d=1}^{\dim_\mathbb C V} \bar \theta^d \otimes \theta^d 
    &\equiv 
    \TMat{\theta} : \End_\mathbb C(M)^* \to \End_\mathbb C(M)^*
    \\
    &\equiv
    \sum_{l_1,\ldots,l_4=1}^{\dim_\mathbb C M} (\mathbb E_\theta)^{l_3l_1}_{l_4l_2} \; m^{l_3} \otimes m_{l_1} \otimes m_{l_4} \otimes m^{l_2} \, ,
\end{split}
\end{equation}
where $(\mathbb E_\theta)^{l_3l_1}_{l_4l_2} := \sum_{d=1}^{\dim_\mathbb C V} (\theta^{l_1}_{l_2})^d \, \overline{(\theta^{l_3}_{l_4})^d}$.
The transfer matrix of any injective MPS tensor admits a single and non-degenerate \emph{right} eigenvalue $\xi \in \mathbb C$ of largest weight. Graphically,
\begin{equation}\label{transferfixedpoint}
    \RightTMatFixed{1}{\theta}{\Xi} = \xi \cdot \RightTMatFixed{2}{\theta}{\Xi}
\end{equation}
where $\Xi \in \End_\mathbb C(M)^*$ is the unique right eigenvector with eigenvalue $\xi$. Without loss of generality, one can normalise the MPS tensor $\theta$ so that $\xi=1$, making $\Xi$ a \emph{fixed point} of $\mathbb E_\theta$. Similarly, one can consider \emph{left} eigenvectors of the transfer matrix, where again $1$ is the unique peripheral eigenvalue and there is a unique left fixed point. Via basis transformations, it is always possible to bring $\theta$ into its \emph{left canonical form} so that this fixed point is the identity map in $\End_\mathbb C(M)$ \cite{SCHOLLWOCK201196}. Graphically,
\begin{equation}
    \LeftTMatFixed{1}{\theta}{{\rm id}_M \;\;} =  \LeftTMatFixed{2}{\theta}{{\rm id}_M \;\;} \, .
\end{equation}
Below, we always assume such a left canonical form. Since all other eigenvalues have magnitude less than $1$, the product of $N$ transfer matrices tends toward a projector in the limit of large $N$:
\begin{equation}
    \label{eq:VecG_transferprojector}
    \productTMat{} = \smallTMatRG{}{\Xi}{\;\;\;{\rm id}_M}
    + O(e^{N \tilde \xi}) \xrightarrow{N \to \infty}
    \smallTMatRG{}{\Xi}{\;\;\;{\rm id}_M} \, ,
\end{equation}
where $\tilde \xi$ denotes the eigenvalue of second highest magnitude. This implies that the right fixed point $\Xi$ is normalized as $\tr_M[\Xi]=1$ because $\Xi=\lim_{N\rightarrow\infty}\mathbb E_\theta^N\Xi=\tr_M[\Xi]\,\Xi$. A special class of MPS are renormalisation group fixed points (see sec.~\ref{sec:VecG_RG}), defined as those for which the transfer matrix $\mathbb E$ is a projector. For such states, one can set $\Xi={\rm id}_M$ by a local change of basis on the physical sites; when we discuss fixed point states in sec.~\ref{sec:VecG_RG}, we will assume we have done this.

Crucial to our analysis is the statement that any injective MPS can be realised as the unique ground state of a gapped symmetric local parent Hamiltonian. Conversely, any ground state of a one-dimensional local gapped quantum lattice model can be efficiently approximated by an MPS whose \emph{bond dimension} $\dim_\mathbb C M$ does not depend on the system size \cite{Perez-Garcia:2006nqo,Hastings:2007iok,PhysRevB.84.165139}. Furthermore, if there is a single vacuum in this gapped phase, the MPS can always be chosen to be injective by blocking sites; otherwise gapped ground states need to satisfy additional \emph{cluster decomposition properties} in order to be efficiently approximated by an injective MPS, so as to rule out for instance \emph{cat-like} states \cite{Perez-Garcia:2006nqo}. 

It is always possible to bring a periodic\footnote{On twisted boundary conditions, off-block-diagonal components of the MPS tensor cannot be eliminated, so the general form is more complicated; we will not consider this case here.} MPS into a \emph{block-diagonal} form, whereby $M = \bigoplus_a M_a$ and $\theta^d$ is a block-diagonal matrix $\bigoplus_a \theta^d_{a}$, where $\theta^d_{a} \in \End_\mathbb C(M_a)$ for every $d \in \{1,\ldots,\dim_\mathbb C V\}$ \cite{Perez-Garcia:2006nqo}. Graphically,
\begin{equation}
    \label{eq:VecG_blockMPS}
    \MPST{\theta}{}{}{}{} \!\!\! = \bigoplus_{a} \!\!\! \MPST{\theta}{a}{}{}{}  \!\!\! \equiv \bigoplus_a \, \theta_a \, ,
\end{equation}
where we introduced $\theta_a : M_a \to V \otimes M_a$ such that $(\theta_a)^d := \theta^d_a \in \End_\mathbb C(M_a)$, for every $d \in \{1,\ldots,\dim_\mathbb C V\}$.
By blocking sites,\footnote{Without blocking sites, the blocks $\theta_a$ can be taken to be irreducible (i.e. lacking invariant subspaces) but not injective. Some irreducible blocks may be \emph{normal}, meaning that they become injective upon blocking. Other irreducible blocks may fail to be normal, indicating spontaneous symmetry breaking of translation symmetry; such blocks split into multiple irreducible ones upon blocking. We always assume we have blocked sites, so that our tensors are block-injective.} we may further assume that each $\theta_a$ is injective. Given that two injective MPSs yield either proportional or orthogonal states in the infinite volume limit \cite{Cirac:2016iqe}, one focuses in the following on block-injective MPSs whose blocks are injective and orthogonal to each other. This implies that the mixed transfer matrix
\begin{equation}\label{mixedtransfer}
    \mathbb E_{\theta_{a_1}\theta_{a_2}}:=\sum_{d=1}^{\dim_\mathbb C V} \bar \theta_{a_1}^d \otimes \theta_{a_2}^d 
\end{equation}
has spectral radius less than $1$ if $a_1\ne a_2$. The full transfer matrix $\mathbb E_\theta$ decomposes as $\mathbb E_\theta=\bigoplus_{a_1,a_2}\mathbb E_{\theta_{a_1}\theta_{a_2}}$ and has a space of right fixed points spanned by the fixed points $\Xi_a$ of $\mathbb E_{\theta_a}=\mathbb E_{\theta_a \theta_a}$.

The same way states in $\mc H_\Lambda$ can be expressed as MPSs, the symmetry operator $\prod_{\msf i \in \Lambda} \rho(g)_\msf i$ associated with the simple object $\mathbb C_g \in \Vect_G$ can be realised as a Matrix Product Operator (MPO) of the form\footnote{Even though the symmetry operator $\prod_{\msf i \in \Lambda} \rho(g)_\msf i = \bigotimes_{\msf i \in \Lambda} \rho(g)_{\{\msf i\}}$ is a tensor product operator, it is useful to treat it as an MPO with a one-dimensional bond space. Moreover, it offers visual support for the corresponding topological line.}
\begin{equation}
    \label{eq:VecG_MPO}
    \MPO{}{g} 
\end{equation}
in terms of a tensor
\begin{equation}
    \label{eq:VecG_MPOT}
    \MPOT{}{g}{}{}{}{} \!\!\! 
    \equiv \sum_{d_1,d_2 = 1}^{\dim_\mathbb C V} \!\!\!\!\! \MPOT{}{g}{g}{g}{d_1}{d_2} v_{d_1} \otimes v^{d_2} \otimes e_g \otimes e^g \equiv \sum_{d_1,d_2 = 1}^{\dim_\mathbb C V}\rho(g)_{d_2}^{d_1} \; v_{d_1} \otimes v^{d_2} \otimes e_g \otimes e^g \, ,
\end{equation}
where $\Hom(\mathbb C_g,\mathbb C) = \mathbb C\{e^g\}$. All these symmetry operators can be organised into a block-injective MPO
\begin{equation}
    \MPOT{}{}{}{}{}{} \!\!\!= \bigoplus_{g \in G} \!\!\! \MPOT{}{g}{}{}{}{} \!\! .
\end{equation}
The fact that $\rho : \mathbb C[G] \to \End_\mathbb C(V)$ is an algebra homomorphism yields the identity
\begin{equation}
    \label{eq:VecG_fusionMPO}
    \fusionMPO{1}{g_1}{g_2}{}{}{}{}{}{} = \sum_{g_3 \in G} \; \fusionMPO{2}{g_1}{g_2}{g_3}{}{}{}{}{} \, ,
\end{equation}
where, for every $g_1,g_2,g_3 \in G$, we introduced tensors $\mathbb C_{g_1} \otimes \mathbb C_{g_2} \to \mathbb C_{g_3}$ and $\mathbb C_{g_3} \to \mathbb C_{g_1} \otimes \mathbb C_{g_2}$ given by
\begin{equation}
    \fusionT{1}{}{g_1}{g_2}{g_3}{}{}{} \!\!\! = \delta_{g_3,g_1g_2} \; e_{g_1g_2} \otimes e^{g_1} \otimes e^{g_2} 
    \q \text{and} \q
    \fusionT{2}{}{g_1}{g_2}{g_3}{}{}{} \!\!\! = \delta_{g_3,g_1g_2} \; e_{g_1} \otimes e_{g_2} \otimes e^{g_3} \, ,
\end{equation}
respectively, such that
\begin{equation}
    \label{eq:VecG_orthoFusion}
    \orthoFusion  = \mathbb I_{\mathbb C_{g_1g_2}}\, .
\end{equation}
Physically, we interpret this operation as a consequence of the topological invariance of the symmetry lines, allowing us to fuse them locally. Clearly, we have the following equality
\begin{equation}
    \label{eq:VecG_Fmove}
    \fusionAssoc{1} \, = \fusionAssoc{2} \, ,
\end{equation}
for every $g_1,g_2,g_3 \in G$, which is the manifestation that the symmetry $G$ is \emph{non-anomalous} and that the associator of $\Vect_G$ evaluates to the identity.

\subsection{Symmetric states\label{sec:VecG_MPS}}

Let us now review the classification of one-dimensional $G$-symmetric gapped phases in terms of block-injective MPSs \cite{PhysRevB.84.165139,PhysRevB.83.035107,Garre-Rubio:2022uum}. Consider a block-injective tensor $\theta  = \bigoplus_a \theta_a$ of the form \eqref{eq:VecG_blockMPS} and the corresponding MPS. Each block labelled by some index $a$ produces an MPS $\Psi(\theta_a)$ in $\mc H_\Lambda$. The block-injective MPS is said to be $G$-symmetric if the subspace $\mathbb C\{\Psi(\theta_a)\}_a$ is closed under the action of $\prod_{\msf i \in \Lambda}\rho(g)_\msf i$, for every $g \in G$ and any system size $|\Lambda|$. Suppose $\Psi(\theta)$ is $G$-symmetric. It turns out that the set of labels $a$ that index the injective blocks is a $G$-set, i.e. it forms a \emph{permutation representation} of $G$ \cite{PhysRevB.84.165139,PhysRevB.83.035107,Garre-Rubio:2022uum}. Explicitly, this means that acting with $\prod_{\msf i \in \Lambda} \rho(g)_\msf i$ on $\Psi(\theta_a)$ for any $a$ results in an injective MPS proportional to $\Psi(\theta_{g(a)})$, where $g(a)$ refers to the image of $a$ under the action of $g$, for every $g \in G$. The fact that this is true for any system size $|\Lambda|$ implies by the \emph{fundamental theorem} of MPS the existence of linear maps $\phi^{g,a} : \mathbb C_g \otimes M_a \to M_{g(a)}$ and $\bar \phi^{g,a} : M_{g(a)} \to \mathbb C_g \otimes M_a$ such that \cite{PhysRevB.84.165139,PhysRevB.83.035107,Garre-Rubio:2022uum}
\begin{equation}
    \label{eq:VecG_localAction}
    \raisebox{16pt}{\actionMPO{1}{}{a}{}{}{g}{}{}} 
    = \raisebox{10pt}{\actionMPO{2}{}{a}{g(a)}{}{g}{\phi^{g,a}}{\bar \phi^{g,a}}}
\end{equation}
and
\begin{equation}
    \label{eq:VecG_orthoAction}
    \orthoAction{a}{g}{g(a)}{g(a)}{\phi^{g,a}}{\bar \phi^{g,a}}  = \mathbb I_{M_{g(a)}} \, ,
\end{equation}
for every label $a$ and $g \in G$. Physically, one can also interpret this local action as a consequence of the topological invariance of the symmetry lines. These relations imply that the fixed points of the MPS transfer matrix \eqref{transferfixedpoint} satisfy
\begin{equation}
    \actionSymFixedPoints{1}{\phi^{g,a}}{\;\;\Xi_a}{g}{a}{g(a)} \!\! = 
    \actionSymFixedPoints{2}{\phi^{g,a}}{\;\;\Xi_a}{g}{a}{g(a)} \!\! = 
    \actionSymFixedPoints{3}{\phi^{g,a}}{\;\;\Xi_a}{g}{a}{g(a)} .
\end{equation}
Since $\Xi_{g(a)}$ is the unique fixed point of $\mathbb E_{\theta_{g(a)}}$, we conclude
\begin{equation}
    \label{eq:VecG_actionFixedPoints}
    \actionSymFixedPoints{1}{\phi^{g,a}}{\;\;\Xi_a}{g}{a}{g(a)} \!\!\! = 
    \actionSymFixedPoints{4}{\phi^{g,a}}{\;\;\Xi_{g(a)}}{g}{a}{g(a)} .
\end{equation}
Now, consider the actions of two symmetry lines associated with simple objects $\mathbb C_{g_1}$ and $\mathbb C_{g_2}$ in $\Vect_G$, respectively. Instead of successively acting with both lines, one can first (locally) fuse them before (locally) acting with the resulting line associated with $\mathbb C_{g_1g_2} \in \Vect_G$. Given that $g_1(g_2(a)) = (g_1g_2)(a)$, for every label $a$ and $g_1,g_2 \in G$, we have the following relation:\footnote{Note that some labels are occasionally omitted for visual clarity; however, these can always be unambiguously deduced from those remaining.}
\begin{equation}
    \assocAction{1} \; = \; \assocAction{2} \, .
\end{equation}
From orthogonality conditions \eqref{eq:VecG_orthoFusion} and \eqref{eq:VecG_orthoAction} follow the existence of complex numbers $\alpha^{\act}(g_1,g_2)(a) \in \mathbb C$ satisfying
\begin{equation}
    \label{eq:VecG_moduleAssoc}
     \moduleAssoc{2} \;=\; \alpha^{\act}(g_1,g_2)(a) \moduleAssoc{1} \, .
\end{equation}
Explicitly, these complex numbers evaluate to 
\begin{equation}
    \Fsym{2} \!\!\! = \alpha^{\act}(g_1,g_2)(a) \; \mathbb I_{M_{g_1(g_2(a))}} \, .
\end{equation}
Now, consider the actions of three symmetry lines associated with simple objects $\mathbb C_{g_1}$, $\mathbb C_{g_2}$ and $\mathbb C_{g_3}$ in $\Vect_G$, respectively. Instead of successively acting with the three lines, one can instead fuse the first two, fuse the resulting line associated with $\mathbb C_{g_1g_2} \in \Vect_G$ with the third one, before acting with the resulting line associated with $\mathbb C_{g_1 g_2 g_3} \in \Vect_G$. Using eq.~\eqref{eq:VecG_Fmove} and eq.~\eqref{eq:VecG_moduleAssoc}, one can relate the resulting combinations of lines in two different ways, which requires that the equation
\begin{equation}
    \label{eq:VecG_modPent}
    \alpha^{\act}(g_2,g_3)(a) \, \alpha^{\act}(g_1,g_2g_3)(a) = \alpha^{\act}(g_1g_2,g_3)(a) \, \alpha^{\act}(g_1,g_2)(g_3(a)) 
\end{equation}
holds for every label $a$ and $g_1,g_2,g_3 \in G$. Note that the linear maps $\{\phi^{g,a}\}_{g \in G,a}$ are only defined up to multiplicative phase factors so that we actually have an equivalence class of complex numbers $\alpha^{\act}(g_1,g_2)(a) \in \mathbb C$ satisfying eq.~\eqref{eq:VecG_modPent} (see below).

It turns out that any collection $\{\phi^{g,a}\}_{g\in G,a}$ of linear maps satisfying \eqref{eq:VecG_moduleAssoc} such that eq.~\eqref{eq:VecG_modPent} holds provides the input datum of a ($\mathbb C$-linear) \emph{module category} over $\Vect_G$ (see app.~\ref{app:algebras} for definition). Moreover, by block-injectivity of $\Psi(\theta)$, this module category is \emph{finite} and \emph{semisimple} so that the set of labels that index the injective blocks provides the set of isomorphism classes of simple objects of the module category. Therefore, the classification of $G$-symmetric gapped phases in terms of $G$-symmetric block-injective MPSs reduces to the classification of \emph{indecomposable} (finite semisimple) module categories over $\Vect_G$. Note that this is compatible with the hypothesis that gapped phases are described at long distances by topological quantum field theories (TFQTs). Indeed, 2d TQFTs that can be coupled to topological lines in $\Vect_G$ are classified by indecomposable (finite semisimple) module categories over $\Vect_G$ \cite{Turaev2010Homotopy}. 

\subsection{Classification \label{sec:VecG_classification}}

Let us now recall the classification of indecomposable module categories over $\Vect_G$. As reviewed in app.~\ref{app:algebras}, module categories over $\Vect_G$ can be constructed from algebra objects in $\Vect_G$. By definition, an algebra in $\Vect_G$ is a $G$-graded algebra, i.e., a $G$-graded vector space $A = \bigoplus_{g \in G}A_g$ that is equipped with a grading preserving multiplication so that $A_{g_1}A_{g_2} \subseteq A_{g_1g_2}$, for every $g_1,g_2 \in G$. For instance, let $H \leq G$ be a subgroup and $\beta$ be a normalised representative of a cohomology class in $H^2(H,\rU(1))$. The twisted group algebra $\mathbb C[H]^\beta$ is $G$-graded, as a result of $H$ being a subgroup of $G$. 
We denote the corresponding algebra object by $A[H,\beta]$. 
For every algebra $\End(V)$ of endomorphisms of a finite-dimensional complex vector space $V$, $\mathbb C[H]^\beta \otimes \End(V)$ also carries the structure of an algebra object in $\Vect_G$, which happens to be \emph{Morita equivalent} to $A[H,\beta]$.
Let $\mc M(H,\beta)$ be the category $\Mod_{\Vect_G}(A[H,\beta])$ of right $A[H,\beta]$-modules in $\Vect_G$. By construction, $\mc M(H,\beta)$ carries the structure of a left $\Vect_G$-module category. 

Concretely, $\mc M(H,\beta)$ is equivalent to the indecomposable finite semisimple $\Vect_G$-module category of $G/H$-graded vector spaces. Let $\{r_a\}_{a=1,\ldots, (G:H)}$ be a choice of representatives in $G$ for the set of left cosets in $G/H$ such that $r_1 = 1_G$. Representatives of isomorphism classes of simple objects in $\mc M(H,\beta)$ are provided by the one-dimensional vector spaces $\{\mathbb C_{r_aH}\}_{a=1,\ldots,(G:H)}$ such that ${(\mathbb C_{r_{a_1}H})}_{r_{a_2}H} = \delta_{a_1,a_2} \mathbb C$.  For every $g \in G$ and $r_aH \in G/H$, there is a unique $g(a) \in \{1,\ldots,(G:H)\}$ such that $g \cdot r_aH = r_{g(a)}H$. 
The simple object $\mathbb C_g \in \Vect_G$ acts on $M = \bigoplus_{a=1}^{(G:H)} M_{r_aH}$ via the action of $G$ on cosets as $\mathbb C_g \act M = \bigoplus_{a=1}^{(G:H)} M_{r_{g^{-1}(a)}H}$. In particular, we have $\mathbb C_g \act \mathbb C_{r_aH} = \mathbb C_{g \cdot r_aH} = \mathbb C_{r_{g(a)}H}$, for every $g \in G$ and $a \in \{1,\ldots,(G:H)\}$. This provides the action bifunctor $\act : \Vect_G \times \mc M(H,\beta) \to \mc M(H,\beta)$. Constructing the module associator $\alpha^{\act}$ relies on a corollary of \emph{Shapiro's lemma} stating that
\begin{equation}
    H^2(H,\rU(1)) \cong H^2(G,{\rm Fun}(G/H,\rU(1))) \, ,
\end{equation}
where the right $G$-module structure of the abelian group ${\rm Fun}(G/H,\rU(1))$ is given by $(f \cdot g)(-) = f(g \cdot -)$, for every $f \in {\rm Fun}(G/H,\rU(1))$ and $g \in G$.  Indeed, since for each $g \in G$ and $r_aH \in G/H$ there is a unique $g(a) \in \{1,\ldots,(G:H)\}$ such that $g \cdot r_aH = r_{g(a)}H$, there is a unique group element $h_{g,a} \in H$ such that $gr_a = r_{g(a)}h_{g,a}$. Associativity of the multiplication in $G$ imposes that 
\begin{equation}
    \label{eq:hgi_multiplication}
    h_{g_1,g_2(a)} \, h_{g_2,a} = h_{g_1g_2,a} \, ,
\end{equation}
for every $g_1,g_2 \in G$ and $r_aH \in G/H$.
The module associator $\alpha^{\act}$ is specified by isomorphisms
\begin{equation}
    \alpha^{\act}(g_1,g_2)(r_aH) \cdot {\rm id}_{\mathbb C_{(g_1g_2) \cdot r_a H}} : (\mathbb C_{g_1} \otimes \mathbb C_{g_2}) \act \mathbb C_{r_aH} 
    \xrightarrow{\simeq} 
    \mathbb C_{g_1} \act (\mathbb C_{g_2} \act \mathbb C_{r_aH}) \, ,
\end{equation}
where the 2-cochain $\alpha^{\act} \in C^2(G,{\rm Fun}(G/H,\rU(1)))$ is defined via
\begin{equation}
    \label{eq:VecG_ModAssoc}
    \alpha^{\act}(g_1,g_2)(r_aH) := \beta(h_{g_1,g_2(a)},h_{g_2,a}) \, ,
\end{equation}
for every $g_1,g_2 \in G$ and $r_{a}H \in G/H$. It then follows from the 2-cocycle satisfied by $\beta$ that $\alpha^{\act} \in Z^2(G,{\rm Fun}(G/H,\rU(1)))$, i.e.
\begin{equation}
    \alpha^{\act}(g_2,g_3)(r_aH) \, \alpha^{\act}(g_1,g_2g_3)(r_aH) = \alpha^{\act}(g_1g_2,g_3)(r_aH) \, \alpha^{\act}(g_1,g_2)(g_3 \cdot r_aH) \, ,
\end{equation}
    for every $g_1,g_2,g_3 \in G$ and $r_aH \in G/H$, which in turn ensures that $\alpha^{\act}$ fulfils the required \emph{pentagon axiom} (see app.~\ref{app:algebras}). This precisely amounts to eq.~\eqref{eq:VecG_modPent} in the case where the set of labels that index the injective blocks of the block-injective MPS is given by $G/H$. 

It turns out that every indecomposable finite semisimple $\Vect_G$-module category is of the form $\mc M(H,\beta) = \Mod_{\Vect_G}(\mathbb C[H]^\beta)$ for some subgroup $H \leq G$ and 2-cocycle $\beta$ in $Z^2(H,\rU(1))$ \cite{2002math......2130O,Naidu23102007, etingof2016tensor}. Conversely, for every simple object $\mathbb C_{r_aH}$ in $\mc M(H,\beta)$, one can verify that the \emph{internal hom} $\underline{\Hom}(\mathbb C_{r_aH}, \mathbb C_{r_aH})$ is provided by $\bigoplus_{h \in H}\mathbb C_{r_ahr_a^{-1}}$, as an object in $\Vect_G$ (see app.~\ref{app:algebras} for definition). The multiplication of the algebra object $\underline{\Hom}(\mathbb C_{r_aH}, \mathbb C_{r_aH})$ is that of $\mathbb C[r_aHr_a^{-1}]$ twisted by the 2-cocycle $\beta^{r_a}$ such that $ \beta^{r_a}(r_ah_1r_a^{-1},r_ah_2r_a^{-1}) = \beta(h_1,h_2) =  \alpha^{\act}(r_ah_1r_a^{-1},r_ah_2r_a^{-1})(r_aH)$, for every $h_1,h_2 \in H$. This algebra is Morita equivalent to $\mathbb C[H]^\beta$, as expected. 

Naturally, one recovers the well-known classification of $G$-symmetric one-dimensional gapped phases \cite{PhysRevB.83.035107,PhysRevB.84.165139}. In particular, $\mc M(\{1_G\},1) = \Vect_G$ corresponds to the phase where the whole symmetry is spontaneously broken, while $\mc M(G,\beta)$ labels a symmetry protected topological (SPT) phase whose edge modes transform according to $\beta$-projective representations of the group. In other words, no topological lines are preserved in the ground state subspace of the gapped phase labelled by $\mc M(\{1_G\},1)$, whereas the unique ground state of the gapped phase labelled by $\mc M(G,\beta)$ preserves an algebra object of topological lines that is Morita equivalent to $A[G,\beta]$. More generally, we state that $\mc M(H,\beta)$ labels the spontaneously symmetry broken phase where each ground state in the  $(G : H)$-dimensional ground state subspace preserves an algebra object of topological lines that is Morita equivalent to $A[H,\beta]$. In particular, we wish to emphasise that we should not think of the symmetry $\Vect_G$ as spontaneously broken to a subsymmetry---or to any symmetry, for that matter---that would be encoded in a fusion category. This does not involve the appropriate level of abstraction, as the amount of symmetry preserved in any ground state is rather captured by an algebra object in $\Vect_G$.
This rephrasing of the usual classification will turn out to be crucial in sec.~\ref{sec:NonInv}.

\subsection{Renormalisation group fixed points\label{sec:VecG_RG}}

Let $H$ be a subgroup of $G$ and $\beta$ a normalised representative of a cohomology class in $H^2(H,\rU(1))$. Suppose we want to construct ground states at some renormalisation group fixed point of the $\Vect_G$-symmetric gapped phase labelled by the $\Vect_G$-module category $\mc M(H,\beta)$. We already know that $\mc M(H,\beta)$ is equivalent to the category of right modules in $\Vect_G$ over an algebra object in $\Vect_G$ that is Morita equivalent to $A[H,\beta]$. We could use this information to build the renormalisation group fixed points as in ref.~\cite{Bhardwaj:2024kvy}; however this choice would not be compatible with the choice of microscopic Hilbert space we made in sec.~\ref{sec:VecG_Hilbert}. Indeed, recall that in this manuscript one realises the symmetry $\Vect_G$ via a choice of object in $\Rep(G)$. Implicitly, we are employing the result that the \emph{Morita dual} $(\Vect_G)^*_\Vect := \Fun_{\Vect_G}(\Vect,\Vect)$ of $\Vect_G$ with respect to the $\Vect_G$-module category $\Vect$ is equivalent to $\Rep(G)$ \cite{Lootens:2021tet} (see app.~\ref{app:algebras} for definitions). It follows that indecomposable left $\Vect_G$-module $\mc M(H,\beta)$ can also be realised as the category of left modules (in $\Vect$) over an indecomposable finite separable algebra object in $\Rep(G)$ \cite{Mombelli2010}. 
This is the realisation that is compatible with our choice of microscopic Hilbert space \cite{Inamura:2021szw,Delcamp:2024wiv}.
To identify a suitable algebra object in $\Rep(G)$, one must first invoke the 2-equivalence (see app.~\ref{app:algebras})
\begin{equation}
    \begin{array}{ccc}
        \Fun_{\Vect_{G}}(\Vect,-) \colon \Mod(\Vect_G) \to \Mod(\Rep(G)^{\rm op}) \, ,
    \end{array}
\end{equation}
which establishes the one-to-one correspondence between $\Vect_G$-module categories and $\Rep(G)$-module categories.
In particular, one can check that $\Fun_{\Vect_G}(\Vect,\mc M(H,\beta))$ is equivalent to the category $\Rep^\beta(H)$ of finite-dimensional $\beta$-projective representations of $H$ (see ref.~\cite{Delcamp:2024cfp} for an explicit derivation), where the $\Rep(G)$-module structure of $\Rep^\beta(H)$ is simply given by the monoidal structure of $\Vect$ via the \emph{restriction functor} $\Res_H^G : \Rep(G) \to \Rep(H)$. It follows that any algebra object $A$ in $\Rep(G)$ such that $\Mod_{\Rep(G)}(A)$ is equivalent to $\Rep^\beta(H)$, as a $\Rep(G)$-module category, is suitable. 

Following ref.~\cite{Ostrik:2001xnt,DAVYDOV2010319}, let us construct such an algebra given a pair $(H,\beta)$. First of all, recall that algebra objects in $\Rep(G)$ are \emph{$G$-equivariant} algebras, i.e., algebras (in $\Vect$) with a (left) $G$-action that is compatible with the multiplication. Let $U$ be a simple object in $\Rep^\beta(H)$ and let $\pi : \mathbb C[H]^\beta \to \End_\mathbb C(U)$ denotes the corresponding algebra homomorphism between the twisted group algebra $\mathbb C[H]^\beta = \mathbb C\{e_h\}_{h \in H}$ and the algebra $\End_\mathbb C(U)$ of endomorphisms of $U$. The algebra homomorphism equips the algebra $\End_\mathbb C(U)$ with an $H$-equivariant structure, whereby $h \cdot f := \pi(e_h)f\pi(e_h)^{-1} \equiv \pi(h)f \pi(h)^{-1}$, for every $h \in H$ and $f \in \End_\mathbb C(U)$. We promote this algebra object in $\Rep(H)$ to an algebra object in $\Rep(G)$ via the \emph{induction functor} $\Ind_H^G : \Rep(H) \to \Rep(G)$:
\begin{equation}
    \label{eq:defAlgRepG}
    A(H,\beta,U) \equiv \Ind_H^G(\End_\mathbb C(U)) 
    := \mathbb C\big\{\sigma \colon G \to \End_\mathbb C(U) \, \big| \, \sigma(gh^{-1}) = h \cdot \sigma(g) \big\}_{g \in G , \, h \in H} \, .
\end{equation}
The algebraic structure of $A(H,\beta,U)$ is provided by \emph{pointwise} multiplication; its $G$-equivariant structure is given by $(g \cdot \sigma)(-) = \sigma(g^{-1}-)$, for every $\sigma \in A(H,\beta,U)$ and $g \in G$. Whenever $\beta=1$, one can always choose $U$ to be the trivial representation, at which point $A(H,1)$ boils down to the algebra $\mathbb C(G/H)$ of functions $G/H \to \mathbb C$ with pointwise multiplication. As an object in $\Rep(G)$, $\mathbb C(G/H)$ is isomorphic to the induced representation $\Ind_H^G(\mathbb C)$.

Let us examine the $G$-equivariant structure of the algebra $A(H,\beta,U)$ in more detail. 
As in sec.~\ref{sec:VecG_classification}, let $\{r_a\}_{a=1,\ldots, (G:H)}$ denote a set of representatives for the left cosets in $G/H$. For each $g \in G$, there is a unique $a \in \{1, \ldots,(G:H)\}$ such that $g \in r_aH$; therefore, there is a unique $h \in H$ such that $g = r_a h$. Given $f \in \End_\mathbb C (U)$, define the function $\sigma_{a,f} : G \to \End_\mathbb C(U)$ via
\begin{equation}
    \sigma_{a,f}(g) =
    \begin{cases}
        \pi(h)^{-1}f \pi(h) \q &\text{if $g \equiv r_ah \in r_aH$} \, ,
        \\
        0 &\text{otherwise} \, ,
    \end{cases}
\end{equation}
for every $g \in G$. By definition, $\sigma_{a,f} \in A(H,\beta,U)$, for every $a \in \{1,\ldots,(G:H)\}$ and $f \in \End_\mathbb C(U)$. For a fixed $a \in \{1,\ldots,(G:H)\}$, consider the  vector space $\mathbb C\{\sigma_{a,f} \, | \, f \in \End_\mathbb C(U) \}$. Clearly, it defines a subalgebra of $A(H,\beta,U)$ which is isomorphic to $\End_\mathbb C(U)$. As a matter of fact, we have 
\begin{equation}
    \mathbb C
    \big\{ \sigma_{a,f} \, \big| \, f \in \End_\mathbb C(U) \big\} \cong 
    \mathbb C \big\{\sigma \in A(H,\beta,U) \, \big| \, \sigma(g) = 0 \; \text{if} \; g \notin r_aH \big\} =: A(H,\beta,U)_{(a)} \, .
\end{equation}
Moreover, for every function $\sigma \in A(H,\beta,U)$ we have the following decomposition:
\begin{equation}
    \sigma = \sum_{a=1}^{(G:H)}\sigma_{a,f_a} \, ,
\end{equation}
where $f_a := \sigma(r_a) \in \End_\mathbb C(U)$.
One can further check that this decomposition is actually unique, which in turn induces the following decomposition of $A(H,\beta,U)$, as an algebra in $\Vect$:
\begin{equation}
    \label{eq:RepGAlgebraDecomp}
    A(H,\beta,U) \cong \bigoplus_{a = 1}^{(G:H)} A(H,\beta,U)_{(a)} \, .
\end{equation}
Therefore, it is enough to examine the action of $G$ on functions $\sigma_{a,f}$, for every $a \in \{1,\ldots,(G:H)\}$ and $f \in \End_\mathbb C(U)$. By definition, 
$(g \cdot \sigma_{a,f})(x) = \sigma_{a,f}(g^{-1}x)$, for every $g,x \in G$, which is non-zero if and only if $x \in g \cdot r_aH = r_{g(a)}H$. 
Moreover, recall that for each $g \in G$ and $r_aH \in G/H$, there is a unique $h_{g,a} \in H$ such that $gr_a = r_{g(a)}h_{g,a}$. Suppose that $x \equiv r_{g(a)}h \in r_{g(a)}H$, we find
\begin{equation}
\begin{split}
    (g \cdot \sigma_{a,f})(x)
    &= \sigma_{a,f}(r_a h_{g,a}^{-1}r_{g(a)}^{-1}r_{g(a)}h)
    \\
    &= (h^{-1}h_{g,a}) \cdot  \sigma_{a,f}(r_a)
    = \pi(h^{-1}h_{g,a})\, f \, \pi(h^{-1}h_{g,a})^{-1} \, ,
\end{split}
\end{equation}
which implies that
\begin{equation}
    \label{eq:GequivRepGAlg}
    g \cdot \sigma_{a,f}
    = \sigma_{g(a),\pi(h_{g,a})f\pi(h_{g,a})^{-1}} \, .
\end{equation}
Given $U=\mathbb C\{u_b\}_{b=1,\ldots,\dim_\mathbb C U}$, we have $\End_\mathbb C(U)= \mathbb C\{u_{b_1} \otimes u^{b_2}\}_{b_1,b_2=1,\ldots,\dim_\mathbb C U}$ and thus $A(H,\beta,U) = \mathbb C\{\sigma_{a,\, u_{b_1} \otimes \, u^{b_2}}\}_{a,b_1,b_2}$. We conclude this presentation of the algebra $A(H,\beta,U)$ by mentioning that it can always be endowed the structure of a \emph{$\Delta$-separable symmetric Frobenius} algebra \cite{Fuchs:2002cm,fuchs2009,kong2019} making in particular $A(H,\beta,U)$ a \emph{coalgebra} whose comultiplication rule descends from the multiplication of $A(H,\beta,U)$.

\bigskip \noindent
We are now ready to confirm that the category $\Mod(A(H,\beta,U))$ of modules over the $G$-equivariant algebra is equivalent to the $\Vect_G$-module category $\mc M(H,\beta)$. Consider a module over $A(H,\beta,U)$. It follows from eq.~\eqref{eq:RepGAlgebraDecomp} and $A(H,\beta,U)_{(a)} \cong \End_\mathbb C(U)$ that this module is $G/H$-graded such that, for every $a \in \{1,\ldots,(G:H)\}$, homogeneous components of degree $a$ span a vector space isomorphic to $U$. In particular, given $u$ such that $|u| = r_aH \in G/H$, each $\sigma \in A(H,\beta,U)$ acts as $\sigma \cdot u = \sigma(r_a)(u)$. Therefore, we already know that $\Mod(A(H,\beta,U))$ is equivalent to $\mc M(H,\beta)$ as a category. We are left to confirm the $\Vect_G$-module structure, which descends from the $G$-equivariant structure of $A(H,\beta,U)$. Let $U_{r_aH}$ be a simple module over $A(H,\beta,U)$ such that ${(U_{r_aH})}_{r_aH} \cong U$. For every $\mathbb C_g = \mathbb C\{e_g\} \in \Vect_G$, we define $\mathbb C_g \act U_{r_aH}$ to be $\mathbb C_g \otimes U_{r_aH}$ as a vector space. The $A(H,\beta,U)$-module structure on $\mathbb C_g \otimes U_{r_aH}$ is provided by
\begin{equation}
    \sigma \cdot (e_g \otimes u) = e_g \otimes (g^{-1} \cdot \sigma) \cdot u \, ,
\end{equation}
for every $u \in U_{r_aH}$ and $\sigma \in A(H,\beta,U)$, so that, in particular, we have
\begin{equation}
    \label{eq:VecGModStructModRepGAlg}
    (g \cdot \sigma_{a,f})(e_g \otimes u) = e_g \otimes f(u) \, . 
\end{equation}
From eq.~\eqref{eq:GequivRepGAlg}, we deduce that $e_g \otimes u$ must be homogeneous of degree $r_{g(a)}H \in G/H$ so that $\mathbb C_g \act U_{r_aH} \cong U_{r_{g(a)}H}$.
Let $\phi^{g,a} : e_g \otimes u \mapsto \pi(h_{g,a})(u)$ implement the isomorphism $\mathbb C_g \act U_{r_aH} \cong U_{r_{g(a)}H}$. It is indeed compatible with eq.~\eqref{eq:VecGModStructModRepGAlg} since, by virtue of eq.~\eqref{eq:GequivRepGAlg},
\begin{equation}
\begin{split}
    (g \cdot \sigma_{a,f})\big(\phi^{g,a}(e_g \otimes u)\big) 
    =(g \cdot \sigma_{a,f})(\pi(h_{g,a})(u)) 
    = \pi(h_{g,a})(f(u)) = \phi^{g,a}\big( e_g \otimes f(u)\big) \, .
\end{split}
\end{equation}
It now remains to compute the module associator. For every $\mathbb C_{g_1}, \mathbb C_{g_2}$ and $U_{r_aH} \in \Mod(A(H,\beta,U))$, isomorphisms $(\mathbb C_{g_1} \otimes \mathbb C_{g_2}) \act U_{r_aH} \cong U_{r_{(g_1g_2)(a)}H}$ and $\mathbb C_{g_1} \act (\mathbb C_{g_2} \act U_{r_aH}) \cong U_{r_{(g_1g_2)(a)}H}$ are implemented by
\begin{equation}
\begin{split}
    e_{g_1} \otimes e_{g_2} \otimes u &\xmapsto{\q} e_{g_1g_2} \otimes u \xmapsto{\phi_{g_1g_2,a}} \pi(h_{g_1g_2,a})(u) 
    \\
    \text{and} \q
    e_{g_1} \otimes e_{g_2} \otimes u&\xmapsto{{\rm id}_{\mathbb C_{g_1}} \otimes \, \phi_{g_2,a}} e_{g_1} \otimes \pi(h_{g_2,a})(u) \xmapsto{\phi_{g_1,g_2(a)} } \pi(h_{g_1,g_2(a)})(\pi(h_{g_2,a})(u)) \, ,
\end{split}
\end{equation}
respectively. It finally follows from eq.~\eqref{eq:hgi_multiplication} that these isomorphisms differ by $\beta(h_{g_1,g_2(a)},h_{g_2,a})$, which reproduces the expected module associator of $\mc M(H,\beta)$ as per eq.~\eqref{eq:VecG_ModAssoc}. Note that choosing a different map implementing the isomorphism $\mathbb C_g \act U_{r_aH} \cong U_{r_{g(a)}H}$ would result in a 2-cocycle in the same cohomology class. This completes the demonstration that $\Mod(A(H,\beta,U)) \simeq \mc M(H,\beta)$, as $\Vect_G$-module categories.

\bigskip \noindent
Exploiting the previous derivation, let us construct the ground state subspace of a renormalisation group fixed point associated with $\mc M(H,\beta)$. Our exposition is closely related to that of ref.~\cite{Kapustin:2016jqm}. Vacua are labelled by simple objects in $\mc M(H,\beta)$, which we just realised as the category of left modules over the algebra $A(H,\beta,U)$. Every such $A(H,\beta,U)$-module also possesses the structure of a left \emph{comodule} over $A(H,\beta,U)^{*,{\rm cop}}$, which is the dual coalgebra $A(H,\beta,U)^*$ with coopposite comultiplication. It follows from $A(H,\beta,U)$ being endowed the structure of a $\Delta$-separable Frobenius algebra that the coalgebra $A(H,\beta,U)^{*,{\rm cop}}$ also possesses the structure of an algebra whose multiplication rule is the opposite of that of the dual Frobenius algebra, thereby inviting us to consider $A(H,\beta,U)^{*,{\rm op},{\rm cop}}$. Therefore, we take ground states to span a subspace of the tensor product space $\bigotimes_{\msf i \in \Lambda} A(H,\beta,U)^*_{\{\msf i\}}$ obtained by choosing the local vector space to be $A(H,\beta,U)^* \equiv \mathbb C\{\sigma_{a,u^{b_1} \otimes u_{b_2}}\}_{a,b_1,b_2}$.\footnote{These manipulations are required by the fact that we defined in sec.~\ref{sec:VecG_Hilbert} the MPS tensors as linear maps of the form $M \to V \otimes M$.} Physically, one typically wants to think of $\bigotimes_{\msf i \in \Lambda} A(H,\beta,U)^*_{\{\msf i\}}$ as being itself a subspace of a microscopic Hilbert space $\mc H_\Lambda = \bigotimes_{\msf i \in \Lambda} V_{\{\msf i\}}$, where $V \in \Rep(G)$ is a faithful representation such that $A(H,\beta,U)^* \hookrightarrow V$. The action of topological lines in $\Vect_G$ on this Hilbert space is provided by the $G$-equivariant structure of $A(H,\beta,U)^*$ so that $\rho : \mathbb C[G] \to \End_\mathbb C(A(H,\beta,U)^*)$ is given by 
\begin{equation}
\begin{split}
    \label{eq:VecG_basisExp}
    \rho(g)(\si{a}{b_2}{b_4} ) := g \cdot \si{a}{b_2}{b_4} 
    &= \sigma_{g(a), h_{g,a}\cdot u^{b_2},h_{g,a}\cdot u_{b_4}}
    \\
    &=
    \sum_{b_1,b_3}
    \pi(h_{g,a}^{-1})^{b_2}_{b_1} \, \pi(h_{g,a})^{b_3}_{b_4} \; \si{g(a)}{b_1}{b_3}    \, ,
\end{split}
\end{equation}
for every $g \in G$, $a \in \{1,\ldots,(G:H)\}$ and $b_2,b_4 \in \{1,\ldots,\dim_\mathbb C U\}$. 
For every simple object $U_{r_aH} \in \Mod(A(H,\beta,U))$, with $a \in \{1,\ldots,(G:H)\}$, let $\theta_a : U_{r_aH} \to A(H,U,\beta)^* \otimes U_{r_aH}$ denote the left $A(H,U,\beta)^{*,{\rm op}, {\rm cop}}$-comodule structure of $U_{r_aH}$. 
For every $a \in \{1,\ldots,(G:H)\}$, we write
\begin{equation}
    \theta_a \equiv \sum_{b_1,b_2=1}^{\dim_\mathbb C U}
    \thet{a}{b_1}{b_2} \; u_{b_1} \otimes u^{b_2} 
\end{equation}
such that $\thet{a}{b_1}{b_2} \in A(H,\beta,U)^*$.
Assuming closed periodic boundary conditions, one associates to the tensor $\theta_a$ the following injective MPS in $\bigotimes_{\msf i \in \Lambda}A(H,\beta,U)^*_{\{\msf i\}}$:
\begin{equation}
\begin{split}
    \Psi(\theta_a) &:= \sum_{b_1,\ldots,b_{|\Lambda|}}\thet{a}{b_1}{b_2}\otimes\thet{a}{b_2}{b_3} \otimes \cdots \otimes\thet{a}{b_{|\Lambda|}}{b_1}  \, .
\end{split}
\end{equation}
But, it follows from the definitions that 
\begin{equation}
    \thet{a}{b_1}{b_2} = \si{a}{b_1}{b_2} \, ,
\end{equation}
for every $a \in \{1,\ldots,(G:H)\}$ and $b_1,b_2 \in \{1,\ldots,\dim_\mathbb C U\}$. Therefore, the previous formula boils down to
\begin{equation}
    \label{eq:VecG_GS}
    \Psi(\theta_a) = \sum_{b_1,\ldots , b_{\Lambda}} \si{a}{b_1}{b_2} \otimes \si{a}{b_2}{b_3} \otimes \cdots \otimes \si{a}{b_{|\Lambda|}}{b_1} \, .
\end{equation}
A topological line labelled by $\mathbb C_g \in \Vect_G$ acts on this MPS as
\begin{equation}
    \begin{split}
    \bigg( \prod_{\msf i \in \Lambda} \rho(g)_\msf i \bigg) \Psi( \theta_a)
    &=
    \sum_{b_1,\ldots , b_{|\Lambda|}} (g \cdot \si{a}{b_1}{b_2}) \otimes \cdots \otimes (g \cdot \si{a}{b_{|\Lambda|}}{b_1})
    \\
    &= \sum_{b_1,\ldots,b_{|\Lambda|}} \si{g(a)}{b_1}{b_2} \otimes \cdots \otimes \si{g(a)}{b_{|\Lambda|}}{b_1} = \Psi(\theta_{g(a)} ) \, ,
    \end{split}
\end{equation}
where we used eq.~\eqref{eq:VecG_basisExp}.
As expected, acting with topological lines $\mathbb C_g \in \Vect_G$ permutes ground states via the left action of the group on $G/H$.
It follows that for each $a \in \{1,\ldots,(G:H)\}$, topological lines labelled by $\mathbb C_{r_ahr_a^{-1}} \in \Vect_G$, for every $r_ahr_a^{-1}$ in the stabiliser $r_aHr_a^{-1}$ of $r_aH \in G/A$, leave the ground state associated with $r_aH \in G/H$ invariant. Finally, it follows from our verification of $\Mod(A(H,\beta,U)) \simeq \mc M(H,\beta)$ that $\Psi(\theta_a)$ preserves the algebra $\mathbb C[r_aHr_a^{-1}]^{\beta^{r_a}}$ of lines in $\Vect_G$. This implies that the resulting $G$-symmetric block-injective MPS defines a renormalisation group fixed point associated with the gapped phase $\mc M(H,\beta)$. 
\section{Algebras of order parameters\label{sec:VecG_operators}}

\emph{Within the framework of the previous section, we explain how to construct irreducible multiplets of string operators associated with simple objects in the so-called Drinfel'd centre $\mc Z(\Vect_G)$ of $\Vect_G$. We then demonstrate that order parameters form Lagrangian algebras in $\mc Z(\Vect_G)$.} 

\subsection{String operators\label{sec:VecG_String}}

We work within the framework of sec.~\ref{sec:VecG_Hilbert}, whereby the microscopic Hilbert space is of the form $\mc H_\Lambda = \bigotimes_{\msf i \in \Lambda} V_{\{\msf i\}}$, where $V$ a faithful representation in $\Rep(G)$ with algebra homomorphism $\rho \colon \mathbb C[G] \to \End_\mathbb C(V)$ so that the action of the topological line $\mathbb C_g \in \Vect_G$ on $\mc H_\Lambda$ is given by $\prod_{\msf i \in \Lambda} \rho(g)_\msf i$. 

Let us begin with a discussion of multiplets of \emph{local} operators. Note that locality of the operator does not necessarily indicate that the support of the corresponding operator in the algebra $\mc A_\Lambda$ is a single site $\msf i \in \Lambda$. Given a simple object $W$ in $\Rep(G)$ with action homomorphism $\varrho : \mathbb C[G] \to \End_\mathbb C(W)$, one can construct a local operator $\mc O_\msf i \in \mc A_\Lambda$ transforming in the representation $W$ if and only if $\Hom_{\Rep(G)}(W \otimes V,V) \equiv \Hom_G(W \otimes V,V)$ is non-trivial. For instance, choosing $V$ to be the \emph{regular} representation $\mathbb C(G)$ guarantees that a local operator in $\mc A_\Lambda$ supported at a single site can be associated to any simple object in $\Rep(G)$. More generally, since every irreducible representation occurs in some finite tensor power of any faithful representation, a local operator in $\mc A_\Lambda$ with finite support can be associated to any simple object in $\Rep(G)$. Henceforth, we assume without loss of generality that $V$ is chosen such that $\Hom_{G}(W \otimes V,V)$ is non-trivial for every $W \in \Irr(\Rep(G))$. 

Given $W \in \Rep(G)$, let $\varphi^{WV}_V \colon W \otimes V \to V$ be a choice of $G$-equivariant map. Let us write $W= \mathbb C\{w_c\}_{c=1,\ldots, \dim_\mathbb C W}$. For every $c \in \{1,\ldots,\dim_\mathbb C \! W\}$, we define $\mc O(w_c)_\msf i$, for any $\msf i \in \Lambda$, as the embedding of $\mc O(w_c) := \varphi^{WV}_V \! (w_c \otimes -) \in \End_\mathbb C(V)$ into $\mc A_\Lambda$. Graphically,
\begin{equation}
    \label{eq:VecG_localOp}
    \localOp{c} \equiv \! \localOpExt{\varphi^{WV}_V}{W}{c} \! \equiv \mc O(w_c) \, .
\end{equation}
By virtue of the $G$-equivariance of $\varphi^{WV}_V$, we have
\begin{equation}
\begin{split}
    \label{eq:VecG_localOpTube}
    \Ad_{\rho(g)}\big(\mc O(w_c)\big) := \rho(g)\,  \mc O(w_c) \, \rho(g)^{-1} 
    = \! \localOpTube{1}{c}{g}{}{} \! 
    &= \! \localOpTube{2}{\varphi^{WV}_V}{g}{W}{c} 
    \\[-2em]
    &= \mc O \big(\varrho(g)(w_c)\big) \, ,
\end{split}
\end{equation}
for every $g \in G$, which in turn implies that $\mc O(w_c)_\msf i$ transforms in the representation $W$ under the action of symmetry lines: 
\begin{equation}
    \bigotimes_{\msf j \in \Lambda} \Ad_{\rho(g)_\msf j}\big( \mc O(w_c)_\msf i\big) = \mc O\big(\varrho(g)(w_c) \big)_\msf i \, .
\end{equation}
Therefore, every $\varphi^{WV}_V$ labels a multiplet $\{\mc O(w_c)_\msf i\}_{c=1,\ldots,\dim_\mathbb C \! W}$ of local operators transforming in the irreducible representation $W$. Notice that while the number of multiplets depends on the choice of Hilbert space, the number of local operators in each multiplet is fixed by the dimension of the corresponding representation. Therefore, depending on the Hilbert space, there may be considerable freedom in finding local operators forming a given multiplet.

The same way we define local operators transforming in a representation $W \in \Rep(G)$, every $G$-equivariant map $\varphi^{VW^*}_W: V \otimes W^*  \to V$ labels a multiplet $\{\mc O(w^c)_\msf i\}_{c=1,\ldots,\dim_\mathbb C W}$ of local operators transforming in the representation $W^* = \mathbb C\{w^c\}_{c=1,\ldots,\dim_\mathbb C W}$. Combining both types of operators, we can consider the operator $\mc O(w_{c_1})_{\msf i} \, \mc O(w^{c_2})_{\msf j}$, for any $c_1,c_2 \in \{1,\ldots,\dim_\mathbb C W \}$ and $\msf i,\msf j \in \Lambda$, which transform in the representation $W \otimes W^* \cong \End_\mathbb C(W)$.

\bigskip \noindent
Let us now examine multiplets of \emph{twisted sector local operators}. Loosely, a twisted sector local operator is a local operator that live at the endpoint of a twisting symmetry defect. Since local operators are labelled by objects in $\Rep(G)$ and symmetry defects are labelled by objects in $\Vect_G$, one expects a twisted sector local operator to be associated with a vector space that is both $G$-equivariant and $G$-graded. But, 
\begin{equation}
    \bigotimes_{\msf j \in \Lambda}\Ad_{\rho(g)_\msf j} (\rho(x)_{\msf i}) = \rho(gxg^{-1})_{\msf i} \, ,
\end{equation}
for every $g,x \in G$ and $\msf i \in \Lambda$, which requires $G$-equivariance and $G$-grading to be compatible with one another.

More formally, define a $G$-equivariant $G$-graded vector space as a $G$-graded vector space $W = \bigoplus_{x \in G}W_x$ together with a collection $\{\varrho(g):W \xrightarrow{\simeq} W\}_{g \in G}$ of automorphisms such that $\varrho(g)(W_{x}) \subseteq W_{gxg^{-1}}$, for every $g,x \in G$, and $\varrho(g_1)\varrho(g_2) = \varrho(g_1g_2)$, for every $g_1,g_2 \in G$. Consider the \emph{non-degenerate braided fusion} category $\Vect_G^G$  of $G$-equivariant $G$-graded vector space and $G$-equivariant grading preserving linear maps (see app.~\ref{app:algebras} for definitions). The monoidal structure is provided by the usual tensor product of $G$-graded vectors spaces with the group acting diagonally. Since conjugacy classes partition the group, the support of an object in $\Vect_G^G$ is a union of conjugacy classes in $\Cl(G)$. Moreover, for every $W \in \Vect_G^G$, we have by definition that $W = \bigoplus_{[x_0] \in \Cl(G)}W_{[x_0]}$ with $W_{[x_0]} := \bigoplus_{x \in [x_0]} W_x$.
It follows that 
\begin{equation}
    \label{eq:VecGG_decomp}
    \Vect_G^G = \bigoplus_{[x_0] \in \Cl(G)} (\Vect_G^G)_{[x_0]} \, ,
\end{equation}
where $(\Vect_G^G)_{[x_0]}$ is the subcategory consisting of objects $W \in \Vect_G^G$ such that $\text{supp}(W) = [x_0]$. Furthermore, one can show that (see e.g. ref.~\cite{DAVYDOV2017149})
\begin{equation}
    \label{eq:VecGG_equiv}
    (\Vect_G^G)_{[x_0]} \simeq \Mod(\mathbb C[Z_G(x_0)]) \, ,
\end{equation}
where $Z_G(x_0)$ denotes the centraliser in $G$ of the representative $x_0$ of $[x_0]$. Let us briefly review the main argument here. On the one hand, given $W_{[x_0]} \in (\Vect_G^G)_{[x_0]}$, $W_{x_0}$ has the structure of a $\mathbb C[Z_G(x_0)]$-module, by definition. On the other hand, given $\hat W = \mathbb C\{w_c\}_{c=1,\ldots,\dim_\mathbb C \! \hat W} \in \Mod(\mathbb C[Z_G(x_0)])$ with algebra homomorphism $\hat \varrho : \mathbb C[Z_G(x_0)] \to \End_\mathbb C(\hat W)$, the induced representation of $\hat W$ in $G$ defined as
\begin{equation}
    W \equiv \Ind_{Z_G(x_0)}^G(\hat W) = \mathbb C\big\{ \varpi : G \to \hat W \, \big| \, \varpi(gz^{-1}) = \hat \varrho(z)(\varpi(g)) \big\}_{g \in G, \, z \in Z_G(x_0)} 
\end{equation}
is a $G$-equivariant $G$-graded vector space with support $\supp(W)=[x_0]$. Indeed, mimicking the derivations of sec.~\ref{sec:VecG_RG}, let $\{r_x\}_{x \in [x_0]}$ be a choice of representatives in $G$ for the left cosets in $G/Z_G(x_0)$.\footnote{Recall from the orbit-stabiliser theorem that $(G : Z_G(x_0)) = |[x_0]|$.} For each $g \in G$, there is a unique $x \in [x_0]$ such that $g \in r_x Z_G(x_0)$. Given $c \in \{1,\ldots,\dim_\mathbb C \hat W\}$, define the function $\varpi_{x,w_c} : G \to \hat W$ via
\begin{equation}
    \varpi_{x,w_c}(g) =
    \begin{cases}
        \hat \varrho(z)^{-1}(w_c) \q &\text{if $g \equiv r_xz \in r_xZ_G(x_0)$} \, ,
        \\
        0 &\text{otherwise} \, ,
    \end{cases}
\end{equation}
for every $g \in G$. One readily verifies that $W = \mathbb C\{\varpi_{x,w_c}\}_{x \in [x_0], c=1,\ldots,\dim_\mathbb C \hat W}$ such that $\varpi_{x,w_c}$ is homogeneous of degree $|\varpi_{x,w_c}|=x=r_x x_0 r_x^{-1} \in G$. As usual, the $G$-equivariance structure is given by $(g \cdot \varpi_{x,w_c})(-) = \varpi_{x,w_c}(g^{-1}-)$. Explicitly, for each $g \in G$ and $r_x Z_G(x_0) \in G / Z_G(x_0)$, there is a unique group element $z_{g,x} \in Z_G(x_0)$ such that $gr_x = r_{gxg^{-1}}z_{g,x}$. Suppose that $y \equiv r_{gxg^{-1}} z \in r_{gxg^{-1}} Z_G(x_0)$, we find
\begin{equation}
\begin{split}
    (g \cdot \varpi_{x,w_c})(y) &= \varpi_{x,w_c}(r_x z_{g,x}^{-1}r_{gxg^{-1}}^{-1} r_{gxg^{-1}}z)
    \\
    &= \hat \varrho(z^{-1}z_{g,x}) \big( \varpi_{x,w_c} \big)
\end{split}
\end{equation}
which implies that 
\begin{equation}
    g \cdot \varpi_{x,w_c} = \varpi_{gxg^{-1},\hat \varrho(z_{g,x})(w_c)} \, ,
\end{equation}
for every $g \in G$. Therefore, $|g \cdot \varpi_{x,w_c}| = gxg^{-1} = g |\varpi_{x,w_c}|g^{-1}$. This confirms that $W \in (\Vect_G^G)_{[x_0]}$. We can further show that the two functors thus constructed are indeed inverse of each other, up to natural isomorphisms, thereby establishing equivalence \eqref{eq:VecGG_equiv}. 
It follows from eq.~\eqref{eq:VecGG_decomp} and $\eqref{eq:VecGG_equiv}$ that simple objects in $\Vect_G^G$ are labelled by pairs $([x_0],\hat W)$ consisting of a conjugacy class $[x_0] \in \Cl(G)$ and a simple object $\hat W \in \Mod(\mathbb C[Z_G(x_0)])$. 

\bigskip\noindent
Given a simple object $([x_0],\hat W)$ in $\Vect_G^G$, let us now construct multiplets of twisted sector local operators. As before, let $W$ denote the induced representation $\Ind_{Z_G(x_0)}^G(\hat W)$ of $\hat W$ in $G$. Let $\varphi^{WV}_V \colon W \otimes V \to V$ be a choice of $G$-equivariant map so that $\varphi^{WV}_V(\varpi_{x,w_c} \otimes -) \in \End_\mathbb C(V)$, for every $x \in [x_0]$ and $c \in \{1,\ldots,\dim_\mathbb C \hat W\}$.\footnote{In order to compute the space of $G$-equivariant maps $\varphi^{WV}_V$, it is useful to recall the `projection formula', which stipulates that $W \otimes V =  \Ind_{Z_G(x_0)}^G(\hat W) \otimes V \cong \Ind_{Z_G(x_0)}^G(\hat W \otimes \Res_{Z_G(x_0)}^G(V))$.} By virtue of the $G$-equivariance of $\varphi^{WV}_V$, we have
\begin{equation}
    \Ad_{\rho(g)}\big( \varphi^{WV}_V(\varpi_{x,w_c} \otimes -)\big) = \varphi^{WV}_V \! \big(\varpi_{gxg^{-1}, \hat \varrho(z_{g,x})(w_c)} \otimes -\big) \, ,
\end{equation}
for every $g \in G$. By construction, as a $G$-graded vector space, $W = \bigoplus_{x \in [x_0]}W_x$ such that $\supp(W) = [x_0]$ and $W_x \cong \hat W$, for every $x \in [x_0]$. It follows that, for every $x \in [x_0]$, the restriction of $\varphi^{WV}_V(\varpi_{x,w_c} \otimes -) \in \End_\mathbb C(V)$ to $Z_G(x_0)$ transforms like $\hat W$. For every $x \in [x_0]$ and $c \in \{1,\ldots,\dim_\mathbb C \hat W\}$, one defines a tensor
\begin{equation}
    \label{eq:VecG_twistedLocalOp}
    \twistedLocalOp{c}{x}{} \!\!\!
    \equiv
    \! \twistedLocalOpExt{\varphi^{WV}_V\!\!\!}{W}{\bar \psi^W_{\mathbb C_x,c}}{x}
    \equiv
    \varphi^{WV}_V(\varpi_{x,w_c} \otimes -) \otimes e^x  \, ,
\end{equation}
where we introduced
\begin{equation}
    \raisebox{-4pt}{\mat{\bar \psi^W_{\mathbb C_x,c}}{W}{x}{}{}} \!\!\!\!\! \equiv \varpi_{x,w_c} \otimes e^x \, ,
\end{equation}
which encodes a choice of basis morphism $\psi^W_{\mathbb C_x,c}$ in $\Hom_{\Vect_G}(W,\mathbb C_x) \cong \hat W$.

For every choice of $G$-equivariant map $\varphi^{WV}_V$, we claim that tensors \eqref{eq:VecG_twistedLocalOp} furnish a multiplet of twisted sector local operators associated with the simple object $([x_0],\hat W)$ in $\Vect_G^G$. First, one needs to clarify how symmetry $\Vect_G$ acts on such twisted sector local operators. It is a standard mathematical result that the (non-degenerate braided fusion) category $\Vect_G^G$ is equivalent to the Drinfel'd centre $\mc Z(\Vect_G)$ of $\Vect_G$ (see app.~\ref{app:centres}). For every fusion category $\mc C$, its Drinfel'd centre $\mc Z(\mc C)$ can be realised as the category $\Mod(\Tu(\mc C))$ of modules over its so-called \emph{tube algebra} $\Tu(\mc C)$ \cite{ocneanu1994chirality,ocneanu2001operator,Izumi2000,MUGER2003159,Neshveyev_Yamashita_2018}. This is via this tube algebra that the symmetry lines act on twisted sector local operators \cite{Lin:2022dhv,Bhardwaj:2023wzd,Bartsch:2023wvv}. As reviewed in app.~\ref{app:centres}, the tube algebra $\Tu(\Vect_G)$ of $\Vect_G$ is equivalent to the \emph{quantum double} of $G$ \cite{Drinfeld:1989st,Dijkgraaf:1989pz,Roche:1990hs}, which is itself equivalent to the \emph{groupoid algebra} $\mathbb C[G/\!/G]$ \cite{Willerton2005TheTD}.
Here, $G/\!/G$ refers to the groupoid with object set $G$, morphisms $x \! \xrightarrow{\,g\,} \! gxg^{-1}$, for every $g,x \in G$, and composition provided by group multiplication.\footnote{Although the equivalence $\Tu(\Vect_G) \cong \mathbb C[G/\!/G]$ is not crucial to our construction, it is useful to introduce the groupoid $G/\!/G$ as its cohomology naturally appears in our construction (see below).} Concretely, $\Tu(\Vect_G) = \mathbb C\{e_g \otimes \varsigma_x\}_{g,x \in G}$ with multiplication (see app.~\ref{app:centres}) 
\begin{equation}
    (e_{g_1} \otimes \varsigma_y) \cdot (e_{g_2} \otimes \varsigma_x) := \delta_{y,g_2xg_2^{-1}} \, (e_{g_1g_2} \otimes \varsigma_x) \, ,
\end{equation}
for every $g_1,g_2,x,y \in G$.
In particular, one immediately recovers from groupoid representation theory that simple objects in $\Vect_G^G$ are labelled by pairs $([x_0],\hat W)$. 
The $\Tu(\Vect_G)$-module associated with the simple objet $([x_0],\hat W)$ is simply given by $\mathbb C\{\varpi_{x,w_c}\}_{x \in [x_0],c=1,\ldots,\dim_
\mathbb C \hat W}$ with (see app.~\ref{app:centres})
\begin{equation}
    (e_g \otimes \varsigma_y) \cdot \varpi_{x,w_c} := \delta_{y,x} \, \varpi_{gxg^{-1},\hat \varrho(z_{g,x})(w_c)} \, ,
\end{equation}
for every $g , y \in G$, $x \in [x_0]$ and $c \in \{1,\ldots, \dim_\mathbb C \hat W\}$.
As reviewed in app.~\ref{app:centres}, $\Tu(\Vect_G) \cong \mathbb C[G/\!/G]$ further possesses the structure of a (quasi-triangular Hopf) $*$-algebra. 

Within our framework, $\Tu(\Vect_G)$ is realised as follows: As a vector space, it is spanned by tensor networks of the form\footnote{The geometrical choices we make in this graphical representation are justified by the derivations of sec.~\ref{sec:VecG_OP}.}
\begin{equation}
    \label{eq:VecG_Tube}
    \Tube{g}{{}^gx}{gx}{x}{}{} \hspace{-5pt}  = \,  \left(\!\! \Tube{g^{-1}}{x}{\;\;\;xg^{-1}}{\!{}^gx}{}{} \right)^\dagger  ,
\end{equation}
for every $g,x \in G$, where ${}^gx := gxg^{-1}$. Henceforth, we refer to such a tensor network as a `tube'. The r.h.s. of eq.~\eqref{eq:VecG_Tube} realises the $*$-structure of $\Tu(\Vect_G)$, which is verifiably an algebra anti-homomorphism. The multiplication rule of $\Tu(\Vect_G)$ reads
\begin{align}
    \nn
    &\Tube{g_1}{{}^{g_1}y}{g_1y}{y}{}{} \hspace{-6pt} \cdot \Tube{g_2}{{}^{g_2}x}{g_2x}{x}{}{} \hspace{-8pt} 
    := \delta_{y,{}^{g_2}x} 
    \TubeMult{g_2}{g_1}{{}^{g_1 g_2}x\;}{\; g_1 {}^{g_2}x}{{}^{g_2}x}{g_2x}{x} 
    \\[-3em] 
    & \q\q\q = \delta_{y,{}^{g_2}x} 
    \Tube{g_1g_2}{{}^{g_1 g_2}x}{g_1g_2x}{x}{}{} \!\!\!  , 
\end{align}
for every $g_1,g_2,x,y \in G$, where we used eq.~\eqref{eq:VecG_fusionMPO} together with eq.~\eqref{eq:VecG_Fmove} and eq.~\eqref{eq:VecG_orthoFusion}. We are now in a position to verify that twisted sector local operators \eqref{eq:VecG_twistedLocalOp} furnish simple modules over $\Tu(\Vect_G)$. Clearly, contracting a tube \eqref{eq:VecG_Tube} with a tensor \eqref{eq:VecG_twistedLocalOp} results in a tensor of the same form. Concretely, acting with the tube labelled by $g,y \in G$ on the tensor labelled by $x \in G$ and $c_2 \in \{1,\ldots,\dim_\mathbb C \hat W\}$ results in   
\begin{equation}
    \label{eq:VecG_twistedLocalOpTube}
    \delta_{y,x} \,
    \twistedLocalOpTube{1}{c_2}{g}{{}^gx}{gx}{x}{}{}
    \hspace{-9pt}
    = 
    \delta_{y,x} \,
    \twistedLocalOpTube{2}{\bar \psi^W_{\mathbb C_x,c_2}}{g}{{}^gx}{gx}{x}{W}{\varphi^{WV}_V}
    \hspace{-9pt}
    =
    \delta_{y,x} 
    \sum_{c_1 = 1}^{\dim_\mathbb C \hat W} \hat \varrho(z_{g,x})^{c_1}_{c_2}
    \twistedLocalOp{c_1}{{}^gx}{} \!\!\! ,
\end{equation}
where we used eq.~\eqref{eq:VecG_localOpTube} before performing all the contractions. This confirms that the collection of tensors of the form \eqref{eq:VecG_twistedLocalOp}, for every $x \in [x_0]$ and $c \in \{1,\ldots,\dim_\mathbb C \hat W\}$, furnish a module over $\Tu(\Vect_G)$ with underlying vector space $\bigoplus_{x \in G}\Hom_{\Vect_G}(\mathbb C_x,W)$, namely that associated with the simple object $([x_0],\hat W)$ in $\Mod(\Tu(\Vect_G)) \simeq \Vect_G^G$ (see app.~\ref{app:centres}). We conclude that for every choice of $G$-equivariant map $\varphi^{WV}_V : W\otimes V \to V$, tensors \eqref{eq:VecG_twistedLocalOp}, for every $x \in [x_0]$ and $c \in \{1,\ldots,\dim_\mathbb C \hat W\}$, furnish a multiplet of twisted sector local operators that transform in the irreducible representation $([x_0],\hat W)$ of $\Tu(\Vect_G)$. Multiplets of twisted local operators associated with any object in $\Vect_G^G$ are constructed similarly.

For every choice of $G$-equivariant map $\varphi^{VW^*}_V \colon V \otimes W^*  \to V$, one can construct another collection of tensors that form the  \emph{Hermitian dual} of the module associated with $([x_0],\hat W)$. Concretely, for every $x \in [x_0]$ and $c \in \{1,\ldots,\dim_\mathbb C \hat W\}$, define
\begin{equation}
    \label{eq:VecG_twistedLocalOpDual}
    \twistedLocalOpDual{c}{x}\!  
    \equiv
    \twistedLocalOpDualExt{\varphi^{VW^*}_V\!\!\!}{W}{\psi^W_{\mathbb C_x,c}}{x}
    \equiv
    e_x \otimes \varphi^{VW^*}_V(- \otimes \varpi^{x,w_c})   \, ,
\end{equation}
where we introduced
\begin{equation}
    \raisebox{-4pt}{\matDual{\psi^W_{\mathbb C_x,c}}{x}{W}{}{}} \!\!\!\!\! \equiv e_x \otimes \varpi^{x,w_c}  .
\end{equation}
In particular, it follows from eq.~\eqref{eq:VecG_Tube} that the Hermitian dual of \eqref{eq:VecG_twistedLocalOpTube} agrees with the result of acting with the Hermitian dual of the tube labelled by $g,y \in G$ on the tensor \eqref{eq:VecG_twistedLocalOpDual}.

Let $([x_0],\hat W)$ label a simple object in $\Vect_G^G$ and choose a group element $x \in G$. Consider twisted sector local operators of the form \eqref{eq:VecG_twistedLocalOp} and $\eqref{eq:VecG_twistedLocalOpDual}$ associated with $c_1$ and $c_2$ in $\{1,\ldots,\dim_\mathbb C \hat W\}$, respectively. Placing the corresponding tensors at sites $\msf i< \msf j$ in $\Lambda$, we contract them via a symmetry MPO \eqref{eq:VecG_MPO} with open boundary conditions. This `truncated' symmetry  MPO implements the action of the symmetry line $\mathbb C_x \in \Vect_G$ on sites $\msf i+1,\msf i+2, \ldots, \msf j-2,\msf j-1$. We denote the embedding in $\mc A_\Lambda$ of the resulting tensor network operator as $\mc S(x,c_1,c_2)_{\msf i,\ldots,\msf j}$. Graphically, 
\begin{equation}
    \label{eq:VecG_StringOp}
    \mc S(x,c_1,c_2)_{\msf i,\ldots,\msf j} \equiv \StringOp{x}{c_1}{c_2} \, .
\end{equation}
In the physics literature, operators of this form are typically referred to as `string operators' \cite{PhysRevB.40.4709,PhysRevLett.100.167202,PhysRevB.86.125441}. However, key aspects set our construction apart from typical definitions. Indeed, a string operator is commonly defined as a product of local operators with a truncated symmetry operator of the form $\prod_{\msf k=\msf i+1}^{\msf j-1} \rho(x)_\msf k$. In general, this approach fails to explicitly capture the $G$-grading structure of twisted sector local operators. Therefore, a naive interpretation of such a definition could yield contradictions. For instance, consider two simple objects in $\Vect_G^G$ labelled by $([x_0],\hat W_1)$ and $([x_0],\hat W_2)$, where $[x_0] \in \Cl(G)$ and $\hat W_1, \hat W_2 \in \Irr(\Mod(\mathbb C[Z_G(x_0)]))$, such that $ \Ind_{Z_G(x_0)}^G(\hat W_1) \cong \Ind_{Z_G(x_0)}^G(\hat W_2)$. If we were to simply require the endpoint local operators to transform in the representation $\Ind_{Z_G(x_0)}^G(\hat W_1) \cong \Ind_{Z_G(x_0)}^G(\hat W_2)$, one would not be able to distinguish the resulting string operators. Instead, one typically requires the endpoint local operators to transform in the representations $\hat W_1$ and $\hat W_2$ when acting with symmetry lines in $\Vect_{Z_G(x_0)}$, respectively, which does produce the desired result. But this only becomes a constructive definition when introducing the $G$-grading structure of the twisted sector local operators. 

Let us examine a couple of special cases. Choosing $[x_0] = [1_G]$, one recovers the situation described at the beginning of this section, whereby the string operator $\mc S(1_G,c_1,c_2)_{\msf i,\ldots,\msf j}$ boils down to $\mc O(w_{c_1})_\msf i \, \mc O(w^{c_2})_{\msf j}$. Choosing $\hat W = \mathbb C \in \Mod(\mathbb C[Z_G(x_0)])$, for any $[x_0] \in \Cl(G)$, it follows from $\Ind^G_{Z_G(x_0)}(\mathbb C) \cong \mathbb C(G/Z_G(x_0))$ that even though $\hat W$ is trivial, the end operators may need to transform non-trivially.\footnote{Whenever $G$ is abelian, since conjugacy classes are all of size one, the end operators always need transform trivially.} In general, one can construct many multiplets of twisted sector local operators that transform adequately. In particular, let us emphasise that $\prod_{\msf k=\msf i+1}^{\msf j-1} \rho(x)_\msf k$, for any $x \in G$ corresponds to a very specific choice of multiplet.

\bigskip\noindent
We conclude with brief comments about `fusion' of twisted sector local operators. Let $W_1$ and $W_2$ be two objects in $\Vect_G^G$. Given $G$-equivariant maps $\varphi^{W_1 V}_V: W_1 \otimes V \to V$ and $\varphi^{W_2V}_V : W_2 \otimes V \to V$, one can construct multiplets of twisted local operators transforming in the representations of $\Tu(\Vect_G)$ associated with $W_1$ and $W_2$, respectively. Given $x_1 \in \supp(W_1)$ and $x_2 \in \supp(W_2)$, let $c_1 \in \{1, \ldots,\dim_\mathbb C (W_1)_{x_1}\}$ and $c_2 \in \{1,\ldots,\dim_\mathbb C (W_2)_{x_2}\}$ label basis vectors of $(W_1)_{x_1}$ and $(W_2)_{x_2}$, respectively. Consider the tensor network
\begin{equation}
    \fusionTwistedLocalOp{x_1}{x_2}{x_1x_2}{c_2}{c_1} \, ,
\end{equation}
where we are adapting eq.~\eqref{eq:VecG_twistedLocalOp} to the case of non-necessarily simple objects in $\Vect_G^G$. Clearly, the collection of such tensors form a new module over $\Tu(\Vect_G)$, and thus a new multiplet of twisted sector local operators. By definition, they transform in the representation associated with $W_1 \otimes W_2 \in \Vect_G^G$.

\subsection{Order parameters\label{sec:VecG_OP}}

Given a $G$-symmetric gapped phase, we are now ready to identify string operators that may serve as \emph{order parameters} in the ground state subspace. We are asking for necessary conditions that the corresponding multiplets of twisted sector local operators must satisfy in order for the string operators to acquire a non-vanishing expectation value in at least one state in the ground state subspace. 

Consider a representative of the gapped phase associated with the indecomposable $\Vect_G$-module category $\mc M(H,\beta)$. Recall that each ground state in the $(G:H)$-dimensional ground state subspace preserves an algebra object of topological lines that is Morita equivalent to $A[H,\beta]$. Specifically, the ground state labelled by $a \in \{1,\ldots,(G:H)\}$ preserves the internal hom $\underline{\Hom}(\mathbb C_{r_aH},\mathbb C_{r_aH})$ in $\Vect_G$, which is isomorphic to $\mathbb C[r_aHr_a^{-1}]^{\beta_{r_a}}$ as an algebra. 

Let us begin by identifying local order parameters. Let $W$ be a simple object in $\Rep(G)$. For every $G$-equivariant map $\varphi^{WV}_V: W \otimes V \to V$, one can construct a multiplet of local operators according to eq.~\eqref{eq:VecG_localOp}, which transform in the representation $W$. Write $\la - \ra_a : W \to \mathbb C$ the operation that consists in computing the expectation value of these local operators in the ground state labelled by $a \in \{1,\ldots,(G:H)\}$, where $\mathbb C$ is here interpreted as the trivial representation of $G$. Since the ground state preserves the algebra $\mathbb C[r_aHr_a^{-1}]^{\beta_{r_a}}$ of symmetry lines---i.e., the corresponding symmetry operators leave the ground state invariant---the map $\la-\ra_a$ must be $H$-equivariant. But $\Hom_{H}(\Res^G_H (W),\mathbb C)$ is non-trivial if and only if $\mathbb C$ is a direct summand of $\Res^G_H(W)$. Therefore, $\mathbb C$ is required to be a direct summand of $\Res^G_H(W)$ for expectation values of local operators transforming in the representation $W$ to be non-vanishing. Note that this is only a necessary condition for acquiring a vacuum expectation value; it is often possible to construct a local operator that contains a singlet under $H$ but nevertheless yields a vanishing expectation value. 

Suppose we found a $G$-equivariant map $\varphi^{WV}_V: W \otimes V \to V$ so that local operators in the resulting multiplet have non-vanishing expectation values. One can conclude that $\mathbb C$ is a direct summand of $\Res^G_H(W)$. Linearly independent $G$-equivariant maps $W \otimes V \to V$ yield inequivalent multiplets of local operators. However, these inequivalent multiplets may still yield linearly dependent expectation values. Computing the number of inequivalent multiplets of local operators transforming in the representation $W$ that yields linearly independent expectation values amounts to computing the multiplicity of $\mathbb C$ in $\Res^G_H(W)$, i.e., the dimension of the vector space $\Hom_H(\Res^G_H(W),\mathbb C)$. By Frobenius reciprocity, $\Hom_H(\Res^G_H(W),\mathbb C) \cong \Hom_G(W,\Ind_H^G(\mathbb C))$. We conclude that local order parameters form a (reducible) multiplet of local operators that transform in the representation $\Ind_H^G(\mathbb C) \cong \mathbb C(G/H)$. The corresponding object in $\Vect_G^G$ is given by $\bigoplus_{W \in \mathbb C(G/H)} \dim_\mathbb C W \cdot ([1_G],W)$. 

Naturally, this is in agreement with the statement that ground states are in one-to-one correspondence with the transitive $G$-set $G/H$. Indeed, it is always possible to construct a local operator that projects onto a specific ground state so that its expectation value is non-vanishing in this ground state only. For instance, consider the renormalisation group fixed points of sec.~\ref{sec:VecG_RG}. For every $a \in \{1,\ldots,(G:H)\}$, defining $\varsigma_a \colon G \to \mathbb C$ via $\varsigma_a(g) = 1$ if $g \in r_aH$, and $0$ otherwise, 
we have $\mathbb C(G/H) = \mathbb C \{ \varsigma_a\}_{a=1,\ldots,(G:H)}$. The map $\mathbb C(G/H) \otimes A(H,\beta,U)^*  \to A(H,\beta,U)^*$ defined by $\varsigma_{a_2} \otimes \si{a_1}{b_1}{b_2} \mapsto \delta_{a_1,a_2} \, \si{a_1}{b_1}{b_2}$, for every $a_1,a_2 \in \{1,\ldots,(G:H)\}$ and $b_1,b_2 \in \{1,\ldots,\dim_\mathbb C U\}$, is clearly $G$-equivariant and defines a (reducible) multiplet of local operators. The local operator $\mc O(\varsigma_a)_\msf i$ acts as a projector onto the ground state labelled by $a$. Finally, local operators can be combined into a single order parameter in this reducible multiplet that distinguishes all the $(G:H)$-many degenerate ground states.

\bigskip \noindent
Let us now study the general scenario. Given a simple object $([x_0],\hat W)$ in $\Vect_G^G$, consider the string operator $\mc S(x,c_1,c_2)_{\msf i,\ldots,\msf j}$ constructed in eq.~\eqref{eq:VecG_StringOp}, for $x \in [x_0]$ and $c_1,c_2 \in \{1,\ldots,\dim_\mathbb C \hat W\}$. The expectation value of this string operator in the ground state labelled by $a \in \{1,\ldots,(G:H)\}$ is provided by 
\begin{equation}
    \label{eq:VecG_StringOpExp}
    \StringOpExpVal{a}{x}{c_1}{c_2} .
\end{equation}
Employing eq.~\eqref{eq:VecG_localAction} at every site $\msf k \in [\msf i+1,\msf j-1]$ together with eq.~\eqref{eq:VecG_orthoAction}, it follows from eq.~\eqref{eq:VecG_transferprojector} that \eqref{eq:VecG_StringOpExp} evaluates to
\begin{equation}
    \label{eq:VecG_expValFact}
    \StringOpExpValLimL{a}{x}{\, \bar \phi^{x,a}}{c_1}{\Xi_a}    
    \cdot
    \StringOpExpValLimR{a}{x}{\, \phi^{x,a}}{c_2}{\Xi_a}  \, ,
\end{equation}
up to terms that decay exponentially in the length of the string. Immediately, one obtains a preliminary constraint. In order for the factors in eq.~\eqref{eq:VecG_expValFact} not to vanish, the line $\mathbb C_x \in \Vect_G$ must belong to the algebra of preserved lines, i.e., $x$ must lie in $r_aHr_a^{-1}$. Thus we assume that there is a group element $h \in H$ such that $x= {}^{r_a}h$. Which additional necessary conditions must the string operator satisfy in order for these factors not to vanish? 

Clearly, the two factors in eq.~\eqref{eq:VecG_expValFact} play symmetric roles. Therefore, it is enough to study one of them. For instance, let us focus on the right-hand-side factor. This quantity is the result of contracting a twisted sector local operator of the form \eqref{eq:VecG_twistedLocalOpDual} with the tensor network: 
\begin{equation}
    \label{eq:VecG_twistedLocalOpEnv}
    \TwisedLocalOpEnv{a}{{}^{r_a}h}{\phi^{r_ahr_a^{-1}\!,a}}{\Xi_a} \, .
\end{equation}
While twisted sector local operators of the form $\eqref{eq:VecG_twistedLocalOpDual}$ are arranged into the Hermitian dual of the simple module over $\Tu(\Vect_G)$ labelled by $([x_0],\hat W)$, we claim that tensor networks of the form \eqref{eq:VecG_twistedLocalOp}, for every $h \in H$ and $a \in \{1,
\ldots,(G:H)\}$, form a reducible module over $\Tu(\Vect_G)$. Acting with the tube \eqref{eq:VecG_Tube} corresponding to $e_g \otimes \varsigma_y \in \Tu(\Vect_G)$ on the tensor network \eqref{eq:VecG_twistedLocalOpEnv} labelled by $h \in H$ and $a \in \{1,\ldots,(G:H)\}$ results in
\begin{equation}
    \delta_{y,{}^{r_a}h}
    \TubeModEnvA{g}{\, {}^{gr_a}h}{g{}^{r_a}h}{{}^{r_a}h}{a}{\phi^{r_ahr_a^{-1}\!,a}}{}{}{\Xi_a} \, . 
\end{equation}
One can simplify this tensor network by utilising eq.~\eqref{eq:VecG_localAction} together with eq.~\eqref{eq:VecG_orthoAction}, before invoking eq.~\eqref{eq:VecG_actionFixedPoints} as well as eq.~\eqref{eq:VecG_moduleAssoc} and its converse:
\begin{equation}
\begin{split}
    \label{eq:VecG_TubeModEnv}
    \delta_{y,{}^{r_a}h}
    \TubeModEnvB{g}{\, {}^{gr_a}h}{g{}^{r_a}h}{{}^{r_a}h}{a}{g(a)}{\phi^{g,a}\;\;\;\;}{\phi^{r_ahr_a^{-1}\!,a}}{\;\;\;\;\bar \phi^{g,a}}
    \!\!\! 
    &=    
    \delta_{y,{}^{r_a}h}   \,
    \frac{\beta(h_{g,({}^{r_a}h)(a)},h_{{}^{r_a}h,a})}{\beta(h_{{}^{gr_a}h,g(a)},h_{g,a})}
    \TwisedLocalOpEnv{g(a)}{{}^{gr_a}h}{\phi^{gr_ah(gr_a)^{-1}\!,a}}{\;\;\;\,\Xi_{g(a)}} 
    \\[-2em]
    &= 
    \delta_{y,{}^{r_a}h} \,
    \frac{\beta(h_{g,a},h)}{\beta(h_{g,a}hh_{g,a}^{-1},h_{g,a})}
    \TwisedLocalOpEnv{g(a)}{{}^{gr_a}h}{\phi^{gr_ah(gr_a)^{-1}\!,g(a)}}{\;\;\;\,\Xi_{g(a)}}
    ,
\end{split}
\end{equation}
where in the last step we used that $h_{g,({}^{r_a}h)(a)} = h_{g,a}$, $h_{{}^{r_a}h,a}=h$ and $h_{{}^{gr_a}h,g(a)} = h_{g,a}h h_{g,a}^{-1}$, for every $g \in G$, $h \in H$ and $a \in \{1,\ldots,(G:H)\}$. Let us confirm that this action does endow the space of tensor networks \eqref{eq:VecG_twistedLocalOpEnv} with the structure of a module over $\Tu(\Vect_G)$. Firstly, it follows from $gr_a = r_{g(a)}h_{g,a}$, which implies ${}^{gr_a}h = r_{g(a)}(h_{g,a}h h_{g,a}^{-1})r_{g(a)}^{-1}$, that \eqref{eq:VecG_TubeModEnv} is indeed of the form \eqref{eq:VecG_twistedLocalOpEnv}, whereby the label in $\{1,\ldots,(G:H)\}$ is chosen to be $g(a)$, while the group element in $H$ is chosen to $h_{g,a}hh_{g,a}^{-1}$. Secondly, let
\begin{equation}
    \label{eq:transgression}
    \beta_{h_2}(h_1) \equiv  \msf t(\beta)(h_2\! \xrightarrow{\,h_1\,} \! h_1h_2h_1^{-1}) := \frac{\beta(h_1,h_2)}{\beta(h_1h_2h_1^{-1},h_1)} \, ,
\end{equation}
for every $h_1,h_2 \in H$. The so-called \emph{transgression} map $\msf t : Z^2 (H,\rU(1)) \to Z^1(H/\!/H,\rU(1))$ maps the group 2-cocycle $\beta$ to the groupoid 1-cocycle $\msf t(\beta)$, which satisfies 
\begin{equation}
\begin{split}
    \label{eq:loopGrpCocycleCond}
        \beta_{h_3}(h_1h_2) &= 
        \msf t(\beta) (h_3 \! \xrightarrow{\, h_1h_2 \,}h_1h_2h_3h_2^{-1}h_1^{-1})
        \\
        &= \msf t(\beta)(h_2h_3h_2^{-1} \! \xrightarrow{\, h_1 \,} \! h_1h_2h_3h_2^{-1}h_1^{-1}) \, \msf t(\beta)(h_3 \! \xrightarrow{\, h_2 \,}\! h_2h_3h_2^{-1})
        = \beta_{h_2h_3h_2^{-1}}(h_1) \, \beta_{h_3}(h_2) \, ,
\end{split}
\end{equation}
for every $h_1,h_2,h_3 \in H$. In particular, notice that for every $h \in H$, the restriction of $\beta_h$ to the centraliser $Z_H(h)$ of $h$ in $H$ defines a one-dimensional representation that we denote by $\mathbb C_{\beta_h}$.
This confirms that the tensor networks \eqref{eq:VecG_twistedLocalOpEnv} arrange into a module over the tube algebra $\Tu(\Vect_G)$. Let us denote this $\Tu(\Vect_G)$-module by $L(H,\beta)$.

What can we infer from these derivations? For every basis vector of the simple module labelled by $([x_0],\hat W)$ and every basis vector of the module $L(H,\beta)$, the right-hand-side factor of \eqref{eq:VecG_expValFact} computes a complex number. As matter of fact, this operation defines an inner product between these $\Tu(\Vect_G)$-modules. By definition \eqref{eq:VecG_Tube} of the $*$-structure of $\Tu(\Vect_G)$, one can further confirm that this inner product is invariant under the action of $\Tu(\Vect_G)$. This implies that, in order for the expectation value \eqref{eq:VecG_StringOpExp} of the string operator $\mc S(x,c_1,c_2)_{\msf i, \ldots, \msf j}$ not to vanish, the simple module $([x_0],\hat W)$ must be direct summand of $L(H,\beta)$. Conversely, suppose that the string operator $\mc S(x,c_1,c_2)_{\msf i,\ldots,\msf j}$ acquires a non-vanishing expectation value, one can conclude that $([x_0],\hat W)$ is a direct summand of $L(H,\beta)$. Now, the definition of the string operator $\mc S(x,c_1,c_2)_{\msf i,\ldots,\msf j}$ requires choosing $G$-equivariant maps $\varphi^{WV}_V : W \otimes V \to V$ and $\varphi^{VW^*}_V : V \otimes W^*  \to V$. Computing the number of inequivalent multiplets of twisted local operators transforming in the module $([x_0],\hat W)$ of $
\Tu(\Vect_G)$ that produce linearly independent expectation values amounts to computing the multiplicity of $([x_0],\hat W)$ in $L(H,\beta)$, thus allowing us to fully reconstruct this module. 

Before continuing, let us briefly comment on the expectation value of the string operators on the block-injective MPS $\Psi(\theta)=\sum_a\Psi(\theta_a)$ as opposed to on its blocks $\Psi(\theta_a)$. Because distinct blocks are orthogonal, the mixed transfer matrices \eqref{mixedtransfer} have spectral radii less than $1$. Therefore, for any finitely supported operator $\mathcal O$, the overlap of $\Psi(\theta_{a_1})$ and $\msf O \Psi(\theta_{a_2})$ vanishes unless $a_1=a_2$. We conclude that the full expectation value is determined by the expectation values in the individual blocks.

\subsection{Lagrangian algebras\label{sec:VecG_Lagrangian}}

Above, we demonstrated that string order parameters for the $G$-symmetric gapped phase associated with the $\Vect_G$-module category $\mc M(H,\beta)$ are such that the corresponding twisted sector local operators form the $\Tu(\Vect_G)$-module $L(H,\beta)$. Let us elucidate further the meaning of this object. 

We begin by computing the $G$-equivariant $G$-graded vector space corresponding to $L(H,\beta)$ under $\Vect_G^G \simeq \Mod(\Tu(\Vect_G))$. Consider the twisted group algebra $\mathbb C[H]^\beta = \mathbb C\{e_h\}_{h \in H}$. It is equipped with an $H$-equivariant structure given by
\begin{equation}
    h_1 \cdot e_{h_2} = \beta_{h_2}(h_1) \,   e_{h_1h_2h_1^{-1}} \, , 
\end{equation}
for every $h_1,h_2 \in H$. We promote the $H$-action of $\mathbb C[H]^\beta$ to a $G$-action via the induction functor:
\begin{equation} \label{eq:LA_definition}
    \Ind_H^G(\mathbb C[H]^\beta) := \mathbb C\big\{ \varsigma : G \to \mathbb C[H]^\beta \, \big| \, \varsigma(gh^{-1}) = h \cdot \varsigma(g)\big\}_{\forall g \in G, \forall h \in H} \, .
\end{equation}
The $G$-equivariant structure, as usual, is given by $(g \cdot \varsigma)(-) = \varsigma(g^{-1}-)$. Moreover, $\Ind_H^G(\mathbb C[H]^\beta)$ is $G$-graded, whereby every function $\varsigma \in \Ind_H^G(\mathbb C[H]^\beta)$ is homogeneous of degree $x \in G$ if and only if $\varsigma(y) \in \mathbb C[H]^\beta$ is homogeneous of degree $y^{-1}xy$, for every $y \in G$. In particular, we have $\Ind_H^G(\mathbb C[H]^\beta)_{1_G} \cong \mathbb C(G/H)$. Furthermore, if $\varsigma$ is homogeneous of degree $x \in G$, then, by definition, $g \cdot \varsigma$ is homogeneous of degree $gxg^{-1}$. Therefore, $\Ind_H^G(\mathbb C[H]^\beta)$ defines an object in $\Vect_G^G$. 
Mimicking the derivations of sec.~\ref{sec:VecG_RG}, we find that
$\Ind_H^G(\mathbb C[H]^\beta) = \mathbb C\{\varsigma_{a,h}\}_{a=1,\ldots,(G:H),h\in H}$, where the function $\varsigma_{a,h} \colon G \to \mathbb C[H]^\beta$ is defined by
\begin{equation}
    \label{eq:LA_basisFunctions}
    \varsigma_{a,h}(g) =
    \begin{cases}
        \beta_h(h_1^{-1}) \,  e_{h_1^{-1}h h_1} \q &\text{if $g \equiv r_ah_1 \in r_aH$} \, ,
        \\
        0 &\text{otherwise} \, ,
    \end{cases}
\end{equation}
for every $g \in G$. By definition, the function $\varsigma_{a,h}$ is homogeneous of degree $|\varsigma_{a,h}| = r_ahr_a^{-1} \in G$.
Let us examine the action of $G$ on $\varsigma_{a,h}$, for every $a \in \{1,\ldots,(G:H)\}$ and $h \in H$. By definition, 
$(g \cdot \varsigma_{a,h})(x) = \varsigma_{a,h}(g^{-1}x)$, for every $g,x \in G$, which is non-zero if and only if $x \in g \cdot r_aH = r_{g(a)}H$. 
Moreover, recall that for each $g \in G$ and $r_aH \in G/H$, there is a unique $h_{g,a} \in H$ such that $gr_a = r_{g(a)}h_{g,a}$. Suppose that $x \equiv r_{g(a)}h_1 \in r_{g(a)}H$, we find
\begin{equation}
\begin{split}
    (g \cdot \varsigma_{a,h})(x)
    &= \varsigma_{a,h}(r_a h_{g,a}^{-1}r_{g(a)}^{-1}r_{g(a)}h_1)
    \\[-.5em]
    &= (h_1^{-1}h_{g,a}) \cdot  \varsigma_{a,h}(r_a)
    = 
    \beta_h(h_1^{-1}h_{g,a}) \, e_{h_1^{-1}h_{g,a}hh_{g,a}^{-1}h_1} \, ,
\end{split}
\end{equation}
where we used in the last step the fact that $\varsigma_{a,h}(r_a) = e_h$.
Invoking the cocycle condition \eqref{eq:loopGrpCocycleCond} finally implies that
\begin{equation}
    \label{eq:GequivLalg}
    g \cdot \varsigma_{a,h}
    = \beta_h(h_{g,a}) \, \varsigma_{g(a), h_{g,a}hh_{g,a}^{-1}} \, .
\end{equation}
At this point, one can explicitly verify that $|g \cdot \varsigma_{a,h}| = g |\varsigma_{a,h}|g^{-1}$. 

As detailed in app.~\ref{app:centres}, the corresponding $\Tu(\Vect_G)$-module is simply given by $\mathbb C\{\varsigma_{a,h}\}_{a,h}$ with
\begin{equation}
    (e_g \otimes \varsigma_y) \cdot \varsigma_{a,h} := \delta_{y,{}^{r_a}h} \, \beta_h(h_{g,a}) \, \varsigma_{g(a), h_{g,a}hh_{g,a}^{-1}} \, ,
\end{equation}
for every $g,y \in G$, $a \in \{1,\ldots,(G:H)\}$ and $h \in H$. This is precisely the $\Tu(\Vect_G)$-module $L(H,\beta)$ we found in sec.~\ref{sec:VecG_OP}. Concretely, the basis vector $\varsigma_{a,h}$ is identified with the tensor network \eqref{eq:VecG_twistedLocalOpEnv} associated with the ground state labelled by $a$ and twisting line $\mathbb C_{|\varsigma_{a,h}|} = \mathbb C_{r_ahr_a^{-1}} \in \Vect_G$.

At this point, it would be useful to have an explicit formula providing the multiplicity of any simple object $\equiv ([x_0],\hat W) \in \Vect_G^G$ in $L(H,\beta)$. As modules over $\Tu(\Vect_G)$, characters of $([x_0],\hat W)$ and $L(H,\beta)$ are given by (see app.~\ref{app:centres}) 
\begin{equation}
    \begin{split}
        \chi^{([x_0],\hat W)}(e_g \otimes \varsigma_x) &= \delta_{x \in [x_0]} \, \delta_{gx,xg} \, \chi^{\hat W}(z_{g,x}) \, ,
        \\
        \chi^{L(H,\beta)}(e_g \otimes \varsigma_x) &= \frac{\delta_{gx,xg}}{|H|} \sum_{y \in G} \delta_{g \in y^{-1}Hy} \, \delta_{x \in y^{-1}Hy} \, \beta_{yxy^{-1}}(ygy^{-1}) \, ,
    \end{split} \label{eq:LA_character}
\end{equation}
respectively, for every $e_g \otimes \varsigma_x \in \Tu(\Vect_G)$. Their inner product reads
\begin{equation}
\begin{split}
    \label{eq:multiplicityL}
    \la \chi^{L(H,\beta)} , \chi^{([x_0],\hat W )}\ra
    &= \frac{1}{|G|} \sum_{g,x \in G} \overline{\chi^{L(H,\beta)}(e_g \otimes \varsigma_x)} \, \chi^{([x_0],\hat W)}(e_g \otimes \varsigma_x)
    \\
    &= \frac{1}{|G||H|} \sum_{g,x,y \in G}
    \delta_{gx,xg} \, \delta_{g \in y^{-1}Hy} \, \delta_{x \in y^{-1}Hy}  \, \delta_{x \in [x_0]} \, \overline{\beta_{yxy^{-1}}(ygy^{-1})} \, \chi^{\hat W}(z_{g,x})
    \\
    &= \frac{1}{|H|} \sum_{g,x \in G} \delta_{gx,xg} \, \delta_{g \in H} \, \delta_{x \in H} \, \delta_{x \in [x_0]} \, 
    \frac{\chi^{\hat W}(z_{g,x})}{\beta_x(g)}
    \\
    &= \frac{1}{|H|} \sum_{x \in [x_0] \cap H} \sum_{g \in Z_H(x)} \frac{\chi^{\hat W}(r_x^{-1}gr_x)}{\beta_x(g)} = \frac{1}{|H|} \sum_{x \in [x_0] \cap H} \sum_{g \in Z_H(x_0)} \frac{\chi^{\hat W}(g)}{\beta_{x_0}(g)}
\end{split}
\end{equation}
where in the last step we used the facts that $\beta_{r_x x_0 r_x^{-1}}(r_x gr_x^{-1}) = \beta_{x_0}(g)$, for every $g \in Z_H(x_0)$ and $x \in [x_0]$. By inspection, we find that 
\begin{equation}
    \la \chi^{L(H,\beta)} , \chi^{([x_0],\hat W )}\ra
    =
    \la \chi^{\mathbb C[H]^\beta}, \chi^{\Res^G_H([x_0],\hat W)} \ra \, ,
\end{equation}
which is consistent with the Frobenius reciprocity condition
\begin{equation}
    \Hom_{\Vect_G^G}\big(L(H,\beta),W\big) \cong \Hom_{\Vect_G^H}\big(\mathbb C[H]^\beta,\Res^G_H(W)\big) \, ,
\end{equation}
where $\Vect_G^H$ is the category of $H$-equivariant $G$-graded vector spaces. 

Let us explore a couple of special cases. On the one hand, choosing $H = \{1_G\}$, one finds that $L(\{1_G\},1) \cong \mathbb C(G) \in \Vect_G^G$, which is consistent with the fact that in the gapped phase where the whole symmetry is spontaneously broken, all the order parameters are local operators. On the other hand, choosing $H = G$, one finds that $L(G,\beta) \cong \mathbb C[G]^\beta \in \Vect_G^G$, which decomposes as 
\begin{equation}
    \label{eq:LA_SPT}
    \mathbb C[G]^\beta \cong \bigoplus_{[x_0] \in \Cl(G)} ([x_0],\mathbb C_{\beta_{x_0}}) \, .
\end{equation}
Notice that the object $L(G,\beta) \in \Vect_G^G$ only depends on $\beta$ through its transgression \eqref{eq:transgression}. 
As such, it is possible for multiplets of twisted sector local operators associated with distinct $G$-symmetric gapped phases to correspond to the same object in $\Vect_G^G$.
This is a phenomenon that is well documented for gapped phases preserving a finite group with non-trivial \emph{Bogomolov multiplier} \cite{PhysRevB.86.125441,10.1063/1.4895764,Kobayashi:2025ykb}.\footnote{A finite group $G$ with non-trivial Bogomolov multiplier produces distinct symmetry protected topological phases whose response theories produce identical topological invariants on the two-torus. This can only happen if $G$ is non-abelian. Indeed, whenever $G$ is abelian, is is enough to compute the topological invariants on the two-torus. Within our context, this means that whenever $G$ is abelian the object $L(H,\beta)$ fully specifies the gapped phase. In other words, we must be able to reconstruct a 2-cocycle $\beta$ from the collection $\{\mathbb C_{\beta_x}\}_{x \in G}$ of one-dimensional representations. Indeed, consider the map $Z^2(G,\rU(1)) \to \Hom(G,\Hom(G,\rU(1)))$ given by $\beta \mapsto (x \mapsto \beta_x)$. It can be verified to be a group homomorphism whose kernel consists of coboundaries, so that $\mathbb C_{\beta_x}$ being trivial for every $x \in G$ implies that $\beta$ is trivial  \cite{deGroot:2021vdi}.} However, this does not necessarily imply that the same string operators may serve as order parameters for distinct gapped phases. Indeed, even though the corresponding twisted sector local operators transform in the same way, they may still be different.

\bigskip \noindent
Mathematically, the resolution of the above conundrum goes as follows: By induction, the object $L(H,\beta) \in \Vect_G^G$ is equipped with the structure of an algebra in $\Vect_G^G$, i.e., a $G$-equivariant $G$-graded algebra. The multiplication rule, which is provided by the pointwise multiplication of functions $G \to \mathbb C[H]^\beta$, reads
\begin{equation}
    \varsigma_{a_1,h_1} \cdot \varsigma_{a_2,h_2} = \delta_{a_1,a_2} \, \beta(h_1,h_2) \, \varsigma_{a_1,  h_1h_2} \, ,
\end{equation}
for every $a_1,a_2 \in \{1,\ldots,(G:H)\}$ and $h_1,h_2 \in H$, where we used the fact that
\begin{equation}
    \beta_{h_1}(h_3) \, \beta_{h_2}(h_3) \,  \beta(h_3h_1h_3^{-1},h_3h_2h_3^{-1}) = \beta_{h_1h_2}(h_3) \, \beta(h_1,h_2) \, , 
\end{equation}
for every $h_1,h_2,h_3 \in H$. As expected, one recovers that $L(G,\beta)$ is equivalent to $\mathbb C[G]^\beta$ as an algebra. Notice that the multiplication rule of $L(H,\beta)$ explicitly depends on the 2-cocycle $\beta$. Therefore, even for two 2-cocycles $\beta_1$ and $\beta_2$ such that $L(H,\beta_1)$ and $L(H,\beta_2)$ would be isomorphic as objects in $\Vect_G^G$, they would still be non-isomorphic algebra objects. As a matter fact, one can readily check that $L(H,\beta)$ is a \emph{Lagrangian algebra} in $\Vect_G^G$, i.e., a connected, indecomposable, separable and commutative algebra such that ${\rm FPdim}(L(H,\beta)) = |G|$ (see app.~\ref{app:algebras}).\footnote{Note that $L(H,\beta)$ is clearly not commutative as an algebra in $\Vect$, but it is commutative as an algebra in $\Vect_G^G$ thanks to the definition of the braiding (see app.~\ref{app:algebras}) \cite{DAVYDOV2017149}.} Crucially, Lagrangian algebras in $\Vect_G^G$ are in one-to-one correspondence with indecomposable module categories over $\Vect_G$ so that each symmetric gapped phase is assigned a unique Lagrangian algebra, up to isomorphisms. Overall, we have shown that string order parameters do organise into the Lagrangian algebra in $\mc Z(\Vect_G)$ associated with the gapped phase \cite{Bhardwaj:2023idu}.

\subsection{Multiplication rule\label{sec:VecG_multiplication}}

Physically, one detects the multiplication rule of the Lagrangian algebra $L(H,\beta)$ by checking the fusion rules of the corresponding twisted sector local order parameters. For instance, consider the renormalisation group fixed point we built in sec.~\ref{sec:VecG_RG} from the algebra $A(H,\beta,U) \in \Rep(G)$.
Let $\varphi^{L(H,\beta)A(H,\beta,U)^*}_{A(H,\beta,U)^*} \colon L(H,\beta) \otimes A(H,\beta,U)^*  \to A(H,\beta,U)^*$ be the linear map defined by
\begin{equation}
    \varsigma_{a_2,h}  \otimes \si{a_1}{b_1}{b_2} 
    \mapsto \delta_{a_1,a_2} \, \sigma_{a_1,u^{b_1} \otimes \, h \cdot u_{b_2}} \, .
    \label{eq:canonicalmap}
\end{equation}
It follows from
\begin{equation}
\begin{split}
    &(g \cdot \varsigma_{a_2,h})  \otimes (g \cdot \si{a_1}{b_1}{b_2}) 
    \\
    & \q \mapsto  
    \delta_{g(a_1),g(a_2)} \,
    \beta_h(h_{g,a}) \, 
    \sigma_{g(a_1),h_{g,a} \cdot u^{b_1} \otimes (h_{g,a}hh_{g,a}^{-1}) \cdot (h_{g,a} \cdot u^{b_2})} 
    \\ 
    & \hspace{2.2em} =     
    \delta_{a_1,a_2} \,
    \beta(h_{g,a},h) \, 
    \sigma_{g(a_1),h_{g,a} \cdot u^{b_1} \otimes (h_{g,a}h) \cdot u^{b_2}} 
    \\ 
    & \hspace{2.2em} =    
    \delta_{a_1,a_2} \,
    \sigma_{g(a_1),h_{g,a} \cdot u^{b_1} \otimes h_{g,a} \cdot (h \cdot u^{b_2})} 
    = \delta_{a_1,a_2} \; g \cdot \sigma_{a_1,(u^{b_1} \otimes  h \cdot u_{b_2})} \, , 
\end{split}
\end{equation} 
that $\varphi^{L(H,\beta)A(H,\beta,U)^*}_{A(H,\beta,U)^*}$ is a $G$-equivariant map. Therefore, one can use it to construct twisted sector local operators. Notice that these twisted sector local operators are special in the sense that they yield string operators that stabilise the ground states. Moreover, from the definition, it is clear that such twisted sector local operators fuse according to the multiplication rule of $L(H,\beta)$, i.e.,
\begin{equation}
    \fusionTwistedLocalOp{{}^{r_{a_1}\!}h_1}{{}^{r_{a_2}\!}h_2}{\q\q{}^{r_{a_1}\!}h_1{}^{r_{a_2}\!}h_2}{a_2}{a_1} = \delta_{a_1,a_2} \fusionTwistedLocalOp{{}^{r_{a_1}\!}h_1}{{}^{r_{a_1}\!}h_2}{\q\;\;\;{}^{r_{a_1}\!}(h_1h_2)}{a_1}{a_1} = \delta_{a_1,a_2} \, \beta(h_1,h_2) \twistedLocalOp{a_1}{\q\;\;\;{}^{r_{a_1}\!}(h_1h_2)}{},  
\end{equation}
for every $h_1,h_2 \in H$ and $a_1,a_2 \in \{1,\ldots,(G:H)\}$.
But there is considerable freedom in finding a multiplet of twisted sector local operators that furnish a copy of the object $L(G,\beta)$ and there is no guarantee a priori that they should all reproduce this multiplication rule. 

\bigskip \noindent
In light of these comments, let us argue that the (canonical) multiplet associated with the map \eqref{eq:canonicalmap} is in fact the one that `survives in the infrared' in the sense described below. Pick any multiplet of twisted sector local operators furnishing the object $L(H,\beta)$. Although these twisted sector local operators transform in the same $\Tu(\Vect_G)$-module as those built from the $G$-equivariant map \eqref{eq:canonicalmap}, they may be built from a different map $\varphi^{L(H,\beta)A(H,\beta,U)^*}_{A(H,\beta,U)^*}$. Recall that $L(H,\beta) = \mathbb C\{\varsigma_{a,h}\}_{a=1,\ldots,(G:H),h \in H}$ with $|\varsigma_{a,h}| = r_ahr_a^{-1}$. The vector space underlying the $\Tu(\Vect_G)$-module associated with $L(H,\beta) = \bigoplus_{g \in G} L(H,\beta)_g$ is provided by $\bigoplus_{g \in G} \Hom_{\Vect_G}(\mathbb C_g,L(H,\beta))$. The dimension of $L(H,\beta)_g$ equals the number of cosets $r_aH \in G/H$ such that $g(a) = a$. Indeed, for every $a \in \{1,\ldots,(G:H)\}$ and $g \in G$ such that $g(a) = a$, we know that there is a unique $h_{g,a} \in H$ such that $g = r_a h_{g,a} r_a^{-1}$. Therefore, for every $g \in G$ and $a \in \{1,\ldots,(G:H)\}$ such that $g(a)=a$, one can construct a twisted sector local operator 
\begin{equation}
    \twistedLocalOp{a}{g}{} \!\!\!
    =  \mc O(\varsigma_{a,h_{g,a}}) \otimes e^g \, ,
\end{equation}
where we introduced the notation
\begin{equation}
    \mc O(\varsigma_{a,h_{g,a}}) := \varphi^{L(H,\beta)A(H,\beta,U)^*}_{A(H,\beta,U)^*}(\varsigma_{a,h_{g,a}} \otimes -) \, .
\end{equation}
Equivalently, for every $a \in \{1,\ldots,(G:H)\}$ and $h \in H$, one constructs a twisted sector local operator with twisting line $\mathbb C_{{}^{r_a}h} \in \Vect_G$. 
Similarly, one picks any multiplet of twisted sector local operators furnishing the Hermitian dual of $L(H,\beta)$. Together, they yield string operators of the form $\mc S(g,a_1,a_2)_{\msf i,\ldots,\msf j}$, for every $a_1,a_2 \in \{1,\ldots,(G:H)\}$ and $g \in G$ such that $g(a_1)=a_1$ and $g(a_2)=a_2$. In order for such a string operator to have a non-vanishing expectation value in the ground state subspace, we already know that the linear map $\mc O(\varsigma_{a_1,h_{g,a_1}})$ must decompose into blocks over $G/H$ because ground state labelled by distinct cosets in $G/H$ are orthogonal in the infrared limit.

Let us now compute what $\mc S(g,a_1,a_2)_{\msf i,\ldots,\msf j}$ 
projects onto in the infrared. We constructed in sec.~\ref{sec:VecG_RG} the MPS tensor $\theta$ of the renormalisation group fixed point associated with the algebra object $A(H,\beta,U) \in \Rep(G)$. After normalisation, the corresponding transfer matrix is provided by
\begin{equation}
    \mathbb E_\theta \equiv \bigoplus_{a=1}^{(G:H)} \mathbb E_{\theta_a} = 
    \frac{1}{\dim_\mathbb C U}\bigoplus_{a=1}^{(G:H)} 
    \TMatRG{a}{\mathbb I_{U^*}\;\;}{\,\mathbb I_U} \, .
\end{equation}
Clearly, both the left and right fixed points of this transfer matrix are provided by the identity map on the virtual space. In the same vein, for every $h \in H$, let us introduce the mixed transfer matrix
\begin{equation}
    \label{eq:VecG_MixedTMat}
    \mathbb E_\theta^h \equiv \bigoplus_{a=1}^{(G:H)} \mathbb E_{\theta_a}^h :=  \bigoplus_{a=1}^{(G:H)} 
    \TMatOp{a}{{}^{r_a}h}{} \, .
\end{equation}
Given $a \in \{1,\ldots,(G:H)\}$ and $h \in H$, one defines the right fixed point $\Xi_a^h$ of the mixed transfer matrix $\mathbb E^h_{\theta_a}$ according to
\begin{equation}
    \TMatOpFixed{1}{a}{{}^{r_a}h}{\Xi^h_a} \! 
    =
    \TMatOpFixed{2}{a}{{}^{r_a}h}{\Xi^h_a} \, .
\end{equation}
Using eq.~\eqref{eq:VecG_localAction}, together with the fact that the transfer matrix $\mathbb E_{\theta_a}$ has a unique (right) fixed point yields
\begin{equation}
    \raisebox{-8pt}{\TMatOpFixed{3}{a}{{}^{r_a}h}{\Xi^h_a}} \!\!\! = 
    \TMatOpFixed{4}{a}{{}^{r_a}h}{\; \mathbb I_{U^*}}\, .
\end{equation}
Remembering from sec.~\ref{sec:VecG_RG} that $\phi^{g,a} \colon e_g \otimes u \mapsto \pi(h_{g,a})(u)$, one easily deduces the space of right fixed points of the mixed transfer matrix $\mathbb E^h_\theta$. One proceeds similarly for the left fixed points. Finally, for every $h \in H$ we introduce
\begin{equation}
    \mathbb E_\theta^{\mc O,a_1,h} \equiv \bigoplus_{a=1}^{(G:H)} \mathbb E_{\theta_a}^{\mc O,a_1, h} :=  \bigoplus_{a=1}^{(G:H)} 
    \TMatOp{a}{{}^{r_a}h}{a_1} \, ,
\end{equation}
where $\mathbb E_{\theta_a}^{\mc O,a_1, h}$ is zero unless $({}^{r_a}h)(a_1) = a_1$.

Now, consider the expectation value of $\mc S(g,a_1,a_2)_{\msf i,\ldots,j}$ in the ground state labelled by $a \in \{1,\ldots,(G:H)\}$. In order to get a non-zero value, we already know that there must be a $h \in H$ such that $g = r_ahr_a^{-1}$. Since we are considering a renormalisation group fixed point model, it immediately reduces to an expression of the form \eqref{eq:VecG_expValFact}. As usual, it is enough to focus on the contribution from a single twisted sector local operator, say the one on the left-hand side. The left-hand `half' of the expectation value loosely takes the form $\mathbb E_{\theta_a} \cdots \mathbb E_{\theta_a} \, \mathbb E_{\theta_a}^{\mc O,a_1,h} \, \mathbb E_{\theta_a}^h \, \mathbb E_{\theta_a}^h \cdots$, such that 
\begin{equation}
    \mathbb E_{\theta_a} \, \mathbb E_{\theta_a}^{\mc O,a_1,h} \, \mathbb E_{\theta_a}^h = \frac{1}{(\dim_\mathbb C U)^2}
    \TMatProj
    \, .
\end{equation}
In particular, one recognises the type of inner product between $\Tu(\Vect_G)$-module one considered in sec.~\ref{sec:VecG_OP}, which enforces $a_1=a$ in order to be non-zero. 
One interprets the virtual operator $\mathbb E_{\theta_a} \, \mathbb E_{\theta_a}^{\mc O,a,h} \, \mathbb E_{\theta_a}^h$, which is the projection of $\mathbb E_{\theta_a}^{\mc O,a,h}$ onto the dominant eigenspace of the transfer matrix, as the part that survives in the infrared.\footnote{Since we are considering a renormalisation group fixed point, no information is actually projected out in the sense that replacing $\mathbb E_{\theta_a}^{\mc O,a,h}$ by $\mathbb E_{\theta_a} \, \mathbb E_{\theta_a}^{\mc O,a,h} \, \mathbb E_{\theta_a}^h$ in the computation of the expectation values actually provide the exact same result, up to normalisation.} By injectivity, for every $a \in \{1,\ldots,(G:H)\}$ and $h \in H$, there is a unique linear map $\widetilde{\mc O}(\varsigma_{a,h})$ such that
\begin{equation}
    \mathbb E^{\widetilde{\mc O},a,h}_{\theta_a} = \mathbb E_{\theta_a} \, \mathbb E_{\theta_a}^{\mc O,a,h} \, \mathbb E_{\theta_a}^h \, .
\end{equation}
which is simply obtained by absorbing the result of the aforementioned inner product into the identity matrix $\mathbb I_{U^*}$. Explicitly, one finds that $\widetilde{\mc O}(\varsigma_{a,h})$ is a linear map that projects onto the block of $A(H,\beta,U)^*$ labelled by $a$, and the operator in each block is proportional to $\mathbb I_{U^*} \otimes \pi(h)$. Bringing everything together, one obtains the projection in the infrared of the multiplet of twisted sector local operators, which readily fuse according to the multiplication rule of $L(H,\beta)$.

What about an arbitrary representative of the gapped phase labelled by $\mc M(H,\beta)$? For any representative, consider the corresponding block-injective tensor $\theta \colon M \to V \otimes M$ whose properties were discussed in sec.~\ref{sec:VecG_MPS}. By construction, $M = \bigoplus_{a=1}^{G:H} M_a$ such that $M_{a_1} \cong M_{a_2}$, for every $a_1,a_2 \in \{1,\ldots,(G:H)\}$. Moreover, it follows from eq.~\eqref{eq:VecG_moduleAssoc} that $M_a$, for every $a \in \{1,\ldots,(G:H)\}$, is equipped with the structure of a $\beta$-projective representation of $H$ in such a way that the whole virtual space $M \otimes M^*$ transforms like $\Ind_H^G(\End_\mathbb C(M_a))$, for any $a \in \{1,\ldots,(G:H)\}$. By definition, the tensor $\theta$ interpreted as a map from the virtual space to the physical one is injective. Moreover, thanks to eq.~\eqref{eq:VecG_localAction}, it is $G$-equivariant. Therefore, the physical vector space $V$ must contain $\Ind_H^G(\End_\mathbb C(M_a))$ as a subrepresentation. This guarantees that the local Hilbert space is large enough to accommodate all the twisted local operators furnishing the multiplet associated with the Lagrangian algebra $L(H,\beta)$. Besides, in light of the above derivations, we posit that, for any gapped representative, any multiplet of string order parameters flows in infrared to a multiplet of twisted local operators that not only transform like $L(H,\beta)$ but also fuse according to its multiplication rule.

\subsection{Example: $\Vect_{\mathbb Z_2 \oplus \mathbb Z_2}$\label{sec:VecG_CS}}

Consider the \emph{Klein four-group} $\mathbb Z_2 \oplus \mathbb Z_2 = \{\sn{()}, \sn{(12)(34)}, \sn{(13)(24)}, \sn{(14)(23)}\}$. As an abelian group, each conjugacy class contains a single group element and the stabiliser subgroup of each conjugacy class is the group $\mathbb Z_2 \oplus \mathbb Z_2$ itself. It admits four one-dimensional (linear) irreducible representations denoted by $\{\mathbb C_{(1,1)}, \mathbb C_{(1,-1)}, \mathbb C_{(-1,1)}, \mathbb C_{(-1,-1)}\}$. The character table is given by
\begin{equation}
    \begin{array}{l|cccc} 
        & \sn{()} & \sn{(12)(34)} & \sn{(13)(24)} & \sn{(14)(23)} \\ 
        \hline
        \mathbb C_{(1,1)}  & 1 & 1 & 1 & 1 \\
        \mathbb C_{(1,-1)} & 1 & 1 & -1 & -1 \\
        \mathbb C_{(-1,1)}  & 1 & -1 & 1 & -1 \\
        \mathbb C_{(-1,-1)} & 1 & -1 & -1 & 1
    \end{array} \, .
\end{equation}
It counts five (conjugacy classes of) subgroups given by $\mathbb Z_2 \oplus \mathbb Z_2$, $\mathbb Z_2 \oplus 0 = \la \sn{(12)(34)}\ra$, $0 \oplus \mathbb Z_2 = \la \sn{(13)(24)}\ra$, the diagonal subgroup $\Delta(\mathbb Z_2 \oplus \mathbb Z_2) = \la \sn{(14)(23)}\ra$ and the trivial subgroup. We employ the same notation for the irreducible representation of the subgroups as for $\mathbb Z_2 \oplus \mathbb Z_2$. The Schur multiplier $H^2(\mathbb Z_2 \oplus \mathbb Z_2,\rm U(1))$ is found to be isomorphic to $\mathbb Z_2$. Given a normalised representative of the non-trivial cohomology class in $H^2(\mathbb Z_2 \oplus \mathbb Z_2,\rU(1))$, the group $\mathbb Z_2 \oplus \mathbb Z_2$ admits a single irreducible projective representations denoted by $U_2$ and satisyfing $U_2 \otimes U_2 \cong \mathbb C_{(1,1)} \oplus \mathbb C_{(1,-1)} \oplus \mathbb C_{(-1,1)} \oplus \mathbb C_{(-1,-1)}$.  

The sixteen simple objects in $\Vect_{\mathbb Z_2 \oplus \mathbb Z_2}^{\mathbb Z_2 \oplus \mathbb Z_2}$ are simply obtained by pairing every group element with every linear irreducible representation, each labelling a multiplet of twisted sector local operators. Finally, using the above data, we find six Lagrangian algebras, namely $L(\mathbb Z_2,1)$, $L(\Delta(\mathbb Z_2 \oplus \mathbb Z_2),1)$, $L(0 \oplus \mathbb Z_2,1)$, $L(\mathbb Z_2 \oplus 0,1)$, $L(\mathbb Z_2 \oplus \mathbb Z_2,1)$ and $L(\mathbb Z_2 \oplus \mathbb Z_2,\beta)$, which we explicitly compute below:

\bigskip\noindent
$\bul$ $L(0,1)$: By definition, every vector in $L(0,1)$ is homogeneous of degree the identity in $\mathbb Z_2 \oplus \mathbb Z_2$ so that ${\rm supp}(L(0,1)) = \sn{()}$. Moreover, as an object in $\Rep(\mathbb Z_2 \oplus \mathbb Z_2)$, it is isomorphic to $\Ind_{\mathbb Z_1}^{\mathbb Z_2 \oplus \mathbb Z_2}(\mathbb C)$. Therefore,
\begin{equation}
    L(\mathbb Z_1,1) \cong (\sn{()},\mathbb C_{(1,1)}) \oplus (\sn{()},\mathbb C_{(1,-1)}) \oplus (\sn{()},\mathbb C_{(-1,1)}) \oplus (\sn{()},\mathbb C_{(-1,-1)}) \, .
\end{equation}

\medskip\noindent
$\bul$ $L(\Delta(\mathbb Z_2 \oplus \mathbb Z_2),1)$: Since $\mathbb Z_2 \oplus \mathbb Z_2$ is abelian, $\varsigma \in L(\Delta(\mathbb Z_2 \oplus \mathbb Z_2),1)$ is homogeneous of degree $y \in \mathbb Z_2 \oplus \mathbb Z_2$ if and only if $\varsigma(x) \in \mathbb C[\Delta(\mathbb Z_2 \oplus \mathbb Z_2)]$ is homogeneous of degree $y$, for every $x \in \mathbb Z_2 \oplus \mathbb Z_2$. Therefore, ${\rm supp}(L(\Delta(\mathbb Z_2 \oplus \mathbb Z_2)),1) = \{\sn{()}, \sn{(14)(23)}\}$ so that $L(\Delta(\mathbb Z_2 \oplus \mathbb Z_2),1)$ decomposes as $L(\Delta(\mathbb Z_2 \oplus \mathbb Z_2),1)_{()} \oplus L(\Delta(\mathbb Z_2 \oplus \mathbb Z_2),1)_{(14)(23)}$. Moreover, as objects in $\Rep(\mathbb Z_2 \oplus \mathbb Z_2)$, both $L(\Delta(\mathbb Z_2 \oplus \mathbb Z_2),1)_{()}$ and $L(\Delta(\mathbb Z_2 \oplus \mathbb Z_2),1)_{(14)(23)}$ are isomorphic to $\Ind_{\Delta(\mathbb Z_2 \oplus \mathbb Z_2)}^{\mathbb Z_2 \oplus \mathbb Z_2}(\mathbb C_{(1,1)}) \cong \mathbb C_{(1,1)} \oplus \mathbb C_{(-1,-1)}$. Bringing everything together, one obtains
\begin{equation}
    L(\Delta(\mathbb Z_2 \oplus \mathbb Z_2),1) \cong (\sn{()},\mathbb C_{(1,1)}) \oplus (\sn{()}, \mathbb C_{(-1,-1)}) \oplus (\sn{(14)(23)}, \mathbb C_{(1,1)}) \oplus (\sn{(14)(23)}, \mathbb C_{(-1,-1)})  \, .
\end{equation}

\medskip \noindent
$\bul$ $L(0 \oplus \mathbb Z_2,1)$: Proceeding as above, we have $L(0 \oplus \mathbb Z_2,1) \cong L(0 \oplus \mathbb Z_2,1)_{()} \oplus L(0 \oplus \mathbb Z_2,1)_{(13)(24)}$. Moreover, as objects in $\Rep(\mathbb Z_2 \oplus \mathbb Z_2)$, both $L(0 \oplus \mathbb Z_2,1)_{()}$ and $L(0 \oplus \mathbb Z_2,1)_{(13)(24)}$ are isomorphic to $\Ind_{0 \oplus \mathbb Z_2}^{\mathbb Z_2 \oplus \mathbb Z_2}(\mathbb C_{(1,1)}) \cong \mathbb C_{(1,1)} \oplus \mathbb C_{(-1,1)}$, so that
\begin{equation}
    L(0 \oplus \mathbb Z_2,1) \cong (\sn{()},\mathbb C_{(1,1)}) \oplus (\sn{()},\mathbb C_{(-1,1)}) \oplus (\sn{(13)(24)},\mathbb C_{(1,1)}) \oplus (\sn{(13)(24)},\mathbb C_{(-1,1)}) \, . 
\end{equation}

\medskip \noindent
$\bul$ $L(\mathbb Z_2 \oplus 0,1)$: Proceeding as above, we have $L(\mathbb Z_2 \oplus 0,1) \cong L(\mathbb Z_2 \oplus 0,1)_{()} \oplus L(\mathbb Z_2 \oplus 0,1)_{(12)(34)}$. Moreover, as objects in $\Rep(\mathbb Z_2 \oplus \mathbb Z_2)$, both $L(\mathbb Z_2 \oplus 0,1)_{()}$ and $L(\mathbb Z_2 \oplus 0,1)_{(12)(34)}$ are isomorphic to $\Ind_{\mathbb Z_2 \oplus 0}^{\mathbb Z_2 \oplus \mathbb Z_2}(\mathbb C_{(1,1)}) \cong \mathbb C_{(1,1)} \oplus \mathbb C_{(1,-1)}$, so that
\begin{equation}
    L(\mathbb Z_2 \oplus 0,1) \cong (\sn{()},\mathbb C_{(1,1)}) \oplus (\sn{()},\mathbb C_{(1,-1)}) \oplus (\sn{(12)(34)},\mathbb C_{(1,1)}) \oplus (\sn{(12)(34)},\mathbb C_{(1,-1)}) \, . 
\end{equation}

\medskip\noindent
$\bul$ $L(\mathbb Z_2 \oplus \mathbb Z_2,1)$: By definition, $L(\mathbb Z_2 \oplus \mathbb Z_2,1) \cong \mathbb C[\mathbb Z_2 \oplus \mathbb Z_2]$ such that all the homogeneous elements transform in the trivial representation of $\mathbb Z_2 \oplus \mathbb Z_2$. It immediately follows that
\begin{equation}
    L(\mathbb Z_2 \oplus \mathbb Z_2,1) \cong (\sn{()},\mathbb C_{(1,1)}) \oplus (\sn{(12)(34)},\mathbb C_{(1,1)}) \oplus (\sn{(13)(24)},\mathbb C_{(1,1)}) \oplus (\sn{(14)(23)},\mathbb C_{(1,1)}) \, .
\end{equation}

\medskip \noindent
$\bul$ $L(\mathbb Z_2 \oplus \mathbb Z_2,\beta)$: Let us now choose $\beta$ to be the normalised representative of the non-trivial cohomology class in $H^2(\mathbb Z_2 \oplus \mathbb Z_2, \rm U(1))$ such that $\beta(g_1,g_2)$ equals $-1$ if $g_1 \in \{\sn{(13)(24)},\sn{(14)(23)}\}$ and $g_2 \in \{\sn{(12)(34)},\sn{(14)(23)}\}$, and equals $+1$ otherwise. By definition, $L(\mathbb Z_2 \oplus \mathbb Z_2,\beta) = \mathbb C[\mathbb Z_2 \oplus \mathbb Z_2]^\beta = \mathbb C\{e_g\}_{g \in \mathbb Z_2 \oplus \mathbb Z_2}$ with the $G$-equivariant structure being provided by $g_1 \cdot e_{g_2} = \frac{\beta(g_1,g_2)}{\beta(g_2,g_1)} \, e_{g_2}$, for every $g_1,g_2 \in \mathbb Z_2 \oplus \mathbb Z_2$. By inspection, $\mathbb C\{e_{(12)(34)}\} \cong \mathbb C_{(1,-1)}$, $\mathbb C\{e_{(13)(24)}\} \cong \mathbb C_{(-1,1)}$ and $\mathbb C\{e_{(14)(23)}\} \cong \mathbb C_{(-1,-1)}$. 
Bringing everything together yields
\begin{equation}
    L(\mathbb Z_2 \oplus \mathbb Z_2,\beta) \cong (\sn{()},\mathbb C_{(1,1)}) \oplus (\sn{(12)(34)},\mathbb C_{(1,-1)}) \oplus (\sn{(13)(24)},\mathbb C_{(-1,1)}) \oplus (\sn{(14)(23)},\mathbb C_{(-1,-1)}) \, .
\end{equation}

\medskip \noindent
Let us discuss some physical consequences of these computations. We focus on the non-trivial SPT and its corresponding Lagrangian algebra $L(\mathbb Z_2 \oplus \mathbb Z_2,\beta)$. Suppose for now that the microscopic Hilbert space of the physical system is provided by $\mc H_\Lambda = \bigotimes_{\msf i \in \Lambda} V_{\{\msf i\}}$, where $V$ is chosen to be $\End_\mathbb C(U_2)^* \cong \mathbb C(\mathbb Z_2 \oplus \mathbb Z_2)$. This is the Hilbert space of the renormalisation group fixed point associated with $\mc M(\mathbb Z_2 \oplus \mathbb Z_2,\beta)$---a.k.a. the \emph{cluster state} \cite{Son:2011qqi,PhysRevA.84.022304}---that we constructed in sec.~\ref{sec:VecG_RG}. Recall that the vector space $\End_\mathbb C(U_2) \cong {\rm Mat}_\mathbb C(2)$ is spanned by the identity matrix $\mathbb I$ and the spin-1/2 \emph{Pauli matrices} $\{\sigma^x , \sigma^z,\sigma^x\sigma^z\}$.  
Besides, let $\pi \colon \mathbb C[\mathbb Z_2 \oplus \mathbb Z_2]^\beta \to \End_\mathbb C(U_2)$. With the above conventions, we have $\pi(\sn{()}) = \mathbb I$, $\pi(\sn{(12)(34)}) = \sigma^x$, $\pi(\sn{(13)(24)}) = \sigma^z$ and $\pi(\sn{(14)(23)}) = \sigma^x \sigma^z$. Recall that the group $G$ acts on $f \in \End_\mathbb C(U_2)$ via $g \cdot f = \pi(g) f \pi(g)^{-1}$, for every $g \in G$.
Since $\End_\mathbb C(U_2) \cong \mathbb C^2 \otimes \mathbb C^2$, as a vector space, it is convenient to think about the local Hilbert space as a pair of \emph{qubits}. Choosing to work in a basis of eigenvectors of $\sigma^z$, one finds that the topological lines $\mathbb C_{(12)(34)}$, $\mathbb C_{(13)(24)}$ and $\mathbb C_{(14)(23)}$ acts on the local Hilbert space $\mathbb C^2 \otimes \mathbb C^2$ as $\sigma^z \otimes \mathbb I$, $\mathbb I \otimes \sigma^z$ and $\sigma^z \otimes \sigma^z$, respectively. We are now ready to construct the string order parameters. The multiplet of twisted sector local operators is defined with respect to the linear map we introduced in eq.~\eqref{eq:canonicalmap}. Here, to every homogeneous element in $L(\mathbb Z_2 \oplus \mathbb Z_2,\beta)$ of degree $g$, this map assigns the operator $\pi(g)$, which acts on $\End_\mathbb C(U_2)$ by postcomposition. Paying attention to the fact that $\sigma^x \sigma^z = - \sigma^z \sigma^x$, in the qubit basis it yields
\begin{equation}
    \label{eq:cluster_OP}
    \twistedLocalOp{1}{\;\;\;\;\;(12)(34)}{} \!\!\! = (\mathbb I \otimes \sigma^x) \otimes e^{(12)(34)} \, , \q
    \twistedLocalOp{1}{\;\;\;\;\;(13)(24)}{} \!\!\! = (\sigma^x \otimes \sigma^z) \otimes e^{(13)(24)} \, ,
\end{equation}
and similarly for the twisted sector local operators of the form \eqref{eq:VecG_twistedLocalOpDual}. It follows from $(\mathbb I \otimes \sigma^x)(\sigma^x \otimes \sigma^z) = - (\sigma^x \otimes \sigma^z)(\mathbb I \otimes \sigma^x)$ that these twisted sector local operators do fuse according to the multiplication rule of $L(\mathbb Z_2 \oplus \mathbb Z_2, \beta)$. Bringing everything together, one recovers for instance the usual string operator associated with the simple object $(\sn{(12)(34)}, \mathbb C_{(1,-1)})$:
\begin{equation}
    \label{eq:VecG_stringCS}
    \mc S(\sn{(12)(34)},1,1)_{\msf i,\ldots,\msf j} = (\mathbb I \otimes \sigma^x)_\msf i \bigg(\prod_{\msf k =\msf i+1}^{\msf j-1}(\sigma^z \otimes \mathbb I)_\msf k \bigg) (\sigma^z \otimes \sigma^x)_{\msf j} \, .
\end{equation}
The other two non-trivial string operators associated with $(\sn{(13)(24)}, \mathbb C_{(1,-1)})$ and $(\sn{(14)(23)}, \mathbb C_{(1,-1)})$, respectively, are constructed similarly.

A more interesting scenario consists in expressing the same multiplet of string order parameters for a spin-1 model where the local Hilbert space would typically be taken to be the three-dimensional spin-1 representation of ${\rm SO}(3)$. This is the setup of the \emph{AKLT model} \cite{PhysRevLett.59.799} and the \emph{antiferromagnetic Heisenberg model}, which are both in the same SPT phase as above, i.e. the \emph{Haldane phase} \cite{PhysRevLett.50.1153,Kennedy:1992ifl,Oshikawa_1992,PhysRevB.81.064439,PhysRevB.83.035107}. In this context, local operators are typically expressed in terms of spin-1 operators $\{S^x,S^y,S^z\}$ in the standard angular momentum basis. But our framework still require defining the microscopic Hilbert space in terms of objects in $\Rep(\mathbb Z_2 \oplus \mathbb Z_2)$. Notice that the restriction of the spin-1 representation of ${\rm SO}(3)$ to $\mathbb Z_2 \oplus \mathbb Z_2$ is isomorphic to $\mathbb C_{(1,-1)} \oplus \mathbb C_{(-1,1)} \oplus \mathbb C_{(-1,-1)}$, which one chooses as our local vector space $V$. Furthermore, one uses this isomorphism to express the spin-1 operators in a basis of $\mathbb C_{(1,-1)} \oplus \mathbb C_{(-1,1)} \oplus \mathbb C_{(-1,-1)}$. At this point, one can verify that the topological lines $\mathbb C_{(12)(34)}$ and $\mathbb C_{(14)(23)}$ acts diagonally on the local Hilbert space as $e^{i \pi S^z}$ and $e^{i \pi S^x}$, respectively (see the treatment of the \emph{Kennedy--Tasaki} duality in ref.~\cite{Lootens:2021tet}). Similarly, the same linear map we used in eq.~\eqref{eq:cluster_OP} now produces the twisted sector local operators $S^x \otimes e^{(12)(34)}$ and $-S^x \otimes e^{(13)(24)}$. It follows that the same abstract string operator associated with the simple object $(\sn{(12)(34)}, \mathbb C_{(1,-1)})$ now acts as 
\begin{equation}
    \mc S(\sn{(12)(34)},1,1)_{\msf i,\ldots,\msf j} 
    = - S^x_\msf i \exp\bigg(i \pi \! \sum_{\msf k=\msf i+1}^{\msf j-1} \! S^z_\msf k \bigg) S^x_\msf j \, ,
\end{equation}
recovering the celebrated string order parameter introduced by Nijs and Rommelse in ref.~\cite{PhysRevB.40.4709}. We can repeat the same operations for $(\sn{(13)(24)}, \mathbb C_{(-1,1)})$ and verify once again that twisted sector local operators in the multiplet fuse according to the multiplication rule of $L(\mathbb Z_2 \oplus \mathbb Z_2,\beta)$. 

\subsection{Example: $\Vect_{\mathbb A_4}$\label{sec:VecG_Example}}

Let $\mathbb A_4$ be the \emph{alternating} group of order $12$. It is isomorphic to $ 
(\mathbb Z_2 \oplus \mathbb Z_2) \rtimes \mathbb Z_3$, where $\mathbb Z_3$ cyclically permutes the three non-identity elements of the $\mathbb Z_2 \oplus \mathbb Z_2$ normal subgroup. It can be interpreted as the group of orientation-preserving symmetries of the tetrahedron, or equivalently, the group of even permutations on $\{\sn{1},\sn{2},\sn{3},\sn{4}\}$. In that spirit, we denote the identity element by $\sn{()}$, the generators of $\mathbb Z_2 \oplus \mathbb Z_2$ by $\sn{(12)}\sn{(34)}$ and $\sn{(13)(24)}$, and the generator of $\mathbb Z_3$ by $\sn{(123)}$.

Let us begin by listing basic properties of $\mathbb A_4$. As a set, it is the disjoint union of four conjugacy classes, which read 
\begin{gather}
    [\sn{()}]=\{\sn{()}\} \, , \q 
    [\sn{(12)(34)}] = \{\sn{(12)(34)},\sn{(13)(24)},\sn{(14)(23)}\} \, ,
    \\
    [\sn{(123)}] = \{\sn{(123)},\sn{(134)},\sn{(142)},\sn{(243)}\} \, , \q
    [\sn{(132)}] = \{\sn{(132)},\sn{(143)},\sn{(124)},\sn{(234)}\} \, .
\end{gather}
The associated centraliser subgroups are $Z_{\mathbb A_4}(\sn{()}) = \mathbb A_4$, $Z_{\mathbb A_4}(\sn{(12)(34)}) = \mathbb Z_2 \oplus \mathbb Z_2 = \la \sn{(12)(34)}, \sn{(13)(24)} \ra$, $Z_{\mathbb A_4}(\sn{(123)}) = \mathbb Z_3 = \la \sn{(123)} \ra$ and $Z_{\mathbb A_4}(\sn{(132)}) = \mathbb Z_3 = \la \sn{(132)}\ra$, respectively. Correspondingly, it admits four (linear) irreducible representations, namely three one-dimensional irreducible representations denoted by $\{\mathbb C_1, \mathbb C_{\omega}, \mathbb C_{\bar \omega}\}$, where $\omega$ stands for $\exp(\frac{2\pi i}{3})$, and a single three-dimensional irreducible representation denoted by $V_3$. The character table is given by
\begin{equation}
    \begin{array}{l|cccc} 
    \label{table:A4charTab}
        & [\sn{()}] & [\sn{(12)(34)}] & [\sn{(123)}] & [\sn{(132)}] \\ 
        \hline
        \mathbb C_1     & 1 & 1 & 1 & 1 \\
        \mathbb C_\omega & 1 & 1 & \omega & \bar \omega \\
        \mathbb C_{\bar \omega}  & 1 & 1 & \bar \omega & \omega \\
        V_3      & 3 & -1 & 0 & 0
    \end{array} \, ,
\end{equation}
while the non-trivial fusion rules read
\begin{equation}
\begin{gathered}
    \mathbb C_\omega \otimes \mathbb C_{\omega} \cong \mathbb C_{\bar \omega} \, , 
    \q 
    \mathbb C_{\bar \omega} \otimes \mathbb C_{\bar \omega} \cong \mathbb C_\omega \, , 
    \q
    \mathbb C_\omega \otimes \mathbb C_{\bar \omega} \cong \mathbb C_1 \, , 
    \q
    \mathbb C_\omega \otimes V_3 \cong  V_3 \cong \mathbb C_{\bar \omega} \otimes V_3  \, ,
    \\
    V_3 \otimes V_3 \cong \mathbb C_1 \oplus \mathbb C_\omega \oplus \mathbb C_{\bar \omega} \oplus 2 \cdot V_3 \, .
\end{gathered}
\end{equation}
For future reference, we choose a basis for $V_3$ such that the representation matrices of $\sn{(123)}$ and $\sn{(12)(34)}$ ar given by
\begin{equation}
    \label{eq:basisVA4}
    \begin{pmatrix}
    0 & 0 & 1 \\
    1 & 0 & 0 \\
    0 & 1 & 0 
    \end{pmatrix}
    \q \text{and} \q
    \begin{pmatrix}
    1 & 0 & 0 \\
    0& -1 & 0 \\
    0 & 0 & -1
\end{pmatrix} \, ,
\end{equation}
respectively.

It counts five conjugacy classes of subgroups represented by $\mathbb A_4$, $\mathbb Z_2 \oplus \mathbb Z_2 = \la \sn{(12)(34)}, \sn{(13)(24)} \ra$, $\mathbb Z_3 = \la \sn{(123)} \ra$, $\mathbb Z_2 = \la \sn{(12)(34)}\ra$ and the trivial subgroup. By analogy, we denote the irreducible representations of $\mathbb Z_2$ by $\{\mathbb C_1, \mathbb C_{-1}\}$, those of $\mathbb Z_2 \oplus \mathbb Z_2$ by $\{\mathbb C_{(1,1)}, \mathbb C_{(1,-1)}, \mathbb C_{(-1,1)}, \mathbb C_{(-1,-1)}\}$, and those of $\mathbb Z_3$ by $\{\mathbb C_{1}, \mathbb C_\omega, \mathbb C_{\bar \omega}\}$. The character table of $\mathbb Z_2 \oplus \mathbb Z_2$ was provided above, while that of $\mathbb Z_3$ is given by 
\begin{equation}
    \begin{array}{l|ccc} 
        & \sn{()} & \sn{(123)} & \sn{(132)}  \\ 
        \hline
        \mathbb C_1     & 1 & 1 & 1  \\
        \mathbb C_\omega & 1 & \omega & \bar \omega  \\
        \mathbb C_{\bar \omega}  & 1 & \bar \omega & \omega
    \end{array} \, .
\end{equation}
They satisfy the obvious fusion rules.

The Schur multiplier $H^2(\mathbb A_4,\rm U(1))$ is found to be isomorphic to $\mathbb Z_2$. Given 
a (normalised) representative of the non-trivial cohomology class in $H^2(\mathbb A_4, \rm U(1))$, the group $\mathbb A_4$ admits three two-dimensional irreducible projective representations denoted by $\{U_1,U_\omega, U_{\bar \omega}\}$. They satisfy the fusion rules
\begin{gather}
    \nn
    \mathbb C_\omega \otimes U_{\omega} \cong U_{\bar \omega} \, , 
    \q 
    \mathbb C_{\bar \omega} \otimes U_{\bar \omega} \cong U_\omega \, , 
    \q
    \mathbb C_\omega \otimes U_{\bar \omega} \cong U_1 \, , 
    \\
    U_1 \otimes V_3 \cong  U_\omega \otimes V_3 \cong U_{\bar \omega} \otimes V_3 \cong U_1 \oplus U_\omega \oplus U_{\bar \omega} \, ,
    \\
    \nn
    U_1 \otimes U_1 \cong U_\omega \otimes U_{\bar \omega} \cong \mathbb C_1 \oplus V_3 \, , \q 
    U_{\bar \omega} \otimes U_{\bar \omega} \cong U_1 \otimes U_{\omega} \cong \mathbb C_\omega \oplus V_3 \, , \q 
    U_\omega \otimes U_\omega \cong U_1 \otimes U_{\bar \omega} \cong \mathbb C_{\bar \omega} \oplus V_3 
    \, .
\end{gather}
Out of the four representatives of conjugacy classes of proper subgroups, only $\mathbb Z_2 \oplus \mathbb Z_2$ admits a non-trivial Schur multiplier, which is also isomorphic to $\mathbb Z_2$. We still denote its single two-dimensional irreducible projective representation by $U_2$.

As reviewed in sec.~\ref{sec:VecG_OP}, simple objects in $\Vect_{\mathbb A_4}^{\mathbb A_4}$ are labelled by pairs $([x_0],\hat W)$ consisting of a conjugacy class $[x_0] \in \Cl(\mathbb A_4)$ and a simple object $\hat W \in \Mod(\mathbb C[Z_{\mathbb A_4}(x_0)])$. Using the above data, we find fourteen simple objects: $([\sn{()}],\mathbb C_1)$, $([\sn{()}],\mathbb C_\omega)$, $([\sn{()}],\mathbb C_{\bar \omega})$, $([\sn{()}],V_3)$, $([\sn{(12)(34)}],\mathbb C_{(1,1)})$, $([\sn{(12)(34)}],\mathbb C_{(1,-1)})$, $([\sn{(12)(34)}],\mathbb C_{(-1,1)})$, $([\sn{(12)(34)}],\mathbb C_{(-1,-1)})$, $([\sn{(123)}],\mathbb C_1)$, $([\sn{(123)}],\mathbb C_{\omega})$, $([\sn{(123)}],\mathbb C_{\bar \omega})$, $([\sn{(132)}],\mathbb C_1)$, $([\sn{(132)}],\mathbb C_\omega)$ and $([\sn{(132)}],\mathbb C_{\bar \omega})$, each labelling a multiplet of twisted sector local operators. Finally, Lagrangian algebras in $\Vect_{\mathbb A_4}^{\mathbb A_4}$ are labelled by (conjugacy classes) of subgroups $H \leq \mathbb A_4$ and cohomology classes in $H^2(H,{\rm U(1)})$. Using the above data, we find seven Lagrangian algebras: $L(0,1)$, $L(\mathbb Z_2,1)$, $L(\mathbb Z_3,1)$, $L(\mathbb Z_2 \oplus \mathbb Z_2,1)$, $L(\mathbb Z_2 \oplus \mathbb Z_2,\beta)$, $L(\mathbb A_4,1)$ and $L(\mathbb A_4,\beta)$, which we explicitly compute below:\footnote{In order to obtain the decomposition into simple objects of each Lagrangian, one could simply employ the multiplicity formula \eqref{eq:multiplicityL}, but it more enlightening to derive it from the definition.}

\bigskip\noindent
$\bul$ $L(0,1)$: Every element of $L(0,1)$ is homogeneous of degree $\sn{()} \in \mathbb A_4$. But, $L(0,1)_{()} \cong \mathbb C(\mathbb A_4)$, as an object in $\Rep(A_4)$. It follows from the decomposition of the regular representation into irreducible representations that
\begin{equation}
    L(0,1) \cong ([\sn{()}],\mathbb C_1) \oplus
    ([\sn{()}],\mathbb C_\omega) \oplus ([\sn{()}],\mathbb C_{\bar \omega}) \oplus 3 \cdot ([\sn{()}],V_3)   
\end{equation}

\medskip \noindent
$\bul$ $L(\mathbb Z_2,1)$: Consider the subgroup $\mathbb Z_2 = \la \sn{(12)(34)} \ra$. Recall that $\varsigma \in L(\mathbb Z_2,1)$ is homogeneous of degree $y \in \mathbb A_4$ if and only if $\varsigma(x) \in \mathbb C[\mathbb Z_2]$ is homogeneous of degree $x^{-1}yx$, for every $x \in G$. Therefore, $\text{supp}(L(\mathbb Z_2,1)) = [\sn{()}] \cup [\sn{(12)(34)}]$. On the one hand, we already know that $L(\mathbb Z_2,1)_{[()]} \cong \mathbb C(\mathbb A_4/\mathbb Z_2)$, which is isomorphic to $\Ind_{\mathbb Z_2}^{\mathbb A_4}(\mathbb C_1)$ as an object in $\Rep(\mathbb A_4)$. From the character table, we read that $\Res_{\mathbb Z_2}^{\mathbb A_4}(\mathbb C_1) \cong \Res_{\mathbb Z_2}^{\mathbb A_4}(\mathbb C_\omega) \cong \Res^{\mathbb A_4}_{\mathbb Z_2}(\mathbb C_{\bar \omega}) \cong \mathbb C_1$ and
\begin{equation}
    \dim_\mathbb C \Hom_{\Rep(\mathbb Z_2)} \big(\Res_{\mathbb Z_2}^{\mathbb A_4}(V_3),\mathbb C_1 \big) = \frac{1}{2}(3-1) = 1 \, .
\end{equation}
By Frobenius reciprocity, one obtains that $\Ind_{\mathbb Z_2}^{\mathbb A_4}(\mathbb C_1) \cong \mathbb C_1 \oplus \mathbb C_\omega \oplus \mathbb C_{\bar \omega} \oplus V_3$. It follows that, as an object in $\Vect_{\mathbb A_4}^{\mathbb A_4}$, we have $L(\mathbb Z_2,1)_{[()]} \cong ([\sn{()}],\mathbb C_1) \oplus ([\sn{()}],\mathbb C_{\omega}) \oplus ([\sn{()}], \mathbb C_{\bar \omega}) \oplus ([\sn{()}],V_3)$. On the other hand, suppose that $\varsigma \in L(\mathbb Z_2,1)$ is homogeneous of degree $\sn{(12)(34)}$, which requires that $\text{supp}(\varsigma)=Z_{G}(\sn{(12)(34)}) \cong \mathbb Z_2 \oplus \mathbb Z_2$. The defining requirement that $\varsigma(gh^{-1}) = h \cdot \varsigma(g)$, for every $g \in \mathbb Z_2 \oplus \mathbb Z_2$ and $h \in \mathbb Z_2$, then implies that $L(\mathbb Z_2,1)_{(12)(34)} \cong \mathbb C((\mathbb Z_2 \oplus \mathbb Z_2)/\mathbb Z_2)$, which is isomorphic to $\Ind_{\mathbb Z_2}^{\mathbb Z_2 \oplus \mathbb Z_2}(\mathbb C_1)$ as an object in $\Rep(\mathbb Z_2 \oplus \mathbb Z_2)$. From the character table of $\mathbb Z_2 \oplus \mathbb Z_2$, we read that $\Res^{\mathbb Z_2 \oplus \mathbb Z_2}_{\mathbb Z_2}(\mathbb C_{(1,1)}) \cong \Res^{\mathbb Z_2 \oplus \mathbb Z_2}_{\mathbb Z_2}(\mathbb C_{(1,-1)}) \cong \mathbb C_1$. By Frobenius reciprocity, one obtains that $\Ind^{\mathbb Z_2 \oplus \mathbb Z_2}_{\mathbb Z_2}(\mathbb C_1) \cong \mathbb C_{(1,1)} \oplus \mathbb C_{(1,-1)}$. Repeating the same derivations for $L(\mathbb Z_2,1)_{(13)(24)}$ and $L(\mathbb Z_2,1)_{(14)(23)}$ yields that, as an object in $\Vect_{\mathbb A_4}^{\mathbb A_4}$, we have $L(\mathbb Z_2,1)_{[(12)(34)]} \cong ([\sn{(12)(34)}],\mathbb C_{(1,1)}) \oplus ([\sn{(12)(34)}],\mathbb C_{(1,-1)})$. Bringing everything together yields
\begin{equation}
    L(\mathbb Z_2,1) \cong 
    ([\sn{()}],\mathbb C_1) \oplus ([\sn{()}],\mathbb C_{\omega}) \oplus ([\sn{()}], \mathbb C_{\bar \omega}) \oplus ([\sn{()}],V_3) \oplus ([\sn{(12)(34)}],\mathbb C_{(1,1)}) \oplus ([\sn{(12)(34)}],\mathbb C_{(1,-1)}) \, .
\end{equation}

\medskip\noindent
$\bul$ $L(\mathbb Z_3,1)$: Consider the subgroup $\mathbb Z_3 = \la \sn{(123)} \ra$. By definition, $\varsigma \in L(\mathbb Z_3,1)$ is homogeneous of degree $y \in \mathbb A_4$ if and only if $\varsigma(x) \in \mathbb C[\mathbb Z_3]$ is homogeneous of degree $x^{-1}yx$, for every $x \in \mathbb A_4$. This means $x^{-1}yx$ must evaluate to an element within $\mathbb Z_3$. Therefore, $\supp(L(\mathbb Z_3,1)) = [\sn{()}] \cup [\sn{(123)}] \cup [\sn{(132)}]$. On the one hand, we have $L(\mathbb Z_3,1)_{[()]} \cong \mathbb C(\mathbb A_4 / \mathbb Z_3)$, which is isomorphic to $\Ind_{\mathbb Z_3}^{\mathbb A_4}(\mathbb C_1)$ as an object in $\Rep(\mathbb A_4)$. From the character table and Frobenius reciprocity, one obtains that $\Ind_{\mathbb Z_3}^{\mathbb A_4}(\mathbb C_1) \cong \mathbb C_1 \oplus V_3$. Therefore, as an object in $\Vect_{\mathbb A_4}^{\mathbb A_4}$, we have $L(\mathbb Z_3,1)_{[()]} \cong ([\sn{()}],\mathbb C_1) \oplus ([\sn{()}],V_3)$. On the other hand, suppose $\varsigma$ is homogeneous of degree $y \in [\sn{(123)}]$. For $\varsigma(x)$ to be non-zero, the condition $x^{-1}yx \in \mathbb Z_3$ must hold. Since conjugation preserves the conjugacy class and $[\sn{(123)}] \cap \mathbb Z_3 = \{\sn{(123)}\}$, it follows that $x^{-1}yx = y$, which implies $x \in Z_{\mathbb A_4}(y)$. The support is therefore exactly the centraliser $Z_{\mathbb A_4}(\sn{(123)}) = \mathbb Z_3$. Given that the index $(Z_{\mathbb A_4}(y) : \mathbb Z_3) = 1$, the equivariance structure on this one-dimensional space corresponds to the trivial representation of the centraliser. By the analogous argument for the class $[\sn{(132)}]$, we obtain $L(\mathbb Z_3,1)_{[(123)]} \cong ([\sn{(123)}], \mathbb C_1)$ and $L(\mathbb Z_3,1)_{[(132)]} \cong ([\sn{(132)}], \mathbb C_1)$. Bringing everything together yields 
\begin{equation}
    L(\mathbb Z_3,1) \cong ([\sn{()}],\mathbb C_1) \oplus ([\sn{()}],V_3) \oplus ([\sn{(123)}],\mathbb C_1) \oplus ([\sn{(132)}],\mathbb C_1) \, .
\end{equation}

\medskip\noindent
$\bul$ $L(\mathbb Z_2 \oplus \mathbb Z_2,1)$: Consider the subgroup $\mathbb Z_2 \oplus \mathbb Z_2 = \la \sn{(12)(34)},\sn{(13)(24)}\ra$. By definition, $\varsigma \in L(\mathbb Z_2 \oplus \mathbb Z_2,1)$, is homogeneous of degree $y \in \mathbb A_4$ if and only if $\varsigma(x) \in \mathbb C[\mathbb C_2 \oplus \mathbb Z_2]$ is homogeneous of degree $x^{-1}yx$, for every $x \in G$. Therefore, $\supp(L(\mathbb Z_2 \oplus \mathbb Z_2,1)) = [\sn{()}] \cup [\sn{(12)(34)}]$. On the one hand, we have $L(\mathbb Z_2 \oplus \mathbb Z_2,1)_{[()]} \cong \mathbb C(\mathbb A_4 / \mathbb Z_2 \oplus \mathbb Z_2)$, which is isomorphic to $\Ind_{\mathbb Z_2 \oplus \mathbb Z_2}^{\mathbb A_4}(\mathbb C_{(1,1)})$ as an object in $\Rep(\mathbb A_4)$. From the character table and Frobenius reciprocity, one obtains that $\Ind_{\mathbb Z_2 \oplus \mathbb Z_2}^{\mathbb A_4}(\mathbb C_{(1,1)}) \cong \mathbb C_1 \oplus \mathbb C_\omega \oplus \mathbb C_{\bar \omega}$. Therefore, as an object in $\Vect_{\mathbb A_4}^{\mathbb A_4}$, we have $L(\mathbb Z_2 \oplus \mathbb Z_2,1)_{[()]} \cong ([\sn{()}],\mathbb C_1) \oplus ([\sn{()}],\mathbb C_\omega) \oplus ([\sn{()}],\mathbb C_{\bar \omega})$. On the other hand, suppose that $\varsigma \in L(\mathbb Z_2 \oplus \mathbb Z_2,1)$ is homogeneous of degree $\sn{(12)(34)}$ so that, for every $x \in \mathbb A_4$, $\varsigma(x) \in \mathbb C[\mathbb Z_2 \oplus \mathbb Z_2]$ must be homogeneous of degree $x^{-1}\sn{(12)(34)}x$. Recall that the basis functions $\varsigma_{a,h}$ we introduced in eq.~\eqref{eq:LA_basisFunctions}, where $a \in \{1,\ldots,(\mathbb A_4 : \mathbb Z_2 \oplus \mathbb Z_2)\}$ and $h \in \mathbb Z_2 \oplus \mathbb Z_2$ are homogeneous of degree $r_a h r_a^{-1}$. But $\mathbb A_4 / (\mathbb Z_2 \oplus \mathbb Z_2) = \{\mathbb Z_2 \oplus \mathbb Z_2 , \sn{(123)}(\mathbb Z_2 \oplus \mathbb Z_2), \sn{(132)}(\mathbb Z_2 \oplus \mathbb Z_2)\}$ and $\sn{(123)}\cdot \sn{(13)(24)} \cdot \sn{(132)} = \sn{(132)}\cdot \sn{(14)(23)} \cdot \sn{(123)} = \sn{(12)(34)}$, so that the space of functions $\varsigma$ that are homogeneous of degree $\sn{(12)(34)}$ is spanned by $\varsigma_{1,(12)(34)}$, $\varsigma_{2,(13)(24)}$ and $\varsigma_{3,(14)(23)}$, whereby $r_1 = \sn{()}$, $r_2 = \sn{(123)}$ and $r_3 = \sn{(132)}$. Similarly, one finds that the nine-dimensional space of homogeneous functions of degree in $[\sn{(12)(34)}]$ is spanned by functions $\varsigma_{a,h}$ such that $a \in \{1,\ldots,(\mathbb A_4 : \mathbb Z_2 \oplus \mathbb Z_2)\}$ and $h \in [\sn{(12)(34)}]$. Now, recall that the $G$-equivariance structure of $L(\mathbb Z_2 \oplus \mathbb Z_2,1)$ is given by $g \cdot \varsigma_{a,h} = \varsigma_{g(a),h_{g,a}hh_{g,a}^{-1}} = \varsigma_{g(a),h}$, for every $g \in \mathbb A_4$ and $h \in [\sn{(12)(34)}]$. Therefore, for each $h \in [\sn{(12)(34)}]$, $\mathbb C\{\varsigma_{a,h}\}_{a=1,\ldots,(\mathbb A_4 : \mathbb Z_2 \oplus \mathbb Z_2)} \cong \mathbb C(\mathbb A_4 / \mathbb Z_2 \oplus \mathbb Z_2)$, so that $L(\mathbb Z_2 \oplus \mathbb Z_2,1)_{[(12)(34)]} \cong 3 \cdot ([\sn{(12)(34)}],\mathbb C_{(1,1)})$.
Bringing everything together yields
\begin{equation}
    L(\mathbb Z_2 \oplus \mathbb Z_2,1) \cong 
    ([\sn{()}],\mathbb C_1) \oplus ([\sn{()}],\mathbb C_{\omega}) \oplus ([\sn{()}], \mathbb C_{\bar \omega}) \oplus 3 \cdot ([\sn{(12)(34)}],\mathbb C_{(1,1)}) \, .
\end{equation}

\medskip \noindent
$\bul$ $L(\mathbb Z_2 \oplus \mathbb Z_2,\beta)$: Let us now choose $\beta$ to be the normalised representative of the non-trivial cohomology class in $H^2(\mathbb Z_2 \oplus \mathbb Z_2,\rm U(1))$ such that $\beta(h_1,h_2)$ equals $-1$ if $h_1 \in \{\sn{(13)(24)},\sn{(14)(23)}\}$ and $h_2 \in \{\sn{(12)(34)},\sn{(14)(23)}\}$, and equals $+1$ otherwise. We have already established that $L(\mathbb Z_2 \oplus \mathbb Z_2,\beta)_{[()]}$ decomposes as $([\sn{()}],\mathbb C_1) \oplus ([\sn{()}],\mathbb C_\omega) \oplus ([\sn{()}],\mathbb C_{\bar \omega})$, as an object in $\Vect_{\mathbb A_4}^{\mathbb A_4}$. We also already know that the space of functions in $L(\mathbb Z_2 \oplus \mathbb Z_2,\beta)$ that are homogeneous of degree in $[\sn{(12)(34)}]$ is spanned by functions $\varsigma_{a,h}$ for $a \in \{1,\ldots,(\mathbb A_1 : \mathbb Z_2 \oplus \mathbb Z_2)\}$ and $h \in [\sn{(12)(34)}]$. The $G$-equivariant structure is now given by $g \cdot \varsigma_{a,h} = \beta_{h}(h_{g,a}) \, \varsigma_{g(a),h}$, where $\beta_h(h_{g,a}) = \beta(h_{g,a},h)/\beta(h,h_{g,a})$. For instance, fix $h$ to be $\sn{(12)(34)}$. Given the above formula for $\beta$, we have $\beta_{(12)(34)}(h_{g,a}) = \beta(h_{g,a},\sn{(12)(34)})$,
For every $g \in \mathbb Z_2 \oplus \mathbb Z_2$, we have $g \cdot \varsigma_{a,(12)(34)} = \beta(h_{g,a},\sn{(12)(34)}) \, \varsigma_{a,(12)(34)}$. In particular, for every $g \in \mathbb Z_2 \oplus \mathbb Z_2$, $g \cdot \varsigma_{1,(12)(34)} = \beta(g,\sn{(12)(34)}) \, \varsigma_{1,(12)(34)}$, which equals  $-\varsigma_{1,(12)(34)}$ if $g \in \{\sn{(13)(24)}, \sn{(14)(23)}\}$,  and equals $\varsigma_{1,(12)(34)}$ otherwise. Therefore, $\mathbb C\{\varsigma_{1,(12)(34)}\} \cong \mathbb C_{(1,-1)}$ as a representation of $\mathbb Z_2 \oplus \mathbb Z_2$. Similarly, it follows from $h_{(12)(34),2} = \sn{(13)(24)}$, $h_{(12)(34),3} = \sn{(14)(23)}$, $h_{(13)(24),2} = \sn{(14)(23)}$ and $h_{(13)(24),3} = \sn{(12)(34)}$ that $\mathbb C\{\varsigma_{2,(12)(34)}\} \cong \mathbb C\{\varsigma_{3,(12)(34)}\} \cong \mathbb C_{(1,-1)}$. A similar computation holds  for $h = \sn{(13)(24)}$ and $h = \sn{(14)(23)}$.
This is enough to imply that, for each $h \in [\sn{(12)(34)}]$, $\mathbb C\{\varsigma_{a,h}\}_{a=1,\ldots,(\mathbb A_4 : \mathbb Z_2 \oplus \mathbb Z_2)} \cong \Ind_{\mathbb Z_2 \oplus \mathbb Z_2}^{\mathbb A_4}(\mathbb C_{(1,-1)})$. Therefore, as an object in $\Vect_{\mathbb A_4}^{\mathbb A_4}$, we have $L(\mathbb Z_2 \oplus \mathbb Z_2,\beta)_{[(12)(34)]} \cong 3 \cdot ([\sn{(12)(34)}],\mathbb C_{(1,-1)})$.
Bringing everything together yields
\begin{equation}
    L(\mathbb Z_2 \oplus \mathbb Z_2,\beta) \cong 
    ([\sn{()}],\mathbb C_1) \oplus ([\sn{()}],\mathbb C_{\omega}) \oplus ([\sn{()}], \mathbb C_{\bar \omega}) \oplus 3 \cdot ([\sn{(12)(34)}],\mathbb C_{(1,-1)}) \, .
\end{equation}

\medskip \noindent
$\bul$ $L(\mathbb A_4,1)$: By definition, $L(\mathbb A_4,1) = \mathbb C[\mathbb A_4] = \mathbb C\{e_g\}_{g \in \mathbb A_4}$ with $g_1 \cdot e_{g_2} = e_{g_1g_2g_1^{-1}}$, for every $g_1,g_2 \in \mathbb A_4$. One immediately concludes that $\mathbb C[\mathbb A_4] \cong \mathbb C(\mathbb A_4/Z_{\mathbb A_4}(\sn{()})) \oplus \mathbb C(\mathbb A_4/Z_{\mathbb A_4}(\sn{(12)(34)})) \oplus \mathbb C(\mathbb A_4/Z_{\mathbb A_4}(\sn{(123)})) \oplus \mathbb C(\mathbb A_4/Z_{\mathbb A_4}(\sn{(132)}))$, as an object in $\Rep(\mathbb A_4)$. Remembering that $\mathbb C(\mathbb A_4 / Z_{\mathbb A_4}(x_0)) \cong \Ind_{Z_{\mathbb A_4}(x_0)}^{\mathbb A_4}(\mathbb C)$, for every $[x_0] \in \Cl(\mathbb A_4)$, one finally obtains 
\begin{equation}
    L(\mathbb A_4,1) \cong ([\sn{()}],\mathbb C_1) \oplus ([\sn{(12)(34)}],\mathbb C_{(1,1)}) \oplus ([\sn{(123)}],\mathbb C_1) \oplus ([\sn{(132)}],\mathbb C_1) \, .
\end{equation}

\medskip \noindent 
$\bul$ $L(\mathbb A_4, \beta)$: Let $\beta$ be the normalised representative of the non-trivial cohomology class in $H^2(\mathbb A_4,\rU(1))$ that restricts to the 2-cocycle of $\mathbb Z_2 \oplus \mathbb Z_2$ utilised in the computation of $L(\mathbb Z_2 \oplus \mathbb Z_2,\beta)$. By definition, $L(\mathbb A_4,\beta) = \mathbb C[\mathbb A_4]^\beta$. We have already established  that it decomposes as \eqref{eq:LA_SPT}, as an object in $\Vect_{\mathbb A_4}^{\mathbb A_4}$, so that we simply need to compute the one-dimensional representations $\mathbb C_{\beta_{x_0}}$, for every $[x_0] \in \Cl(\mathbb A_4)$. Simply mimicking the derivation of $L(\mathbb Z_2 \oplus \mathbb Z_2,\beta)$ yields  
\begin{equation}
    L(\mathbb A_4,\beta) \cong ([\sn{()}],\mathbb C_1) \oplus ([\sn{(12)(34)}],\mathbb C_{(1,-1)}) \oplus ([\sn{(123)}],\mathbb C_1) \oplus ([\sn{(132)}],\mathbb C_1) \, .
\end{equation}

\bigskip \noindent
Let us make some general remarks about these Lagrangian algebras. First, we notice that out of the fourteen total simple objects of $\mc Z(\Vect_{\mathbb A_4})$, eight simple objects appear in Lagrangian algebras. One can construct multiplets of order parameters associated with each one of these eight simple objects. However, inspecting how these Lagrangian algebras intersect, we find that not all of them are required to identify each one of the seven gapped phases. As a matter of fact, it is enough to consider three order parameters only.
 
We begin by noticing that $([\sn{()}], \mathbb C_\omega)$ and $([\sn{()}], \mathbb C_{\bar \omega})$ always come in pair, and so do $([\sn{(123)}], \mathbb C_1)$ and $([\sn{(132)}], \mathbb C_1)$. Moreover, these two pairs of objects are mutually exclusive, thereby dividing the seven gapped phases into two sets. Together, these pairs of objects signal whether the algebra $\mathbb C[\mathbb Z_3]$ of lines is preserved in the ground state subspace. The object $([\sn{(123)}], \mathbb C_1) \oplus ([\sn{(132)}], \mathbb C_1)$ appears in three Lagrangian algebras indicating that the algebra $\mathbb C[\mathbb Z_3]$ of lines is preserved in the three corresponding gapped phases. Further inspection reveals that these three gapped phases are distinguished by their corresponding Lagrangian algebras, each containing a distinct three-dimensional simple object: $([\sn{()}],V_3)$, $([\sn{(12)(34)}],\mathbb C_{(1,1)})$, and $([\sn{(12)(34)}],\mathbb C_{(1,-1)})$, respectively. The presence of the simple object $([\sn{()}],V_3)$, which corresponds to a multiplet of charged local operators signals that the algebra $\mathbb C[\mathbb Z_2 \oplus \mathbb Z_2]$ of lines cannot be preserved, at least not in its entirety. On the other hand, since $[\sn{(12)(34)}]$ contains all the non-trivial elements of $\mathbb Z_2 \oplus \mathbb Z_2$, the presence of the simple objects $([\sn{(12)(34)}],\mathbb C_{(1,1)})$ and $([\sn{(12)(34)}],\mathbb C_{(1,-1)})$ signals that the algebras $\mathbb C[\mathbb Z_2 \oplus \mathbb Z_2]$ and $\mathbb C[\mathbb Z_2 \oplus \mathbb Z_2]^\beta$, respectively, must be at least partially preserved. The remaining four gapped phases where the algebra of lines $\mathbb C[\mathbb Z_3]$ is not preserved can also be distinguished in terms of these three simple objects: each appears individually in a single Lagrangian algebra, while the three of them simultaneously appears in the last one. Combining these remarks, we can detect any $\mathbb A_4$-symmetric gapped phase by measuring the expectation values of three order parameters, namely two local operators in the multiplets associated with $([\sn{()}],\mathbb C_\omega)$ and $([\sn{()}],V_3)$, respectively, as well as a string operator in the multiplet associated with $([\sn{(12)(34)}],\mathbb C_{(1,1)})$.

Let us consider a couple of examples. Suppose the three order parameters mentioned above acquire non-vanishing expectation values in the ground state subspace. Together, they signal that the algebra $\mathbb C[\mathbb Z_3]$ of lines cannot be preserved, and although the algebras $\mathbb C[\mathbb Z_2 \oplus \mathbb Z_2]$ or $\mathbb C[\mathbb Z_2 \oplus \mathbb Z_2]^{\beta}$ may not be fully preserved either, the whole symmetry $\Vect_{\mathbb A_4}$ cannot be spontaneously broken. There is only one possibility, namely the gapped phase with Lagrangian algebra $L(\mathbb Z_2,1)$. The presence of the simple object $([\sn{(12)(34)}],\mathbb C_{(1,-1)})$ may seem surprising---this type of terms being typical of non-trivial SPTs---but it is consistent with the fact that the restriction of the 2-cocycle to $\mathbb Z_2$ is trivial. Suppose instead that all the order parameters yield vanishing expectation values in the ground state subspace. We know from the local charged operators that all the symmetry lines are preserved in the ground state subspace; and since the string operators associated with $([\sn{(12)(34)}], \mathbb C_{(1,1)})$ does not gain an expectation value either, then they must form the algebra $\mathbb C[\mathbb A_4]^\beta$. The only possibility is the gapped phase associated with the Lagrangian algebra $L(\mathbb A_4,\beta)$.

We conclude with a comment echoing previous remarks. We are accustomed to the idea that string order parameters for non-trivial SPTs may require the endpoint twisted sector local operators to transform non-trivially. In the present example, this is reflected by the appearance of the simple objects $([\sn{(12)(34)}],\mathbb C_{(1,-1)})$---which depends implicitly on the 2-cocycle $\beta$---in the Lagrangian algebra $L(\mathbb A_4,\beta)$. For non-abelian groups, however, such dressing can also be necessary for trivial SPTs. Consider, for instance, the fixed point of the trivial SPT associated with the Lagrangian algebra $L(\mathbb A_4,1)$, constructed from the algebra $A(\mathbb A_4, 1, V_3)$. Observe the presence of the simple object $([\sn{(123)}],\mathbb C_1)$, which, as an object in $\Rep(A_4)$, is isomorphic to $\Ind_{\mathbb Z_3}^{\mathbb A_4}(\mathbb C_1) \cong \mathbb C_1 \oplus V_3$. In particular, this implies that the space of multiplets is quite large. While a generic string operator in a generic multiplet would acquire a non-vanishing expectation value, choosing a multiplet so that the string operator would boil down to the ordinary \emph{disorder operators} would result in a vanishing expectation value, by virtue of the representation matrix of $\sn{(123)}$ in the representation $V_3$ being traceless. Instead, the canonical multiplet defined via eq.~\eqref{eq:canonicalmap} would have a non-vanishing expectation value.

\section{Non-invertible symmetry\label{sec:NonInv}}

\emph{In this section, we extend the techniques we employed in the invertible case to the case of a finite non-invertible symmetry. Although we focus on the case of symmetry lines furnishing the fusion category of representations of a finite group $G$, we argue that the techniques apply more generally.}

\subsection{Microscopic Hilbert space\label{sec:RepG_Hilbert}}

Consider any one-dimensional quantum theory with a symmetry $\Vect_G$. Gauging the symmetry results in a new (effective) one-dimensional quantum theory with a non-invertible symmetry \cite{Frohlich:2009gb,Tachikawa:2017gyf,Bhardwaj:2017xup}.
The topological lines of this new theory are encoded into the fusion category $\Rep(G)$ of finite-dimensional representations of $G$ and $G$-equivariant maps. More generally, as for invertible symmetries, there are many ways to realise a symmetry $\Rep(G)$ in one-dimensional quantum lattice models  \cite{Lootens:2021tet,Lootens:2022avn,Jones:2024lws}. Throughout this section, we work within the following setup: Let $K = \bigoplus_{g \in G} K_g = \mathbb C\{k_r\}_{r=1,\ldots,\dim_\mathbb C K}$ be an object in $\Vect_G$ and $\Lambda$ be a finite subset of the lattice $\mathbb Z$. To every `site' $\msf i \in \Lambda$, we assign a copy of the vector space $K$ together with a copy of the algebra ${\rm Mat}_\mathbb C(\dim_\mathbb C K)$ so that we identify $\mc A_\Lambda := \bigotimes_{\msf i \in \Lambda} {\rm Mat}_\mathbb C(\dim_\mathbb C K)_{\{\msf i\}}$ with the algebra of bounded operators acting on the microscopic Hilbert space $\mc H_\Lambda := \bigotimes_{\msf i \in \Lambda} K_{\{\msf i\}}$. 
As in sec.~\ref{sec:VecG_Hilbert}, any state in $\mc H_\Lambda$ can be expressed as an MPS. 

Consider the topological line labelled by a simple object $V = \mathbb C\{v_d\}_{d=1,\ldots,\dim_\mathbb C V}$ in $\Rep(G)$ with algebra homomorphism $\rho \colon \mathbb C[G] \to \End_\mathbb C(V)$. Let us construct the corresponding symmetry MPO. Note that for every $r \in \{1,\ldots,\dim_\mathbb C K\}$, the basis vector $k_r \in K$ decomposes uniquely as a sum $\sum_{g \in G} k_{r,g}$ of homogeneous components such that $|k_{r,g}| = g \in G$. In other words, recall that $\Vect_G \simeq \Comod(\mathbb C[G])$, such that the \emph{coaction} $K \to \mathbb C[G] \otimes K$ of the $\mathbb C[G]$-comodule associated with $K \in \Vect_G$ is provided by $k_r \mapsto \sum_{g \in G} e_g \otimes k_{r,g}$, for every $g \in G$ and $r \in \{1,\ldots,\dim_\mathbb C K\}$. Assuming closed boundary conditions, the MPO labelled by $V$ is provided by
\begin{equation}
    \sum_{r_1,\ldots,r_{|\Lambda|}} \sum_{g_1,\ldots,g_{|\Lambda|}} {\rm tr}_V \big[\rho(g_1^{-1}) \rho(g_2^{-1}) \cdots \rho(g_{|\Lambda|}^{-1}) \big] \; k_{r_1,g_1} \otimes \cdots \otimes k_{r_{|\Lambda|},g_{|\Lambda|}} \otimes k^{r_1,g_1} \otimes \cdots \otimes k^{r_{|\Lambda|},g_{|\Lambda|}} \, . 
\end{equation}
Finally, every $G$-grading preserving map $K^{\otimes N} \to K^{\otimes N}$ gives rise to a symmetric operator in $\mc A_{\Lambda}$. 

The symmetry MPO labelled by any $V \in \Irr(\Rep(G))$ can be conveniently depicted as
\begin{equation}
    \label{eq:RepG_MPO}
    \MPO{}{V} \, ,
\end{equation}
where we introduced the tensor\footnote{Notice that in the case where $K$ would be chosen of be $\mathbb C[G]$, the tensor \eqref{eq:RepG_MPOT} boils down to a rotated version of the tensor \eqref{eq:VecG_MPOT} we defined previously. In sec.~\ref{sec:VecG_Hilbert}, we realised the symmetry $\Vect_G$ implicitly invoking the equivalence $(\Rep(G))^*_\Vect \simeq \Vect_G$, which makes $\Vect$ an invertible $(\Rep(G),\Vect_G)$-bimodule category. We are now realising the symmetry $\Rep(G)$ invoking the equivalence $(\Vect_G)^*_\Vect \simeq \Rep(G)$, which makes $\Vect$ an invertible $(\Vect_G,\Rep(G))$-bimodule category. This discrepancy explains why the group is required to act from the right. Note that we could have avoided this convention at the cost of reversing some arrows but our choice allows us to treat both cases---and ultimately the case of an arbitrary symmetry fusion category---in a similar way. Moreover, our convention is that dictated by the equivalences $\Mod(\Tu(\Vect_G)) \simeq \mc Z(\Vect_G) \simeq \Vect_G^G \simeq \mc Z(\Rep(G)) \simeq \Mod(\Tu(\Rep(G)))$ (see sec.~\ref{sec:RepG_OP}).}
\begin{equation}
\begin{split}
    \label{eq:RepG_MPOT}
    \MPOT{}{V}{}{}{}{} \!\!\! 
    &\equiv 
    \sum_{d_1,d_2=1}^{\dim_\mathbb C V} \sum_{r =1}^{\dim_\mathbb C K} \MPOT{}{V}{d_1}{d_2}{r}{r} k_r \otimes k^r \otimes v_{d_1} \otimes v^{d_2}
    \\
    &\equiv 
    \sum_{d_1,d_2=1}^{\dim_\mathbb C V} \sum_{r=1}^{\dim_\mathbb C K} \sum_{g \in G} \rho(g^{-1})^{d_1}_{d_2} \, k_{r,g} \otimes k^{r,g} \otimes v_{d_1} \otimes v^{d_2} \, .
\end{split}
\end{equation}
The various symmetry operators can be organised into a block-injective MPO
\begin{equation}
    \MPOT{}{}{}{}{}{} \!\!\!= \!\!\!\!\! \bigoplus_{V \in \Irr(\Rep(G))} \!\!\!\!\! \MPOT{}{V}{}{}{}{} \, .
\end{equation}
For every $V_1,V_2,V_3 \in \Irr(\Rep(G))$, let $N^{V_1V_2}_{V_3}$ be the non-negative integer such that $V_1 \otimes V_2 \cong \bigoplus_{V_3 \in \Irr(\Rep(G))}N^{V_1V_2}_{V_3} V_3$. For each $i \in \{1,\ldots,N^{V_1V_2}_{V_3}\}$, let us denote by $\varphi^{V_1V_2}_{V_3,i}$ and $\bar \varphi^{V_1V_2}_{V_3,i}$ the linear maps furnishing bases for $\Hom_G(V_1 \otimes V_2,V_3)$ and $\Hom_G(V_3,V_1 \otimes V_2)$, respectively. By definition of the monoidal structure in $\Rep(G)$, for every $V_1 = \mathbb C\{v_{d_1}\}_{d_1},V_2 = \mathbb C\{v_{d_2}\}_{d_2} \in \Irr(\Rep(G))$, we have
\begin{equation}
    \label{eq:RepG_fusionMPO}
    \fusionMPO{1}{V_1}{V_2}{}{}{} = \sum_{\substack{V_3 \in V_1 \otimes V_2\\ 1 \leq i \leq N^{V_1 \! V_2}_{V_3}}} \fusionMPO{2}{V_1}{V_2}{V_3}{\varphi^{V_1 \! V_2}_{V_3,i}}{\bar \varphi^{V_1 \! V_2}_{V_3,i}} \, ,
\end{equation}
where, for every $V_3 = \mathbb C\{v_{d_3}\}_{d_3} \in V_1 \otimes V_2$ and $i \in \{1,\ldots,N^{V_1 V_2}_{V_3}\}$, we introduced tensors 
\begin{equation*}
\begin{split}
    \fusionT{1}{\varphi^{V_1\!V_2}_{V_3,i}}{V_1}{V_2}{V_3}{}{}{} 
    &\equiv \!\!\!\!
    \sum_{\substack{1 \leq d_1 \leq \dim_\mathbb C V_1 \\ 1 \leq d_2 \leq \dim_\mathbb C V_2 \\ 1 \leq d_3 \leq \dim_\mathbb C V_3}} 
    \fusionT{1}{\varphi^{V_1\!V_2}_{V_3,i}}{V_1}{V_2}{V_3}{d_1}{d_2}{d_3} \, 
    v_{d_3} \otimes v^{d_1} \otimes v^{d_2} 
    \equiv \!\!\!\!
    \sum_{\substack{1 \leq d_1 \leq \dim_\mathbb C V_1 \\ 1 \leq d_2 \leq \dim_\mathbb C V_2 \\ 1 \leq d_3 \leq \dim_\mathbb C V_3}} 
    \overline{\CC{V_1}{V_2}{V_3}{d_1}{d_2}{d_3}{i}} \, 
    v_{d_3} \otimes v^{d_1} \otimes v^{d_2} 
    \\
    \fusionT{2}{\bar \varphi^{V_1\!V_2}_{V_3,i}}{V_1}{V_2}{V_3}{}{}{}  \!\!\!
    &\equiv \!\!\!\!
    \sum_{\substack{1 \leq d_1 \leq \dim_\mathbb C V_1 \\ 1 \leq d_2 \leq \dim_\mathbb C V_2 \\ 1 \leq d_3 \leq \dim_\mathbb C V_3}} 
    \fusionT{2}{\bar \varphi^{V_1\!V_2}_{V_3,i}}{V_1}{V_2}{V_3}{d_1}{d_2}{d_3} \, 
    v_{d_1} \otimes v_{d_2} \otimes v^{d_3} 
    \equiv \!\!\!\!
    \sum_{\substack{1 \leq d_1 \leq \dim_\mathbb C V_1 \\ 1 \leq d_2 \leq \dim_\mathbb C V_2 \\ 1 \leq d_3 \leq \dim_\mathbb C V_3}} 
    \CC{V_1}{V_2}{V_3}{d_1}{d_2}{d_3}{i} \, 
    v_{d_1} \otimes v_{d_2} \otimes v^{d_3} \, ,
\end{split}
\end{equation*}
which evaluate to \emph{Clebsch--Gordan coefficients} satisfying the orthogonality condition
\begin{equation}
    \label{eq:RepG_orthoFusion}
    \sum_{\substack{1 \leq d_1 \leq \dim_\mathbb C V_1 \\ 1 \leq d_2 \leq \dim_\mathbb C V_2}}
    \CCb{V_1}{V_2}{V_3}{d_1}{d_2}{d_3}{i_1}
    \CC{V_1}{V_2}{V_4}{d_1}{d_2}{d_4}{i_2} = \delta_{V_3,V_4} \, \delta_{d_3,d_4} \, \delta_{i_1,i_2} \, .
\end{equation}
As in the invertible case, we interpret this operation as a consequence of the topological invariance of the symmetry lines, allowing us to fuse them locally. Moreover, it follows from the orthogonality condition \eqref{eq:RepG_orthoFusion} that 
\begin{equation}
    \label{eq:RepG_Fmove}
    \fusionAssocRepG{2} = 
    \sum_{V_5 \in V_1 \otimes V_2} \sum_{\substack{ 1 \leq i_1 \leq N^{V_1 \! V_2}_{V_5} \\ 1 \leq i_2 \leq N^{V_5\! V_3}_{V_4}}}     \big(F^{V_1V_2V_3}_{V_4}\big)^{V_5,i_1i_2}_{V_6,i_3i_4}  \, \fusionAssocRepG{1} \, ,
\end{equation}
for every $V_1,\ldots,V_4,V_6 \in \Irr(\Rep(G))$, $i_3 \in \{1,\ldots,N^{V_2V_3}_{V_5}\}$ and $i_4 \in \{1,\ldots,N^{V_1 V_6}_{V_4}\}$ 
where we introduced the $F$-symbols
\begin{equation}
    \big(F^{V_1V_2V_3}_{V_4}\big)^{V_5,i_1i_2}_{V_6,i_3i_4} 
    := \frac{1}{\dim_\mathbb C V_4} \sum_{d_1,\ldots,d_6}
    \CCb{V_2}{V_3}{V_6}{d_3}{d_3}{d_6}{i_3}
    \CCb{V_1}{V_6}{V_4}{d_1}{d_6}{d_4}{i_4}
    \CC{V_1}{V_2}{V_5}{d_1}{d_2}{d_5}{i_1}
    \CC{V_5}{V_3}{V_4}{d_5}{d_3}{d_4}{i_2} \, .
\end{equation}
Eq.~\eqref{eq:RepG_Fmove} merely encodes a change of basis of $\Hom_G(V_1 \otimes V_2 \otimes V_3 , V_4)$. These $F$-symbols are subject to the \emph{Biedenharn--Elliott} identity, which encodes the pentagon axiom of the monoidal associator of $\Rep(G)$.

\subsection{Symmetric gapped phases\label{sec:RepG_Phases}}

Let us now review the classification of one-dimensional $\Rep(G)$-symmetric gapped phases in terms of block-injective MPSs. As in sec.~\ref{sec:VecG_MPS}, one could proceed by  showing that a $\Rep(G)$-symmetric block-injective MPS furnish a $\Rep(G)$-module category, classify indecomposable $\Rep(G)$-module categories, and then construct gapped phases representatives associated with indecomposable $\Rep(G)$-module categories \cite{Garre-Rubio:2022uum}. Since we have already commented on the classification of $\Rep(G)$-module categories, we shall directly construct the corresponding representatives. 

In sec.~\ref{sec:VecG_RG}, we reviewed that (finite semisimple) indecomposable module categories over $\Rep(G)$ and in one-to-one correspondence with those over $\Vect_G$. For every subgroup $H \leq G$ and normalised representative $\beta$ of a cohomology class $[\beta] \in H^2(H,\rU(1))$, recall that the corresponding indecomposable $\Rep(G)$-module category is provided by the category $\Rep^\beta(H)$ of finite-dimensional projective representation of $H$ such that the $\Rep(G)$-module structure is provided by the tensor product of representations via the restriction functor $\Res^G_H : \Rep(G) \to \Rep(H)$. Let us construct a representative of the gapped phase associated with $\Rep^\beta(H)$. Consider a block-injective MPS built from a tensor $\theta \colon M \to K \otimes M = \sum_{l_1,l_2 = 1}^{\dim_\mathbb C M} \sum_{r=1}^{\dim_\mathbb C K} (\theta^{l_1}_{l_2})^r \; k_r \otimes m_{l_1} \otimes m^{l_2}$ such that $\theta^r := \sum_{l_1,l_2}^{\dim_\mathbb C M}(\theta^{l_1}_{l_2})^r \; m_{l_1} \otimes m^{l_2} \equiv \bigoplus_{U \in \Irr(\Rep^\beta(H))} \theta^r_U$, where $\theta^r_U \in \End_\mathbb C(M_U)$, for every $r \in \{1,\ldots,\dim_\mathbb C K\}$. Graphically,\footnote{Here, $U \in \Irr(\Rep^\beta(H))$ merely labels a block of the block-injective MPS tensor, while the corresponding bond space is provided by $M_U$. In the case of renormalisation group fixed points (see sec.~\ref{sec:RepG_RG}), these may coincide, but they generically do not.}
\begin{equation}
    \label{eq:RepG_blockMPS}
    \MPST{\theta}{}{}{}{} \!\!\! = \hspace{-5pt} \bigoplus_{U \in \Irr(\Rep^\beta(H))} \hspace{-20pt} \MPST{\theta}{U}{}{}{}  \!\!\!.  
\end{equation}
This block-injective MPS is assumed to be $\Rep(G)$-symmetric, in the sense that the subspace of $\mc H_\Lambda$ spanned by its constitute injective MPSs is closed under the action of symmetry MPOs \eqref{eq:RepG_MPO}, for every $V \in \Irr(\Rep(G))$. For every $V \in \Irr(\Rep(G))$ and $U_1,U_2 \in \Irr(\Rep^\beta(H))$, let $N^{VU_1}_{U_2}$ be the non-negative integer such that $V \act U_1 := \Res^G_H(V) \otimes U_1 \cong \bigoplus_{U_2 \in \Irr(\Rep^\beta(H))} N^{VU_1}_{U_2} \, U_2$. For every $i \in \{1,\ldots,N^{VU_1}_{U_2}\}$, there are three-valent tensors $\phi^{VU_1}_{U_2,i}$ and $\bar \phi^{VU_1}_{U_2,i}$ such that 
\begin{equation}
    \label{eq:RepG_localAction}
    \raisebox{16pt}{\actionMPORepG{1}{}{U_1}{}{}{V}{}{}} 
    = \!\!\! \sum_{\substack{U_2 \in V \act U_1 \\ 1 \leq i \leq N^{VU_1}_{U_2}}} \!\!\!
    \raisebox{8pt}{\actionMPORepG{2}{}{U_1}{U_2}{}{V}{\phi^{VU_1}_{U_2,i}}{\bar \phi^{VU_1}_{U_2,i}}}
\end{equation}
and 
\begin{equation}
    \label{eq:RepG_orthoAction}
    \orthoActionRepG{U_1}{V}{U_2}{U_3}{\phi^{VU_1}_{U_2,i_1}}{\bar \phi^{V U_1}_{U_3,i_2}} \; = \delta_{U_2,U_3} \, \delta_{i_1,i_2} \, \mathbb I_{M_{U_2}}\, .
\end{equation}
These relations imply that the fixed points of the MPS transfer matrix satisfy an analogue of \eqref{eq:VecG_actionFixedPoints}.

Now, consider the actions of two symmetry lines associated with simple objects $U_1$ and $U_2$ in $\Rep(G)$, respectively. Instead of successively acting with both lines, one can first (locally) fuse them before (locally) acting with the resulting line, which gives rise so the following relation:
\begin{equation}
    \sum_{\substack{V_3 \in V_1 \otimes V_2 \\ U_2 \in V_3 \act U_1 \\ i_1,i_2}} \hspace{-8pt}
    \assocActionRepG{1}  \!\!\! = \sum_{\substack{U_3 \in V_2 \act U_1 \\ U_2 \in V_1 \act U_3 \\ i_3,i_4}}  \assocActionRepG{2} \, .
\end{equation}
From orthogonality conditions \eqref{eq:RepG_orthoFusion} and \eqref{eq:RepG_orthoAction} follow the existence of complex $\F{\act}$-symbols satisfying
\begin{equation}
    \label{eq:RepG_moduleAssoc}
     \moduleAssocRepG{2} \;=\; 
     \sum_{\substack{V_3 \in V_1 \otimes V_2 \\ i_1,i_2}} \!\!
     \big(\F{\act}^{V_1V_2U_1}_{U_2}\big)^{V_3,i_1i_2}_{U_3,i_3i_4} \; \moduleAssocRepG{1} \, ,
\end{equation}
where
\begin{equation}
    \big(\F{\act}^{V_1V_2U_1}_{U_2}\big)^{V_3,i_1i_2}_{U_3,i_3i_4}  \; \delta_{U_2,U_4} \, \mathbb I_{M_{U_2}}
    =
    \FsymRepG \, .  
\end{equation}
Explicitly, the complex $\F{\act}$-symbols evaluate to
\begin{align}    
    \big(\F{\act}^{V_1V_2U_1}_{U_2}\big)^{V_3,i_1i_2}_{U_3,i_3i_4} 
    = \frac{1}{\dim_\mathbb C U_2} \sum_{\substack{d_1,d_2,d_3 \\ b_1,b_2,b_3}}
    \CCb{V_2}{U_1}{U_3}{d_2}{b_1}{b_3}{i_3}
    \CCb{V_1}{U_3}{U_2}{d_1}{b_3}{b_2}{i_4}
    \CC{V_1}{V_2}{V_3}{d_1}{d_2}{d_3}{i_1}
    \CC{V_3}{U_1}{U_2}{d_3}{b_1}{b_2}{i_2} \, ,
\end{align}
where we introduced the Clebsch--Gordan coefficients associated with the action bifunctor $\act : \Rep(G) \times \Rep^\beta(H) \to \Rep^\beta(H)$. Now, consider the actions of three symmetry lines. Instead of successively acting with the three lines, one can instead fuse the first two, fuse the resulting line with the third one, before acting with the resulting line. Using eq.~\eqref{eq:RepG_Fmove} and eq.~\eqref{eq:RepG_moduleAssoc}, one can relate the resulting combinations of lines in two different ways, which requires the mixed Biedenharn--Elliott identity 
\begin{equation}
\begin{split}
    \label{eq:BEid}
    &\sum_{i_7=1}^{N^{V_5U_3}_{U_2}}
    \big(\F{\act}^{V_1V_2U_3}_{U_2}\big)^{V_5,i_1i_7}_{U_4,i_5i_6} \,
    \big(\F{\act}^{V_5V_3U_1}_{U_2}\big)^{V_3,i_2i_3}_{U_3,i_2i_7} 
    \\
    & \q =
    \sum_{V_6 \in V_2 \otimes V_3}
    \sum_{i_8=1}^{N^{V_2V_3}_{V_6}}
    \sum_{i_9=1}^{N^{V_1V_6}_{V_4}}
    \sum_{i_{10}=1}^{N^{V_6 U_1}_{U_4}}
    \big(\F{\act}^{V_2V_3U_1}_{U_4}\big)^{V_6,i_8i_{10}}_{U_3,i_4i_5}
    \big(\F{\act}^{V_1V_6U_1}_{U_2}\big)^{V_4,i_9i_3}_{U_4,i_{10}i_6}
    \big(\F{}^{V_1V_2V_3}_{V_4}\big)^{V_5,i_1i_2}_{V_6,i_8i_9}
\end{split}
\end{equation}
to hold, for every $V_1,\ldots,V_5 \in \Irr(\Rep(G))$, $U_1,\ldots,U_4 \in \Rep^{\beta}(H)$, $i_1 \in \{1,\ldots, N^{V_1V_2}_{V_5}\}$ and multiplicity labels $i_1,\ldots,i_6$.
This identity guarantees that the pentagon axiom of the module associator of $\Rep^\beta(H)$ is satisfied. 

\bigskip \noindent
Above, we constructed a representative of the $\Rep(G)$-symmetric gapped phase associated with the $\Rep(G)$-module category $\Rep^\beta(H)$ such that the set of vacua is in one-to-one correspondence with $\Irr(\Rep^\beta(H))$. Pick a simple object $U \in \Irr(\Rep^\beta(H))$. Recall that the internal hom $\underline{\Hom}(U,U) \in \Rep(G)$ is such that we have an isomorphism $\Hom_{\Rep^\beta(H)}(X \act U,U) \cong \Hom_{\Rep(G)}(X,\underline{\Hom}(U,U))$ that is natural in $X$, for every $X \in \Rep(G)$ (see app.~\ref{app:algebras}). By Frobenius reciprocity, 
\begin{align}
    \Hom_{\Rep^\beta(H)}(X \act U,\, U)
    &=
    \Hom_{\Rep^\beta(H)}(\Res^G_H(X) \otimes U, U )
    \\ \nn
    &\cong 
    \Hom_{\Rep^\beta(H)}(\Res^G_H(X),\End_\mathbb C(U))
    \cong
    \Hom_{\Rep(G)}(X, \Ind^{G}_{H}(\End_\mathbb C(U))) \, ,    
\end{align}
for every $X \in \Rep(G)$. An application of the \emph{Yoneda lemma} yields $\underline{\Hom}(U,U) \cong \Ind_H^G(\End_\mathbb C(U))$. The $\Rep(G)$-module structure of $\Rep^\beta(H)$ endows $\underline{\Hom}(U,U)$ with the structure of an algebra in $\Rep(G)$ (see app.~\ref{app:algebras}). One can further verify that $\underline{\Hom}(U,U)$ is isomorphic to the algebra object $A(H,\beta,U)$ one constructed in sec.~\ref{sec:VecG_RG}. Finally, one recovers the $\Rep(G)$-module category $\Rep^\beta(H)$ as $\Mod_{\Rep(G)}(A(H,\beta,U))$ (see app.~\ref{app:algebras}). In particular, this implies that for any two $U_1,U_2 \in \Irr(\Rep^\beta(H))$, the algebras $A(H,\beta,U_1)$ and $A(H,\beta,U_2)$ are Morita equivalent.

Physically, the internal hom $\underline{\Hom}(U,U)$ is the algebra of lines that is preserved in the ground state labelled by $U \in \Irr(\Rep^\beta(H))$. By definition, the algebra $\underline{\Hom}(U,U)$ is constituted of lines $X \in \Irr(\Rep(G))$ for which the hom-space $\Hom_{\Rep^\beta(H)}(X \act U,U)$ is non-trivial. Therefore, we say that a symmetry line is preserved in the ground state labelled by $U \in \Irr(\Rep^\beta(U))$ if acting with it on the ground state results in a state with non-zero overlap. We often find a stronger condition in the literature, namely that the resulting state is proportional to the initial one. This stronger condition typically ensures that the preserved lines are closed under the tensor product, in such a way that they may form a fusion subcategory. As we already commented in sec.~\ref{sec:VecG_classification}, this does not accurately capture spontaneous breaking of invertible or non-invertible symmetries. Indeed, the amount of symmetry preserved in the ground state labelled by any $U \in \Irr(\Rep^\beta(H))$ is captured by an algebra object, namely $\underline{\Hom}(U,U) \cong A(H,\beta,U)$ in $\Rep(G)$. Embracing this viewpoint is crucial for two reasons: (i) This guarantees that algebras of preserved lines in any two ground states are Morita equivalent, i.e., they preserve equivalent amounts of symmetry. (ii) This is required to ensure that we are not excluding any order parameters (see sec.~\ref{sec:RepG_OP}).

At this point, it is interesting to observe that for distinct $U_1,U_2 \in \Irr(\Rep^\beta(H))$, the algebra objects $A(H,\beta,U_1)$ and $A(H,\beta,U_2)$---though Morita equivalent---may be drastically different, e.g., whenever $U_1$ and $U_2$ do not share the same dimension. Let us study an example. Let $G=\mathbb A_4$ be the alternating group of order 12 explored in sec.~\ref{sec:VecG_Example}.
Consider the gapped phase associated with the $\Rep(\mathbb A_4)$-module category provided by $\Rep(\mathbb A_4)$ itself. Algebras of preserved lines in the ground states labelled by the one-dimensional irreducible representation $\mathbb C_1$, $\mathbb C_\omega$ and $\mathbb C_{\bar \omega}$ are all trivial, while that in the ground state labelled by the three-dimensional representation $V_3$ is given by $A(\mathbb A_4,1,V_3) \cong \End_\mathbb C(V_3)$, which decomposes as $\mathbb C_1 \oplus \mathbb C_\omega \oplus \mathbb C_{\bar \omega} \oplus 2 \cdot V_3$. Therefore, in spite of the symmetry $\Rep(G)$ being spontaneously broken, some symmetry is preserved in the ground state labelled by $V_3$, which translates into more entanglement \cite{Lootens:2024gfp}. In that regard, the various ground states are not on equal footing \cite{Bhardwaj:2023idu,Lootens:2024gfp,Chung:2025ulc}.
Were we to only include lines that strictly preserve the ground state labelled by $V_3$---i.e., those labelled by the one-dimensional irreducible representations---we would obtain an algebra that is not Morita equivalent to the trivial one, which would ultimately lead to contradictions as we must still be able to reconstruct the whole gapped phase from any ground state. To further illustrate our remarks, let us also consider the gapped phase associated with the $\Rep(\mathbb A_4)$-module category $\Rep(\mathbb  Z_3)$. The algebra of lines that is preserved in the ground state labelled by $\mathbb C_{1/\omega/\bar \omega} \in \Irr(\Rep(\mathbb Z_3))$ is provided by $A(\mathbb Z_3,1,\mathbb C_{1/\omega/\bar \omega}) \cong \Ind_{\mathbb Z_3}^{\mathbb A_4}(\mathbb C_{1}) \cong \mathbb C_{1} \oplus V_3$. Clearly, this collection of lines is not closed under the tensor product of $\Rep(\mathbb A_4)$, but they do form an algebra, as required.

\subsection{Renormalisation group fixed points\label{sec:RepG_RG}}

Let $H$ be a subgroup of $G$ and $\beta$ a normalised representative of a cohomology class in $H^2(H,\rU(1))$. Suppose we want to construct the ground state subspace at some renormalisation group fixed point of the $\Rep(G)$-symmetric gapped phase labelled by the $\Rep(G)$-module category $\Rep^\beta(H)$. Above, we commented that $\Rep^\beta(H)$ is equivalent to the category $\Mod_{\Rep(G)}(A(H,\beta,U))$ of right modules in $\Rep(G)$ over the algebra object $A(H,\beta,U)$, for any $U \in \Irr(\Rep^\beta(H))$. We could exploit this equivalence in order to build renormalisation group fixed points, but, as in our treatment of the symmetry $\Vect_G$, this would not be compatible with the choice of microscopic Hilbert space we made in sec.~\ref{sec:RepG_Hilbert}.\footnote{The realisation of the symmetry $\Rep(G)$ alluded to here differs drastically from the one considered in this manuscript, as the microscopic Hilbert space is typically not a tensor product space.} Indeed, recall that our microscopic Hilbert space requires a choice of object in $\Vect_G$. Here, $\Vect_G$ is realised as the Morita dual $(\Rep(G))^*_\Vect$ of $\Rep(G)$ with respect to $\Vect$. From this Morita equivalence follows that $\Rep^\beta(H)$ can also be realised as the category of left modules in $\Vect$ over an indecomposable (finite separable) algebra object in $\Vect_G$. A suitable choice is provided by the algebra object $A[H,\beta]$, which also possesses the structure of a $\Delta$-separable symmetric Frobenius algebra. Recall that $A[H,\beta]$ is the algebra object with underlying object $\bigoplus_{h \in H} \mathbb C_h$ whose algebraic structure is provided by the twisted group algebra $\mathbb C[H]^\beta = \mathbb C\{e_h\}_{h \in H}$ such that $|e_h| = h \in G$ and $e_{h_1} \cdot e_{h_2} = \beta(h_1,h_2) \, e_{h_1h_2}$, for every $h,h_1,h_2 \in G$. Indeed, we already know that $\Rep^\beta(H) \simeq \Mod(\mathbb C[H]^\beta)$.

Let us exploit the equivalence $\Mod(\mathbb C[H]^\beta) \simeq \Rep^\beta(H)$ to construct the ground state subspace at a renormalisation group fixed point associated with the $\Rep(G)$-module category $\Rep^\beta(H)$. As in sec.~\ref{sec:VecG_RG}, our conventions dictate that rather than the $\Delta$-separable symmetric Frobenius algebra $A[H,\beta]$, one should consider $A[H,\beta]^{*,{\rm op}, {\rm cop}} = \mathbb C\{e^h\}_{h \in H}$ such that $|e^h| = h^{-1}$,\footnote{The evaluation pairing $A[H,\beta]^* \otimes A[H,\beta] \to \mathbb C, e^h \otimes e_h \mapsto e^h(e_h)=1$ being a linear map in $\Vect_G$ of degree the identity in $H$, it requires $|e^h|=h^{-1} \in H$.} for every $h \in H$, and exploit the fact that every left module over $A[H,\beta]$ is also a left comodule over $A[H,\beta]^{*,{\rm op}, {\rm cop}}$. Therefore, we take ground states to span a subspace of the tensor product space $\bigotimes_{\msf i \in \Lambda} A[H,\beta]^*_{\{\msf i\}}$, which may itself be the subspace of a larger microscopic Hilbert space. The action of topological lines in $\Rep(G)$ on $\bigotimes_{\msf i \in \Lambda}A[H,\beta]^*_{\{\msf i\}}$ is provided by the $G$-graded structure of $A[H,\beta]^* = \mathbb C\{e^h\}_{h \in G}$ via definition \eqref{eq:RepG_MPOT} of the MPO tensors. Concretely, given $V \in \Rep(G)$ with $\rho : \mathbb C[G] \to \End_\mathbb C(V)$, recall from eq.~\eqref{eq:RepG_MPOT} that the MPO tensor actually depends on the inverse of the grading of the basis element in the $G$-graded vector space $K = \bigoplus_{g \in G}K_g$. Since $|e^{h}|=h^{-1} \in H$, for every $h \in H$, the MPO reads 
\begin{equation}
    \sum_{h_1,\ldots,h_{|\Lambda|} \in H} {\rm tr}_V \big[\rho(h_1) \rho(h_2) \cdots \rho(h_{|\Lambda|}) \big] \; e^{h_1} \otimes \cdots \otimes e^{h_{|\Lambda|}} \otimes e_{h_1} \otimes \cdots \otimes e_{h_{|\Lambda|}} \, . 
\end{equation}
Let $U = \mathbb C\{u_b\}_{b=1,\ldots,\dim_\mathbb C U}$ be a simple object in $\Mod(\mathbb C[H]^\beta)$ with algebra homomorphism $\pi \colon \mathbb C[H]^\beta \to \End_\mathbb C (U)$. For every $b_1,b_2 \in \{1,\ldots,\dim_\mathbb CU\}$, one defines an algebra element\footnote{\label{foot:insanity}Let us spell out these manipulations in some detail. Given that the multiplication of $A[H,\beta] = \mathbb C\{e_h\}_{h \in H}$ is provided by $e_{h_1} \cdot e_{h_2} = \beta(h_1,h_2) \, e_{h_1h_2}$, for every $h_1,h_2 \in H$, the comultiplication that makes $A[H,\beta]$ a $\Delta$-separable symmetric Frobenius algebra is given by $\Delta(e_h) = |H|^{-1}\sum_{y \in H} \beta(y,y^{-1}h)^{-1}\, e_y \otimes e_{y^{-1}h}$. By definition of the dual, the comultiplication of $A[H,\beta]^* = \mathbb C\{e^h\}_{h \in H}$ is defined via the multiplication of $A[H,\beta]$. After renormalisation, it reads $\Delta^*(e^h) = |H|^{-1}\sum_{y \in H} \beta(y,y^{-1}g) \, e^y \otimes e^{y^{-1}h}$. The coopposite comultiplication is then given by $\Delta^{*,{\rm cop}}(e^h) = |H|^{-1}\sum_{y \in H} \beta(y,y^{-1}g) \, e^{y^{-1}h} \otimes e^y$. The compatible multiplication rule, which is the opposite of that $A[H,\beta]^*$, is then provided by $e^{h_1} \cdot e^{h_2} = \beta(h_2,h_1)^{-1} \, e^{h_2h_1}$.
Now, given $U = \mathbb C\{u_b\} \in \Mod(A[H,\beta])$, one can verify that $u_{b_2} \mapsto \sum_{h \in H} e^h \otimes e_h \cdot u_{b_2} = \sum_{h \in H}\sum_{b_1} \pi(h)^{b_1}_{b_2} \, e^h \otimes u_{b_1} \equiv \sum_{b_1} \pi^{b_1}_{b_2} \otimes u_{b_1}$ equips $U$ with the structure of a left comodule over $A[H,\beta]^{*,{\rm op},{\rm cop}}$.}
\begin{equation}
    \pi^{b_1}_{b_2} := \sum_{h \in H} \pi(h)^{b_1}_{b_2} \, e^{h} \in A[H,\beta]^* \, .
\end{equation}
Assuming closed periodic boundary conditions, one associates to this collection of algebra elements the following injective MPS in $\bigotimes_{\msf i \in \Lambda}A[H,\beta]^*_{\{\msf i\}}$:
\begin{equation}
    \label{eq:RepG_RGT}
\begin{split}
    \Psi(\pi) 
    &:= \sum_{b_1,\ldots,b_{|\Lambda|}}\pi^{b_1}_{b_2} \otimes \pi^{b_2}_{b_3} \otimes \cdots \otimes \pi^{b_{|\Lambda|}}_{b_1} 
    \\
    &=  \sum_{h_1,\ldots,h_{|\Lambda|}} {\rm tr}_U \big[\pi(h_1) \pi(h_2) \cdots \pi(h_{|\Lambda|}) \big] \; e^{h_1} \otimes e^{h_2} \otimes \cdots \otimes e^{h_{|\Lambda|}}
    \, .
\end{split}
\end{equation}
Acting with the topological line $V \in \Irr(\Rep(G))$ with $\rho \colon \mathbb C[G] \to \End_\mathbb C (V)$ on the MPS $\Psi(\pi_1)$ with $\pi_1 \colon \mathbb C[H]^\beta \to \End_\mathbb C(U_1)$ results in  $\sum_{U_2 \in V \act U_1} N^{VU_1}_{U_2} \Psi(\pi_2)$, where we are using the fact that
\begin{equation}
\begin{split}
    \label{eq:RepG_basisExp}
    \rho(h)^{d_1}_{d_2} \, \pi_1(h)^{b_1}_{b_2} = \sum_{\substack{U_2 \in V \act U_1 \\ 1 \leq i \leq N^{VU_1}_{U_2}}} \sum_{b_3,b_4 =1}^{\dim_\mathbb C U_2} \CC{V}{U_1}{U_2}{d_1}{b_1}{b_3}{i} \,
    \pi_2(h)^{b_3}_{b_4} \, 
    \CCb{V}{U_1}{U_2}{d_2}{b_2}{b_4}{i} \, ,
\end{split}
\end{equation}
for every $h \in H$, $b_1,b_2 \in \{1,\ldots,\dim_\mathbb C U\}$ and $d_1,d_2 \in \{1,\ldots,\dim_\mathbb C V\}$. 
As expected, the topological line $V \in \Irr(\Rep(G))$ acts on the ground state subspace according to $\Res^G_H(V) \otimes -$, thereby confirming that the $\Rep(G)$-symmetric block-injective MPS furnished by all the irreducible representations in $\Mod(\mathbb C[H]^\beta)$ defines a renormalisation group fixed point associated with the gapped phase $\Rep^\beta(H)$. In this scenario, it is especially easy to build gapped parent Hamiltonians in terms of $G$-grading preserving linear maps $e^{h_1} \otimes e^{h_2} \mapsto \frac{1}{|H|}\sum_{y \in H} \frac{\beta(y,y^{-1}h_2)}{\beta(h_1,y)} \, e^{h_1y} \otimes e^{y^{-1}h_2}$. As a matter of fact, this map is built from the multiplication and the comultiplication of the $\Delta$-separable symmetric Frobenius algebra $A[H,\beta]^{*,{\rm op},{\rm cop}}$. The resulting local operators are projectors thanks to separability, and commute with one another thanks to the Frobenius property \cite{Inamura:2021szw}.

\subsection{String operators\label{sec:RepG_String}}

We work within the framework of sec.~\ref{sec:RepG_Hilbert}, whereby the microscopic Hilbert space is of the form $\mc H_\Lambda = \bigotimes_{\msf i \in \Lambda} K_{\{\msf i\}}$, where $K = \bigoplus_{g \in G}K_g = \mathbb C\{k_r\}_{r=1,\ldots,\dim_\mathbb C K}$ is an object in $\Vect_G$, so that the action of the topological line $V \in \Irr(\Rep(G))$ on $\mc H_\Lambda$ is given by the MPO \eqref{eq:RepG_MPO}.  Given a symmetry $\Vect_G$, we explained in sec.~\ref{sec:VecG_String} that twisted sector local operators furnish representations of the tube algebra $\Tu(\Vect_G)$, whose category $\Mod(\Tu(\Vect_G))$ is equivalent to the Drinfel'd centre $\mc Z(\Vect_G)$ of $\Vect_G$. More generally, given a non-invertible symmetry encoded in a fusion category $\mc C$, every multiplet of twisted sector local operators corresponds to an object in the Drinfel'd centre $\mc Z(\mc C)$ of $\mc C$, which can be realised as $\Mod(\Tu(\mc C))$. It follows from $\Vect_G$ and $\Rep(G)$ being Morita equivalent that $\mc Z(\Vect_G) \simeq \mc Z(\Rep(G))$ (see app.~\ref{app:algebras}). In sec.~\ref{sec:VecG_String}, we further exploited the fact that $\mc Z(\Vect_G) \simeq \Vect_G^G$. Therefore, we also have $\mc Z(\Rep(G)) \simeq \Vect_G^G$. As a matter of fact, the equivalence $\mc Z(\Vect_G) \simeq \mc Z(\Rep(G))$ naturally factors through $\Vect_G^G$ (see app.~\ref{app:centres}). 

For every a simple object in $\Vect_G^G$, let us construct multiplets of twisted sector local operators for the symmetry $\Rep(G)$. As in sec.~\ref{sec:VecG_String}. given $[x_0] \in \Cl(G)$ and $\hat W \in \Irr(\Mod(\mathbb C[Z_G(x_0)]))$, we denote the corresponding simple object in $\Vect_G^G$ by $([x_0],\hat W)$. Recall that, as an object in $\Vect_G$, the simple object $([x_0],\hat W)$ is isomorphic to $W = \bigoplus_{x \in [x_0]} W_x$ with $W_x \cong \hat W$, for every $x \in [x_0]$. On the other hand,  as an object in $\Rep(G)$, it is isomorphic to the induced representation $W  = \Ind_{Z_G(x_0)}^G(\hat W) = \mathbb C\{\varpi_{x,w_c}\}_{x \in [x_0], c=1,\ldots,\dim_\mathbb C \hat W}$ such that $|\varpi_{x,w_c}| = x \in G$ (see sec.~\ref{sec:VecG_String}). Let $\psi^{WK}_K : W \otimes K \to K$ be a choice of $G$-grading preserving map. For every $w \in W$, the embedding of $\psi^{WK}_K(w \otimes -) \in \End_\mathbb C (T)$ in $\mc A_\Lambda$ is a local operator. In particular, suppose $w = \varpi_{x,w_c}$, then $\psi^{WK}_K(\varpi_{x,w_c} \otimes -)(T_g) \subseteq T_{gx}$. 
In general, $W = \Ind_{Z_G(x_0)}^G(\hat W)$ is not a simple object in $\Rep(G)$. Let $S = \mathbb C\{s_f\}_{f=1,\ldots,\dim_\mathbb C S} \in \Irr(\Rep(G))$ be a direct summand of $W$ with multiplicity $N^W_S \in \mathbb N$. Given $f \in \{1,\ldots,\dim_\mathbb C S\}$, note that the basis vector $s_f$ is typically not homogeneous. For every $i \in \{1,\ldots,N^W_S\}$, let $\varphi^W_{S,i} : W \to S$ be a choice of basis vector in $\Hom_G(W,S)$. For every $S \in \Irr(\Rep(G))$ and $j \in \{1,\ldots,N^W_S \}$, one defines a tensor
\begin{equation}
    \label{eq:RepG_twistedLocalOp}
    \twistedLocalOp{j}{S}{} \!\!\!
    \equiv
    \! \twistedLocalOpExt{\psi^{W \! K}_K\!\!\!}{W}{\bar \varphi^W_{S,j}}{S}
    \equiv
    \sum_{f=1}^{\dim_\mathbb C S}\psi^{WK}_K \big( \bar \varphi^{W}_{S,j}(s_f)\otimes - \big) \otimes s^f  \, ,
\end{equation}
where
\begin{equation}
    \label{eq:RepG_mat}
    \raisebox{-2pt}{\mat{\bar \varphi^W_{S,j}}{W}{S}{}{}} \!\!\!\!\! \equiv \sum_{f=1}^{\dim_\mathbb C S} \bar \varphi^W_{S,j}(s_f) \otimes s^f 
    \equiv \sum_{f=1}^{\dim_\mathbb C S} \sum_{x \in [x_0]} \sum_{c=1}^{\dim_\mathbb C \hat W} \CCd{W}{S}{x,c}{f}{j} \; \varpi_{x,w_c} \otimes s^f \, .
\end{equation}
In eq.~\eqref{eq:RepG_mat}, we introduced a notation for the coefficients of the $G$-equivariant maps $\varphi^{W}_{S,i}$, which satisfy the orthogonality condition
\begin{equation}
    \sum_{x \in [x_0]} \sum_{c=1}^{\dim_\mathbb C \hat W} \CCdb{W}{S_1}{x,c}{f_1}{j_1} \CCd{W}{S_2}{x,c}{f_2}{j_2} = \delta_{S_1,S_2} \, \delta_{f_1,f_2} \, \delta_{j_1,j_2} \, .
\end{equation}
For every choice of $G$-grading preserving map $\psi^{WK}_K$, we claim that tensors \eqref{eq:RepG_twistedLocalOp} furnish a multiplet of twisted sector local operators associated with the simple object $([x_0], \hat W)$ in $\Vect_G^G \simeq \mc Z(\Rep(G))$. Recall the symmetry $\Rep(G)$ acts on twisted sector local operators via its tube algebra $\Tu(\Rep(G))$. Given an object in $\Vect_G^G$, we explain in app.~\ref{app:centres_RepG} how to construct the corresponding object in $\mc Z(\Rep(G))$, and subsequently the corresponding module in $\Mod(\Tu(\Rep(G)))$. Our goal is to verify that tensors \eqref{eq:RepG_twistedLocalOp} transform in this module.

Within our framework, $\Tu(\Rep(G))$ is realised as follows: As a vector space, it is spanned by tensor networks of the form
\begin{equation}
    \label{eq:RepG_Tube}
    \Tube{V}{S_3}{S_2}{S_1}{\bar \varphi^{VS_1}_{S_2,i_1}}{\varphi^{S_3V}_{S_2,i_2}} 
    \hspace{-10pt}  = \,  
    \left(\!\!  \TubeDag{V}{S_3}{S_2}{S_1}{\bar \varphi^{S_3V}_{S_2,i_2}}{\varphi^{VS_1}_{S_2,i_1}} \!\! \right)^\dagger  ,
\end{equation}
for every $V,S_1,S_2,S_3 \in \Irr(\Rep(G))$, $i_1 \in \{1,
\ldots,N^{VS_1}_{S_2}\}$ and $i_2 \in \{1,\ldots,N^{S_3V}_{S_2}\}$ such that $\Hom_G(V \otimes S_1,S_2) = \mathbb C\{\varphi^{VS_1}_{S_2,i_1}\}_{i_1}$ and $\Hom_G(S_3 \otimes V,S_2) = \mathbb C\{\varphi^{S_3V}_{S_2,i_2}\}_{i_2}$. The r.h.s. of eq.~\eqref{eq:RepG_Tube} realises the $*$-structure of $\Tu(\Rep(G))$. The multiplication rule of $\Tu(\Rep(G))$ reads
\begin{align}
    \nn
    &\Tube{V_1}{S_6}{S_5}{S_4}{\bar \varphi^{V_1S_4}_{S_5,i_3}}{\varphi^{S_6V_1}_{S_5,i_4}} \hspace{-16pt} \cdot \Tube{V_2}{S_3}{S_2}{S_1}{\bar \varphi^{V_2S_1}_{S_3,i_1}}{\varphi^{S_3V_2}_{S_2,i_2}} \hspace{-16pt} 
    := \delta_{S_3,S_4} 
    \TubeMultRepG 
    \\[-1em]
    & \q\q\q \equiv  
    \sum_{\substack{V_3 \in V_1 \otimes V_2 \\ S_7 \in V_3 \otimes S_1}} 
    \sum_{\substack{1 \leq i_5 \leq N^{V_3 S_1}_{S_7} \\ 1 \leq i_6 \leq N^{S_6 V_3}_{S_7}}}
    \Gamma^{V_1,V_2,S_1,\ldots,S_6,i_1,\ldots,i_4}_{V_3,S_7,i_5,i_6}
    \Tube{V_3}{S_6}{S_7}{S_1}{\bar \varphi^{V_3S_1}_{S_7,i_5}}{\varphi^{S_6V_3}_{S_7,i_6}} \!\!\!  , 
\end{align}
for every $V_1,V_2,S_1,\ldots,S_6 \in \Irr(\Rep(G))$, $i_1 \in \{1,\ldots,N^{V_2 S_1}_{S_2}\}$, $i_2 \in \{1,\ldots,N^{S_3 V_2}_{S_2}\}$,  $i_3 \in \{1,\ldots,N^{V_1 S_4}_{S_5}\}$ and $i_4 \in \{1,\ldots,N^{S_6 V_1}_{S_5}\}$, where the structure constants of the algebra are given by 
\begin{equation}
    \label{eq:RepG_TubeCsts}
    \Gamma^{V_1,V_2,S_1,\ldots,S_6,i_1,\ldots,i_4}_{V_3,S_7,i_5,i_6}
    :=   \delta_{S_3,S_4}  \sum_{\substack{1 \leq i_7 \leq N^{V_1 S_2}_{S_7} \\  1 \leq i_8 \leq N^{S_5 V_2}_{S_7} \\ 1 \leq i_9 \leq N^{V_1V_2}_{V_3} }}\big(F^{V_1S_3V_2}_{S_7}\big)^{S_5,i_3i_8}_{S_2,i_2i_7} \, 
    \big(\bar F^{S_6 V_1 V_2}_{S_7}\big)^{V_3,i_9i_6}_{S_5,i_4i_8} \, 
    \big(\bar F^{V_1 V_2 S_1}_{S_7}\big)^{V_3,i_9i_5}_{S_2,i_1i_7} \, .
\end{equation}
In order to obtain eq.~\eqref{eq:RepG_TubeCsts}, we used eq.~\eqref{eq:RepG_fusionMPO} together with eq.~\eqref{eq:RepG_Fmove} and eq.~\eqref{eq:RepG_orthoFusion}. 

We are now in a position to verify that twisted sector local operators \eqref{eq:RepG_twistedLocalOp} furnish simple modules over $\Tu(\Rep(G))$. Given $([x_0],\hat W) \in \Irr(\Vect_G^G)$, contracting the tube labelled by $V,S_1,S_2,S_3 \in \Irr(\Rep(G))$, $i_1 \in \{1,\ldots,N^{VS_1}_{S_2}\}$ and $i_2 \in \{1,\ldots,N^{S_2 V}_{S_3}\}$ with the tensor \eqref{eq:RepG_twistedLocalOp} labelled by $S_5 \in \Irr(\Rep(G))$ and $j_2 \in \{1,\ldots,N^W_{S_1}\}$ results in
\begin{equation}
\begin{split}
    \label{eq:RepG_twistedLocalOpTube}
    \delta_{S_1,S_5} \,
    \twistedLocalOpTube{1}{j_2}{V}{S_3}{S_2}{S_1}{\bar \varphi^{V S_1}_{S_2,i_1}}{\varphi^{S_3 V}_{S_2,i_2}}
    \hspace{-14pt}
    &= 
    \delta_{S_1,S_5} \,
    \twistedLocalOpTube{3}{}{V}{S_3}{S_2}{S_1}{W}{\psi^{WK}_K}
    \\
    &\equiv
    \sum_{\substack{S_4 \in W \\ 1 \leq j_1 \leq N^W_{S_4}}} \big( \Upsilon^{S_1 V S_3}_{S_2,i_1i_2} \big)_{S_5,j_2}^{S_4,j_1}
    \twistedLocalOp{j_1}{S_4}{} \!\!\! ,
\end{split}
\end{equation}
where
\begin{align}
    \label{eq:RepG_TubeModCsts}
    \big( \Upsilon^{S_1 V S_3}_{S_2,i_1 i_2} \big)_{S_5,j_2}^{S_4,j_1}  = \frac{\delta_{S_1,S_5} \, \delta_{S_3,S_4}}{\dim_\mathbb C S_3} \,  
    \tr_W \! \left[ \ModuleCsts \right] \, .
\end{align}
Explicitly, the $\Upsilon$-symbols evaluate to
\begin{align}
    \label{eq:RepG_UpsSymbols}
    \big( \Upsilon^{S_1 V S_3}_{S_2,i_1 i_2} \big)^{S_4,j_1}_{S_5,j_2} =
    \frac{\delta_{S_1,S_5} \, \delta_{S_3,S_4}}{\dim_\mathbb C S_3}  \!\!\!\!\!\!\!
    \sum_{\substack{x \in [x_0] \\ 1 \leq c \leq \dim_\mathbb C \hat W}} \!
    \sum_{\substack{1 \leq f_1 \leq \dim_\mathbb C S_1 \\ 1 \leq f_2 \leq \dim_\mathbb C S_2 \\ 1 \leq f_3 \leq \dim_\mathbb C S_3 \\ 1 \leq d_1,d_2 \leq \dim_\mathbb C V}} \!\!
    \CCdb{W}{S_3}{x,c}{f_3}{j_1} \,
    \CCb{S_3}{V}{S_2}{f_3}{d_1}{f_2}{i_2}
     \,
    \rho(x^{-1})^{d_1}_{d_2} \, 
    \CC{V}{S_1}{S_2}{d_2}{f_1}{f_2}{i_1} \,
    \CCd{W}{S_1}{x,c}{f_1}{j_2} \! ,  
\end{align}
where $\rho : \mathbb C[G] \to \End_\mathbb C (V)$. 
In order to obtain eq.~\eqref{eq:RepG_TubeModCsts}, we used the fact that $\psi^{WK}_K$ is $G$-grading preserving, together with eq.~\eqref{eq:RepG_MPOT} before performing all the contractions. We affirm that this action endows the space of tensors \eqref{eq:RepG_twistedLocalOp} with the structure of a $\Tu(\Rep(G))$-module, i.e., that the symbols \eqref{eq:RepG_TubeModCsts} coincide with the matrix coefficients of a $\Tu(\Rep(G))$-module. Indeed,
\begin{equation}
\begin{split}
    \label{eq:RepG_ModStr}
    &\sum_{\substack{S_8 \in W \\ 1 \leq j_2 \leq N^W_{S_8}}} \!\!\! 
    \big( \Upsilon^{S_4 V_1 S_6}_{S_5,i_3 i_4} \big)_{S_8,j_2}^{S_7,j_1} 
    \big( \Upsilon^{S_1 V_2 S_3}_{S_2,i_1 i_2} \big)_{S_9,j_3}^{S_8,j_2} 
    \\[-1em ]
    & \q = \frac{\delta_{S_3,S_4} \, \delta_{S_6,S_7} \, \delta_{S_1,S_9} \, \delta_{S_4,S_8}}{\dim_\mathbb C S_6} \;
    \tr_{W} \! \left[ \RepGTubeModStr \right]
    \\
    & \q =
    \sum_{\substack{V_3 \in V_1 \otimes V_2 \\ S_7 \in V_1 \otimes S_2}} 
    \sum_{\substack{1 \leq i_5 \leq N^{V_3 S_1}_{S_7} \\ 1 \leq i_6 \leq N^{S_6 V_3}_{S_7}}}
    \Gamma^{V_1,V_2,S_1,\ldots,S_6,i_1,\ldots,i_4}_{V_3,S_7,i_5,i_6} \, 
    \big( \Upsilon^{S_1 V_3 S_6}_{S_7,i_5 i_6} \big)_{S_9,j_3}^{S_7,j_1} \, , 
\end{split}
\end{equation}
for every $V_1,V_2,S_1,\ldots,S_6 \in \Irr(\Rep(G))$, $i_1 \in \{1,\ldots,N^{V_2 S_1}_{S_2}\}$, $i_2 \in \{1,\ldots,N^{S_3 V_2}_{S_2}\}$,  $i_3 \in \{1,\ldots,N^{V_1 S_4}_{S_5}\}$, $i_4 \in \{1,\ldots,N^{S_6 V_1}_{S_5}\}$, $j_1 \in \{1,\ldots,N^W_{S_7}\}$ and $j_3 \in \{1,\ldots,N^W_{S_9}\}$.
In the first step, we used the fact that\footnote{This formula follows from thinking about the closed tensor network on the l.h.s. as a morphism in $\Hom_G(\mathbb C,\mathbb C)$ in such a way that the sum over $j_2 \in \{1,\ldots,N^W_{S_3}\}$ amounts to resolving an identity morphism. In deriving eq.~\eqref{eq:RepG_ModStr}, we used this property after expressing the $\Upsilon$-symbols in terms of the corresponding tensor networks, where one should think of the map $\bar \varphi^{W}_{S_3,j_2}$ depicted on the l.h.s. as a piece of the former tensor network.}
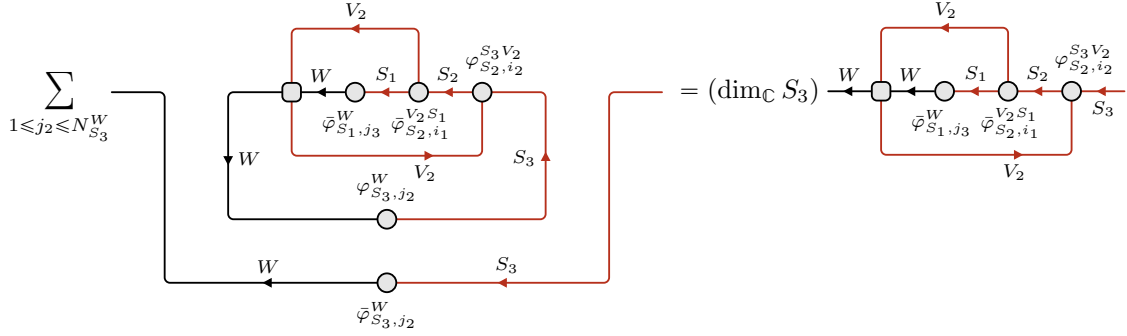
\begin{equation*}
    \sum_{1 \leq j_2 \leq N^W_{S_3}} \!\!\! \raisebox{-28pt}{\sesameTrick{1}} = (\dim_\mathbb C S_3) \; \sesameTrick{2}
\end{equation*}
followed by
\begin{equation}
    \rho_1(x^{-1})^{d_1}_{d_2} \, \rho_2(x^{-1})^{d_3}_{d_4} = \sum_{\substack{V_3 \in V_1 \otimes V_2 \\ 1 \leq i \leq N^{V_1 \! V_2}_{V_3}}} \sum_{d_5,d_6 =1}^{\dim_\mathbb C V_3} \CC{V_1}{V_2}{V_3}{d_1}{d_3}{d_5}{i} \,
    \rho_3(x^{-1})^{d_5}_{d_6} \, 
    \CCb{V_1}{V_2}{V_3}{d_2}{d_4}{d_6}{i} \, .
\end{equation}
Finally, invoking $\Mod(\Tu(\Rep(G))) \simeq \mc Z(\Rep(G))$, one confirms in app.~\ref{app:centres_RepG} that the module furnished by tensors \eqref{eq:RepG_twistedLocalOp} with underlying vector space $\bigoplus_{S \in \Irr(\Rep(G))}\Hom_{G}(S,W)$ is that associated with the simple object $([x_0],\hat W)$ in $\Vect_G^G \simeq \Mod(\Tu(\Rep(G)))$. We conclude that for every choice of $G$-grading preserving map $\psi^{WK}_K : W \otimes K \to K$, tensors \eqref{eq:RepG_twistedLocalOp}, for every simple direct summand $S$ of $W$ and $i \in \{1,\ldots,N^W_S\}$, furnish a multiplet of twisted sector local operators that transform in the irreducible module $([x_0],\hat W)$ of $\Tu(\Rep(G))$. Multiplets of twisted local operators associated with any object in $\Vect_G^G$ are constructed similarly.

\bigskip\noindent
For every choice of $G$-grading preserving map $\psi^{KW^*}_K \colon K \otimes W^* \to K$, one can construct another collection of tensors that form the Hermitian dual of the module associated with $([x_0],\hat W)$ . Concretely, for every simple direct summand $S$ of $W$ and $j \in \{1,\ldots,N^W_S\}$, define
\begin{equation}
    \label{eq:RepG_twistedLocalOpDual}
    \twistedLocalOpDual{j}{S}\!  
    \equiv
    \twistedLocalOpDualExt{\psi^{KW^*}_K\!\!\!}{W}{\varphi^{W}_{S,j}}{S}
    \equiv   
    \sum_{x \in [x_0]} \sum_{c=1}^{\dim_\mathbb C \hat W} \varphi^{W}_{S,j}(\varpi_{x,w_c}) \otimes \psi^{K W^*}_K (- \otimes \varpi^{x,w_c} ) \, ,
\end{equation}
where
\begin{equation}
    \raisebox{-3pt}{\matDual{\bar \varphi^{W}_{S,j}}{S}{W}{}{}} \!\!\!\!\! 
    \equiv \sum_{x \in [x_0]} \sum_{c=1}^{\dim_\mathbb C \hat W} \varphi^W_{S,j}(\varpi_{x,w_c}) \otimes \varpi^{x,w_c} 
    \equiv \sum_{f=1}^{\dim_\mathbb C S} \sum_{x \in [x_0]} \sum_{c=1}^{\dim_\mathbb C \hat W} \CCdb{W}{S}{x,c}{f_4}{j} \; s_f \otimes \varpi^{x,w_c} \, .
\end{equation}
Given definition \eqref{eq:RepG_Tube} of the $*$-structure of $\Tu(\Rep(G))$, the Hermitian dual of \eqref{eq:RepG_twistedLocalOpTube} agrees with the result of acting with the Hermitian dual of the tube on the tensor \eqref{eq:RepG_twistedLocalOpDual}.

Let $([x_0],\hat W)$ label a simple object in $\Vect_G^G$ and choose a simple direct summand $S \in \Irr(\Rep(G))$ of $W$. Consider twisted sector local operators of the form \eqref{eq:RepG_twistedLocalOp} and $\eqref{eq:RepG_twistedLocalOpDual}$ associated with $j_1$ and $j_2$ in $\{1,\ldots,N^W_S\}$, respectively. Placing the corresponding tensors at sites $\msf i< \msf j$ in $\Lambda$, we contract them via a symmetry MPO \eqref{eq:RepG_MPO} with open boundary conditions. We denote the embedding in $\mc A_\Lambda$ of the resulting tensor network operator as $\mc S(S,j_1,j_2)_{\msf i,\ldots,\msf j}$. Graphically, 
\begin{equation}
    \label{eq:RepG_StringOp}
    \mc S(S,j_1,j_2)_{\msf i,\ldots,\msf j} \equiv \StringOp{S}{j_1}{j_2} \, .
\end{equation}
These are the string operators for $\Rep(G)$-symmetric gapped phases. 

\bigskip \noindent
Let us examine some special cases. Given $[x_0] \in \Cl(G)$, choose $\hat W = \mathbb C$. Recall that as an object in $\Vect_G$, $([x_0],\mathbb C)$ is isomorphic to $W= \bigoplus_{x \in [x_0]} \mathbb C$. On the other hand, as an object in $\Rep(G)$, it is isomorphic to the induced representation $\Ind^G_{Z_G(x_0)}(\mathbb C) \cong \mathbb C(G/Z_G(x_0)) = \mathbb C\{\varpi_x\}_{x \in [x_0]}$ such that $|\varpi_x|=x \in G$, for every $x \in [x_0]$. By Frobenius reciprocity, the trivial representation $\mathbb C$ is always a direct summand of $\mathbb C(G/Z_{G}(x_0))$ such that $N^{\mathbb C(G/Z_G(x_0))}_\mathbb C = 1$. Choose $S = \mathbb C$. By definition, $\bar \varphi^{\mathbb C(G/Z_{G}(x_0))}_\mathbb C(1) = \sum_{x \in [x_0]} \varpi_x$. Suppose that $\mc H_\Lambda = \bigotimes_{\msf i \in \Lambda} \mathbb C[G]^*$, where $\mathbb C[G]^* = \bigoplus_{g \in G} \mathbb C_{g^{-1}}  = \mathbb C\{e^g\}_{g \in G}\in \Vect_G$. Consider the multiplet of local operators defined via the $G$-grading preserving map
\begin{equation}
    \label{eq:RepG_canonicalMap}
\begin{array}{rrcl}
     \psi^{W\mathbb C[G]^*}_{\mathbb C[G]^*} : &W \otimes \mathbb C[G]^* &\to &\mathbb C[G]^*  \\
      {} & \varpi_x \otimes e^g  & \mapsto & e^{gx^{-1}} \, .
\end{array} 
\end{equation}
It follows that the local operator labelled by $S= \mathbb C$ acts by right translation as 
\begin{equation}
    R^{[x_0]} \equiv \sum_{x \in [x_0]} R^x := \sum_{x \in [x_0]} \sum_{g \in G} e^{gx^{-1}} \otimes e_g \, .    
\end{equation}
Similarly, $\psi^{\mathbb C[G]^* W^*}_{\mathbb C[G]^*} : e^g \otimes \varpi^x \mapsto e^{xg}$ would yield $L^{[x_0]} \equiv \sum_{x \in [x_0]} L^x := \sum_{x \in [x_0]}\sum_{g \in G}e^{xg} \otimes e_g$. The remaining operators in the multiplet, which are twisted sector local operators, are constructed analogously. 

For instance, choose $G = \mathbb A_4$, and consider the simple object $([\sn{(12)(34)}],\mathbb C_{(1,1)})$. As an object in $\Vect_{\mathbb A_4}$, it is isomorphic to $\bigoplus_{x \in [(12)(34)]} \mathbb C_{(1,1)}$; as an object in $\Rep(\mathbb A_4)$, it is isomorphic to $W = \Ind_{\mathbb Z_2 \oplus \mathbb Z_2}^{\mathbb A_4}(\mathbb C_{(1,1)}) \cong \mathbb C_1 \oplus \mathbb C_\omega \oplus \mathbb C_{\bar \omega}$. Choose $S = \mathbb C_\omega$. By definition, $\bar \varphi^W_{\mathbb C_\omega}(1) = \varpi_{(12)(34)} + \omega \, \varpi_{(13)(24)} + \bar \omega \, \varpi_{(14)(23)}$. One can now construct the corresponding twisted sector local operator in the multiplet associated with the map \eqref{eq:RepG_canonicalMap}, and ultimately the string operator. Since $S = \mathbb C_\omega$ is a one-dimensional irreducible representation, it remains possible to write down the result very explicitly, namely\footnote{Similar operators appeared for instance in ref.~\cite{Bhardwaj:2024kvy}.}
\begin{equation}
    \label{eq:RepG_ExStringSign}
    \mc S(\mathbb C_\omega,1,1)_{\msf i,\ldots,\msf j} =  R^{[x_0],\omega}_\msf i
    \bigg(\prod_{\msf k=\msf i+1}^{\msf j -1} \chi^\omega_{\msf k} \bigg) L_{\msf j}^{[x_0],\omega} \, ,
\end{equation}
where $\chi^\omega := \sum_{g \in G} \chi^\omega(g) \, e^g \otimes e_g$, $R^{[x_0],\omega} := R^{(12)(34)} + \omega R^{(13)(24)} + \bar \omega R^{(14)(23)}$ and $L^{[x_0],\omega} := L^{(12)(34)} + \bar \omega L^{(13)(24)} + \omega L^{(14)(23)}$. 

Let us now study a different type of multiplet. Still choosing $G = \mathbb A_4$, consider the simple objects $([\sn{()}],V_3)$ and $([\sn{(12)(34)}],\mathbb C_{(1,-1)})$ in $\Vect_{\mathbb A_4}^{\mathbb A_4}$. Since $\Ind_{\mathbb Z_2 \oplus \mathbb Z_2}^{\mathbb A_4}(\mathbb C_{(1,-1)}) \cong V_3$, there is a single string operator in every multiplet associated with these simple objects. For both simple objects, the twisting line is always given by $V_3$, but the twisted sector local operators do differ. Given that every basis vector in $([\sn{()}],V_3)$ is homogeneous of degree the identity, the resulting string operator amounts to summing the symmetry MPO labelled by $V_3$ over all possible basis open boundary conditions. In contrast, basis vectors in the $\mathbb A_4$-equivariant $\mathbb A_4$-graded vector space $([\sn{(12)(34)}],\mathbb C_{(1,-1)})$ are homogeneous of degrees $\sn{(12)(34)}$, $\sn{(13)(24)}$ and $\sn{(14)(23)}$, respectively. Let $\mathbb C_{(1,-1)} = \mathbb C\{w_1\}$, $W = \Ind_{\mathbb Z_2 \oplus \mathbb Z_2}^{\mathbb A_4}(\mathbb C_{(1,-1)}) = \mathbb C\{\varpi_{x,w_1}\}_{x \in [(12)(34)]}$ and $V_3 = \mathbb C\{s_1,s_2,s_3\}$ be the basis given in eq.~\eqref{eq:basisVA4}. It follows that the $G$-equivariant map $\bar \varphi^{W}_{V_3} \colon V_3 \to W$ is taken to act as $s_1 \mapsto \varpi_{(12)(34),w_1}$, $s_2 \mapsto \varpi_{(13)(24),w_1}$ and $s_3 \mapsto \varpi_{(14)(23),w_1}$. Given that $|\varpi_{x,w_1}| = x \in \mathbb A_4$, for every $x \in [\sn{(12)(34)}]$, the twisted sector local operator on $\mathbb C[A_4]^*$ defined via eq.~\eqref{eq:RepG_canonicalMap} reads
\begin{equation}
    \label{eq:RepG_ExEndPoint}
    \twistedLocalOp{1}{V_3}{} \!\!\! = R^{(12)(34)} \otimes s^1 + R^{(13)(24)} \otimes s^2 + R^{(14)(23)} \otimes s^3 \, .
\end{equation}
Finally, the string operator is obtained applying the construction depicted in eq.~\eqref{eq:RepG_StringOp}.

\subsection{Order parameters\label{sec:RepG_OP}}   

Given a $\Rep(G)$-symmetric gapped phase, we now would like to identify string operators that may serve as \emph{order parameters} in the ground state subspace. As in the invertible case we are asking for necessary conditions that the corresponding multiplets of twisted sector local operators must satisfy in order for the string operators to acquire a non-vanishing expectation value in at least one state in the ground state subspace. Though more technical, the derivation mimics that of sec.~\ref{sec:VecG_OP}. As a matter of fact the conclusion is the same, namely that suitable irreducible multiplets of twisted sector local order parameters must furnish a direct summand of a Lagrangian algebra in $\mc Z(\Rep(G))$. In particular, by virtue of $\mc Z(\Rep(G)) \simeq \Vect_G^G \simeq \mc Z(\Vect_G)$, which follows from the Morita equivalence between $\Rep(G)$ and $\Vect_G$, we have already computed in sec.~\ref{sec:VecG_Lagrangian} these Lagrangian algebras.

Consider a representative of the gapped phase associated with the indecomposable $\Rep(G)$-module category $\Rep^\beta(H)$. Recall that  ground states in the ground state subspace preserve algebra objects of topological lines that are two-by-two Morita equivalent. Specifically, the ground state labelled by $U \in \Irr(\Rep^\beta(U))$ preserves the internal hom $\underline{\Hom}(U,U) \cong A(H,U,\beta)$ in $\Rep(G)$. Our goal is to demonstrate that in order for a string operator to serve as an order parameter, its constitutive irreducible multiplet of twisted sector local operators must furnish a direct summand of the Lagrangian algebra $L(H,\beta)$. 

In order to appreciate most of the technical subtleties that are specific to a non-invertible symmetry, it is enough to focus on the $\Rep(G)$-symmetric gapped phase associated with $\Rep^\beta(G)$. Given a simple object $([x_0],\hat W)$ in $\Vect_G^G$, consider the string operator $\mc S(S,j_1,j_2)_{\msf i, \ldots, \msf j}$ constructed in eq.~\eqref{eq:RepG_StringOp}, for $S \in \Irr(\Rep(G))$ a direct summand of $W= \Ind_{Z_G(x_0)}^G(\hat W)$ and $j_1,j_2 \in \{1,\ldots,N^W_S\}$. The expectation value of this string operator in the ground state labelled by $U \in \Irr(\Rep^\beta(G))$ is given by 
\begin{equation}
    \label{eq:RepG_StringOpExp}
    \StringOpExpVal{U}{S}{j_1}{j_2} .
\end{equation}
Employing eq.~\eqref{eq:RepG_localAction} followed by eq.~\eqref{eq:RepG_orthoAction} at every site $\msf k \in [\msf i+1,\msf j-1]$ together, it follows from eq.~\eqref{eq:VecG_transferprojector} that \eqref{eq:RepG_StringOpExp} evaluates to 
\begin{equation}
    \label{eq:RepG_expValFact}
    \sum_{i=1}^{N^{SU}_U}
    \StringOpExpValLimL{U}{S}{\, \bar \phi^{SU^{\phantom{I^l}}}_{U,i}}{j_1}{\Xi_U}
    \cdot
    \StringOpExpValLimR{U}{S}{\, \phi^{SU^{\phantom{I^l}}}_{U,i}}{j_2}{\Xi_U}  \, ,
\end{equation}
up to terms that decay exponentially in the length of the string.
As in the invertible case, one immediately obtains a preliminary constraint, namely that the simple $S$ object must be a direct summand of the algebra object $A(H,\beta,U)$ of preserved lines in the ground state labelled by $U$. Focusing on the right-hand-side factor in eq.~\eqref{eq:RepG_expValFact}, consider the collection of tensors networks
\begin{equation}
    \label{eq:RepG_twistedLocalOpEnv}
    \TwisedLocalOpEnv{U}{S}{\phi^{SU^{\phantom{I^l}}}_{U,i}}{\Xi_U} 
\end{equation}
over $U \in \Irr(\Rep^\beta(G))$ and basis vector $\phi^{SU}_{U,i}$ in $\Hom_{\Rep^\beta(G)}(S \otimes U , U)$ labelled by $i \in \{1,\ldots,N^{SU}_U\}$. Let us show that this collection furnishes a (reducible) module over $\Tu(\Rep(G))$. Acting with the tube \eqref{eq:RepG_Tube} labelled by $V,S_1,S_2,S_3 \in \Irr(\Rep(G))$, $i_1 \in \{1,\ldots,N^{VS_1}_{S_2}\}$ and $i_2 \in \{1,\ldots,N^{S_2V}_{S_3}\}$ on the tensor network \eqref{eq:RepG_twistedLocalOpEnv} labelled by $U_1 \in \Rep^\beta(G)$, $S_4 \in \End_\mathbb C(U_1)$ and $j_2 \in \{1,\ldots,N^{S_4U_1}_{U_1}\}$ results in
\begin{equation}
    \delta_{S_1,S_4}
    \TubeModEnvA{V}{S_3}{S_2}{S_1}{U_1}{\phi^{S_1U_1^{\phantom{I^{l^l}}}}_{U_1,j_2}}{\bar \varphi^{VS_1}_{S_2,i_1}}{\varphi^{S_3V}_{S_2,i_2}}{\;\;\Xi_{U_1}} \, . 
\end{equation}
One can simplify this tensor network by utilising eq.~\eqref{eq:RepG_localAction} together with eq.~\eqref{eq:RepG_orthoAction}, before invoking eq.~\eqref{eq:RepG_moduleAssoc} and its converse:
\begin{align}
    \label{eq:RepG_TubeModEnv}
    &\delta_{S_1,S_4}
    \sum_{\substack{U_2 \in V \act U_1 \\ 1 \leq j_4 \leq N^{VU_1}_{U_2}}} 
    \RepGTubeModEnvB
    \\[-2em] \nn
    & \q =    
    \delta_{S_1,S_4}  
    \sum_{\substack{U_2 \in V \act U_1 \\ 1 \leq j_4 \leq N^{VU_1}_{U_2}}} 
    \sum_{\substack{ 1 \leq j_3 \leq N^{S_2U_1}_{U_2} \\ 1 \leq j_1 \leq N^{S_3U_2}_{U_2}}}
    \big(\Fbar{\act}^{S_3VU_1}_{U_2}\big)^{U_2,j_4j_1}_{S_2,i_2j_3} \, \big(\F{\act}^{VS_1U_1}_{U_2}\big)^{S_2,i_1j_3}_{U_1,j_2j_4}
    \TwisedLocalOpEnv{U_2}{S_3}{\phi^{S_3U_2^{\phantom{I^l}}}_{U_2,j_1}}{\;\;\Xi_{U_2}} \, .
\end{align}
Using the mixed Biedenharn--Elliott identity \eqref{eq:BEid}, one can show that
\begin{align}
    &\delta_{S_3,S_4} 
    \sum_{\substack{V_3 \in V_1 \otimes V_2 \\ S_7 \in V_3 \otimes S_1}} 
    \sum_{\substack{1 \leq i_5 \leq N^{V_3 S_1}_{S_7} \\ 1 \leq i_6 \leq N^{S_6 V_3}_{S_7}}}
    \sum_{\substack{U_3 \in V_3 \act U_1 \\ 1 \leq j_9 \leq N^{V_3 U_1}_{U_2}}} 
    \sum_{\substack{ 1 \leq j_8 \leq N^{S_2U_1}_{U_3} \\ 1 \leq j_1 \leq N^{S_3U_3}_{U_3}}}
    \big(\Fbar{\act}^{S_6V_3U_1}_{U_3}\big)^{U_3,j_9j_1}_{S_7,i_6j_8} \; \big(\F{\act}^{V_3S_1U_1}_{U_3}\big)^{S_7,i_5j_8}_{U_1,j_3j_9} \;
    \\[-2em] \nn
    & \hspace{21.5em} \Gamma^{V_1,V_2,S_1,\ldots,S_6,i_1,\ldots,i_4}_{V_3,S_7,i_5,i_6}  
    \\[.7em] \nn 
    & \q =  \delta_{S_3,S_4} \!\!
    \sum_{\substack{U_2 \in V_2 \act U_1 \\ 1 \leq j_5 \leq N^{V_2U_1}_{U_2}}} 
    \sum_{\substack{ 1 \leq j_4 \leq N^{S_2U_1}_{U_2} \\ 1 \leq j_2 \leq N^{S_3U_2}_{U_2}}} \sum_{\substack{U_3 \in V_1 \act U_2 \\ 1 \leq j_7 \leq N^{V_1 U_2}_{U_3}}} \sum_{\substack{1 \leq j_6 \leq N^{S_5 U_2}_{U_3} \\ 1 \leq j_1 \leq N^{S_3 U_3}_{U_3}}}
    \big(\Fbar{\act}^{S_3V_2U_1}_{U_2}\big)^{U_2,j_5j_1}_{S_2,i_2j_4} \, \big(\F{\act}^{V_2S_1U_1}_{U_2}\big)^{S_2,i_1j_4}_{U_1,j_3j_5}
    \\[-2em] \nn
    & \hspace{24.5em}
    \big(\Fbar{\act}^{S_6V_1U_2}_{U_3}\big)^{U_3,j_7j_2}_{S_5,i_4j_6} \, \big(\F{\act}^{V_1S_4U_2}_{U_3}\big)^{S_5,i_3j_6}_{U_2,j_2j_7} \, ,
\end{align}
for every for every $V_1,V_2,S_1,\ldots,S_8 \in \Irr(\Rep(G))$, $i_1 \in \{1,\ldots,N^{V_2 S_1}_{S_2}\}$, $i_2 \in \{1,\ldots,N^{S_3 V_2}_{S_2}\}$,  $i_3 \in \{1,\ldots,N^{V_1 S_4}_{S_5}\}$, $i_4 \in \{1,\ldots,N^{S_6 V_1}_{S_5}\}$ and $j_3 \in \{1,\ldots,N^{S_8 U_1}_{U_1}\}$. 
This confirms that the action \eqref{eq:RepG_TubeModEnv} does endow the space of tensor networks \eqref{eq:RepG_twistedLocalOpEnv} with the structure of a module over $\Tu(\Rep(G))$.
In app.~\ref{app:centres_RepG}, one proves that this module is that associated with the Lagrangian algebra $L(G,\beta)$ under the equivalence $\Vect_G^G \simeq \mc Z(\Rep(G))  \simeq \Mod(\Tu(\Rep(G)))$. 
As in the invertible case, this proves that in order for an irreducible multiplet of twisted sector local operators to provide string operators that may have non-vanishing expectation values in the $\Rep(G)$-symmetric gapped phases associated with the $\Rep(G)$-module category $\Rep^\beta(G)$, the corresponding simple object in $\mc Z(\Rep(G)) \simeq \Vect_G^G$ must be a direct summand of the Lagrangian algebra $L(G,\beta)$. Conversely, by inspecting the space of twisted sector local operators yielding non-vanishing expectation values, one can reconstruct the Lagrangian algebra object. Almost as easily, one can verify that in the gapped phase associated with the $\Rep(G)$-module category $\Rep^\beta(H)$, for any $H \leq G$ and $[\beta] \in H^2(H,\rU(1))$, irreducible multiplets of twisted sector local operators must furnish direct summands of the Lagrangian algebra $L(H,\beta)$. 

\bigskip \noindent
A couple of physical remarks are in order. Consider the gapped phase associated with the module category $\Rep(G)$ over itself. This is the gapped phase where the whole symmetry $\Rep(G)$ is spontaneously broken in the sense that the dimension of the ground state subspace equals the number of elements in $\Irr(\Rep(G))$. Moreover, there is a preferred vacuum---that associated with the identity object in $\Rep(G)$---such that acting on it by any simple non-trivial line in $\Rep(G)$ results in an orthogonal vacuum. In that sense, this is analogous to the spontaneously symmetry broken phase associated with an invertible symmetry. However, as we have already clarified in sec.~\ref{sec:RepG_Phases}, all the vacua are not quite on equal footing since the algebras of lines that are preserved in these ground states may be very different---in spite of being Morita equivalent. In particular, that associated with the preferred vacuum mentioned above is trivial. This implies that order parameters in this ground state are all strictly local. In contrast, ground states labelled by higher-dimensional irreducible representations of $G$ admit genuine string operators. Mathematically, this follows from the fact that simple direct summands in the Lagrangian algebra $L(G,1) \cong \bigoplus_{[x_0] \in \Cl(G)} ([x_0],\mathbb C)$ typically include non-trivial irreducible representations of $G$ as objects in $\Rep(G)$.\footnote{Note that the same Lagrangian algebra $L(G,\beta)$ was associated with an SPT for the symmetry $\Vect_G$. This duality can be traced back to the fact that $(\Vect_G)^*_\Vect \simeq \Rep(G)$, which encodes the notion that gauging the symmetry $\Vect_G$ results in a theory with symmetry $\Rep(G)$.} 

Now, suppose that the Bogomolov multiplier of $G$ is non-trivial and let $\beta$ be a normalised representative of a non-trivial class in it. By definition, for every $g_1 \in G$ and $g_2 \in Z_G(g_1)$, $\beta(g_1,g_2) = \beta(g_2,g_1)$. This implies that every conjugacy class of $G$ is \emph{$\beta$-regular} so that the number of simple objects in $\Rep^\beta(G)$ equals the number of conjugacy classes of $G$ and thus the number of simple objects in $\Rep^\beta(G)$. We commented in sec.~\ref{sec:VecG_Lagrangian} that the corresponding Lagrangian algebra $L(G,\beta) = \bigoplus_{[x_0] \in \Cl(G)} ([x_0], \mathbb C_{\beta_{x_0}})$ is equivalent to $L(G,1)$ as an object in $\Vect_G^G$, but not as an algebra. For instance, consider the renormalisation group fixed point of sec.~\ref{sec:RepG_RG} for the gapped phase associated with $\Rep(G)$-module category $\Rep^\beta(G)$. Recall that the microscopic Hilbert space is provided by $\mc H_\Lambda = \bigotimes_{\msf i \in \Lambda} A[G,\beta]^*$, where $A[G,\beta]^* = \mathbb C\{e^g\}_{g \in G}$, such that MPS tensors are provided by the left $A[G,\beta]^{*,{\rm op}, {\rm cop}}$-comodule structure of the left $A[G,\beta]$-modules. Consider the multiplet of twisted sector local operators furnishing $L(G,\beta)$ associated with the $G$-grading preserving map
\begin{equation}
\begin{array}{rrcl}
     \psi^{L(G,\beta) A[G,\beta]^*}_{A[G,\beta]^*} : &L(G,\beta) \otimes A[G,\beta]^* &\to &A[G,\beta]^*  \\
      {} & e_{g_1} \otimes e^{g_2}  & \mapsto & \beta(g_2g_1^{-1},g_1^{-1}) \; e^{g_2g_1^{-1}}
\end{array} \, .
\end{equation}
According to foot.~\ref{foot:insanity}, this map essentially implements the comultiplication rule of $A[H,\beta]^{*,{\rm op},{\rm cop}}$. It follows that twisted sector local operators in this multiplet fuse according to the multiplication rule of $L(G,\beta)$. Mimicking the derivations of sec.~\ref{sec:VecG_Lagrangian}, one could show that, in the infrared, multiplets of order parameters furnishing $L(G,\beta)$ are further required to be defined in terms of such $G$-grading preserving maps so as to yield non-vanishing expectation values, thereby distinguishing the gapped phases $\Rep(G)$ and $\Rep^\beta(G)$. But a stronger statement applies here. Indeed, only in the gapped phase $\Rep(G)$ does there exist a ground state in which only local operators admit a non-trivial expectation value.

\subsection{Example: $\Rep(\mathbb A_4)$}

Consider the non-invertible symmetry encoded into the fusion category $\Rep(\mathbb A_4)$ of finite-dimensional representations of the alternating group $\mathbb A_4$ of order $12$ (see sec.~\ref{sec:VecG_Example}). We focus on the two SPT phases associated with the $\Rep(\mathbb A_4)$-module category $\Rep(\mathbb Z_1)$ and $\Rep^\beta(\mathbb Z_2 \oplus \mathbb Z_2)$ with $\beta$ the normalised representative of the non-trivial cohomology class in $H^2(\mathbb Z_2 \oplus \mathbb Z_2,\rm U(1))$ \cite{you2025symmetricentanglersnoninvertiblespt}. 
The relevant Lagrangian algebras $L(0,1)$ and $L(\mathbb Z_2 \oplus \mathbb Z_2,\beta)$ were computed in sec.~\ref{sec:VecG_Example}. By inspection, we find that the multiplets $([\sn{()}], V_3)$ and $([\sn{(12) (34)}], \mathbb{C}_{(1,-1)})$ of twisted sector local operators distinguish these two phases. 

In sec.~\ref{sec:RepG_RG}, we explained how to construct renormalisation group fixed points for these two phases. The input algebras are provided by $A[\mathbb C,1] \cong \mathbb C$ and $A[\mathbb Z_2 \oplus \mathbb Z_2,\beta]^{*,{\rm op},{\rm cop}} \cong (\mathbb C[\mathbb Z_2 \oplus \mathbb Z_2]^{\beta})^{*,{\rm op},{\rm cop}}$, respectively, and the MPS are given by eq.~\eqref{eq:RepG_RGT}. It is convenient to embed both models into the microscopic Hilbert space $\bigotimes_{\msf i \in \Lambda} \mathbb C[\mathbb A_4]^*_{\{\msf i\}}$, at which point the unique string operator in any multiplet associated with $([\sn{()}], V_3)$ and $([\sn{(12) (34)}], \mathbb{C}_{(1,-1)})$ are of the same form as those we constructed in sec.~\ref{sec:RepG_String}. 

First, consider the trivial SPT associated with the $\Rep(\mathbb A_4)$-module category $\Rep(\mathbb Z_2) = \Vect$.  The unique ground state is the product state $e^{1_{\mathbb A_4}} \otimes e^{1_{\mathbb A_4}} \otimes \cdots \otimes e^{1_{\mathbb A_4}}$. In sec.~\ref{sec:RepG_String}, we explained that for any multiplet associated with $([\sn{()}],V_3)$, the unique string operator acts diagonally in the computational basis. Furthermore, the ground state is clearly an eigenvector of the symmetry operator labelled by $V_3$. Together, this guarantees that the expectation value of this string operator is non-zero. In contrast, the string operator in any multiplet associated with $(\sn{(12)(34)},\mathbb C_{(1,-1)})$ would have a vanishing expectation value as it would act at the endpoints by right or left translations by a non-identity group element

Now, consider the non-trivial SPT associated with the $\Rep(\mathbb A_4)$-module category $\Rep^\beta(\mathbb Z_2 \oplus \mathbb Z_2)$. The unique ground state in this phase can be verified to be the cluster state that we already encountered in sec.~\ref{sec:VecG_CS} in the context of the invertible symmetry $\Vect_{\mathbb Z_2 \oplus \mathbb Z_2}$. Recall that $\Res^{V_3}_{\mathbb Z_2 \oplus \mathbb Z_2}(V_3) \cong \mathbb C_{(1,-1)} \oplus \mathbb C_{(-1,1)} \oplus \mathbb C_{(-1,-1)}$. 
Up to a choice of basis, it follows that the non-invertible symmetry operator labelled by $V_3$ acts on the cluster state as the sum of three operators associated with the non-trivial invertible lines in $\Vect_{\mathbb Z_2 \oplus \mathbb Z_2}$. Choosing the multiplet $([\sn{(12)(34)}],\mathbb C_{(1,-1)})$ associated with the $\mathbb A_4$-grading preserving map \eqref{eq:RepG_canonicalMap}, it follows from eq.~\eqref{eq:RepG_ExEndPoint} that the string operator acts in the same way as the sum of the three non-trivial string operators we constructed in sec.~\ref{sec:VecG_CS}. Therefore, the expectation value is non-vanishing.\footnote{Within our framework, it essentially follows from the fact that both the MPS tensor and the action tensor encoding the local action of the symmetry line $V_3$ via the restriction functor $\Res^{V_3}_{\mathbb Z_2 \oplus \mathbb Z_2}$ evaluate to the Pauli matrices. When computing the expectation value, the non-trivial action of the endpoint local operator produces a virtual tensor that precisely cancels with the action tensor.} As a matter of fact, different choices of $\mathbb A_4$-grading preserving maps allows us to isolate each one of these string operators. For instance, one can choose $\psi^{W \mathbb C[\mathbb A_4]^*}_{\mathbb C[\mathbb A_4]^*}$, where $W = \Ind^{\mathbb A_4}_{\mathbb Z_2 \oplus \mathbb Z_2}(\mathbb C_{(1,-1)})$, such that $\varpi_{(12)(34),w_1} \otimes - \mapsto R^{(12)(34)}$, and zero otherwise. The resulting string operator then coincides with eq.~\eqref{eq:cluster_OP}. The fact there are three linearly independent $\mathbb A_4$-grading preserving maps one can choose, each reproducing one of the non-trivial string operators of the cluster state we constructed in sec.~\ref{sec:VecG_CS} explains the multiplicity of the simple object $([\sn{(12)(34)}],\mathbb C_{(1,-1)})$ in the Lagrangian algebra.  This shows that string order parameters can distinguish different gapped phases much more easily than by computing the corresponding $\F{\act}$-symbols \cite{you2025symmetricentanglersnoninvertiblespt}.

\subsection{Generalisation}

For non-invertible symmetry lines furnishing the fusion category $\Rep(G)$, one demonstrated in this section that in order for an irreducible multiplet of twisted sector local operators to provide string order parameters in the gapped phase associated with the $\Rep(G)$-module category $\Rep^\beta(H)$, they must transform in a direct summand of the Lagrangian algebra $L(H,\beta)$. The crux of our argument relies on the observation that the expectation value of any string operator in the ground state subspace explicitly depends on the $\Tu(\Rep(G))$-module associated with $L(H,\beta) \in \Vect_G^G$ via the equivalence $\Mod(\Tu(\Rep(G))) \simeq \mc Z(\Rep(G)) \simeq \Vect_G^G$. We established this correspondence in app.~\ref{app:centres_RepG} by explicitly computing the Lagrangian algebra as an object in $\mc Z(\Rep(G))$, before expressing its half-braiding natural isomorphism in terms of the  $\F{\act}$- and $\Fbar{\act}$-symbols appearing in eq.~\eqref{eq:RepG_TubeModEnv}. 

We claim that our approach generalises to any fusion category. Given a fusion category $\mc C$, consider the gapped phase associated with the (indecomposable finite semisimple) $\mc C$-module category $\mc M$. It is easy to imagine how the derivations of sec.~\ref{sec:RepG_OP} would generalise to this case. In particular, one would find that in order for string operators to acquire non-vanishing expectation values, they must be built from irreducible multiplets of twisted sector local operators that transform in a simple direct summand of a certain module over $\Tu(\mc C)$. The defining $\Tu(\mc C)$-action of this module would be expressed in terms of $\F{\act}$- and $\Fbar{\act}$-symbols, as in eq.~\eqref{eq:RepG_basisExp}. One would then demonstrate that this $\Tu(\mc C)$-module precisely coincides with a Lagrangian algebra object in $\mc Z(\mc C)$ under the equivalence $\mc Z(\mc C) \simeq \Mod(\Tu(\mc C))$. More concretely, indecomposable finite semisimple $\mc C$-module categories $\mc M$ are in one-to-one correspondence with Lagrangian algebras in $\mc Z(\mc C)$. Given such an $\mc M$, the corresponding Lagrangian algebra can be constructed as the \emph{full centre} $\mc Z(\mc M)$ of $\mc M$ in $\mc Z(\mc C)$ \cite{DAVYDOV2010319,Davydov:2011kb}. Crucially, this construction establishes that the half-braiding natural isomorphism of the Lagrangian algebra in $\mc Z(\mc C)$ associated with $\mc M$ can be expressed in terms of the module associator of $\mc M$, in a way that precisely coincides with the $\F{\act}$- and $\Fbar{\act}$-symbols appearing in the definition of the $\Tu(\mc C)$-module structure (see app.~\ref{app:algebras}).\footnote{Consider $\Rep^\beta(G) \in \Mod(\Rep(G))$. The full centre construction relates the half-braiding isomorphism $R_{V,L(G,\beta)} : V \otimes L(G,\beta) \cong L(G,\beta) \otimes V$ of the Lagrangian algebra $L(G,\beta) \in \mc Z(\Rep(G))$ associated with $\Rep^\beta(G)$ to the components $\alpha^{\act}_{V,L(G,\beta),U_1}$ and $\bar \alpha^{\act}_{L(G,\beta),V,U_1}$ of the module associator $\alpha^{\act}$ and its inverse of $\Rep^\beta(G)$, for every irreducible linear representation $V$ and irreducible $\beta$-projective representation $U_1$. Decomposing $L(G,\beta)$ into simple objects in $\Rep(G)$ and computing the matrix coefficients of $R_{V,L(G,\beta)}$ precisely recovers the combinations of $\F{\act}$- and $\Fbar{\act}$-symbols found in eq.~\eqref{eq:RepG_TubeModEnv} (see app.~\ref{app:centres_RepG}).}

\section{Phases of symmetric mixed states\label{sec:mixed}}

\emph{In this section, we briefly demonstrate how our approach further generalises to the case of symmetric mixed states with respect to invertible symmetries. We explain how string order parameters furnish condensable algebras in the Drinfel'd centre of the symmetry fusion category.}

\subsection{Matrix product density operators}

The study of symmetric phases of matter can be extended from pure states to mixed states. In this context, decoherence both destroys some patterns of entanglement and enables new ones, leading to a richer phase classification. Previous work on symmetric mixed states has studied their SPT phases \cite{deGroot:2021vdi,Ma_2023,Lee_2025,Zhang_2026,Ma_2025} (including some \emph{intrinsically mixed} SPT phases which are obstructed from arising in pure states) as well as phases that exhibit an intrinsically mixed form of symmetry breaking called \emph{strong-to-weak spontaneous symmetry breaking} (SWSSB) \cite{Lee_2023,Ma_2025,PRXQuantum.6.010344,sala2024spontaneousstrongsymmetrybreaking}. The representation of these states with matrix product tensor networks has also been investigated \cite{PRXQuantum.6.010348,Xue:2024bkt,PhysRevX.15.021060}. Mixed state phases can be studied from the perspective of symmetry TFT \cite{Qi:2025tal,Schafer-Nameki:2025fiy,luo2025topologicalholographymixedstatephases}. Consistent with these results, we find that string order parameters furnish (typically non-Lagrangian) \emph{condensable algebras} in the Drinfel'd centre of the symmetry fusion category. Specifically, we find that expectation values $\tr[\mc S \Psi]$ of string operators $\mc S$ in a mixed state $\Psi$ are enough to reconstruct the condensable algebra as an object, which ultimately implies that it suffices to compute such expectation values to distinguish phases.\footnote{In particular, knowledge of R\'enyi-2 correlators $\tr[\mc S\Psi\mc S'\Psi]$ \cite{Lee_2023,PRXQuantum.6.010344} and more general R\'enyi-1 correlators $\tr[\mc S\sqrt{\Psi}\mc S'\sqrt{\Psi}]$ \cite{Weinstein_2025} are not necessary to distinguish a phase from all others. For example, the $\mathbb Z_2$ strong-to-weak symmetry breaking phase, associated with the condensable algebra $C=0$, is completely characterised by the vanishing of all string expectation values \cite{PRXQuantum.6.010348}.}

\bigskip \noindent
We work within the framework of sec.~\ref{sec:VecG_Hilbert}, whereby the microscopic Hilbert space is defined to be $\mc H_\Lambda = \bigotimes_{\msf i \in \Lambda} V_{\{i\}}$, where $V \in \Rep(G)$ is a faithful representation. Any \emph{mixed state} in $\mc H_\Lambda$ can be expressed as a matrix product density operator (MPDO). Here, we focus on MPDOs with finite bond dimension in the thermodynamic limit; physically, this condition is meant to exclude criticality and is analogous to the gapped condition on pure states.

We begin with a few remarks on MPDOs. These form a special class of MPOs, which up to this point have been used to parametrise symmetry operators. 
Since they are now meant to describe (mixed) states rather than operators, it is natural to view an MPO as an MPS where the physical space $V$ has been replaced by $\End_\mathbb C(V)$. In general, a translation invariant MPO in $\mc H_{\Lambda}$ is defined via a choice of linear map $\theta : M \otimes V \to V \otimes M$:
\begin{equation}
    \theta \equiv \sum_{l_1,l_2=1}^{\dim_\mathbb C M} \theta^{l_1}_{l_2} \;  m_{l_1} \otimes m^{l_2} \equiv \sum_{l_1,l_2 = 1}^{\dim_\mathbb C M} \sum_{d_1,d_2=1}^{\dim_\mathbb C V} (\theta^{l_1}_{l_2})^{d_1}_{d_2} \; v_{d_1} \otimes v^{d_2} \otimes m_{l_1} \otimes m^{l_2}
    \, ,   
\end{equation}
such that $\theta^{l_1}_{l_2} \in \End_\mathbb C(V)$, for every $l_1,l_2 \in \{1,\ldots,\dim_
\mathbb C M\}$, and $(\theta^{l_1}_{l_1})^{d_1}_{d_2} \in \mathbb C$, for every $d_1,d_2 \in \{1,\ldots,\dim_\mathbb C V\}$. We depict this tensor as
\begin{equation}
    \theta \equiv \!\!\! \DOT{}{}{}{}{} \!\!\!
    \equiv 
    \sum_{l_1,l_2 = 1}^{\dim_\mathbb C M}
    \sum_{d=1}^{\dim_\mathbb C V} \! \DOT{}{l_1}{l_2}{d_1}{d_2} v_{d_1} \otimes v^{d_2} \otimes m_{l_1} \otimes m^{l_2} \, .
\end{equation}
One further defines $\theta^{d_1}_{d_2} := \sum_{l_1,l_2=1}^{\dim_\mathbb C M} (\theta^{l_1}_{l_2})^{d_1}_{d_2} \; m_{l_1} \otimes m^{l_2} \in \End_\mathbb C(M)$, for every $d_1,d_2 \in \{1,\ldots,\dim_\mathbb C V\}$. Assuming \emph{closed periodic} boundary conditions, the MPO $\mathcal O(\theta) \in \End_\mathbb C(\mc H_\Lambda)$ associated with the map $\theta$ reads
\begin{equation}
    \mathcal O(\theta) := \sum_{l_1,\ldots,l_{|\Lambda|}} \theta^{l_1}_{l_2} \otimes \theta^{l_2}_{l_3} \otimes \cdots \otimes \theta^{l_{|\Lambda|}}_{l_1} \, .
\end{equation}
Alternatively, the same MPO can be written as 
\begin{equation}
    \mathcal O(\theta) = \sum_{\substack{d_1,\ldots,d_{|\Lambda|}\\d'_1,\ldots,d'_{|\Lambda|}}} {\rm tr}_M \big[\theta^{d_1}_{d_1'} \, \theta^{d_2}_{d_2'} \cdots \theta^{d_{|\Lambda|}}_{d'_{|\Lambda|}} \big] \; v_{d_1}\otimes v^{d_1'} \otimes v_{d_2} \otimes v^{d_2'} \otimes \cdots \otimes v_{d_{|\Lambda|}} \otimes v^{d'_{|\Lambda|}} \, . 
\end{equation}
Graphically,
\begin{equation}
    \DO{}{} \, .
\end{equation}
Just as for an MPS, an MPO is said to be \emph{injective} if the linear map $\End_\mathbb C(M) \to \End_\mathbb C(V), m_{l_1} \otimes m^{l_2} \mapsto \theta^{l_1}_{l_2}$ is injective. And just as for MPS, it is always possible to bring a periodic MPO into a block-diagonal form, whereby $M = \bigoplus_a M_a$ and $\theta^{d_1}_{d_2}$ is a block-diagonal matrix $\bigoplus_a \thet{a}{d_1}{d_2}$, where $\thet{a}{d_1}{d_2} \in \End_\mathbb C(M_a)$ for every $d_1,d_2 \in \{1,\ldots,\dim_\mathbb C V\}$ \cite{Perez-Garcia:2006nqo}. Graphically,
\begin{equation}
    \DOT{}{}{}{}{} \!\!\! = \bigoplus_{a=1} \!\!\! \DOT{a}{}{}{}{} \!\! \! \equiv \bigoplus_{a=1} \theta_a \, ,
\end{equation}
where we introduced $\theta_a : M_a \otimes V \to V \otimes M_a$ such that $(\theta_a)^{d_1}_{d_2} := \thet{a}{d_1}{d_2} \in \End_\mathbb C(M_a)$, for every $d_1,d_2 \in \{1,\ldots,\dim_\mathbb C V\}$. By blocking sites, we may once again further assume that each $\theta_a$ is injective. Each injective block gives rise to an operator depicted as
\begin{equation}
    \DO{}{a} \, .
\end{equation}
We now specialise to MPDOs. An MPDO $\Psi(\theta)$ is an MPO that satisfies the conditions of \emph{self-adjointness} $\Psi(\theta)^\dagger=\Psi(\theta)$ and \emph{positivity}. These conditions constrain the topological invariants that the MPDO can realise \cite{PRXQuantum.6.010348,Xue:2024bkt}. In the context of MPDOs, the injectivity condition is known as \emph{strong injectivity}. 

The transfer matrix $\mathbb E_\theta$ of the MPDO tensor $\theta$ is defined as\footnote{This transfer matrix is distinguished from the transfer matrix of the MPO tensor realized as an MPS tensor. The latter transfer matrix involves involves two copies of the tensor.}
\begin{equation}
\begin{split}
    \mathbb E_\theta := \sum_{d=1}^{\dim_\mathbb C V} \theta^d_d 
    &\equiv 
    \DOTM{} \in \End_\mathbb C(M)
    \\
    &\equiv
    \sum_{l_1,l_2=1}^{\dim_\mathbb C M} (\mathbb E_\theta)^{l_1}_{l_2} \; m_{l_1} \otimes m^{l_2} \, ,
\end{split}
\end{equation}
where $(\mathbb E_\theta)^{l_1}_{l_2} := \sum_{d=1}^{\dim_\mathbb C V} (\theta^{l_1}_{l_2})^d_d$.
An MPDO is said to be \emph{weakly injective} if the transfer matrix admits a single and non-degenerate eigenvalue $\xi \in \mathbb C$ of largest weight. Graphically,
\begin{equation}
    \DOFptR{1}{}{}{g}{\Xi_{\rm r}}{} \! = \xi \cdot \DOFptRT{}{\Xi_{\rm r}}
    \, , \q
    \DOFptL{1}{}{}{g}{\Xi_{\rm l}\! } \, = \xi \cdot  \DOFptLT{}{\Xi_{\rm l}} \, ,
\end{equation}
where $\Xi_{\rm r}, \Xi_{\rm l} \in M$ are the unique right and left eigenvectors with eigenvalue $\xi$. Without loss of generality, one can normalise the MPDO tensor $\theta$ so that $\xi=1$, making $\Xi_{\rm r}$ and $\Xi_{\rm l}$ right and left \emph{fixed points} of $\mathbb E_\theta$. Since all other eigenvalues have magnitude less than $1$, the product of $N$ transfer matrices tends toward a projector in the limit of large $N$:
\begin{equation}
    \label{eq:mixed_transferprojector}
    \productDOTM{} = \DOFptLR{}{\Xi_{\rm l}}{\Xi_{\rm r}}
    + O(e^{N \tilde \xi}) \xrightarrow{N \to \infty}
    \DOFptLR{}{\Xi_{\rm l}}{\Xi_{\rm r}} \, ,
\end{equation}
where $\tilde \xi$ denotes the eigenvalue of second highest magnitude. A special class of MPDOs are renormalisation group fixed points, defined as those for which the transfer matrix $\mathbb E$ is a projector.

It is worth noting that the blocks $\Psi(\theta_a)$ of an MPDO $\Psi(\theta)$ may themselves fail to be MPDOs because they fail positivity.\footnote{For example, in the two-block MPDO $\mathbb I\cdots\mathbb I+\sigma^x\cdots \sigma^x$, the block corresponding to $\sigma^x\cdots \sigma^x$ is not itself an MPDO.} It is believed that strong injectivity and positivity imply weak injectivity, but this claim is yet to be proven.\footnote{The claim is stated as a conjecture in ref.~\cite{PRXQuantum.6.010348}. It is proven for \emph{locally purified} MPDOs in ref.~\cite{PhysRevX.15.021060}.} We will assume in the following that at least one of the positive blocks is weakly injective, and this will turn out to imply that many other blocks are weakly injective too. The transfer matrix $\mathbb E_\theta$ of a block injective MPDO tensor decomposes as $\mathbb E_\theta=\oplus_a\mathbb E_{\theta_a}$ and has a space of fixed points spanned by the fixed points $\Xi_a$ of $\mathbb E_{\theta_a}$ of the blocks. 

The expectation value of an operator $\mc O$ supported on a finite interval has the form
\begin{equation}
    \DOvevO{}{}\, .
\end{equation}
In the thermodynamic limit, it involves an infinite number of copies of the transfer matrix. Therefore on a block whose transfer matrix has spectral radius less than $1$, the expectation value vanishes.\footnote{Two examples to have in mind are the following: In the strong-to-weak symmetry breaking state $\mathbb I\cdots\mathbb I+\sigma^x \cdots \sigma^x$, the block corresponding to $\sigma^x \cdots \sigma^x$ has transfer matrix $\mathbb E=0$ and does not contribute. In the pure symmetry breaking state $(|0\cdots 0\rangle+|1\cdots 1\rangle)(\langle 0\cdots0|+\langle 1\cdots 1|)$, the two off-diagonal blocks do not contribute.} Note, however, that these blocks still affect expectation values of non-local operators like symmetry operators.

\subsection{Classification}

We now review the classification of phases of symmetric mixed states, as formulated in ref.~\cite{Qi:2025tal,davydov2009modularinvariantsgrouptheoreticalmodular,Wen:2023otf,PRXQuantum.6.010348,Xue:2024bkt,PhysRevX.15.021060}.\footnote{While this classification has appeared in the literature, its explicit realisation in terms of block-injective MPDOs has not, to our knowledge, been spelled out in the full generality treated here.} Let $H \leq G$ be subgroup of $G$ and $N \lhd H$ a normal subgroup of $H$. Pick a cocycle $\beta\in Z^2(N,\rU(1))$ and a function $\epsilon \colon N \times H \to \rU(1)$ satisfying
\begin{align}
    \label{eq:condensableEpsA}
    \epsilon_{n}(h_1h_2) &= \epsilon_{h_2nh_2^{-1}}(h_1) \, \epsilon_{n}(h_2) \, ,
    \\
    \label{eq:condensableEpsB}
    \epsilon_{n_1}(h) \, \epsilon_{n_2}(h) \,  \beta(hn_1h^{-1},hn_2h^{-1}) &= \epsilon_{n_1n_2}(h) \, \beta(n_1,n_2) \, , 
    \\
    \label{eq:condensableEpsC}
    \frac{\beta(n_1,n_2)}{\beta(n_1n_2n_1^{-1},n_1)} &= \epsilon_{n_2}(n_1) \, ,
\end{align}
for every $h,h_1,h_2 \in H$ and $n,n_1,n_2 \in N$. Each quadruple $(H,N,\beta,\epsilon)$, up to simultaneously shifting $\beta$ by a coboundary ${\rm d}\lambda$ and $\epsilon_n(h)$ by $\lambda(h)/\lambda(nhn^{-1})$, labels a phase of $G$-symmetric mixed states.
Below we explain how to construct a representative of the phase of $G$-symmetric mixed states associated with the quadruple $(H,N,\beta,\epsilon)$. But first, we need to explain the physical meaning of this classification. Throughout, we borrow the notation of sec.~\ref{sec:VecG_classification}.

\bigskip \noindent
Consider a block-injective tensor $\theta=\bigoplus_a\theta_a$ and the corresponding MPDO. Each block labelled by some index $a$ produces an MPDO $\Psi(\theta_a)$ in $\End_\mathbb C (\mc H_\Lambda)$. The block-injective MPDO is said to be $G$-symmetric if the subspace $\mathbb C\{\Psi(\theta_a)\}_a$ is closed under the left action of the symmetry lines $\prod_{i\in\Lambda}\rho(g)_i$, for every $g\in G$ and any system size $|\Lambda|$, and is also closed under the right action of the symmetry lines. Suppose the MPDO is symmetric. Then, regarding the MPDO as an MPS with physical space $\End_\mathbb C(V)$, the arguments from before imply that the MPDO define a $\Vect_{G\times G}$-module category; in particular, the set of labels $a$ that index the injective blocks is a $(G\times G)$-set such that $(g_{\rm l}, g_{\rm r})(a) = g_{\rm l}(g_{\rm r}(a)) = g_{\rm r}(g_{\rm l}(a))$, for every $(g_{\rm l},g_{\rm r}) \in G \times G$. Explicitly, acting with the left symmetry on $\Psi(\theta_a)$ for any $a$ results in an injective MPS proportional to $\Psi(\theta_{g_{\rm l}(a)})$, whereas acting with the right symmetry results in a state proportional to $\Psi(\theta_{g_{\rm r}(a)})$. However, not all $\Vect_{G \times G}$-module categories are compatible with the constraints of self-adjointness and positivity of the MPDO \cite{PRXQuantum.6.010348,Xue:2024bkt}, and it is only after imposing these constraints that the aforementioned classification is obtained.

As in the pure state case, an indecomposable module category has blocks labelled by cosets of the subgroup of block-fixing symmetry lines in the full group of symmetry lines. In the present setting, this means the blocks are labelled by $a\in \{1,\ldots,(G\times G:S)\}$, where $S$ is the subgroup that stabilises blocks. The subgroup $H \leq G$ consists of the lines whose diagonal action fixes all blocks, i.e., lines $\mathbb C_h \in \Vect_G$ for which $h_lh_r(a)=a$ for all $a$. The subgroup $N\lhd H$ consists of lines whose left action fixes all blocks; one can verify that $N$ is normal in $H$ and, since $\Psi(\theta)$ is self-adjoint, is equal to the subgroup of lines whose right action fixes all blocks. Thus, $S$ is the subgroup $\{(nh,h)\in G\times G \, | \, (n,h) \in N \times H\}$. This means that the number of blocks is $(G:H)(G:N)$. We will refer to $H$ and $N$, respectively, as the subgroups of \emph{weakly and strongly preserved lines}.\footnote{In the language typically used to describe mixed state phases, the block-injective MPDOs we consider are \emph{strongly symmetric} with respect to the full group $G$ of symmetries. $H$ is the group of \emph{not-completely-broken symmetries}, while $N$ is the group of \emph{not-at-all-broken symmetries}. The quotient $K=H/N$ consists of \emph{strong-to-weak-broken} symmetries. No generality has been lost in taking all symmetries to be strong: to realize a strong-to-weak broken symmetry as a weak symmetry of the block-injective MPDO, one can remove blocks from the MPDO so that the left action of that symmetry line is no longer a symmetry; for example, the $\mathbb Z_2$ strong-to-weak symmetry breaking state $\mathbb I\cdots\mathbb I+\sigma^x\cdots \sigma^x$ \cite{Ma_2025,PRXQuantum.6.010344} becomes the weakly symmetric state $\mathbb I\cdots\mathbb I$ upon removing the block corresponding to $\sigma^x \cdots \sigma^x$.}

Let us now argue that only $(G:H)$ of these blocks contribute to local observables. It was shown above that a block contributes if and only if its transfer matrix has a spectral radius of $1$. Fix a basepoint block $a_0$, which we assume is weakly injective. Any other block is related to the basepoint by the action of some element $(g_1,g_2) \in G\times G$. Therefore, the transfer matrix $\mathbb E_{\theta_a}$ of a block $a$ is given by $\tr_V[\theta_a]=\tr_V[U_{g_2}^{-1}\theta_{a_0}U_{g_1}]=\tr_V[U_{g_1g_2^{-1}}\theta_{a_0}]$. If $g_1g_2^{-1}\in N$, the transfer matrix of the block $a$ has unique peripheral eigenvalue $1$; in fact, it is related to the transfer matrix of the basepoint $a_0$ by conjugation by $\phi^{r_{a_0}g_1g_2^{-1}r_{a_0}^{-1},a_0}$, defined below in eq. \eqref{eq:mixed_strongAction}. On the other hand, if $g_1g_2^{-1}\notin N$, we believe that the spectral radius is strictly less that $1$; this has been shown when the MPDO represents a pure state \cite{Bridgeman_2017} and the argument straightforwardly extends to large class of MPDOs, namely those with local purifications. If $g_1g_2^{-1}=n\in N$, then $g_2^{-1}=g_1^{-1}n$, and since $n$ fixes the block $a$, this means that the block $a$ is related to the basepoint $a_0$ by the action of $(g_1,g_1)\in G\times G$. Since the intersection of the diagonal subgroup $\Delta(G):=\{(g,g)\in G\times G \, | \, g \in G\}$ with the stabiliser $S$ is $\Delta(H)$, we conclude that the contributing blocks are labelled by cosets in $\Delta(G) /( \Delta(G)\cap S) \cong G/H$. Note that the space of contributing blocks is closed under conjugation. Note also that since the basepoint $a_0$ is weakly injective by assumption and the contributing blocks are related to it by conjugation, they are weakly injective as well. In the following, we restrict to these contributing blocks.

The block-injective built from $\theta$ is weakly $G$-symmetric in the sense that the space spanned by its constitutive injective MPDOs is closed under the action of $\bigotimes_{\msf i \in \Lambda} {\rm Ad}_{\rho(g)_\msf i}$, for every $g \in G$. For every $g \in G$ and $a \in \{1,\ldots,(G:H)\}$, there are four-valent tensors $\eta^{g,a}$ and $\bar \eta^{g,a}$ such that  
\begin{equation}
    \label{eq:mixed_weakLocalAction}
    \actionDOWeak{1}{}{a}{}{}{g}{}{}
    \, = \, 
    \actionDOWeak{2}{}{a}{g(a)}{}{g}{\eta^{g,a}}{\bar \eta^{g,a}}
\end{equation}
and
\begin{equation}
    \label{eq:mixed_weakOrtho}
    \orthoActionWeak{a}{g}{g(a)}{g(a)}{\eta^{g,a}}{\bar \eta^{g,a}}  = \mathbb I_{M_{g(a)}} \, .
\end{equation}
Consider the adjoint actions of two symmetry lines labelled by simple objects $\mathbb C_{g_1}$ and $\mathbb C_{g_2}$ in $\Vect_G$, respectively. As in the pure state case, one can either successively act with both lines, or first locally fuse them before acting with the resulting line. However, in the mixed state case, this does not give rise to a non-trivial complex phase due to the constraints of self-adjointness and positivity on the MPDO \cite{PRXQuantum.6.010348,Xue:2024bkt}.

By construction, for every $a \in \{1,\ldots,(G:H)\}$, the MPDO labelled by $a$ is weakly $r_aHr_a^{-1}$-symmetric in the sense that is invariant under the action of $\bigotimes_{\msf i \in \Lambda} {\rm Ad}_{\rho(r_ahr_a^{-1})_\msf i}$, for every $h \in H$. The MPDO is also strongly $N$-symmetric in the sense that it is an eigenstate of $\bigotimes_{\msf i \in \Lambda} \rho(r_anr_a^{-1})_\msf i$, for every $n \in N$. For every $n \in N$ and $a \in \{1,\ldots,(G:H)\}$, there are three-valent tensors $\phi^{{}^{r_a}n,a}$ and $\bar \phi^{{}^{r_a}n,a}$ such that 
\begin{equation}    
    \label{eq:mixed_strongAction}
    \raisebox{12pt}{\actionDOStrong{1}{}{a}{}{}{{}^{r_a}n}{}{}} 
    \! = \! \raisebox{8pt}{\actionDOStrong{2}{}{a}{a}{}{{}^{r_a}n}{\phi^{r_anr_a^{-1},a}}{\bar \phi^{r_anr_a^{-1},a}}}
\end{equation}
and
\begin{equation}
    \label{eq:mixed_strongOrtho}
    \orthoAction{a}{{}^{r_a}n}{a}{a}{\phi^{r_anr_a^{-1}\!,a}}{\bar \phi^{r_anr_a^{-1}\!,a}}  = \mathbb I_{M_a} \, .
\end{equation}
In analogy with the pure state case, we have
\begin{equation}
     \moduleAssocStrong{2} \hspace{-6pt} = \beta(n_1,n_2) \moduleAssocStrong{1} \, ,
\end{equation}
which implies that in every block, there is an algebra of strongly preserved lines that is Morita equivalent to $\mathbb C[N]^\beta$.

In addition, the strong and the weak symmetry conditions are compatible in the following way: 
\begin{equation}    
    \label{eq:mixed_Assoc}
    \assocDO{1} = \epsilon_{r_anr_a^{-1}}(h_{g,a}) \assocDO{2}
\end{equation}
where we used the fact that ${}^{gr_a}n = {}^{r_{g(a)}h_{g,a}}n$.

These relations imply that the right fixed points $\Xi_{{\rm r},a}$ of the transfer matrix satisfy
\begin{equation}
    \DOFptSym{1} = \DOFptSym{2} = \DOFptSym{3} \, ,
\end{equation}
where in the last step we used eq.~\eqref{eq:mixed_weakLocalAction} together with eq.~\eqref{eq:mixed_weakOrtho}.
Then, since $\Xi_{{\rm r},a}$ is the unique right fixed point of $\mathbb E_{\theta_a}$, it can be concluded that
\begin{equation}
    \label{eq:mixed_actionFixedPoints}
    \DOFptR{2}{a}{g(a)}{g}{\;\;\Xi_{{\rm r},a}}{\eta^{g,a}} \! = \, \DOFptRT{g(a)}{\;\;\;\;\;\;\Xi_{{\rm r},g(a)}} \, .
\end{equation}
Analoguous relations hold for left fixed points.

\subsection{Order parameters \label{sec:Mixed_OP}}

Given a  phase of $G$-symmetric mixed states, we are now ready to identify string operators that may serve as \emph{order parameters}. Consider a representative of the phase associated with the quadruple $(H,N,\beta,\epsilon)$.
Given a simple object $([x_0],\hat W)$ in $\Vect_G^G$, let $\mc S(x,c_1,c_2)_{\msf i,\ldots,\msf j}$ be the string operator constructed in eq.~\eqref{eq:VecG_StringOp}, for $x \in [x_0]$ and $c_1,c_2 \in \{1,\ldots,\dim_\mathbb C \hat W\}$. Let us compute the expectation value of this string operator in state labelled by $a \in \{1,\ldots,(G:H)\}$. By definition, the twisting line $\mathbb C_x$ needs to be part of the algebra of \emph{strongly} preserved lines for the expectation value not to vanish, i.e. it must lie in $r_aNr_a^{-1}$. Therefore, we assume there is a group element $n \in N$ such that $x = {}^{r_a}n$. Employing eq.~\eqref{eq:mixed_strongAction} at every site $\msf k \in [\msf i+1,\msf j-1]$ together with eq.~\eqref{eq:mixed_strongOrtho}, it follows from eq.~\eqref{eq:mixed_transferprojector} that the expectation value evaluates to
\begin{equation}
    \label{eq:mixed_expValFact}
    \MixedStringOpExpValLimL{a}{{}^{r_a}n}{\, \bar \phi^{r_anr_a^{-1}\!,a}}{c_1}{\Xi_{{\rm l},a}\;\;}{\;\;\,\Xi_{{\rm r},a}}    
    \cdot 
    \MixedStringOpExpValLimR{a}{{}^{r_a}n}{\, \phi^{r_anr_a^{-1}\!,a}}{c_2}{\Xi_{{\rm l},a}\;\;}{\;\;\,\Xi_{{\rm r},a}}  \, ,
\end{equation}
up to terms that decay exponentially in the length of the string. As usual, let us focus on the right-hand-side factor. This quantity is the result of contracting a twisted sector local operator of the form \eqref{eq:VecG_twistedLocalOpDual} with the tensor network: 
\begin{equation}
    \label{eq:mixed_twistedLocalOpEnv}
    \MixedTwistedLocalOpEnv{a}{{}^{r_a}n}{\, \phi^{r_anr_a^{-1}\!,a}}{}{\Xi_{{\rm l},a}\;\;}{\;\;\,\Xi_{{\rm r},a}}  \, .
\end{equation}
We claim that tensor networks of the form \eqref{eq:mixed_twistedLocalOpEnv}, for every $n \in N$ and $a \in \{1,\ldots,(G:H)\}$, form a reducible module over $\Tu(\Vect_G)$. Acting with the tube \eqref{eq:VecG_Tube} corresponding to $e_g \otimes \varsigma_y \in \Tu(\Vect_G)$ on the tensor network \eqref{eq:mixed_twistedLocalOpEnv} labelled by $n \in N$ and $a \in \{1,\ldots,(G:H)\}$ results in
\begin{equation}
    \delta_{y,{}^{r_a}n} \cdot 
    \MixedTubeModEnvA{g}{\, {}^{gr_a}n}{g{}^{r_a}n}{{}^{r_a}n}{a}{\phi^{r_anr_a^{-1}\!,a}}{}{\Xi_{{\rm l},a}\;\;}{\;\;\,\Xi_{{\rm r},a}}
    \!\! =
    \delta_{y,{}^{r_a}n}  \cdot
    \MixedTubeModEnvB{g}{\, {}^{gr_a}n}{g{}^{r_a}n}{{}^{r_a}n}{a}{\phi^{r_anr_a^{-1}\!,a}}{}{\Xi_{{\rm l},a}\;\;}{\;\;\,\Xi_{{\rm r},a}}
    \, . 
\end{equation}
One can simplify this tensor network by utilising eq.~\eqref{eq:mixed_weakLocalAction}, before invoking eq.~\eqref{eq:mixed_actionFixedPoints} as well as eq.~\eqref{eq:mixed_Assoc} and eq.~\eqref{eq:mixed_weakOrtho}:
\begin{equation}
    \label{eq:mixed_TubeModEnv}
    \delta_{y,{}^{r_a}n} \cdot
    \MixedTubeModEnvC{g}{\, {}^{gr_a}n}{g{}^{r_a}n}{{}^{r_a}n}{a}{\phi^{r_anr_a^{-1}\!,a}}{}{\Xi_{{\rm l},a}\;\;}{\;\;\;\;\;\,\Xi_{{\rm  r},g(a)}}
    = 
    \delta_{y,{}^{r_a}n} \;
    \epsilon_n(h_{g,a}) \cdot
    \MixedTwistedLocalOpEnv{g(a)}{{}^{gr_a}n}{\, \phi^{gr_an(gr_a)^{-1}\!,g(a)}\;\;\;\;\;\;\;\;}{}{\Xi_{{\rm l},g(a)}\;\;\;\;\;}{\;\;\;\;\;\;\,\Xi_{{\rm r},g(a)}} \, .
\end{equation}
It remains to verify that this action does endow the space of tensor networks \eqref{eq:mixed_twistedLocalOpEnv} with the structure of a module over $\Tu(\Vect_G)$. Firstly, it follows from $gr_a = r_{g(a)}h_{g,a}$, which implies ${}^{gr_a}n = r_{g(a)}(h_{g,a}nh_{g,a}^{-1})r_{g(a)}^{-1}$, that \eqref{eq:mixed_TubeModEnv} is indeed of the form \eqref{eq:mixed_twistedLocalOpEnv}, whereby the label in $\{1,\ldots,(G:H)\}$ is chosen to be $g(a)$, while the group element in $N \lhd H$ is chosen to be $h_{g,a}nh_{g,a}^{-1}$. Together, it confirms that the tensor networks \eqref{eq:mixed_twistedLocalOpEnv} arrange into a module over the tube algebra $\Tu(\Vect_G)$. Let us denote this $\Tu(\Vect_G)$-module by $C(H,N,\beta,\epsilon)$. Invoking the same arguments as in sec.~\ref{sec:VecG_OP} and sec.~\ref{sec:RepG_OP}, one deduces that in order for the string operator $\mc S(x,c_1,c_2)_{\msf i,\ldots,\msf j}$, the simple object $([x_0],\hat W)$ must be a direct summand of the object in $\Vect_G^G$ associated with $C(H,N,\beta,\epsilon)$.

\subsection{Condensable algebras}

Let us shed some light on the $\Tu(\Vect_G)$-module $C(H,N,\beta,\epsilon)$ introduced in the previous section. 
We begin by computing the $G$-equivariant $G$-graded vector space corresponding to $C(H,N,\beta,\epsilon)$ via $\Vect_G^G \simeq \Mod(\Tu(\Vect_G))$. Let $\mathbb C[N]^\beta_\epsilon = \mathbb C\{e_n\}_{n \in N}$ denote the twisted group algebra $\mathbb C[N]^\beta$ that is equipped with the $H$-equivariant structure given by
\begin{equation}
    h \cdot e_{n} = \epsilon_{n}(h) \,   e_{hnh^{-1}} \, , 
\end{equation}
for every $h \in H$ and $n \in N$. We promote the $H$-action of $\mathbb C[N]^\beta_\epsilon$ to a $G$-action via the induction functor:
\begin{equation} 
    \Ind_H^G(\mathbb C[N]^\beta_\epsilon) := \mathbb C\big\{ \varsigma : G \to \mathbb C[N]^\beta_\epsilon \, \big| \, \varsigma(gh^{-1}) = h \cdot \varsigma(g)\big\}_{\forall g \in G, \forall h \in H} \, .
\end{equation}
The $G$-equivariant structure, as usual, is given by $(g \cdot \varsigma)(-) = \varsigma(g^{-1}-)$. Moreover, $\Ind_H^G(\mathbb C[N]^\beta_\epsilon)$ is $G$-graded, whereby every function $\varsigma \in \Ind_H^G(\mathbb C[N]^\beta_\epsilon)$ is homogeneous of degree $x \in G$ if and only if $\varsigma(y) \in \mathbb C[N]^\beta_\epsilon$ is homogeneous of degree $y^{-1}xy \in N$, for every $y \in G$. In particular, we have $\Ind_H^G(\mathbb C[N]^\beta_\epsilon)_{1_G} \cong \mathbb C(G/H)$. Furthermore, if $\varsigma$ is homogeneous of degree $x \in G$, then, by definition, $g \cdot \varsigma$ is homogeneous of degree $gxg^{-1}$. Therefore, $\Ind_H^G(\mathbb C[N]^\beta_\epsilon)$ defines an object in $\Vect_G^G$. 
Mimicking the derivations of sec.~\ref{sec:VecG_RG}, we find that
$\Ind_H^G(\mathbb C[N]^\beta_\epsilon) = \mathbb C\{\varsigma_{a,n}\}_{a=1,\ldots,(G:H),n\in N}$, where the function $\varsigma_{a,n} \colon G \to \mathbb C[N]^\beta_\epsilon$ is defined by
\begin{equation}
    \varsigma_{a,n}(g) =
    \begin{cases}
        \epsilon_n(h^{-1}) \,  e_{h^{-1}n h} \q &\text{if $g \equiv r_ah \in r_aH$} \, ,
        \\
        0 &\text{otherwise} \, ,
    \end{cases}
\end{equation}
for every $g \in G$. By definition, the function $\varsigma_{a,n}$ is homogeneous of degree $|\varsigma_{a,n}| = r_anr_a^{-1} \in G$.
Let us examine the action of $G$ on $\varsigma_{a,n}$, for every $a \in \{1,\ldots,(G:H)\}$ and $n \in N$. By definition, 
$(g \cdot \varsigma_{a,n})(x) = \varsigma_{a,n}(g^{-1}x)$, for every $g,x \in G$, which is non-zero if and only if $x \in g \cdot r_aH = r_{g(a)}H$. 
Moreover, recall that for each $g \in G$ and $r_aH \in G/H$, there is a unique $h_{g,a} \in H$ such that $gr_a = r_{g(a)}h_{g,a}$. Suppose that $x \equiv r_{g(a)}h \in r_{g(a)}H$, we find
\begin{equation}
\begin{split}
    (g \cdot \varsigma_{a,n})(x)
    &= \varsigma_{a,n}(r_a h_{g,a}^{-1}r_{g(a)}^{-1}r_{g(a)}h)
    \\[-.5em]
    &= (h^{-1}h_{g,a}) \cdot  \varsigma_{a,n}(r_a)
    = 
    \epsilon_n(h^{-1}h_{g,a}) \, e_{h^{-1}h_{g,a}nh_{g,a}^{-1}h} \, ,
\end{split}
\end{equation}
where we used in the last step the fact that $\varsigma_{a,n}(r_a) = e_n$.
Invoking condition \eqref{eq:condensableEpsA} finally implies that
\begin{equation}
    g \cdot \varsigma_{a,n}
    = \epsilon_n(h_{g,a}) \, \varsigma_{g(a), h_{g,a}nh_{g,a}^{-1}} \, .
\end{equation}
At this point, one can explicitly verify that $|g \cdot \varsigma_{a,n}| = g |\varsigma_{a,n}|g^{-1}$. The corresponding $\Tu(\Vect_G)$-module is simply given by $\mathbb C\{\varsigma_{a,n}\}_{a,n}$ with
\begin{equation}
    (e_g \otimes \varsigma_y) \cdot \varsigma_{a,n} := \delta_{y,{}^{r_a}n} \, \epsilon_n(h_{g,a}) \, \varsigma_{g(a), h_{g,a}nh_{g,a}^{-1}} \, ,
\end{equation}
for every $g,y \in G$, $a \in \{1,\ldots,(G:H)\}$ and $n \in N$. This is precisely the $\Tu(\Vect_G)$-module $C(H,N,\beta,\epsilon)$ we found in sec.~\ref{sec:Mixed_OP}. Concretely, the basis vector $\varsigma_{a,n}$ is identified with the tensor network \eqref{eq:mixed_twistedLocalOpEnv} associated with the ground state labelled by $a$ and twisting line $\mathbb C_{|\varsigma_{a,n}|} = \mathbb C_{r_anr_a^{-1}} \in \Vect_G$. Mimicking the derivations of sec.~\ref{sec:VecG_Lagrangian}, one can derive an explicit formula providing the multiplicity of any simple object $([x_0],\hat W) \in \Vect_G^G$ in $C(H,N,\beta,\epsilon)$ \cite{DAVYDOV2017149}.

By induction, the object $C(H,N,\beta,\epsilon) \in \Vect_G^G$ is equipped with the structure of an algebra in $\Vect_G^G$, i.e., a $G$-equivariant $G$-graded algebra. The multiplication rule, which is provided by the pointwise multiplication of functions $G \to \mathbb C[N]^\beta$, reads
\begin{equation}
    \varsigma_{a_1,n_1} \cdot \varsigma_{a_2,n_2} = \delta_{a_1,a_2} \, \beta(n_1,n_2) \, \varsigma_{a_1,  n_1n_2} \, ,
\end{equation}
for every $a_1,a_2 \in \{1,\ldots,(G:H)\}$ and $n_1,n_2 \in N$, where we used eq.~\eqref{eq:condensableEpsB}. One can further check that $C(H,N,\beta,\epsilon)$ is a \emph{condensable algeba} in $\Vect_G^G$,
i.e., a connected, indecomposable, separable and commutative algebra (see app.~\ref{app:algebras}). The notion that condensable algebras should describe order parameters for mixed state gapped phases already appeared in ref.~\cite{Qi:2025tal,Schafer-Nameki:2025fiy}. 

Finally, let us remark that Lagrangian algebras are special cases of  condensable algebras, namely those whose square of the Frobenius--Perron dimension equals the Frobenius--Perron dimension of the braided fusion category they belong to (see app.~\ref{app:algebras}).
In particular, whenever $H=N$ and $\epsilon = \beta$, it follows from eq.~\eqref{eq:condensableEpsC} that $C(H,H,\beta,\beta) \cong L(H,\beta)$. This is in agreement with the notion that if the algebras of weakly preserved and strongly preserved lines are identical, then it reduces to the pure state case.

\newpage
\titleformat{name=\section}[display]
{\normalfont}
{\footnotesize\centering {APPENDIX \thesection}}
{0pt}
{\large\bfseries\centering}
\appendix
\section{Algebras\label{app:algebras}}

\emph{In this appendix, we collect the definitions of category theoretic concepts that are used throughout this manuscript, together with basic results. For additional details, we encourage the reader to consult ref.~\cite{etingof2016tensor}, on which we rely extensively.}

\bigskip \noindent
We assume the reader is familiar with the notion of braided fusion category; otherwise, we recommend consulting ref.~\cite{etingof2010fusion,Drinfeld2010}. Nonetheless, for convenience, we recall below the list of defining properties of a fusion category over the field of complex numbers:
\begin{definition}[Fusion category]
    A fusion category is a finite semisimple $\mathbb C$-linear abelian rigid monoidal category with finitely many isomorphism classes, and such that the space of endomorphisms of the unit is isomorphic to $\mathbb C$.
\end{definition}
\noindent
Throughout, all the categories are assumed to be finite, semisimple and $\mathbb C$-linear. Given a fusion category $\mc C$, the set of representatives of its finitely many isomorphism classes of simple objects is denoted by $\Irr(\mc C)$. The monoidal structure is designated by $(\otimes, 1_\mc C, \alpha)$, where $\otimes : \mc C \times \mc C \to \mc C$ is the tensor product bifunctor, $1_\mc C$ the unit object, and $\alpha : (- \otimes -) \otimes - \xrightarrow{\sim} - \otimes (- \otimes -)$ the monoidal associator.\footnote{Without loss of generality, we always assume the unit object to be strict.} Components of the monoidal associator are written as $\alpha_{X_1,X_2,X_3} : (X_1 \otimes X_2) \otimes X_3 \xrightarrow{\sim} X_1 \otimes (X_2 \otimes X_3)$. For an object $X \in \mc C$, let $X^*$ denotes its dual, and write the corresponding evaluation and coevaluation morphisms as ${\rm ev}_X \colon X^* \otimes X \to 1_\mc C$ and ${\rm coev}_X \colon 1_\mc C \to X^* \otimes X$, respectively. Finally, $\mc C^{\rm op}$ denotes the fusion category opposite to $\mc C$ with monoidal structure $(\otimes^{\rm op},1_\mc C,\alpha^{\rm op})$ such that $X_1 \otimes^{\rm op} X_2 := X_2 \otimes X_1$ and $\alpha^{\rm op}_{X_1,X_2,X_3} := \alpha^{-1}_{X_3,X_2,X_1}$, for every $X_1,X_2,X_3 \in \mc C$.

First, one requires a notion of algebra in a fusion category:

\begin{definition}[Algebra object] An algebra in a fusion category $\mc C \equiv (\mc C, \otimes, 1_\mc C, \alpha)$ is a triple $(A, \mu, \eta)$ consisting of an object $A$ in $\mc C$ and morphisms $\mu: A \otimes A \to A$ and $\eta: 1_\mc C \to A$ in $\mc C$, which are required to satisfy associativity and unitality conditions involving the monoidal structure of $\mc C$. The morphisms $\mu$ and $\eta$ are referred to as the `multiplication' and the `unit' of $A$, respectively.
\end{definition}

\noindent
One recovers the ordinary notion of finite-dimensional unital associative algebra over the field $\mathbb C$ by considering algebras in the fusion category $\Vect$ of complex vector spaces.
For any fusion category $\mc C$, the unit trivially carries the structure of an algebra object. More generally, for any $X \in \mc C$, the object $A = X \otimes X^*$ carries the structure of an algebra, with multiplication ${\rm id_X} \otimes {\rm ev}_X \otimes {\rm id}_{X^*}$ and unit ${\rm coev}_X$. 

The same way one considers modules over ordinary algebras, there is a notion of module over an algebra in a fusion category:

\begin{definition}[Module object] A right module over an algebra $A \equiv (A,\mu,\eta)$ (or right $A$-module) in a fusion category $(\mc C, \otimes , 1_\mc C, \alpha)$ is pair $(M,p)$ consisting of an object $M$ in $\mc C$ and an `action' morphism $p : M \otimes A \to M$ in $\mc C$, which is required to satisfy a pentagon axiom involving the monoidal structure of $\mc C$ and the algebraic structure of $A$, as well as a unitality condition.
\end{definition}
\noindent
For instance, consider the algebra $A = X \otimes X^*$ associated with any object $X \in \mc C$; for every $Y \in \mc C$, the object $M = Y \otimes X^*$ has the structure of a right module over $A$ with ${\rm id}_Y \otimes {\rm ev}_X \otimes {\rm id}_{X^{\! *}} : M \otimes A \to M$. In general, given two modules $(M_1,p_1)$ and $(M_1,p_2)$ over an algebra $A$ in a fusion category $\mc C$, a module homomorphism between them is a morphism $M_1 \to M_2$ between the
corresponding objects that is compatible with the action morphisms $p_1$ and $p_2$. Module homomorphisms between $(M_1,p_1)$ and $(M_2,p_2)$ form a subspace of $\Hom_\mc C(M_1,M_2)$ that is stable under composition. It follows that, given an algebra $A$ in $\mc C$, module objects over it form a finite category denoted by $\Mod_\mc C(A)$.\footnote{Whenever $\mc C = \Vect$, we simply write $\Mod(A) \equiv \Mod_{\Vect}(A)$.}

Left modules over an algebra object $A$ in $\mc C$ (or left $A$-modules) are defined similarly. An object $M$ in a fusion category $\mc C$ is said to be a bimodule over an algebra object $A$ (or $(A,A)$-bimodule) if it is equipped with both a left $A$-module structure and a right $A$-module structure such that the action morphisms are compatible. Given two $(A,A)$-bimodules in $\mc C$, one defines a bimodule homomorphism between them as a morphism in $\mc C$ that is simultaneously a homomorphism of left $A$-modules and of right $A$-modules. We are now ready to introduce an important family of algebra objects: 
\begin{definition}[Separable algebra] 
    An algebra $(A,\mu,\eta)$ in a fusion category $\mc C$ is said to be `separable' if the multiplication $\mu$ admits a `section' $\Delta : A \to A \otimes A$ such $\mu \circ \Delta = {\rm id}_A$ as a homomorphism of $(A,A)$-bimodules.  
\end{definition}
\noindent
The finite category $\Mod_\mc C(A)$ of modules over any \emph{separable} algebra $A$ in a fusion category $\mc C$ can be shown to be semisimple. Moreover, $\Mod_\mc C(A)$ turns out to have the structure of a (finite semisimple) \emph{module category} over $\mc C$:

\begin{definition}[Module category]
    A (left) module category over a fusion category $\mc C$ (or left $\mc C$-module category) is a triple $(\mc M,\act,\alpha^{\act})$ consisting of a (finite semisimple $\mathbb C$-linear) category $\mc M$, an `action' bifunctor $\act : \mc C \times \mc M \to \mc M$, and a natural isomorphism $\alpha^{\act} : (- \otimes - )\act - \xrightarrow{\sim} - \act (- \act - )$, which is required to satisfy a pentagon axiom involving the monoidal structure of $\mc C$.\footnote{We always assume the action of the unit object $1_\mc C \in \mc C$ to be strict so that $1_\mc C \act M = M$, for every $M \in \mc M$.} The natural isomorphism $\alpha^{\act}$ is referred to as the `module associator' of $\mc M$.
\end{definition}
\noindent
Right module categories are defined similarly.
First of all, every fusion category $\mc C$ is a module category over itself---referred to as the `regular' module category---such that the action bifunctor is given  by the tensor product bifunctor. A $\mc C$-module category is said to be `indecomposable' whenever it cannot be written as a direct of $\mc C$-module categories. 
As alluded to above, the category $\Mod_\mc C(A)$ or right modules over any algebra object $A$ in a fusion category $\mc C$ carries the structure of a left $\mc C$-module category. The action bifunctor $\act : \mc C \times \Mod_\mc C(A) \to \Mod_\mc C(A)$ is simply given by the monoidal structure of $\mc C$. Indeed, for any $X \in \mc C$ and $(M,p) \in \Mod_\mc C(A)$, the object $X \act M := X \otimes M$ together with the action morphism
\begin{equation}
    (X \otimes M) \otimes A \xrightarrow{\alpha_{X,M,A}} X \otimes (M \otimes A) \xrightarrow{{\rm id}_X \otimes p} X \otimes M
\end{equation}
defines another object in $\Mod_\mc C(A)$. Components of the module associator $\alpha^{\act}$ are then given by $\alpha^{\act}_{X_1,X_2,M} = \alpha_{X_1,X_2,M}$, for every $X_1,X_2 \in \mc C$ and $M \in \mc M$, and can be shown to be isomorphisms of $A$-modules, which completes the argument. For instance, the category of modules over the trivial algebra object in any fusion category $\mc C$ is equivalent to the regular module category, i.e., $\Mod_\mc C(1_\mc C) \simeq \mc C$.

The above construction can be used to introduce a notion of Morita equivalence between algebras objects:
\begin{definition}[Algebraic Morita equivalence]
    Two algebras $(A_1,\mu_1,\eta_1)$ and $(A_2,\mu_2,\eta_2)$ in a fusion category $\mc C$ are said to be `Morita equivalent' if their categories of modules $\Mod_\mc C(A_1)$ and $\Mod_\mc C(A_2)$ are equivalent as $\mc C$-module categories.
\end{definition}
\noindent
For instance, one can easily verify that the algebra $A = X \otimes X^*$ associated with any object $X \in \mc C$ is Morita equivalent to the trivial algebra associated with the unit object. 

We already know that considering categories of modules over algebras in a fusion category $\mc C$ is a general way of constructing module categories over $\mc C$.
As a matter of fact, every \emph{finite} module category over a fusion category $\mc C$ is equivalent to $\Mod_\mc C(A)$ for a suitable algebra object $A$ \cite{Ostrik:2001xnt}. Given a module category over $\mc C$, finding such an algebra requires the following concept:
\begin{definition}[Internal hom]
    Let $\mc M \equiv (\mc M, \act ,\alpha^{\act})$ be a module category over a  fusion category $\mc C$. For every objects $M_1,M_2 \in \mc M$, there exist an object $\underline{\Hom}(M_1,M_2) \in \mc C$, referred to as the `internal hom' from $M_1$ to $M_2$, and an isomorphism $\underline{\alpha}_{X,M_1,M_2}: \Hom_\mc M(X \act M_1,M_2) \xrightarrow{\sim} \Hom_\mc C(X,\underline{\Hom}(M_1,M_2))$ that is natural in $X$, for every $X \in \Irr(\mc C)$.
\end{definition}
\noindent
Given two objects $M_1$ and $M_2$ in a $\mc C$-module category $\mc M$, consider the isomorphism $\underline{\alpha}_{\underline{\Hom}(M_1,M_2),M_1,M_2}$, and define the `internal evaluation' morphism ${\rm ev}_{M_1,M_2}$ as the preimage of ${\rm id}_{\underline{\Hom}(M_1,M_2)}$, i.e.,
\begin{equation}
    {\rm ev}_{M_1,M_2} := \underline{\alpha}_{\underline{\Hom}(M_1,M_2),M_1,M_2}^{-1}({\rm id}_{\underline{\Hom}(M_1,M_2)}) : \underline{\Hom}(M_1,M_2) \act M_1 \to M_2 \, .
\end{equation}
Given three objects $M_1,M_2,M_3 \in \mc M$, one can then use these internal evaluation morphisms to construct a canonical map of the form
\begin{equation}
    \circ_{M_1,M_2,M_3} : \underline{\Hom}(M_2,M_3) \otimes \underline{\Hom}(M_1,M_2) \to \underline{\Hom}(M_1,M_3) 
\end{equation}
as the image under $\underline{\alpha}_{\underline{\Hom}(M_2,M_3) \otimes \underline{\Hom}(M_1,M_2),M_1,M_3}$ of the composition
\begin{equation}
\begin{aligned}
    &\big(\underline{\Hom}(M_2,M_3) \otimes \underline{\Hom}(M_1,M_2)\big) \act M_1
    \\
    & \q \xrightarrow{\alpha^{\act}_{\underline{\Hom}(M_2,M_3),\underline{\Hom}(M_1,M_2),M_1}}
    \underline{\rm Hom}(M_2,M_3) \act \big(\underline{\rm Hom}(M_1,M_2) \act M_1\big)
    \\
    & \q 
    \xrightarrow{{\rm id}_{\underline{\Hom}(M_2,M_3)} \act {\rm ev}_{M_1,M_2}}
    \underline{\Hom}(M_2,M_3) \act M_2
    \xrightarrow{{\rm ev}_{M_2,M_3}}
    M_3 \, .
\end{aligned}
\end{equation}

\noindent
In particular, the map $\circ_{M_1,M_1,M_2}$ endows any internal hom $\underline{\Hom}(M_1,M_1)$ with the structure of an algebra in $\mc C$, while the map $\circ_{M_1,M_1,M_2}$ endows any internal hom $\underline{\Hom}(M_1,M_2)$ with the structure of a right $\underline{\Hom}(M_1,M_1)$-module in $\mc C$. Given a finite $\mc C$-module category $\mc M$, one can then demonstrate that for every $M \in \mc M$, the category $\Mod_\mc C(\underline{\Hom}(M,M))$ is equivalent to $\mc M$, as a $\mc C$-module category \cite{etingof2016tensor}. For instance, consider the regular $\mc C$-module category; for every $X \in \mc C$, we have $\underline{\Hom}(X,X) \cong X \otimes X^*$, as expected.

One requires a notion of functors between module categories:
\begin{definition}[Module functors]
    Let $(\mc M,\act,\alpha^{\act})$ and $(\mc N, \act', \alpha^{\act'})$ be two module categories over a fusion category $\mc C$. A $\mc C$-module functor from $\mc M$ to $\mc N$ is a pair $(F,\omega)$ consisting of a functor $F : \mc M \to \mc N$ and a natural isomorphism $\omega : F(- \act -) \xrightarrow{\sim} - \act' F(-)$, which is required to satisfy a pentagon axiom involving the module structures of both $\mc M$ and $\mc N$.
\end{definition}
\noindent
Similarly, one defines a notion of module natural transformation between module functors, which form the morphisms of the category $\Fun_\mc C(\mc M,\mc N)$ of $\mc C$-module functors between two $\mc C$-module categories $\mc M$ and $\mc N$. Given a fusion category $\mc C$, $\mc C$-module categories, $\mc C$-module functors and $\mc C$-module natural transformations are organised into a 2-category $\Mod(\mc C)$.

Crucially, given an indecomposable (finite semisimple) $\mc C$-module category $\mc M$, the category $\mc C^*_\mc M := \Fun_\mc C(\mc M,\mc M)$ of $\mc C$-module endofunctors of $\mc M$ carries the structure of a fusion category, where the monoidal structure is provided by the composition of $\mc C$-module functors. We refer to $\mc C^*_\mc M$ as the `Morita dual' of $\mc C$ with respect to $\mc M$:

\begin{definition}[Categorical Morita equivalence]
    Two fusion categories $\mc C$ and $\mc D$ are said to be `Morita equivalent' if there is an indecomposable (finite semisimple) $\mc C$-module category $\mc M$ such that ${\mc C}^*_{\mc M} \simeq \mc D^{\rm \, op}$.
\end{definition}
\noindent
This notion of categorical Morita equivalence does categorify the previous notion of algebraic Morita equivalence. In particular, module categories over a fusion category $\mc C$ are in one-to-one correspondence with module categories over its Morita dual $\mc C^*_\mc M$ with respect to any (indecomposable finite semisimle) $\mc C$-module category $\mc M$. Specifically, given a $\mc C$-module category $\mc N$, $\Fun_\mc C(\mc M,\mc N)$ carries  the structure of a module category over $(\mc C^*_\mc M)^{\rm op}$. As a matter of fact, the 2-functor $\Fun_\mc C(\mc M,-) : \Mod(\mc C) \to \Mod((\mc C^*_\mc M)^{\rm op})$ is a 2-equivalence. This is a 2-categorical invariant of Morita equivalence. Another invariant of Morita equivalence is provided by the Drinfel'd centre construction \cite{Majid1991,JOYAL199143}:
\begin{definition}[Drinfel'd centre]
    The `Drinfel'd centre' $\mc Z(\mc C)$ of a fusion category $\mc C$ is the monoidal category whose objects are pairs $(Z,R_{-,Z})$ consisting of an object $Z \in \mc C$ and a `half-braiding' natural isomorphism $R_{-,Z} : - \otimes Z \xrightarrow{\sim} Z \otimes -$, which is required to satisfy a hexagon axiom involving the monoidal structure of $\mc C$. A morphism in $\mc Z(\mc C)$ between two objects $(Z_1,R_{-,Z_1})$ and $(Z_2,R_{-,Z_2})$ is a morphism $f \in \Hom_\mc C(Z_1,Z_2)$ such that $(f \otimes {\rm id}_X) \circ R_{X,Z_1} = R_{X,Z_2} \circ ({\rm id}_X \otimes f)$. The tensor product of these two objects in given by $(Z_1 \otimes Z_2,R_{-,Z_1 \otimes Z_2})$, where $R_{-,Z_1 \otimes Z_2}$ is defined via a hexagonal commutative diagram involving the half-braiding isomorphisms as well as the monoidal structure of $\mc C$.
\end{definition}
\noindent 
In addition to being a monoidal category, the Drinfel'd centre $\mc Z(\mc C)$ of a fusion category $\mc C$ can be verified to carry a fusion structure. Moreover, it is braided, where the braiding natural isomorphism is constructed from the individual half-braiding natural isomorphisms. Furthermore, it is `non-degenerate', in the sense that the full subcategory of objects $(Z,R_{-,Z}) \in \mc Z(\mc C)$ satisfying $R_{X,Z} \circ R_{Z,X} = {\rm id}_{X \otimes Z}$, for every $X \in \mc C$, is equivalent to $\Vect$ \cite{MUGER2003159,Drinfeld2010}. 
For every indecomposable module category $\mc M$ over a fusion category $\mc C$, we can then show that $\mc Z(\mc C) \simeq \mc Z(\mc C^*_\mc M)$, as non-degenerate braided fusion categories \cite{MUGER200381}. The proof consists in first establishing that $\mc Z(\mc C)$ is equivalent to $(\mc C \boxtimes \mc C^*_\mc M)^*_\mc M$, which is symmetric in $\mc C$ and $\mc C^*_\mc M$.

Whenever dealing with a braided fusion category $\mc B$ with braiding $R$, one can require an algebra object $(A,\mu,\eta)$ in $\mc B$ to be `commutative', i.e., $\mu = \mu \circ R_{A,A}$, and subsequently `\'etale':
\begin{definition}[\'Etale algebra] 
    An algebra $A$ in a braided fusion category $\mc B$ is said to be `\'etale' if it is indecomposable, separable and commutative. 
\end{definition}
\noindent
In general, an algebra object is said to be connected whenever the hom-space between the underlying object and the unit object is isomorphic to $\mathbb C$. In the physics literature, connected \'etale algebras are often referred to as `condensable' algebras. We are now ready to introduce  the following family of algebra objects in a braided fusion category:
\begin{definition}[Lagrangian algebra]
    A `Lagrangian' algebra in a braided fusion category $\mc B$ is a connected \'etale algebra $L$ such that ${\rm FPdim}(L)^2 = {\rm FPdim}(\mc B)$.
\end{definition}
\noindent
In the main text, we always consider Lagrangian algebras in the Drinfel'd centre of a fusion category $\mc C$, where ${\rm FPdim}(\mc Z(\mc C)) = {\rm FPdim}(\mc C)^2$. In this case, it is always possible to construct a Lagrangian algebra $L$ in $\mc Z(\mc C)$ from the data of an indecomposable separable algebra in $\mc C$ via the `full centre' construction \cite{DAVYDOV2010319}:
\begin{definition}[Full centre of an algebra]
    Let $(A,\mu,\eta)$ be an algebra in a fusion category $\mc C$. Consider pairs $((Z,R_{-,Z}),\zeta)$ consisting of objects $(Z,R_{-,Z})$ in $\mc Z(\mc C)$ and morphisms $\zeta : Z \to A$ in $\mc C$ such that the following diagram commutes:
    \begin{align}
    \begin{tikzcd}[ampersand replacement=\&, column sep=3em, row sep=2em]
    |[alias=A1]| A \otimes Z
    \& {} \&
    |[alias=A3]| Z \otimes A
    \\
    |[alias=B1]| A \otimes A
    \&
    |[alias=B2]| A
    \&
    |[alias=B3]| A \otimes A
    \arrow[from=A1,to=A3,"R_{A,Z}"]  
    \arrow[from=A1,to=B1,"{\rm id}_A \otimes \zeta"'] 
    \arrow[from=A3,to=B3,"\zeta \otimes {\rm id}_A"] 
    \arrow[from=B1,to=B2,"\mu"'] 
    \arrow[from=B3,to=B2,"\mu"] 
    \end{tikzcd} \, .
    \end{align}
    The `full centre' $\mc Z(A)$ of $A$ is an object in $\mc Z(\mc C)$ equipped with a morphism $\mc Z(A) \to A$ such that for every such pair $((Z,R_{-,Z}),\zeta)$ there is a unique morphism $Z \to \mc Z(A)$ in $\mc Z(\mc C)$ making the diagram
    \begin{align}
        \begin{tikzcd}[ampersand replacement=\&, column sep=2.5em, row sep=2em]
            |[alias=A1]| Z \& {} \& |[alias=A2]| \mc Z(A)
            \\
            {} \& |[alias=B]| A \& {}
            \arrow[from=A1,to=A2,""]
            \arrow[from=A1,to=B,"\zeta"']
            \arrow[from=A2,to=B,""]
        \end{tikzcd}
    \end{align}
    commutes.
\end{definition}

\noindent
As alluded to above, the full centre $\mc Z(A)$ of an indecomposable separable algebra $A$ in a fusion category $\mc C$ can be shown to be a Lagrangian algebra in $\mc Z(\mc C)$ \cite{DAVYDOV2017149}. In a similar vein, one can define the centre of a module category over a fusion category $\mc C$:
\begin{definition}[Full centre of a module category]
    Let $(\mc M,\act, \alpha^{\act})$ be a module category over a fusion category $\mc C$. Consider pairs $((Z,R_{-,Z}),\zeta)$ consisting of objects $(Z,R_{-,Z})$ in $\mc Z(\mc C)$ and collections $\zeta$ of morphisms $\zeta_M : Z \act M \to M$ in $\mc M$ that are natural in $M \in \mc M$ such that the following diagram commutes for every $M \in \mc M$ and $X \in \mc C$: 
    \begin{align}
    \begin{tikzcd}[ampersand replacement=\&, column sep=3em, row sep=2em]
    |[alias=A1]| (X \otimes Z) \act M
    \& {} \&
    |[alias=A3]| (Z \otimes X) \act M
    \\
    |[alias=B1]| X \act (Z \act M)
    \&
    |[alias=B2]| X \act M
    \&
    |[alias=B3]| Z \act (X \act M)
    \arrow[from=A1,to=A3,"R_{X,Z} \act {\rm id}_M"]  
    \arrow[from=A1,to=B1,"\alpha^{\act}_{X,Z,M}"'] 
    \arrow[from=A3,to=B3,"\alpha^{\act}_{Z,X,M}"] 
    \arrow[from=B1,to=B2,"{\rm id}_X \act \zeta_M"'] 
    \arrow[from=B3,to=B2,"\zeta_{X \act M}"] 
    \end{tikzcd} \, .
    \end{align}
    The `full centre' $\mc Z(\mc M)$ of $\mc M$ is an object in $\mc Z(\mc C)$ equipped with a collection of morphisms $\mc Z(\mc M) \act M \to M$ such that for every such pair $((Z,R_{-,Z}),\zeta)$ there is a unique morphism $Z \to \mc Z(\mc M)$ in $\mc Z(\mc C)$ making the diagram
    \begin{align}
        \begin{tikzcd}[ampersand replacement=\&, column sep=2.5em, row sep=2em]
            |[alias=A1]| Z \act M \& {} \& |[alias=A2]| \mc Z(\mc M) \act M
            \\
            {} \& |[alias=B]| M \& {}
            \arrow[from=A1,to=A2,""]
            \arrow[from=A1,to=B,"\zeta_M"']
            \arrow[from=A2,to=B,""]
        \end{tikzcd}
    \end{align}
    commutes, for every $M \in \mc M$.
\end{definition}

\noindent
For any algebra $A$ in a fusion category $\mc C$, one can then verify that the centre $\mc Z(\Mod_\mc C(A))$ of the $\mc C$-module category $\Mod_\mc C(A)$ coincides with the full centre $\mc Z(A)$ of $A$ \cite{DAVYDOV2010319}. The $\mc C$-module categories of modules over Morita equivalent algebras being equivalent, this makes the full centre construction an invariant of Morita equivalence. Therefore, Morita equivalent indecomposable separable algebras yield equivalent Lagrangian algebras.

\section{Drinfel'd centres and tube algebras\label{app:centres}}

\emph{Numerous results in the main text exploits the equivalences between the category of $G$-equivariant $G$-graded vector spaces, the Drinfel'd centres of the category of $G$-graded vectors spaces and the category of $G$-representations, as well as the category of modules over the tube algebras of the category of $G$-graded vector spaces and the category of $G$-representations. Here, we briefly sketch these equivalences before deriving results that are used in the main text.}

\bigskip \noindent
Before proceeding with our examples, let us briefly sketch the general case. Let $\mc C$ be a (unitary) fusion category. Consider the vector space 
\begin{equation}
    \Tu(\mc C) \equiv \!\!\! \bigoplus_{X_1,X_2 \in \Irr(\mc C)} \!\!\! \Tu(\mc C)^{X_1}_{X_2} \, , \q
    \Tu(\mc C)^{X_1}_{X_2} := \bigoplus_{Y \in \Irr(\mc C)} \Hom_\mc C(X_2 \otimes Y, Y \otimes X_1) \, .
\end{equation}
For every $X_1,X_2 \in \Irr(\mc C)$, one thinks about the vector space $\Tu(\mc C)^{X_1}_{X_2}$ as the space of formal $\mathbb C$-linear combinations of equivalence classes of \emph{string diagrams} on the cylinder for incoming simple boundary condition $X_2$ and outgoing simple boundary condition $X_1$. This vector space has the structure of a $*$-algebra with the multiplication of two elements defined by `gluing' the corresponding equivalence classes of string diagrams and the $*$-operation obtained by flipping the string diagrams upside down, inverting the orientation of all the strings, and taking the inverse of all the morphisms. We refer to this $*$-algebra as the \emph{tube algebra} $\Tu(\mc C)$ of $\mc C$. More concretely, given $Y_1,Y_2,X_1,X_2,X_3,X_4 \in \Irr(\mc C)$, let $\varphi^{X2Y_2}_{Y_2X_1} \in \Tu(\mc C)^{X_1}_{X_2}$ and $\varphi^{X_4Y_1}_{Y_1X_3} \in \Tu(\mc C)^{X_3}_{X_4}$, which we regard as elements of $\Tu(\mc C)$. The multiplication rule reads
\begin{align}
    \nn
    \varphi^{X_4Y_1}_{Y_1X_3} \cdot \varphi^{X_2 Y_2}_{Y_2X_1} 
    &:= \delta_{X_2,X_3} \!\!\! \sum_{\substack{Y_3 \in Y_1 \otimes Y_2 \\ 1 \leq i \leq N^{Y_1Y_2}_{Y_3}}} ({\rm id}_{X_4} \otimes \bar \varphi^{Y_1Y_2}_{Y_3,i}) \circ (\varphi^{X_4Y_1}_{Y_1X_2} \otimes {\rm id}_{Y_2}) \circ ({\rm id}_{Y_1} \otimes \varphi^{X_2Y_2}_{Y_1X_1}) \circ (\varphi^{Y_1Y_2}_{Y_3,i} \otimes {\rm id_{X_1}})
    \\[-.2em]
    &\in \Tu(\mc C)^{X_1}_{X_4} \, . 
\end{align}
In order to extract the structure constants of this algebra, one typically works in the basis provided by the decomposition
\begin{equation}
    \Tu(\mc C)^{X_1}_{X_3} = \bigoplus_{Y,X_2  \in \Irr(\mc C)} \Hom_\mc C (X_3 \otimes Y,X_2) \otimes \Hom_\mc C(X_2,Y \otimes X_1) \, .
\end{equation}
At this point, one recovers the framework of the main text and the structure constants can be expressed in terms of the $F$-symbols of the monoidal structure (see e.g. eq.~\eqref{eq:RepG_TubeCsts}). Similarly, one can conveniently expand the result of the $*$-operation mentioned above in this basis.
By virtue of being a finite-dimensional $*$-algebra, $\Tu(\mc C)$ is semisimple, and so is its category $\Mod(\Tu(\mc C))$ of modules. It is well known that $\Mod(\Tu(\mc C)) \simeq \mc Z(\mc C)$ \cite{ocneanu1994chirality,ocneanu2001operator,Izumi2000,MUGER2003159,Neshveyev_Yamashita_2018}. In particular, given an object $(Z,R_{-,Z}) \in \mc Z(\mc C)$, the vector space underlying the corresponding module in $\Mod(\Tu(\mc C))$ is given by $\bigoplus_{X \in \Irr(\mc C)}\Hom_\mc C(X,Z)$ and the action of the tube element $\varphi^{X_2Y}_{YX_1} \in \Tu(\mc C)^{X_1}_{X_2}$ on the module element $\varphi^{X}_Z \in \Hom_\mc C(X,Z)$ is given by
\begin{equation}
\begin{split}
    \varphi^{X_2Y}_{YX_1} \cdot \varphi^X_Z 
    :=
    \delta_{X_1,X} \, &({\rm id_Z} \otimes {\rm ev}^*_Y) \circ (R_{Y,Z} \otimes {\rm id}_{Y^*}) \circ ({\rm id}_Y \otimes \varphi^{X}_Z \otimes {\rm id}_{Y^*}) 
    \\
    &\circ (\varphi^{X_2Y}_{YX} \otimes {\rm id}_{Y^*}) \circ ({\rm id}_{X_2} \otimes {\rm coev}^*_{Y}) \in \Hom_\mc C(X_2,Z) \, .
\end{split}
\end{equation}
This action is best expressed graphically in the same vein as eq.~\eqref{eq:RepG_twistedLocalOpTube}.

\subsection{Drinfel'd centre of the category of group graded vector spaces}\label{app:centres_vecggeqs}

Consider the fusion category $\Vect_G$ of $G$-graded vector spaces. The Drinfel'd centre $\mc Z(\Vect_G)$ of $\Vect_G$ is equivalent, as a braided fusion category, to the category $\Vect_G^G$ of $G$-equivariant $G$-graded vector spaces. Recall that an object in $\Vect_G^G$ consists of a $G$-graded vector spaces $W = \bigoplus_{x \in G}W_x$ together with a compatible $G$-equivariant structure such that $g \cdot W_x \subseteq W_{gxg^{-1}}$, for every $g ,x \in G$. Let $(W,R_{-,W})$ be an object in $\mc Z(\Vect_G)$; it consists of a $G$-graded vector space $W = \bigoplus_{x \in G}W_x$ and a collection of $G$-grading preserving isomorphisms $R_{K,W} \colon K \otimes W \xrightarrow{\sim} W \otimes K$ that are natural in $K$, for every $K \in \Vect_G$. In particular, for every $g \in G$, we have $R_{\mathbb C_g,W} \colon \mathbb C_g \otimes W \xrightarrow{\sim} W \otimes \mathbb C_g$. 
It follows from  $\mathbb C_g \otimes W \cong \bigoplus_{x \in G} W_{g^{-1}x}$ and $W \otimes \mathbb C_g \cong \bigoplus_{x \in G}W_{xg^{-1}}$ that $R_{\mathbb C_g,W}$ boils down to a collection of isomorphisms $W_x \xrightarrow{\sim}W_{gxg^{-1}}$, for every $g,x \in G$. The natural isomorphism $R_{-,W}$ is subject to a hexagon axiom, which imposes
\begin{equation}
    (R_{\mathbb C_{g_1},W} \otimes {\rm id}_{\mathbb C_{g_2}}) \circ ({\rm id}_{\mathbb C_{g_1}} \otimes R_{\mathbb C_{g_2},W})= R_{\mathbb C_{g_1g_2},W} \, ,
\end{equation}
for every $g_1,g_2 \in G$. It follows that the isomorphisms $W_x \xrightarrow{\sim}W_{gxg^{-1}}$ do equip $W$ with a $G$-equivariant structure that is compatible with the $G$-grading. Conversely, given an object $W \in \Vect_G^G$, the corresponding object in $\mc Z(\Vect_G)$ has the same underlying $G$-graded vector space and the half-braiding natural isomorphism is provided by
\begin{equation}
    R_{K,W}(k \otimes w) = g \cdot w \otimes k \, ,
\end{equation}
for $K = \bigoplus_{g \in G}K_g \in \Vect_G$, $k \in K_g$ and $w \in W$. This is the equivalence $\mc Z(\Vect_G) \simeq \Vect_G^G$.

Let us consider a couple of special cases. Given $[x_0] \in \Cl(G)$ and $\hat W \in \Irr(\Rep(Z_G(x_0)))$, let $W \equiv ([x_0],\hat W)$ be the corresponding simple object in $\Vect_G^G$. Recall that, as a $G$-graded vector space, $W = \bigoplus_{x \in [x_0]} W_x$, such that $W_x \cong \hat W$, for every $x \in [x_0]$, and as a $G$-representation, $W \cong \Ind_{Z_G(x_0)}^G(\hat W) = \mathbb C\{\varpi_{x,w_c}\}_{x \in [x_0], c=1,\ldots,\dim_\mathbb C \hat W}$ such that $g \cdot \varpi_{x,w_c} = \varpi_{gxg^{-1},z_{g,x} \cdot w_c}$, where $gr_x = r_{gxg^{-1}} z_{g,x}$, for every $g \in G$ (see sec.~\ref{sec:VecG_OP}). Therefore, the half-braiding natural isomorphism of the corresponding object in $\mc Z(\Vect_G)$ is provided by
\begin{equation}
    R_{\mathbb C_g,W}(e_g \otimes \varpi_{x,w_c}) = \varpi_{gxg^{-1},z_{g,x} \cdot w_c} \otimes e_g \, ,
\end{equation}
for every $g \in G$, $x \in [x_0]$ and $c \in \{1,\ldots,\dim_\mathbb C \hat W\}$.

Diagrammatically, 
\begin{equation}
    \halfBraidingVecG{1} \equiv R_{\mathbb C_g,W} \, ,
\end{equation}
which can be fully specified by the collection of so-called $\Omega$-symbols satisfying (see notation in sec.~\ref{sec:VecG_OP}) 
\begin{equation} 
    \halfBraidingVecG{1} \equiv 
    \sum_{x \in [x_0]} \sum_{c_1,c_2 =1}^{\dim_\mathbb C \hat W} 
    \Omega^{W \mathbb C_g}_{\mathbb C_x \mathbb C_{gx} \mathbb C_{gxg^{-1}},c_1c_2}  \; 
    \halfBraidingVecG{2} \, .
\end{equation}
It follows from the orthogonality conditions satisfied by the $G$-grading preserving linear maps that
\begin{equation}
    \Omega^{W \mathbb C_g}_{\mathbb C_x \mathbb C_{gx} \mathbb C_{gxg^{-1}},c_1c_2}  \, \mathbb I_{\mathbb C_{gx}} 
    = \halfBraidingCstsVecG \, .
\end{equation}
Explicitly, the $\Omega$-symbols evaluate to
\begin{equation}
    \Omega^{W \mathbb C_g}_{\mathbb C_x \mathbb C_{gx} \mathbb C_{gxg^{-1}},c_1c_2}  = \hat \varrho(z_{g,x})^{c_1}_{c_2} \, ,
\end{equation}
where $\hat \varrho : \mathbb C[Z_G(x_0)] \to \End_\mathbb C(\hat W)$. As another example, consider the Lagrangian algebra $L(H,\beta)=\Ind_H^G(\mathbb C[H]^\beta) \in \Vect^G_G$ defined in sec.~\ref{sec:VecG_Lagrangian}. Writing $L(H,\beta) = \mathbb C\{\varsigma_{a,h}\}_{a=1,\ldots,(G:H),h \in H}$, the $G$-equivariant $G$-graded vector space structure is defined via $|\varsigma_{a,h}| = r_ahr_a^{-1}$ and $g \cdot \varsigma_{a,h} = \beta_h(h_{g,a})\,\varsigma_{g(a),h_{g,a}hh_{g,a}^{-1}}$, for every $g \in G$, where $g r_a = r_{g(a)} h_{g,a}$ and $\beta_-(-)$ was defined in eq.~\eqref{eq:transgression}. Therefore, the half-braiding natural isomorphism of the corresponding object in $\mc Z(\Vect_G)$ is provided by
\begin{equation}
    R_{\mathbb C_g,L(H,\beta)}(e_g \otimes \varsigma_{a,h}) = \beta_h(h_{g,a}) \, \varsigma_{g(a),h_{g,a}hh_{g,a}^{-1}} \otimes e_g \, ,
\end{equation}
for every $g \in G$, $h \in H$ and $a \in \{1,\ldots,(G:H)\}$.

\bigskip \noindent
Let us now briefly review the equivalence $\Vect_G^G \simeq \Tu(\Vect_G)$. Let $\mathbb C[G] = \mathbb C \{e_g\}_{g \in G}$ be the group algebra with multiplication rule $e_{g_1} \cdot e_{g_2} = e_{g_1g_2}$, for every $g_1,g_2 \in G$, and $\mathbb C(G) = \mathbb C \{\varsigma_x\}_{x \in G}$ be the algebra of functions $G \to \mathbb C$ with multiplication rule $\varsigma_x \cdot \varsigma_y = \delta_{x,y} \, \varsigma_{x,y}$, for every $x,y \in G$. By definition, the linear space underlying $\Tu(\Vect_G)$ is $\bigoplus_{g,x \in G} \Hom_{\Vect_G}(\mathbb C_{gxg^{-1}} \otimes \mathbb C_g, \mathbb C_g \otimes \mathbb C_x)$. Clearly, as a vector space, $\Tu(\Vect_G) \cong \mathbb C[G] \otimes \mathbb C(G) = \mathbb C\{e_g \otimes \varsigma_x\}_{g,x \in G}$. In this basis, the multiplication rule of $\Tu(\Vect_G)$ obtained by stacking the string diagrams reads
\begin{equation}
    (e_{g_1} \otimes \varsigma_y) \cdot (e_{g_2} \otimes \varsigma_x) := \delta_{y,g_2xg_2^{-1}} \, (e_{g_1g_2} \otimes \varsigma_x) \, ,
\end{equation}
for every $g_1,g_2,x,y \in G$, while the $*$-structure is provided by $(e_g \otimes \varsigma_x)^* = e_{g^{-1}} \otimes \varsigma_{gxg^{-1}}$, for every $x,g \in G$. 

Let $W$ be a module over $\Tu(\Vect_G)$. Since both $\mathbb C[G]$ and $\mathbb C(G)$ form subalgebras of $\Tu(\Vect_G)$, $W$ is equipped with the structure of a (left) module over both $\mathbb C[G]$ and $\mathbb C(G)$. Since $\Mod(\mathbb C[G]) \simeq \Rep(G)$ and $\Mod(\mathbb C(G)) \simeq \Vect_G$, $W$ is both a $G$-graded vector space and a $G$-representation. More explicitly, we have $W = \bigoplus_{x \in G} W_x$ such that $W_x := (e_{1_G} \otimes \varsigma_x) W$, for every $x \in G$. For every $x \in G$ and $w \in W_x$, define $g \cdot w := (e_g \otimes \varsigma_x) \cdot w$. It follows from
\begin{equation}
    (e_{1_G} \otimes \varsigma_{gxg^{-1}}) \cdot (e_g \otimes \varsigma_x) 
    = 
    (e_g \otimes \varsigma_x)  \cdot (e_{1_G} \otimes \varsigma_x) \, ,
\end{equation}
for every $g,x \in G$, that $|g \cdot w| = g |w|g^{-1} = gxg^{-1}$, for every $w \in W_x$. Therefore, $g \cdot W_x \subseteq W_{gxg^{-1}}$ and $W \in \Vect_G^G$. Conversely, given an object $W \in \Vect_G^G$, its $\Tu(\Vect_G)$-module structure is given by 
\begin{equation}
    (e_g \otimes \varsigma_y) \cdot w := \delta_{x,y} \; g \cdot w
\end{equation}
for $g,x,y \in G$ and $w \in W_x$, in such a way that $g \cdot w = \sum_{x \in G}(e_g \otimes \varsigma_x) \cdot w$. In particular, given a simple object $([x_0],\hat W)$ in $\Vect_G^G$ with $W \cong \Ind_{Z_G(x_0)}^G(\hat W) = \mathbb C\{\varpi_{x,w_c}\}_{x \in [x_0]. c=1,\ldots,\dim_\mathbb C \hat W}$, its $\Tu(\Vect_G)$-module structure is given by 
\begin{equation}
    (e_g \otimes \varsigma_y) \cdot \varpi_{x,w_c} := \delta_{y,x} \, \varpi_{gxg^{-1}, z_{g,x} \cdot w_c} \, ,
\end{equation}
for every $g , y \in G$, $x \in [x_0]$ and $c \in \{1,\ldots, \dim_\mathbb C \hat W\}$. As another example, consider the Lagrangian algebra $L(H,\beta)$. Using the same notation as above, 
\begin{equation}
    (e_g\otimes \varsigma_y)\cdot \varsigma_{a,h}
    =
    \delta_{y,\,r_a h r_a^{-1}}\,
    \beta_h(h_{g,a})\, \varsigma_{g(a),h_{g,a}hh_{g,a}^{-1}} \, ,
\end{equation}
for every $g,y \in G$, $a \in \{1,\ldots,(G:H)\}$ and $h \in H$.

We are now ready to make an important remark: Let $([x_0],\hat W_1)$ and $([x_0],\hat W_2)$ be two simple objects in $\Vect_G^G$. Suppose $\hat W_1$ and $\hat W_2$ are non-isomorphic. It is still possible that $\Ind_{Z_G(x_0)}^G(\hat W_1) \cong \Ind_{Z_G(x_0)}^G(\hat W_2)$. But the simple objects $([x_0],\hat W_1)$ and $([x_0],\hat W_2)$ are well and truly non-isomorphic, precisely because their corresponding $\Tu(\Vect_G)$-module structures `remember' the grading. Physically, this is the reason why it is crucial to treat the endpoints of string operators as twisted sector local operators that transform like modules over $\Tu(\Vect_G)$, rather than local operators dressing a truncated symmetry operator (see sec.~\ref{sec:VecG_String}).

Finally, let us derive the character formulas used in the main text. Given $W \in \Mod(\Tu(\Vect_G))$, let $\chi^W$ be the corresponding character. By definition, we have $\chi^W(e_g \otimes \varsigma_x) = \delta_{gx,xg} \, \tr_{W_x}[g]$. For the simple object $([x_0],\hat W)$, this becomes
\begin{equation} 
    \chi^{([x_0],\hat W)}(e_g \otimes \varsigma_x) = \delta_{x \in [x_0]} \, \delta_{gx,xg} \, \chi^{\hat W}(z_{g,x}) \, ,
\end{equation}
where $z_{g,x}\in Z_G(x_0)$ is defined by $g r_x = r_{g x g^{-1}} z_{g,x}$. Similarly, let us compute the character of $L(H,\beta)$. By definition,
\begin{equation}
\begin{split}
    \chi^{L(H,\beta)}(e_g \otimes \varsigma_x) 
    &=
    \sum_{a=1}^{(G:H)} \sum_{h \in H} \delta_{x,r_ahr_a^{-1}} \, \beta_h(h_{g,a}) \, \varsigma^{a,h}\big(\varsigma_{g(a),h_{g,a}hh_{g,a}^{-1}}\big)
    \\
    &=
    \sum_{a=1}^{(G:H)} \sum_{h \in H} \delta_{x,r_ahr_a^{-1}} \, \delta_{g(a),a} \, \delta_{h_{g,a}hh_{g,a}^{-1},h} \, \beta_h(h_{g,a}) \, . 
\end{split}
\end{equation}
Since $gr_a = r_{g(a)}h_{g,a}$, $g(a)=a$ implies $g =r_a h_{g,a}r_a^{-1}$. Together with $x = r_ahr_a^{-1}$ and $h_{g,a}h = h h_{g,a}$, this implies $gx=xg$. Therefore,
\begin{equation}
\begin{split}
    \chi^{L(H,\beta)}(e_g \otimes \varsigma_x) 
    &= \delta_{gx,xg} \sum_{a=1}^{(G:H)} \sum_{h \in H} \delta_{x,r_ahr_a^{-1}} \, \delta_{g(a),a} \, \beta_{r_a^{-1}xr_a}(r_a^{-1}gr_a)
    \\
    &= \delta_{gx,xg} \sum_{a=1}^{(G:H)} \delta_{x \in r_aHr_a^{-1}} \, \delta_{g \in r_aHr_a^{-1}} \, \beta_{r_a^{-1}xr_a}(r_a^{-1}gr_a)
    \, .
\end{split}
\end{equation}
Replacing the sum over cosets by a sum group elements in $G$ finally yields
\begin{equation}
    \chi^{L(H,\beta)}(e_g \otimes \varsigma_x) = \frac{\delta_{gx,xg}}{|H|} \sum_{y \in G} \delta_{g \in y^{-1}Hy} \, \delta_{x \in y^{-1}Hy} \, \beta_{yxy^{-1}}(ygy^{-1}) \, ,
\end{equation}
which is the formula used in the main text.

\subsection{Drinfel'd centre of the category of group representations\label{app:centres_RepG}}

Consider the category $\Rep(G)$ of representations of $G$. Let us sketch the equivalence $\mc Z(\Rep(G)) \simeq \Vect_G^G$. Let $(W,R_{-,W})$ be an object in $\mc Z(\Rep(G))$; it consists of an object $W \in \Rep(G)$ and a collection of $G$-equivariant isomorphisms $R_{V,W} \colon V \otimes W \xrightarrow{\sim} W \otimes V$ that are natural in $V$, for every $V \in \Rep(G)$. In particular, we have an isomorphism $R_{\mathbb C[G],W} \colon \mathbb C[G] \otimes W \xrightarrow{\sim} W \otimes \mathbb C[G]$, where $\mathbb C[G]$ is treated as the regular representation. For any $w \in W$, consider $R_{\mathbb C[G],W}(e_{1_G} \otimes w)$.
One can expand $R_{\mathbb C[G],W}(e_{1_G} \otimes w) \in W \otimes \mathbb C[G]$ in the basis of $\mathbb C[G]$, $R_{\mathbb C[G],W}(e_{1_G} \otimes w) \equiv \sum_{x \in G} w_{x^{-1}} \otimes e_x$, which defines $w_{x^{-1}} \in W$ uniquely. 
Let $\epsilon \colon \mathbb C[G] \to \mathbb C$ be the $G$-equivariant map $e_g \mapsto 1$, for every $g \in G$. By naturality of the half-braiding, 
\begin{equation}
    ({\rm id}_W \otimes \epsilon) \circ R_{\mathbb C[G],W} = R_{\mathbb C,W} \circ (\epsilon \otimes {\rm id}_W) \, .
\end{equation}
Applying both sides to $e_{1_G} \otimes w$ yields
\begin{equation}
    \sum_{x \in G} w_{x^{-1}} \otimes \epsilon(e_x) = \sum_{x \in G} w_x \otimes 1 = R_{\mathbb C,W}(1 \otimes w) = w \otimes 1 \, ,
\end{equation}
from which follows  that $w = \sum_{x \in G} w_x$.
Defining $W_x := \{w \in W \, | \, R_{\mathbb C[G],W}(e_{1_G} \otimes w) = w \otimes e_{x^{-1}} \}$, we obtain $W = \bigoplus_{x \in G}W_x$. It remains to check that the $G$-equivariant structure is compatible with this $G$-grading. 
Let $w\in W_x$. By definition, $R_{\mathbb C[G],W}(e_{1_G}\otimes w) = w\otimes e_{x^{-1}}$.
Since $R_{\mathbb C[G],W}$ is a map in $\Rep(G)$, we have 
\begin{equation}
    \begin{split}
        R_{\mathbb C[G],W}(e_g\otimes g\cdot w) = R_{\mathbb C[G],W}(g\cdot(e_{1_G}\otimes w)) = g\cdot R_{\mathbb C[G],W}(e_{1_G}\otimes w) = g\cdot w \otimes e_{g x^{-1}} \, .
    \end{split}
\end{equation}
Let $R_g \colon \mathbb C[G] \to \mathbb C[G], e_x \mapsto e_{xg^{-1}}$.
By naturality of the half-braiding,  
\begin{equation}
    ({\rm id}_W \otimes R_g) \circ R_{\mathbb C[G],W} = R_{\mathbb C[G],W} \circ (R_g \otimes {\rm id}_W) \, . 
\end{equation}
Applying both sides to $e_{1_G} \otimes w$ yields $w \otimes e_{x^{-1}g^{-1}} = R_{\mathbb C[G],W}(e_{g^{-1}} \otimes w)$ so that
\begin{equation}
    R_{\mathbb C[G],W}(e_{1_G} \otimes g \cdot w)  = g \cdot R_{\mathbb C[G],W}(e_{g^{-1}} \otimes w) = g \cdot w \otimes e_{gx^{-1}g^{-1}} \, .
\end{equation}
Therefore, for every $w \in W_x$, $g \cdot w \in W_{gxg^{-1}}$, as required. Bringing everything together, $W \in \Vect_G^G$. Conversely, given $W \in \Vect_G^G$, the corresponding object in $\mc Z(\Rep(G))$ has  the same underlying $G$-representation and the half-braiding natural isomorphism is provided by
\begin{equation}
    R_{V,W}(v \otimes w) = w \otimes x^{-1} \cdot v \, ,
\end{equation}
for every $V \in \Rep(G)$, $v \in V$ and $w \in W_x$.

Let us consider a couple of special cases. Let $W = ([x_0],\hat W)$ be a simple object in $\Vect_G^G$. As a $G$-graded vector space $W = \bigoplus_{x \in G} W_x$ such that $W_x = \mathbb C\{\varpi_{x,w_c}\}_{x \in [x_0], c=1,\ldots,\dim_\mathbb C \hat W} \cong \hat W$, for every $x \in [x_0]$. Therefore, the half-braiding of the corresponding object in $\mc Z(\Rep(G))$ is given by
\begin{equation}
    R_{V,W}(v \otimes \varpi_{x,w_c}) = \varpi_{x,w_c} \otimes x^{-1} \cdot v \, ,
\end{equation}
for every $V \in \Rep(G)$ and $v \in V$. 
Diagrammatically,
\begin{equation}
    \halfBraidingRepG{1} \equiv R_{V,W} \, .
\end{equation}
As in the $\Vect_G$, the half-braiding can be fully specified by a collection of $\Omega$-symbols defined via
\begin{equation}
    \label{eq:RepG_OmegaSymbols}
    \halfBraidingRepG{1} \equiv 
    \sum_{\substack{S_1,S_3 \in W \\ S_2 \in V \act S_3}} 
    \sum_{\substack{1 \leq j_1 \leq N^W_{S_1} \\ 1 \leq j_2 \leq N^W_{S_3}}}
    \sum_{\substack{1 \leq i_1 \leq N^{VS_1}_{S_2} \\ 1 \leq i_2 \leq N^{S_3V}_{S_2}}}
    \Omega^{WV}_{S_1S_2S_3,i_1i_2j_1j_2}  \; 
    \halfBraidingRepG{2} \, .
\end{equation}
Using the orthogonality conditions satisfied by the $G$-equivariant linear maps, we obtain
\begin{equation}
    \Omega^{WV}_{S_1S_2S_3,i_1i_2j_1j_2} \, \mathbb I_{S_2}  
     = \halfBraidingCstsRepG \, .
\end{equation}
Explicitly, the $\Omega$-symbols evaluate to
\begin{equation}
    \Omega^{WV}_{S_1S_2S_3,i_1i_2j_1j_2} =
    \frac{1}{\dim_\mathbb C S_2}  \!\!\!\! \sum_{\substack{x \in [x_0] \\ 1 \leq c \leq \dim_\mathbb C \hat W}}
    \sum_{\substack{1 \leq f_1 \leq \dim_\mathbb C S_1 \\ 1 \leq f_2 \leq \dim_\mathbb C S_2 \\ 1 \leq f_3 \leq \dim_\mathbb C S_3 \\ 1 \leq d_1,d_2 \leq \dim_\mathbb C V}} 
    \CCdb{W}{S_3}{x,c}{f_3}{j_1}
    \CCb{S_3}{V}{S_2}{f_3}{d_1}{f_2}{i_2}\,
    \rho(x^{-1})^{d_1}_{d_2} \, 
    \CC{V}{S_1}{S_2}{d_2}{f_1}{f_2}{i_1} 
    \CCd{W}{S_1}{x,c}{f_1}{j_2} \, ,
\end{equation}
where $\rho \colon \mathbb C[G] \to \End_\mathbb C(V)$. As another example, consider the Lagrangian algebra $L(H,\beta)$. Recall that $L(H,\beta) = \mathbb C\{\varsigma_{a,h}\}_{a=1,\ldots,(G:H),h \in H}$ such that $|\varsigma_{a,h}| = r_a h r_a^{-1}$. Therefore, $R_{V,L(H,\beta)}(v \otimes \varsigma_{a,h})  = \varsigma_{a,h} \otimes ({}^{r_a}h^{-1}) \cdot v$, for every $v \in V$.

\bigskip \noindent
We are now ready to perform the main computation of this appendix. Specialising to the Lagrangian algebra $L(G,\beta)$, we need to compute the $\Omega$-symbols of the corresponding object in $\mc Z(\Rep(G))$. As commented above, the half-braiding natural isomorphism of $L(G,\beta)$ is easily expressed in the basis $\mathbb C\{\varsigma_{1,g}\}_{g \in G}$ given that $\varsigma_{1,g}$ is homogeneous of degree $|\varsigma_{1,g}| = g$. Thus, the difficulty stems from decomposing $L(G,\beta)$ into simple objects in $\Rep(G)$. Recall that as an object in $\Rep(G)$, $L(G,\beta)$ is isomorphic to $\mathbb C[G]^\beta = \mathbb C\{e_g\}_{g \in G}$, where $g_1 \cdot e_{g_2} = \beta_{g_2}(g_1)\, e_{g_1g_2g_1^{-1}}$, for every $g_1,g_2 \in G$. Therefore, 
\begin{equation}
    \label{eq:decompGrpAlg}
    \mathbb C[G]^\beta \cong 
    \!\!\! \bigoplus_{U \in \Irr(\Rep^\beta(G))} 
    \!\!\!
    \End_\mathbb C(U) \, ,
\end{equation}
as an object in $\Rep(G)$.
Indeed, by Fourier transform, the matrix unit basis of $\End_\mathbb C(U)$, for every $U \in \Irr(\Rep^\beta(U))$, can be expressed in terms of $\{e_g\}_g$ as
\begin{equation}
    u_{b_1} \otimes u^{b_2} = \sqrt{\frac{\dim_\mathbb C U}{|G|}} 
    \sum_{g \in G} \overline{\pi(g)^{b_1}_{b_2}}\; e_g \, ,
\end{equation}
where $\pi : \mathbb C[G]^\beta \to \End_\mathbb C(U)$. Recall that 
\begin{equation}
    \overline{\pi(g)^{b_1}_{b_2}} = \frac{1}{\beta(g,g^{-1})} \pi(g^{-1})^{b_2}_{b_1} \, ,
\end{equation}
for every $g \in G$ and $b_1,b_2 \in \{1,\ldots,\dim_\mathbb C U\}$. It follows from the orthogonality condition
\begin{equation}
    \frac{1}{|G|} \sum_{g \in G} \pi_1(g)^{b_1}_{b_2} \; \overline{\pi_2(g)^{b_3}_{b_4}} = \frac{\delta_{U_1,U_2}}{\dim_\mathbb C U_1} \, 
    \delta_{b_1,b_3} \, \delta_{b_2,b_4} \, ,
\end{equation}
where $\pi_1 : \mathbb C[G]^\beta \to \End_\mathbb C(U_1)$ and $\pi_2 : \mathbb C[G]^\beta \to \End_\mathbb C(U_2)$, for every $U_1,U_2 \in \Irr(\Rep^\beta(G))$, $b_1,b_2 \in \{1,\ldots,\dim_\mathbb C U_1\}$ and $b_3,b_4 \in \{1,\ldots,\dim_\mathbb C U_2\}$, that the inverse map is given by\footnote{Note that this is the transform that guarantees that both the basis of $\mathbb C[G]^\beta$ and $\End_\mathbb C(U)$ are orthonormal. In order to establish eq.~\eqref{eq:decompGrpAlg} as an isomorphism of $G$-equivariant algebra, a different normalisation is required.} 
\begin{equation}
    e_g = \!\!\! 
    \sum_{U \in \Irr(\Rep^\beta(G))} \sum_{b_1,b_2=1}^{\dim_\mathbb C U} \Bigg( \sqrt{\frac{\dim_\mathbb C U}{|G|}} \,
    {\pi(g)^{b_1}_{b_2}} \Bigg) u_{b_1} \otimes u^{b_2} \, .
\end{equation}
In order to obtain the decomposition of $\mathbb C[G]^\beta$ into simple objects. one further needs to decompose each (linear) representation $\End_\mathbb C(U)$, for every $U \in \Irr(\Rep^\beta(G))$. Given $U = \mathbb C\{u_{b}\}_{b} \in \Irr(\Rep^\beta(G))$, let $S = \mathbb C\{s_{f}\}_{f} \in \Irr(\Rep(G))$ be a direct summand of $\End_\mathbb C(U)$ and write
\begin{equation}
    \Hom_G(\End_\mathbb C(U),S) = \mathbb C\{\varphi^{UU^*}_{S,j}\}_{j} \, .
\end{equation}
Recall that by our convention
\begin{equation}
    \varphi^{U U^*}_{S,j}(u_{b_1} \otimes u^{b_2}) = \sum_{f=1}^{\dim_\mathbb C S}
    \CCb{U}{U^*}{S}{b_1}{b_2}{f}{j} \;  s_{f} \, .
\end{equation}
Besides, note that $\Hom_G(\End_\mathbb C(U),S) \cong \Hom_G(U,S \otimes U)$, which is implemented by\footnote{As a consistency check, consider the orthogonality conditions satisfied by the two sets of Clebsch--Gordan coefficients.} 
\begin{equation}
    \CCb{U}{U^*}{S}{b_1}{b_2}{f_1}{j} 
    = \sqrt{\frac{\dim_\mathbb C S}{\dim_\mathbb C U}} \; 
    \CC{S}{U}{U}{f_1}{b_2}{b_1}{j} \, .
\end{equation}
Therefore, given any basis vector $s_{f}$ in the simple direct summand $S$ of $\End_\mathbb C(U)$ associated with multiplicity label $j \in \{1,\ldots,N^{UU^*}_{S}\}$, one identifies
\begin{equation}
    s_{f}
    = \sqrt{\frac{\dim_\mathbb C S}{\dim_\mathbb C U}}
    \sum_{b_1,b_2 = 1}^{\dim_\mathbb C U}
    \CCb{S}{U}{U}{f_1}{b_2}{b_1}{j} u_{b_1} \otimes u^{b_2} \, .
\end{equation}
Together with the Fourier transform spelt out above, this yields
\begin{equation}
    s_{f}
    = \sqrt{\frac{\dim_\mathbb C S}{|G|}}
    \sum_{b_1,b_2 = 1}^{\dim_\mathbb C U} \sum_{g \in G}
    \Bigg( \CCb{S}{U}{U}{f}{b_2}{b_1}{j} \, \overline{\pi(g)^{b_1}_{b_2}} \Bigg) e_g \, .
\end{equation}
Conversely, for every $g \in G$, we have
\begin{equation}
    \label{eq:invDecompLagAlg}
    e_g = \!\!\! 
    \sum_{\substack{U \in \Irr(\Rep^\beta(G)) \\ S \in U \otimes U^* \\ 1 \leq j \leq N^{UU^*}_S }}
    \sum_{b_1,b_2=1}^{\dim_\mathbb C U} 
    \sum_{f=1}^{\dim_\mathbb C S} 
    \Bigg( \sqrt{\frac{\dim_\mathbb C S}{|G|}} \,
    \CC{S}{U}{U}{f}{b_2}{b_1}{j} \, 
    {\pi(g)^{b_1}_{b_2}} \Bigg) s_f\, .
\end{equation}
We are now ready to compute the $\Omega$-symbols. Let $V = \mathbb C\{v_{d}\}_{d} \in \Irr(\Rep(G))$, $U_1 = \mathbb C\{u_{b_1}\}_{b_1} \in \Irr(\Rep^\beta(G))$ and $S_1 = \mathbb C\{s_{f_1}\}_{f_1}$ the simple direct summand of $\End_\mathbb C(U_1)$ associated with multiplicity label $j_2 \in \{1,\ldots,N^{U_1U_1^*}_{S_1}\}$. Let us begin by computing the image of the basis vector $v_{d_2} \otimes s_{f_1}$ under the half-braiding isomorphism $R_{V,L(G,\beta)} : V \otimes L(G,\beta) \xrightarrow{\sim} L(G,\beta) \otimes V$. By definition, 
\begin{equation}
    R_{V,L(G,\beta)}(v_{d_2} \otimes s_{f_1}) = 
    \sqrt{\frac{\dim_\mathbb C S_1}{|G|}}
    \sum_{b_1,b_2 = 1}^{\dim_\mathbb C U_1} \sum_{g \in G}
    \Bigg( \CCb{S_1}{U_1}{U_1}{f_1}{b_2}{b_1}{j_2} \overline{\pi_1(g)^{b_1}_{b_2}}  \Bigg) e_g \otimes (g^{-1} \cdot v_{d_2}) \, .
\end{equation}
In order to simplify this expression, one begins by spelling out $g^{-1} \cdot v_{d_2} = \sum_{d_1=1}^{\dim_\mathbb C V} \overline{\rho(g)^{d_2}_{d_1}} \, v_{d_1}$ before employing eq.~\eqref{eq:invDecompLagAlg}, which results in
\begin{align}
    \nn
    R_{V,L(G,\beta)}(v_{d_2} \otimes s_{f_1}) = \!\!\!  
    \sum_{\substack{U_2 \in \Irr(\Rep^\beta(G)) \\ S_3 \in U_2 \otimes U_2^* \\ 1 \leq j_1 \leq N^{U_2U_2^*}_{S_3} }}
    \sum_{\substack{1 \leq d_1 \leq \dim_\mathbb C V \\ 1 \leq b_1,b_2 \leq \dim_\mathbb C U_1 \\ 1 \leq b_3,b_4 \leq \dim_\mathbb C U_2 \\ 1 \leq f_3 \leq \dim_\mathbb C S_3 \\ g \in G}}
    \Bigg(
    &\frac{\sqrt{\dim_\mathbb C S_1 \dim_\mathbb C S_3}}{|G|} \,
    \CCb{S_1}{U_1}{U_1}{f_1}{b_1}{b_2}{j_2} \, \CC{S_3}{U_2}{U_2}{f_3}{b_3}{b_4}{j_1}
    \\[-3em]
    &\overline{\rho(g)^{d_2}_{d_1}}  \, \overline{\pi_1(g)^{b_2}_{b_1}} \, \pi_2(g)^{b_4}_{b_3} \Bigg) s_{f_3} \otimes v_{d_1} \, .
\end{align}
Invoking 
\begin{equation}
    \frac{1}{|G|}
    \sum_{g \in G}
    \overline{\rho(g)^{d_2}_{d_1}} \, 
    \overline{\pi_1(g)^{b_2}_{b_1}} \,
    {\pi_2(g)^{b_4}_{b_3}}
    =
    \frac{1}{\dim_\mathbb C U_2}
    \sum_{1 \leq j_4 \leq N^{VU_1}_{U_2}}
    \CCb{V}{U_1}{U_2}{d_2}{b_2}{b_4}{j_4}\,
    \CC{V}{U_1}{U_2}{d_1}{b_1}{b_3}{j_4} \, ,
\end{equation}
it boils down to
\begin{align}
    \nn
    R_{V,L(G,\beta)}(v_{d_2} \otimes s_{f_1}) = \!\!\!  
    \sum_{\substack{U_2 \in \Irr(\Rep^\beta(G)) \\ S_3 \in U_2 \otimes U_2^* \\ 1 \leq j_1 \leq N^{U_2U_2^*}_{S_3} \\ 1 \leq j_4 \leq N^{VU_1}_{U_2}}}
    \sum_{\substack{1 \leq d_1 \leq \dim_\mathbb C V \\ 1 \leq b_1,b_2 \leq \dim_\mathbb C U_1 \\ 1 \leq b_3,b_4 \leq \dim_\mathbb C U_2 \\ 1 \leq f_3 \leq \dim_\mathbb C S_3}}
    \Bigg(
    &\frac{\sqrt{\dim_\mathbb C S_1 \dim_\mathbb C S_3}}{\dim_\mathbb C U_2} \,
    \CCb{S_1}{U_1}{U_1}{f_1}{b_1}{b_2}{j_2} \, \CC{S_3}{U_2}{U_2}{f_3}{b_3}{b_4}{j_1}
    \\[-3em]
    &\CC{V}{U_1}{U_2}{d_1}{b_1}{b_3}{j_4}\,
    \CCb{V}{U_1}{U_2}{d_2}{b_2}{b_4}{j_4}\Bigg) s_{f_3} \otimes v_{d_1} \, .
\end{align}
We are left to express the induced maps $V \otimes S_1 \to S_3 \otimes V$ in the basis
\begin{equation}
    \mathbb C\big\{\bar \varphi^{VS_1}_{S_2,i_1} \otimes \varphi^{S_3 V}_{S_2,i_2}\big\}_{S_2,i_1,i_2} = \bigoplus_{S_2 \in V \otimes S_1} \Hom_{G}(V \otimes S_1,S_2) \otimes \Hom_G(S_2, S_3 \otimes V) 
\end{equation}
so as to obtain
\begin{align}
    \nn
    \Omega^{L(G,\beta)V}_{S_1S_2S_3,i_1i_2j_1j_2} = \!\!\!  
    \sum_{\substack{U_1,U_2 \in \Irr(\Rep^\beta(G)) \\ 1 \leq j_4 \leq N^{VU_1}_{U_2}}}
    \sum_{\substack{1 \leq d_1,d_2 \leq \dim_\mathbb C V \\ 1 \leq b_1,b_2 \leq \dim_\mathbb C U_1 \\ 1 \leq b_3,b_4 \leq \dim_\mathbb C U_2 \\ 1 \leq f_1 \leq \dim_\mathbb C S_1 \\ 1 \leq f_2 \leq \dim_\mathbb C S_2 \\ 1 \leq f_3 \leq \dim_\mathbb C S_3}}
    \Bigg(
    &\frac{\sqrt{\dim_\mathbb C S_1 \dim_\mathbb C S_3}}{\dim_\mathbb C U_2 \dim_\mathbb C S_2} \,
    \CCb{S_1}{U_1}{U_1}{f_1}{b_1}{b_2}{j_2} \, \CC{S_3}{U_2}{U_2}{f_3}{b_3}{b_4}{j_1}
    \\[-2.5em] \nn
    &\CC{V}{S_1}{S_2}{d_2}{f_1}{f_2}{i_1}\, 
    \CC{V}{U_1}{U_2}{d_1}{b_1}{b_3}{j_4} \, 
    \CCb{V}{U_1}{U_2}{d_2}{b_2}{b_4}{j_4}
    \CCb{S_3}{V}{S_2}{f_3}{d_1}{f_2}{i_2}
    \Bigg)  \, ,
\end{align}
By definition of the $\F{\act}$-symbols, we have
\begin{equation}
    \sum_{b_2 = 1}^{\dim_\mathbb C U_1}
    \CCb{V}{U_1}{U_2}{d_2}{b_2}{b_4}{j_4} \, 
    \CCb{S_1}{U_1}{U_1}{f_1}{b_1}{b_2}{j_2}
    = \sum_{\substack{S_4 \in V \otimes S_1 \\ 1 \leq i_3 \leq N^{VS_1}_{S_4} \\ 1 \leq j_3 \leq N^{S_4 U_1}_{U_2}}} 
    \sum_{f_4 = 1}^{\dim_\mathbb C S_4}
    \big(\F{\act}^{VS_1U_1}_{U_2}\big)^{S_4,i_3j_3}_{U_1,j_2j_4} \, 
    \CCb{V}{S_1}{S_4}{d_2}{f_1}{f_4}{i_3} \, 
    \CCb{S_4}{U_1}{U_2}{f_4}{b_1}{b_4}{j_3} \, .
\end{equation}
Together with the orthogonality condition
\begin{equation}
    \sum_{\substack{1 \leq d_2 \leq \dim_\mathbb C V \\ 1 \leq f_1 \leq \dim_\mathbb C S_1}}
    \CCb{V}{S_1}{S_4}{d_2}{f_1}{f_4}{i_3} \, 
    \CC{V}{S_1}{S_2}{d_2}{f_1}{f_2}{i_1} = \delta_{S_2,S_4} \, \delta_{f_2,f_4} \, \delta_{i_1,i_3} \, ,
\end{equation}
it allows us to rewrite the previous expression for the $\Omega$-symbols as
\begin{align}
    \nn
    \Omega^{L(G,\beta)V}_{S_1S_2S_3,i_1i_2j_1j_2} = \!\!\!  
    \sum_{\substack{U_1,U_2 \in \Irr(\Rep^\beta(G)) \\ 1 \leq j_3 N^{S_2U_1}_{U_2} \\ 1 \leq j_4 \leq N^{VU_1}_{U_2}}}
    \sum_{\substack{1 \leq d_1 \leq \dim_\mathbb C V \\ 1 \leq b_1 \leq \dim_\mathbb C U_1 \\ 1 \leq b_3,b_4 \leq \dim_\mathbb C U_2  \\ 1 \leq f_2 \leq \dim_\mathbb C S_2 \\ 1 \leq f_3 \leq \dim_\mathbb C S_3}}
    \Bigg(
    &\frac{\sqrt{\dim_\mathbb C S_1 \dim_\mathbb C S_3}}{\dim_\mathbb C U_2 \dim_\mathbb C S_2} \,
    \big(\F{\act}^{VS_1U_1}_{U_2}\big)^{S_4,i_3j_3}_{U_1,j_2j_4}
    \, \CC{S_3}{U_2}{U_2}{f_3}{b_3}{b_4}{j_1}
    \\[-2.5em] \nn
    &\CCb{S_2}{U_1}{U_2}{f_2}{b_1}{b_4}{j_3}\, 
    \CC{V}{U_1}{U_2}{d_1}{b_1}{b_3}{j_4}
    \CCb{S_3}{V}{S_2}{f_3}{d_1}{f_2}{i_2}
    \Bigg)  \, ,
\end{align}
Finally, it follows from 
\begin{align}    
    \big(\Fbar{\act}^{S_3VU_1}_{U_2}\big)^{U_2,j_4j_1}_{S_2,i_2j_3} 
    = \frac{1}{\dim_\mathbb C U_2}
    \sum_{\substack{1 \leq d_1 \leq \dim_\mathbb C V \\ 1 \leq b_1 \leq \dim_\mathbb C U_1 \\ 1 \leq b_3,b_4 \leq \dim_\mathbb C U_2  \\ 1 \leq f_2 \leq \dim_\mathbb C S_2 \\ 1 \leq f_3 \leq \dim_\mathbb C S_3}}
    \CC{S_3}{U_2}{U_2}{f_3}{b_3}{b_4}{j_1} \, 
    \CC{V}{U_1}{U_2}{d_1}{b_1}{b_3}{j_4} \,
    \CCb{S_2}{U_1}{U_2}{f_2}{b_1}{b_4}{j_3}\, 
    \CCb{S_3}{V}{S_2}{f_3}{d_1}{f_2}{i_2} \, ,
\end{align}
that 
\begin{align}
    \Omega^{L(G,\beta)V}_{S_1S_2S_3,i_1i_2j_1j_2} = 
    \frac{\sqrt{\dim_\mathbb C S_1 \dim_\mathbb C S_3}}{ \dim_\mathbb C S_2} \!\!\!
    \sum_{\substack{U_1,U_2 \in \Irr(\Rep^\beta(G)) \\ 1 \leq j_3 N^{S_2U_1}_{U_2} \\ 1 \leq j_4 \leq N^{VU_1}_{U_2}}} \!\!\!
    \big(\F{\act}^{VS_1U_1}_{U_2}\big)^{S_4,i_3j_3}_{U_1,j_2j_4}
    \, \big(\Fbar{\act}^{S_3VU_1}_{U_2}\big)^{U_2,j_4j_1}_{S_2,i_2j_3}  \, .
\end{align}

\bigskip \noindent
Let us conclude this appendix with some comments about the equivalence $\Mod(\Tu(\Rep(G))) \simeq \mc Z(\Rep(G))$. At the beginning of this appendix, we mentioned that the vector space underlying the $\Tu(\Rep(G))$-module associated with $W \in \mc Z(\Rep(G))$ is given by $\bigoplus_{S \in \Irr(\mc C)} \Hom_\mc C(S,W)$. In sec.~\ref{sec:RepG_String}, we constructed tensors that span this vector space and we explicitly computed how $\Tu(\Rep(G))$ acts on this vector space. In particular, for every simple object in $\mc Z(\Rep(G))$, we computed in eq.~\eqref{eq:RepG_UpsSymbols} the matrix coefficients---referred to in the main text as $\Upsilon$-symbols--- of every element of $\Tu(\Rep(G))$ in the corresponding module. These are immediately related to the $\Omega$-symbols of the half-braiding. By inspection, we find that they differ by a factor of $\frac{\dim_\mathbb C S_2}{\dim_\mathbb C S_3}$. As a consistency check, inserting in eq.~\eqref{eq:RepG_twistedLocalOpTube} definition \eqref{eq:RepG_OmegaSymbols} of the $\Omega$-symbols readily recovers the $\Upsilon$-symbols. This is the essence of the equivalence $\Mod(\Tu(\Rep(G))) \simeq \mc Z(\Rep(G))$.

We are now able to confirm that the $\Tu(\Rep(G))$-module computed in eq.~\eqref{eq:RepG_TubeModEnv} is isomorphic to that associated with $L(G,\beta)$ under the equivalence $\mc Z(\Rep(G)) \simeq \Mod(\Tu(\Rep(G)))$. From our treatment of this equivalence for the simple objects, we know $\Omega$-symbols and $\Upsilon$-symbols of the corresponding $\Tu(\Rep(G))$-module differ by a factor of $\frac{\dim_\mathbb C S_2}{\dim_\mathbb C S_3}$. However, this presupposes that the vector space underlying the $\Tu(\Rep(G))$-module is of the form $\bigoplus_{S \in \Irr(\Rep(G))}\Hom_G(S,W)$, where $W \in \Vect_G^G \simeq \mc Z(\Rep(G))$. Applied to $W = L(G,\beta)$,  this would be $\bigoplus_{S} \bigoplus_{U} \Hom_G(S,\End_\mathbb C(U))$. 
Rather, in the main text, the underlying vector space is taken to be isomorphic to $\bigoplus_S \bigoplus_U \Hom_G(S \otimes U,U)$.   
As spelt out above, the isomorphism $\Hom_G(S, U \otimes U^*) \cong \Hom_G(S \otimes U, U)$ rescales the orthonormal basis elements by a factor of $\sqrt{\dim_\mathbb C S}/\sqrt{\dim_\mathbb C U}$.
Furthermore, in our derivation, the module is realised via tensor network operators evaluated on the physical Hilbert space. In a tensor network, closing a diagram inherently performs a partial trace over the internal degrees of freedom. This physical trace operation weights each module element by additional quantum dimensions. Because the basis elements extracted from the tensor network are normalized with respect to the physical Hilbert--Schmidt inner product, the matrix coefficients evaluated on the lattice (the $\Upsilon$-symbols) further differ from those we would associate to the $\Omega$-symbols by multiplicative factors provided by these quantum dimensions.

\newpage

\renewcommand*\refname{\hfill References \hfill}
\bibliographystyle{alpha}
\bibliography{ref}

\end{document}